\documentclass[12pt,headings=big,numbers=noenddot,DIV=14,a4paper]{article}%

\pdfoutput=1

\usepackage[margin=1 in]{geometry}

\usepackage{graphicx}  
\usepackage{dcolumn}   
\usepackage{bm,relsize}        
\usepackage{amsfonts,amsmath,amssymb,amsthm,mathtools}
\usepackage{slashed} 
\usepackage[normalem]{ulem}
\usepackage{placeins}
\usepackage{lscape}
\usepackage{multirow}
\usepackage{lipsum, color}
\usepackage[usenames,dvipsnames,svgnames,table]{xcolor}
\usepackage[linktoc=page,bookmarks=false,colorlinks=true,linkbordercolor=RoyalBlue,citebordercolor=ForestGreen,urlbordercolor=CornflowerBlue]{hyperref}
\hypersetup{
    colorlinks=true,
    linkcolor=ForestGreen,
    filecolor=ForestGreen,      
    urlcolor=ForestGreen,
    citecolor=ForestGreen,
}
\usepackage[sort&compress,numbers,merge]{natbib}
\usepackage{breakcites}

\renewcommand{\over}[2]{\left(\frac{#1}{#2}\right)}

\def\beq{\begin{equation}}
\def\eeq{\end{equation}}
\def\bea{\begin{eqnarray}}
\def\eea{\end{eqnarray}}

\begin{document}

\begin{flushright}
\footnotesize
DESY 19-204\\
\end{flushright}
\color{black}

\begin{center}


{\LARGE \bf
Beyond the Standard Models \\
\vspace{.3 cm}
 with Cosmic Strings
}

\medskip
\bigskip\color{black}\vspace{0.5cm}

{
{\large Yann Gouttenoire}$^{a,b}$,
{\large G\'eraldine Servant}$^{a,c}$,
{\large Peera Simakachorn}$^{a,c}$
}
\\[7mm]

{\it \small $^a$ DESY, Notkestra{\ss}e 85, D-22607, Hamburg, Germany}\\
{\it \small $^b$ LPTHE,  CNRS \& Sorbonne Universit\'e, 4 Place Jussieu, F-75252, Paris, France}\\
{\it \small $^c$ II. Institute of Theoretical Physics, Universit\"{a}t  Hamburg, D-22761, Hamburg, Germany}\\
\end{center}

\bigskip

\centerline{\bf Abstract}

\bigskip

We examine which information on the early cosmological history can be extracted from the potential  measurement by third-generation gravitational-wave observatories  of a stochastic gravitational wave background (SGWB) produced by cosmic strings.  We consider a variety of cosmological scenarios breaking the  scale-invariant properties of the spectrum, such as early long matter or kination eras,  short intermediate matter and  inflation periods inside a radiation era, and their specific signatures on the  SGWB. This requires to go beyond the usually-assumed scaling regime, to take into account the transient effects during the change of equation of state of the universe.  
We compute  the time evolution of the string network parameters and thus the loop-production efficiency
 during the transient regime, and derive the corresponding shift in the turning-point frequency.   We consider the impact of particle production on the gravitational-wave emission by loops.
We estimate the reach of future interferometers LISA, BBO, DECIGO, ET and CE and radio telescope SKA to probe the new physics energy scale at which the universe has experienced changes in its expansion history. We find that a given interferometer may be sensitive to very different energy scales, depending on the nature and duration of the non-standard era, and the value of the string tension. 
It is fascinating that by exploiting the data from different GW observatories associated with distinct frequency bands, we may be able to reconstruct the full spectrum and therefore extract the values of fundamental physics parameters.

\clearpage
\noindent\makebox[\linewidth]{\rule{\textwidth}{1pt}} 
\tableofcontents
\noindent\makebox[\linewidth]{\rule{\textwidth}{1pt}}

\newpage
\section{Introduction}

The Standard Model of particle physics needs to be completed 
to address observational facts such as the matter antimatter asymmetry and the dark matter of the universe, as well as the origin of inflation. These, together with a number of other fundamental theoretical puzzles associated with e.g. the flavour structure of the matter sector and the ultra-violet properties of  the Higgs scalar field, motivate extensions of the Standard Model featuring new degrees of freedom and new energy scales. In turn, such new physics can substantially impact the expansion history in the early universe and leads to deviations  with respect to the standard cosmological model. Any deviations in the Friedmann equation occurring at  temperatures above the MeV remain to date  essentially unconstrained.

In the standard cosmological model,   primordial inflation is  followed by a long period of radiation domination until the more recent transitions to matter and then dark energy domination. Evidence for this picture comes primarily from observations of the Cosmic Microwave Background (CMB) and the successful predictions of Big-Bang Nucleosynthesis (BBN), which on the other hand, do not allow to test  cosmic temperatures above ${\cal O}$(MeV). 

An exciting prospect for deciphering the pre-BBN universe history and therefore high energy physics unaccessible by particle physics experiments, comes from the possible detection of a stochastic background of gravitational waves (SGWB), originating either from cosmological phase transitions, from cosmic strings or from inflation \cite{Caprini:2018mtu}.

Particularly interesting are cosmic strings (CS), which act as a long-lasting source of gravitational waves (GW) from the time of their production, presumably very early on, until today. The resulting frequency spectrum therefore encodes information from the almost entire cosmic history of our universe, and could possibly reveal precious details about the high energy particle physics responsible for a modified universe expansion.

There has been a large literature on probes of  a non-standard cosmology through the nearly-scale invariant primordial GW spectrum generated during inflation \cite{ Giovannini:1998bp, Giovannini:2009kg, Riazuelo:2000fc, Sahni:2001qp, Seto:2003kc, Tashiro:2003qp, Nakayama:2008ip, Nakayama:2008wy, Durrer:2011bi, Kuroyanagi:2011fy, Kuroyanagi:2018csn, Jinno:2012xb, Lasky:2015lej, Li:2016mmc, Saikawa:2018rcs, Caldwell:2018giq, Bernal:2019lpc, Figueroa:2019paj, DEramo:2019tit}. In contrast, little efforts have been invested to use the scale-invariant GW spectrum generated by CS  \cite{Cui:2017ufi, Cui:2018rwi, Auclair:2019wcv, Guedes:2018afo, Ramberg:2019dgi, Chang:2019mza}  while there has been intense activity on working out predictions for the SGWB produced by CS in standard cosmology \cite{Vilenkin:1981bx, Hogan:1984is, Vachaspati:1984gt, Accetta:1988bg, Bennett:1990ry, Caldwell:1991jj, Allen:1991bk, Battye:1997ji, DePies:2007bm, Siemens:2006yp, Olmez:2010bi, Regimbau:2011bm, Sanidas:2012ee, Sanidas:2012tf, Binetruy:2012ze, Kuroyanagi:2012wm, Kuroyanagi:2012jf,Sousa:2016ggw,Sousa:2020sxs}.

In this paper, we propose to use the detection of a SGWB from local cosmic strings to test the 
existence of alternative stages of cosmological expansion between the end of inflation and the end of the radiation era.
Particularly well-motivated is a stage of early-matter domination era induced by  a heavy cold particle dominating the universe and decaying before BBN. Another possibility is a stage of kination triggered by the fast rolling evolution of a scalar field down its potential, e.g. \cite{Spokoiny:1993kt, Joyce:1996cp} for the pioneering articles. Finally, supercooled confining phase transitions \cite{Creminelli:2001th, Randall:2006py,Nardini:2007me,Konstandin:2011dr,vonHarling:2017yew,Iso:2017uuu,Bruggisser:2018mrt,Baratella:2018pxi,Agashe:2019lhy,DelleRose:2019pgi,vonHarling:2019gme} can induce some late short stages of inflation inside a radiation era.  The latter were motivated at the TeV scale but the properties of the class of scalar potentials naturally inducing a short  inflationary era can be applied to any other scale. We will consider these various possibilities and their imprints on the GW spectrum from cosmic strings.

The dominant source of GW emission from a cosmic string network comes from loops which are continuously formed during the network fragmentation.
We thus primarily need to compute the loop-production efficiency during the non-standard  transition eras.
This is crucial for a precise prediction of the turning-point frequency as a signature of the non-standard era.
The temperature of the universe at the end of the non-standard era can be deduced from the measurement of these turning point frequencies.

The observational prospects for measuring the SGWB from cosmic string networks at LISA was recently reviewed in \cite{Auclair:2019wcv}.
Besides, the  effect  of particle production on the loop distribution and thus on the SGWB was recently discussed \cite{Matsunami:2019fss,Auclair:2019jip} where it was however concluded that the expected cutoff is outside the range of current and planned detectors (see also \cite{Kawasaki:2011dp}). 
Our paper integrates these  recent developments and goes beyond in  several directions:

\begin{itemize}

\item We go beyond the so-called  scaling regime by computing  the time evolution of the string network parameters (long string mean velocity and correlation length) and thus the loop-production efficiency during modifications of the equation of state of the universe, see right panel of Fig.~\ref{figure_preceding_NS_era}. Including these transient effects results in a turning-point frequency smaller by ${\cal O}$(20) compared to the prediction from the scaling regime.\footnote{ The turning-point frequency can even be smaller by ${\cal O}$(400) if in a far-future, a precision of the order of $1\%$ can be reached in the measurement of the SGWB, c.f. Eq.~\eqref{turning_point_general_scaling_app}.} As a result, the energy scale of the universe associated with the departure from the standard radiation era that can be probed is correspondingly larger than the one predicted from scaling networks,  see e.g. Fig.~\ref{fig:contour_all_f_n_sum}.

\item We investigate a large variety of non-standard cosmologies, in particular models where a non-standard era can be rather short inside the radiation era, due for instance to some cold particle temporarily dominating the energy density (short matter era, see Fig.~\ref{intermediate_matter_spectrum}) \cite{Gouttenoire:2019rtn} or some very short stage of inflation (for a couple of efolds) due to a high-scale supercooled confinement phase transition,  see Fig.~\ref{figure_spect_inflation}.
Such inflationary stages occurring at scales up to $10^{14}$ GeV could be probed, see Fig.~\ref{fig:contour_power_inflation2}. Even 1 or 2 e-folds could lead to observable features, see Fig.~\ref{figure_spect_inflation}.

\item  We also consider longer low-scale inflation models. For instance,  an intermediate inflationary era lasting for ${\cal O}(10)$ efolds, the SGWB from cosmic strings completely looses its scale invariant shape and has a peak structure instead,  see Fig.~\ref{fig:beautifulpeaks}.  A TeV scale inflation era can lead to a broad peak either in the LISA or BBO band or even close to the SKA band, depending on the precise value of the string tensions $G\mu$, and the number of efolds $N_e$.

\item We include high-frequency cutoff effects from particle production which can limit observations for small value of the string tension $G\mu \lesssim 10^{-15}$ and high-frequency cutoff from thermal friction,  see Fig.~\ref{sketch_scaling} and top left panel of Fig.~\ref{ST_vos_scaling}, as well as low-frequency cutoff from unstable CS networks,  see Fig.~\ref{fig:peak_spectrum_formation}.

\item We provide the relations between the observed frequency of a given spectral feature and 
the energy scale of the universe for  different physical effects,  see Fig.~\ref{fig:turning_points_lines}: 
i)  the end of a non-standard matter or kination era; ii) the time when particle emission starts to dominate; iii) the time at which the CS network re-enters the horizon after an intermediate inflation era.

{  \item We discuss how to read information about the small-scale structure of CS from the high-frequency tail of the GW spectrum, see App.~\ref{sec:study_impact_mode_nbr}.}

\item We discuss the comparison between local and global string networks, see App.~\ref{app:global_strings}.

\end{itemize}

The plan of the paper is the following.
 In Sec.~\ref{cs_chapter}, we recap the key features of CS networks, their cosmological evolution, decay channels and the pulsar timing array constraints on the string tension. 
 Sec.~\ref{sec:GWCS} reviews the computation of the SGWB from Nambu-Goto CS. We first discuss the underlying assumptions on the small-scale structure and  on the loop distribution and then derive the master formula of the GW frequency spectrum.
 An important discussion concerns the non-trivial frequency-temperature relation and how it depends on the cosmological scenario. 
 Sec.~\ref{sec:VOS} is devoted to the derivation  of the loop production efficiency beyond the scaling regime, taking into account transient effects from the change in the equation of state of the universe. We apply this to predict the SGWB in the standard cosmological model in Sec.~\ref{sec:standard}. We then move to discuss non-standard cosmological histories, a long-lasting matter or kination era before the radiation era in Sec.~\ref{sec:long_NS_era}, a short intermediate matter era inside the radiation era in  Sec.~\ref{sec:interm_matter}, and an intermediate inflationary era in Sec.~\ref{sec:inflation}.
 We discuss the specific spectral features in each of these cases and their observability by future instruments. 
 In Sec.~\ref{sec:peakspectrum}, we illustrate the possibilities  for the GW spectrum to exhibit different types of peak structures due to the presence of both a high and a low-frequency cutoff.
 Sec.~\ref{sec:detectability} summarises proposed approaches to test different scenarios and the physics reach of each experiment.  We conclude in Sec.~\ref{sec:conclusion}.
 Additional details are moved to appendices, such as non-GW constraints on the string tension $G\mu$ in App.~\ref{app:phenoCS}, a step-by-step derivation of the GW spectrum as well as the values of its slopes in App.~\ref{app:derivationGWspectrum}, the formulae of the various turning-point frequencies in 
App.~\ref{derive_turning_points},  the derivation of the equations which govern the evolution of the long-string network in the Velocity-dependent One-Scale (VOS) model in App.~\ref{sec:VOS_proof},  a discussion of the extensions to the original VOS model in App.~\ref{app:VOScalibration}, 
  the prediction of the GW spectrum from global strings in App.~\ref{app:global_strings}, the impact of the cosmology on the size of loops at formation in App.~\ref{app:impactcosmo-loop-size}, and the calculation of the integrated power-law sensitivity curves for each experiment in App.~\ref{app:sensitivity_curves}.

 \section{Recap on Cosmic Strings}\label{cs_chapter}
 
Cosmic strings have been the subject of numerous studies since the pioneering paper \cite{Kibble:1976sj}, see  \cite{Hindmarsh:1994re, Vilenkin:2000jqa, Vachaspati:2015cma} for reviews. 
\subsection{String field theory}
\paragraph{A topological defect:}
CS can originate as fundamental or composite objects in string theory \cite{Witten:1985fp, Dvali:2003zj,Copeland:2003bj, Polchinski:2004ia, Sakellariadou:2008ie, Davis:2008dj, Sakellariadou:2009ev, Copeland:2009ga} or as topological defects from spontaneous symmetry breaking (SSB) when the vacuum manifold $\mathcal{M}$ has a non-trivial first homotopy group $\pi_1(\mathcal{M})$.
Any theory with spontaneous breaking of a $U(1)$ symmetry  has a string solution, since $\pi_1(U(1)) = \mathbb{Z}$. 
More complex vacuum manifolds with string solutions can appear in various grand unified theories \cite{Jeannerot:2003qv, Sakellariadou:2007bv,Buchmuller:2013lra, Dror:2019syi}, e.g. $SO(10) \rightarrow SO(5)\times \mathbb{Z}_2$.
\paragraph{The abelian-Higgs model:}
The standard example of field theories with a string-liked solution is the Abelian-Higgs (AH) model, a field theory with a complex scalar field $\phi$ charged under a $U(1)$ gauge interaction. Note that the symmetry can also be global. The resulting strings solutions corresponding to local and global symmetries are called local and global strings, respectively.
CS correspond to lines where the scalar field sits on the top of its \textit{mexican hat potential} $V(\phi)$ and approaches its vacuum expectation value (VEV) at large distance, the Nielsen-Olesen vortex \cite{Nielsen:1973cs}. When following a closed path around the string, the phase of the complex scalar field returns to its original value after winding around the mexican hat an integer $n$ number of times.
The energy per unit of length, also known as the \textit{string tension} reads \cite{Vilenkin:2000jqa}
\begin{equation}
\mu\approx2\pi \eta^2\,n\,\times
\begin{cases}
1&\hspace{2em}\textrm{for local strings},\\
\ln\left(\frac{m_{\phi}}{H}\right)&\hspace{2em}\textrm{for global strings},\\
\end{cases}
\label{tension_string_exp}
\end{equation}
with $\eta$ the scalar field VEV. The Hubble horizon $H^{-1}$ and the string core width $m_{\phi}^{-1}$ play the role of IR and UV cut-offs.
The logarithmic divergence of the tension of global strings is due to the existence of a long-range interaction mediated by the massless Goldstone mode (the complex phase of $\phi$).

\subsection{Cosmic-string network formation and evolution}

\paragraph{Kibble mechanism:} The formation of cosmic strings occurs during a cosmological phase transition associated with spontaneous symmetry breaking, occurring at a temperature, approximately given by the VEV acquired by the scalar field
\begin{equation}
\label{eq:network_formation}
T_{\rm p}\sim 10^{11} ~ \text{GeV}  \left( \frac{G\mu}{10^{-15}} \right)^{1/2}.
\end{equation} 
CS are randomly distributed and form a network characterized by its correlation length $L$, which can be defined as
\begin{equation}
L \equiv \sqrt{\mu / \rho_\infty},
\end{equation}
where $\mu$ is the string tension, the energy per unit length, and  $\rho_\infty$ is the energy density of long strings. More precisely, long strings form infinite random walks \cite{Scherrer:1986sw} which can be visualized as collections of segments of length $L$.

\paragraph{Loop chopping:}  Each time two segments of a long string cross each other, they inter-commute, with a probability $P$ and form a loop. Loop formation is the main energy-loss mechanism of the long string network.
In numerical simulations \cite{Shellard:1987bv} and analytical modelling \cite{Eto:2006db}, the probability of inter-commutation has been found to be $P=1$ but in some models it can be lower. This is the case of models with extra-dimensions \cite{Dvali:2003zj, Jackson:2004zg}, strings with junctions \cite{Copeland:2006if} or peeling \cite{Laguna:1989hn}, or the case of highly relativistic strings \cite{Achucarro:2006es}. 
\paragraph{Scaling regime:} Just after the network is formed, the strings may interact strongly with the thermal plasma such that their motion is damped. When the damping stops, cosmic strings oscillate and enter the phase of \textit{scaling} evolution.
During this phase, the network experiences two competing dynamics:
\begin{enumerate}
\item Hubble stretching: the correlation length scale stretches due to the cosmic expansion, $L\sim a$.
\item Fragmentation of long strings into loops: a loop is formed after each segment crossing. Right after their formation, loops evolve independently of the network and start to decay through gravitational radiation and/or particle production. 
\end{enumerate}
It is known since a long time ago \cite{Kibble:1984hp, Bennett:1987vf, Bennett:1989ak, Albrecht:1989mk, Allen:1990tv}, that out of the two competing dynamics, Hubble expansion and loop fragmentation, there is an attractor solution, called the \textit{scaling regime}, where the correlation length scales as the cosmic time, 
\begin{equation}
L\sim t .
\label{eq:scaling_def}
\end{equation}
Note however that in the case of global-string network, it has been claimed that the scaling property in Eq.~\eqref{eq:scaling_def}, is logarithmically violated due to the dependence of the string tension on the Hubble horizon \cite{Klaer:2017ond, Gorghetto:2018myk, Kawasaki:2018bzv, Vaquero:2018tib, Buschmann:2019icd, Martins:2018dqg}. More recently, an opposite conclusion has been drawn in \cite{Hindmarsh:2019csc}. 
\paragraph{Number of strings:} During the scaling regime, the number of strings per Hubble patch is conserved
\begin{equation}
\frac{\rho_{\infty} H^{-3}}{\mu L} = \rm constant.
\end{equation}
Moreover, the energy density of the long-string network, which scales as $\rho_\infty\sim  \mu/t^2$, has the same equation-of-state as the main background cosmological fluid $\rho_\textrm{bkg}\sim a^{-n}$, 
\begin{equation}
\frac{\rho_\infty}{\rho_\textrm{bkg}}\sim \frac{a^{n}}{t^2} \sim \textrm{constant},
\end{equation}
where we used $a=t^{2/n}$. Hence, the long-string energy density redshifts as matter during matter domination and as radiation during radiation domination. The scaling regime allows cosmic strings not to dominate the energy density of the universe, unlike other topological defects.
The scaling property of a string network has been checked some fifteen years ago in numerical Nambu-Goto simulations \cite{Ringeval:2005kr, Vanchurin:2005pa, Martins:2005es, Olum:2006ix} and more recently with larger simulations \cite{BlancoPillado:2011dq}. 
During the scaling regime, the loop production function is scale-free, with a power-law shape, meaning that loops are produced at any size between the Hubble horizon $t$ and the scale $\sim \Gamma \, G \mu \, t,$ below which the strings have been smoothened by the gravitational backreaction and there is no further segment crossing.
\paragraph{A scale-invariant  SGWB:} An essential outcome is the scale-invariance of the Stochastic GW Background generated by loops during the scaling regime \cite{Vilenkin:1981bx, Hogan:1984is, Vachaspati:1984gt, Accetta:1988bg, Bennett:1990ry, Caldwell:1991jj,Allen:1991bk, Battye:1997ji, DePies:2007bm, Siemens:2006yp, Olmez:2010bi, Regimbau:2011bm, Sanidas:2012ee, Sanidas:2012tf, Binetruy:2012ze, Kuroyanagi:2012wm, Kuroyanagi:2012jf}.  We construct the GW spectrum in Sec.~\ref{sec:SGWB} and give more details in App.~\ref{app:derivationGWspectrum}. Remarkably, the spectrum generated by loops produced during radiation domination is flat, $\propto f^0$, whereas an early matter domination or an early kination-domination era turns the spectral index from $f^0$ to respectively $f^{-1/3}$ or $f^1$. As recently pointed out by \cite{Blasi:2020wpy}, in presence of an early matter, the slope $f^{-1}$ predicted by \cite{Cui:2017ufi, Cui:2018rwi}, is changed to $f^{-1/3}$ due to the high-k modes. We give more details on the impact of high-k modes on the GW spectrum in the presence of a decreasing slope due to an early matter era, a second period of inflation, particle production, thermal friction or network formation in App.~\ref{sec:study_impact_mode_nbr}.  Hence, the detection of the SGWB from CS by LIGO \cite{Aasi:2014mqd}, DECIGO, BBO \cite{Yagi:2011wg}, LISA \cite{Audley:2017drz}, Einstein Telescope \cite{Hild:2010id, Punturo:2010zz} or Cosmic Explorer \cite{Evans:2016mbw} would offer an unique observation window on the  equation of state of the Universe at the time when the CS loops responsible for the detected GW are formed. In Secs.~\ref{sec:long_NS_era}, \ref{sec:interm_matter} and \ref{sec:inflation}, we study the possibility for probing particular non-standard cosmological scenario: long matter/kination era, intermediate matter and intermediate inflation, respectively.

\subsection{Decay channels of Cosmic Strings}	

Cosmic strings can decay in several ways, as we discuss below.

\paragraph{GW radiation from long strings:}
Because of their topological nature,  straight infinitely-long strings are stable against decay. However, 
small-scale structures of wiggly long strings can generate gravitational radiation. Intuitively, a highly wiggly string can act as a gas of small loops. The GW emission from long strings can be neglected compared to the GW emission from loops, as loops live much longer than a Hubble time \cite{Allen:1991bk, Vilenkin:2000jqa}. Indeed, the GW signal emitted by loops is enhanced by the large number of loops (continuously produced).  Nambu-Goto numerical simulations have shown that the loop energy density is at least $100$ times larger than the long-string energy density \cite{Blanco-Pillado:2013qja}. Only for global strings where loops are short-lived due to efficient Goldstone production, the GW emission from long strings can give a major contribution to the SGWB \cite{Krauss:1991qu, JonesSmith:2007ne, Fenu:2009qf, Figueroa:2012kw}. In what follows, we only consider the emission from loops.
	
\begin{figure}[]
				\centering
				\includegraphics[width=9.5cm]{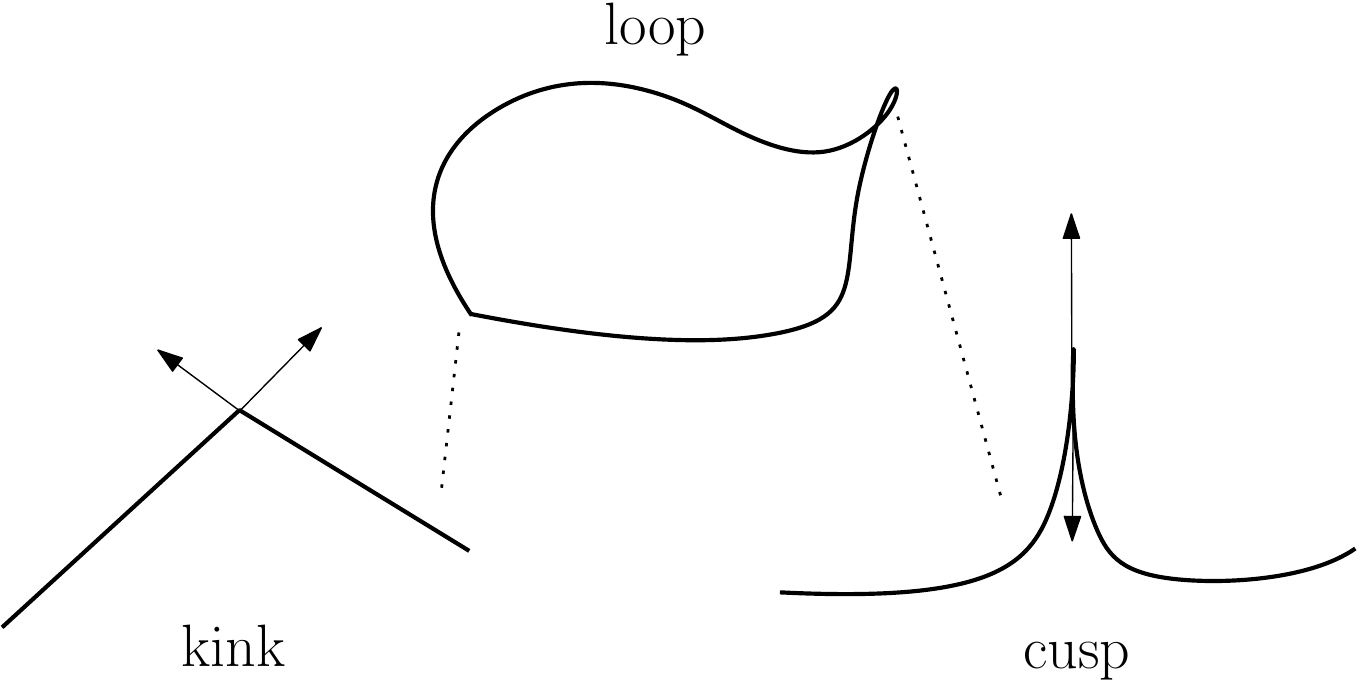}
				\caption{\it \small Cartoon showing the geometry of a kink and a cusp which are singular structures formed on loops. The arrows denote the tangent vectors of the string segments.}
				\label{kink_cusp_cartoon}
\end{figure}

\paragraph{GW radiation from loops (local strings):} 
\label{sec:masslessradiation}
In contrast to long strings, loops do not contain any topological charge and are free to decay into GW. The GW radiation power is found to be \cite{Vilenkin:2000jqa}
\begin{equation}
P_\textrm{GW}=\Gamma G \mu^2,
\label{eq:power_GW_0}
\end{equation}
where the total GW emission efficiency $\Gamma$ is determined from Nambu-Goto simulations, $\Gamma\simeq 50$ \cite{Blanco-Pillado:2017rnf}. Note that the gravitational power radiated by a loop is independent of its length. This can be understood from the quadrupole formula $P = G/5 (Q''')^2$ \cite{maggiore2008gravitational,Vachaspati:1984gt} where the triple time derivative of the quadrupole, $Q''' \propto {\rm mass \, (length)^2 / (time) ^3 \propto \mu}$, is indeed independent of the length. 
The resulting GW are emitted at frequencies \cite{Kibble:1982cb,Hindmarsh:1994re}
\begin{equation}
\label{eq:GWspecfreq_0}
\tilde{f}=\frac{2k}{l}, \qquad  k\in\mathbb{Z}^{+},
\end{equation}  
corresponding to the proper modes $k$ of the loop.  The tilde is used to distinguished the frequency emitted at $\tilde{t}$ from the frequency today 
\begin{equation}
f = a(\tilde{t})/a(t_0)~ \tilde{f}.
\end{equation}
 The frequency dependence of the power spectrum $P_\textrm{GW}(k)$ relies on the nature of the loop small-scale structures \cite{ Damour:2001bk, Ringeval:2017eww}, e.g. kinks or cusps, c.f. Fig.~\ref{kink_cusp_cartoon}.
More precisely, the spectrum of the gravitational power emitted from one loop reads
\begin{equation}
\label{eq:one-loop-spectrum}
P_\textrm{GW}^{(k)}= \Gamma^{(k)} G \mu^2, \qquad \text{with} \quad \Gamma^{(k)} \equiv \frac{\Gamma \, k^{-n} }{ \sum_{p=1}^{\infty}p^{-n}} \ , \  \ n= 
\begin{cases}
4/3 & \mbox{cusps } \\
5/3 & \mbox{kinks} \\
2 &  \mbox{kink-kink collisions }\\
\end{cases}
\end{equation}
where the spectral index $n={4/3}$ when the small-scale structure is dominated by cusps \cite{Vachaspati:1984gt, Burden:1985md,Olmez:2010bi}, $n={5/3}$ for kink domination \cite{Olmez:2010bi}, or $n={2}$ for kink-kink collision domination \cite{Damour:2001bk, Ringeval:2017eww}. { A discussion on how to read information about the small-scale structure of CS from the GW spectrum, is given in App.~\ref{sec:study_impact_mode_nbr}. In particular, we show that the high-frequency slope of the GW spectrum in the presence of an early matter era, a second period inflation, particle production or network formation, which is expected to be $f^{-1}$ from the fundamental, $k=1$, GW spectrum alone, is actually given by $f^{1-n}$.}
Immediately after a loop gets created, at time $t_i$ with a length $\alpha\,t_i$, its length $l(\tilde{t})$ shrinks through emission of GW with a rate $\Gamma G \mu$
\begin{equation}
\label{eq:CSlength0}
l(\tilde{t}) = \alpha t_i -\Gamma G \mu(\tilde{t}-t_i).
\end{equation}
Consequently, the string lifetime due to decay into GW is given by 
\begin{equation}
\label{eq:GWlifetime}
\tau_{\rm GW} = \frac{\alpha \, t_i }{\Gamma G\mu}.
\end{equation}
The superposition of the GW emitted from all the loops formed since the creation of the long-string network generates a Stochastic GW Background. 
Also, cusp formations can emit high-frequency, short-lasting GW bursts \cite{Damour:2000wa, Damour:2001bk, Siemens:2006yp, Olmez:2010bi, Ringeval:2017eww}. If the rate of such events is lower than their frequency, they might be subtracted from the SGWB.

\paragraph{Goldstone boson radiation (global strings):} For global strings, the massless Goldstone  particle production is the main decay channel. The radiation power has been estimated \cite{Vilenkin:2000jqa}
\begin{equation}
P_{\rm Gold}=\Gamma_{\rm Gold}\, \eta^2,
\label{eq:power_goldstone}
\end{equation}
where $\eta$ is the scalar field VEV and $\Gamma_{\rm Gold} \approx 65$ \cite{Vilenkin:1986ku, Chang:2019mza}. We see that the GW emission power in Eq.~\eqref{eq:power_GW_0} is suppressed by a factor $G\mu$ with respect to the Goldstone emission power in Eq.~\eqref{eq:power_goldstone}. Therefore, for global strings, the loops decay into Goldtone bosons after a few oscillations before having the time to emit much GW \cite{Vilenkin:2000jqa, Saurabh:2020pqe}. However, as shown in App.~\ref{app:global_strings}, the SGWB from global string is detectable for large values of the string scale, $\eta \gtrsim 10^{14}$~GeV. Other recent studies of GW spectrum from global strings in standard and non-standard cosmology include \cite{Bettoni:2018pbl, Ramberg:2019dgi, Chang:2019mza}. 
 A well-motivated example of global string is the axion string coming from the breaking of a $U(1)$ Peccei-Quinn symmetry \cite{Vilenkin:1982ks, Vilenkin:1984ib, Vilenkin:1986ku, Sikivie:1982qv}. Ref.~\cite{Ramberg:2019dgi}  shows the detectability of the GW from the axionic network of QCD axion Dark Matter (DM), after introducing an early-matter era which dilutes the axion DM abundance and increases the corresponding Peccei-Quinn scale $\eta$.


\paragraph{Massive particle radiation:} 
When the string curvature size is larger than the string thickness, one expects the quantum field nature of the CS, like the possibility to radiate massive particles, to give negligible effects and one may instead consider the CS as an infinitely thin 1-dimensional classical object with tension $\mu$: the Nambu-Goto (NG) string. 
However, due to the presence of small-scale structures on the strings, regions with curvature comparable to the string core size can develop and the Nambu-Goto  approximation breaks down. In that case, massive radiation can be emitted during processes known as cusp annihilation \cite{BlancoPillado:1998bv} or kink-kink collisions \cite{Matsunami:2019fss}. We discuss massive particle emission in more details in Sec.~\ref{sec:massive_radiation}.

\subsection{Constraints on  the string tension $G \mu$ from GW emission}
\label{subsec:Gmuconstraints}

The observational signatures of Nambu-Goto cosmic strings are mainly gravitational. The GW emission can be probed by current and future pulsar timing arrays and GW interferometers, while the static gravitational field around the string can be probed by CMB, 21 cm, and lensing observables, see app.~\ref{app:phenoCS} for more details on non-GW probes.
The strongest constraints come from pulsar timing array EPTA, $G \mu \lesssim 8 \times 10^{-10}$ \cite{Lentati:2015qwp}, and NANOGrav,  $G \mu \lesssim 5.3 \times 10^{-11}$ \cite{Arzoumanian:2018saf}. Comparison  with the theoretical predictions from the SGWB from cosmic strings leads to  $G\mu \lesssim 2 \times 10^{-11}$ 
\cite{Blanco-Pillado:2017rnf, Cui:2018rwi} or $G\mu \lesssim 10^{-10}$ \cite{Ringeval:2017eww}, even though it can be relaxed to $G\mu \lesssim 5 \times 10^{-7}$ \cite{Sanidas:2012ee}, after taking into account uncertainties on the loop size at formation and on the number of emitting modes. Note that it can also be strengthened by decreasing the inter-commutation probability \cite{Sakellariadou:2004wq, Damour:2004kw, Binetruy:2012ze}. 

By using the EPTA sensitivity curve derived in \cite{Breitbach:2018ddu},
we obtain the upper bound on $G\mu$, one order of magnitude higher, $2 \times 10^{-10}$, instead of $2 \times 10^{-11}$, c.f. Fig.~\ref{ST_vos_scaling}.  This bound 
becomes {$\sim5 \times 10^{-11}$} by using the NANOGrav sensitivity curve derived in \cite{Breitbach:2018ddu}.
Another large source of uncertainty is the nature of the GW spectrum generated by a loop, which depends on the assumption on the loop small-scale structure (e.g. the number of cusps, kinks and kink-kink collisions per oscillations) \cite{Binetruy:2012ze,Ringeval:2017eww}. For instance, the EPTA bound can be strengthened to $G\mu \lesssim 6.7 \times 10^{-14}$ if the loops are very kinky \cite{Ringeval:2017eww}. 
CS can also emit highly-energetic and short-lasting GW bursts due to cusp formation \cite{Damour:2000wa, Damour:2001bk, Siemens:2006yp, Olmez:2010bi, Ringeval:2017eww}. From the non-observation of such events with LIGO/VIRGO \cite{Abbott:2017mem, Abbott:2019prv}, one can constrain $G\mu \lesssim 4.2 \times 10^{-10}$ with the loop distribution function from \cite{Lorenz:2010sm}. However, the constraints are completely relaxed with the loop distribution function from \cite{Blanco-Pillado:2013qja}.

\section{Gravitational waves from cosmic strings}
\label{sec:GWCS}

In the main text of this work, we do not consider the case of global strings where the presence of a massless Goldstone in the spectrum implies that particle production is the main energy loss so that GW emission is suppressed~\cite{Vilenkin:2000jqa}. However, we give an overview of the GW spectrum from global strings in App.~\ref{app:global_strings}, which can be detectable for string scales $\eta \gtrsim 10^{14}~$GeV. Other studies of the sensitivity  of  next generation GW interferometers to GW from global strings are \cite{Bettoni:2018pbl, Ramberg:2019dgi, Chang:2019mza}.

There has been a long debate in the community whether local cosmic strings mainly loose their energy via GW emission or by particle production. We summarise the arguments and clarify the underlying assumptions below.

\subsection{Beyond the Nambu-Goto approximation}
\label{sec:NG}

\paragraph{Quantum field string simulations:}

Quantum field string (Abelian-Higgs) lattice simulations run by Hindmarsh et al. \cite{Vincent:1997cx, Hindmarsh:2008dw, Hindmarsh:2017qff} have shown that decay into massive radiation is the main energy loss and is sufficient to lead to scaling. Then, loops decay within one Hubble time into scalar and gauge boson radiation before having the time to emit GW. It is suggested that the presence of small-scale structures, kinks and cusps, at the string core size are responsible for the energy loss into particle production. In these regions of large string curvature, the Nambu-Goto  approximation, which considers CS as infinitely thin 1-dimensional classical objects, is no longer valid.

However, Abelian-Higgs simulations run by \cite{Moore:1998gp, Olum:1999sg, Moore:2001px} have claimed the opposite result, that energy loss into massive radiation is exponentially suppressed when the loop size is large compared to the thickness of the string.

\paragraph{Small-scale structure:}
	
At formation time, loops are not smooth but made of straight segments linked by kinks \cite{Blanco-Pillado:2015ana}. Kinks are also created in pairs after each string intercommutation, see \cite{Matsunami:2019fssVIDEO} or Fig.~$2.1$ in \cite{Rocha:2008de}. The presence of straight segments linked by kinks prevents the formation of cusps. However, backreaction from GW emission smoothens the shapes, hence allowing for the formation of cusps \cite{Blanco-Pillado:2015ana} (see Fig.~\ref{kink_cusp_cartoon}). Because of the large hierarchy between the gravitational backreaction scale and the cosmological scale $H$, the effects of the gravitational backreaction on the loop shape are not easily tractable numerically. The  effects of backreaction from particle emission are shown in \cite{Matsunami:2019fssVIDEO}. Nevertheless, it has  been proposed since long \cite{Quashnock:1990wv} that the small-scale structures are smoothened below the gravitational backreaction scale $\sim \Gamma \, G \mu \, t,$. Particularly, based on analytical modelling on simple loop models, it has been shown in \cite{Blanco-Pillado:2018ael,Blanco-Pillado:2019nto} that due to gravitational backreaction, kinks get rounded off, become closer to cusps and then cusps get weakened. In earlier works, the same authors \cite{Wachter:2016rwc, Wachter:2016hgi} claimed that whether the smoothening has the time to occur within the loop life time strongly depends on the initial loop shape. In particular, for a four-straight-segment loop, the farther from the square shape, the faster the smoothening, whereas for more general loop shapes, the smoothening may not always occur. 

To summarise the last two paragraphs, the efficiency of the energy loss into massive radiation depends on the nature of the small-scale structure, which can be understood as a correction to the Nambu-Goto approximation. The precise nature of the small-scale structure, its connection with the gravitational backreaction scale and the conflict between Nambu-Goto and Abelian Higgs simulations remain to be explained. Moreover, the value of the gravitational backreaction scale itself, see Sec.~\ref{sec:Ringeval} is matter of debate.  
For our study, we follow the proposal of \cite{Auclair:2019jip} for investigating how the GW spectrum is impacted for two benchmark scenarios: when the small-scale structures are dominated by cusps or when they are dominated by kinks. We give more details in the next paragraph. {
In App.~\ref{sec:study_impact_mode_nbr}, we show that if the high-frequency slope of the fundamental, $k=1$, GW spectrum is $f^{-1}$, as expected in presence of an early matter era or in presence of an Heavide cut-off in the loop formation time, then the existence of the high-$k$ modes, turns it to $f^{-1} \rightarrow f^{1-n}$, where $n$, defined in Eq.~\eqref{eq:one-loop-spectrum}, depends on the small-scale structure. We can therefore read information about the small-scale structure of CS from the high-frequency GW spectrum.}

\paragraph{Massive radiation emission:}
\label{sec:massive_radiation}

In the vicinity of a cusp, the topological charge vanishes where the string cores overlap. Hence, the corresponding portions of the string can decay into massive radiation. The length of the overlapping segment has been estimated to be $\sqrt{r\,l}$ \cite{BlancoPillado:1998bv, Olum:1998ag} where $r \simeq \mu^{-1/2}$ is the string core size and $l$ is the loop length. Hence, the energy radiated per cusp formation is $\mu \sqrt{r l}$, from which we deduce the power emitted from a loop
\begin{equation}
P_{\rm cusp}^{\rm part} \simeq N_{\rm c} \frac{\mu^{3/4}}{l^{1/2}},
\label{eq:power_cusp}
\end{equation}
where $ N_{\rm c}$ is the average number of cusps per oscillation, estimated to be $ N_{\rm c} \sim 2$  \cite{Blanco-Pillado:2015ana}. Note that the consideration of pseudo-cusps, pieces of string moving at highly relativistic velocities, might also play a role \cite{Elghozi:2014kya, Stott:2016loe}.

Even without the presence of cusps, Abelian-Higgs simulations \cite{Matsunami:2019fss} have shown that kink-kink collisions produce particles with a power per loop
\begin{equation}
P_{\rm kink}^{\rm part} \simeq  N_{\rm kk} \frac{\epsilon}{l},
\label{eq:power_kink}
\end{equation}
where $ N_{\rm kk}$ is the average number of kink-kink collisions per oscillation. 
Values possibly as large as $N_{\rm kk} \sim O(10^3)$ have been considered in \cite{Ringeval:2017eww} or even as large as $10^6$ for the special case of strings with junctions 
\cite{Binetruy:2010cc}, due to kink proliferations \cite{Binetruy:2010bq}.
In contrast to the cusp case, the energy radiated per kink-kink collision, $\epsilon$, is independent of the loop size $l$ and we expect $\epsilon \sim \mu^{1/2}$.

Upon comparing the power of GW emission in Eq.~\eqref{eq:power_GW_0} with either Eq.~\eqref{eq:power_cusp} or Eq.~\eqref{eq:power_kink}, one expects gravitational production to be more efficient than particle production when loops are larger than \cite{Auclair:2019jip}
\begin{equation}
\label{eq:length_cusps}
l \gtrsim l_c  \equiv \beta_c \, \frac{\mu^{-1/2}}{(\Gamma G \mu)^2},
\end{equation}
for small-scale structures dominated by cusps, and
\begin{equation}
\label{eq:length_kinks}
l \gtrsim l_k  \equiv \beta_k \, \frac{\mu^{-1/2}}{\Gamma G \mu},
\end{equation}
for kink-kink collision domination. $\beta_{\rm c}$ and $\beta_{\rm k}$ are numbers which depend on the precise refinement. We assume $\beta_{\rm c}, \, \beta_{\rm k} \sim O(1)$. Therefore, loops with length smaller than the critical value in Eq.~\eqref{eq:length_cusps} or Eq.~\eqref{eq:length_kinks} are expected to decay into massive radiation before they have time to emit GW, which means that they should be subtracted when computing the SGWB. Equations (\ref{eq:length_cusps}) and (\ref{eq:length_kinks}) are crucial to determine the cutoff frequency, as we discuss in Sec.~\ref{UVcutoff}.

The cosmological and astrophysical consequences of the production of massive radiation and the corresponding constraints on CS from different experiments are presented in Sec.~\ref{sec:particle_prod_pheno}

\subsection{Assumptions on the loop distribution}
\label{sec:Ringeval}

The SGWB resulting from the emission by CS loops strongly relies on the distribution of loops. In the present section, we introduce the loop-formation efficiency and discuss the assumptions on the loop-production rate, inspired from Nambu-Goto simulations. The loop-formation efficiency is computed later, in Sec.~\ref{sec:VOS}.

\paragraph{Loop-formation efficiency:}
The SGWB resulting from the emission by CS loops strongly relies on the assumption for the distribution of loops which we now discuss. 
The equation of motion of a Nambu-Goto string in a expanding universe implies the following evolution equation for the long string energy density, c.f. Sec.~\ref{sec:VOS_proof}
\begin{equation}
\label{eq:longstringdensity_eq}
\frac{d\rho_{\infty}}{dt}= -2H(1+\bar{v}^2) \rho_{\infty} - \left.\frac{d\rho_{\infty}}{dt}\right|_\textrm{loop},
\end{equation}
where $\bar{v}$ is the long string mean velocity. The energy loss into loop formation can be expressed as \cite{Vilenkin:2000jqa}
\begin{equation}
\label{eq:energyloss_loops}
\left.\frac{d\rho_{\infty}}{dt}\right|_\textrm{loop} \equiv \mu \int_{0}^{\infty} l f(l,t)dl \equiv \frac{\mu}{t^3} \tilde{C}_{\rm eff},
\end{equation}
with $f(l,t)$ the number of loops created per unit of volume, per unit of time $t$ and per unit of length $l$ and where we introduced the loop-formation efficiency $\tilde{C}_{\rm eff}$.
 The loop-formation efficiency $\tilde{C}_{\rm eff}$ is related to the notation introduced in \cite{Cui:2017ufi, Cui:2018rwi} by
\begin{equation}
\tilde{C}_{\rm eff} \equiv \sqrt{2} \,C_{\rm eff}.
\end{equation}
In Sec.~\ref{sec:VOS}, we compute the loop-formation efficiency $C_{\rm eff}$ as a function of the long string network parameters $\bar{v}$ and $L$, which themselves are solutions of the Velocity-dependent One-Scale (VOS) equations.

\paragraph{Only loops produced at the horizon size contribute to the SGWB:}
\label{sec:mainAssumptions}
As pointed out already a long time ago by \cite{Bennett:1989ak, Quashnock:1990wv} and more recently in large Nambu-Goto simulations \cite{Blanco-Pillado:2013qja}, the most numerous loops are the ones of the size of the gravitational backreaction scale 
\begin{equation}
\Gamma G \mu \times t,
\end{equation} 
which acts as a cut-off below which, small-scale structures are smoothened and such that smaller loops can not be produced below that scale. However, it has been claimed that only large loops are relevant for GW \cite{Hogan:2006we, Siemens:2006yp, Blanco-Pillado:2013qja}. In particular, Nambu-Goto  numerical simulations realized by Blanco-Pillado et al. \cite{Blanco-Pillado:2013qja} have shown that a fraction $\mathcal{F}_{\alpha}\simeq 10\%$ of the loops are produced with a length equal to a fraction $\alpha \simeq 10\%$ of the horizon size, and with a Lorentz boost factor $\gamma \simeq \sqrt{2}$. The remaining $90\%$ of the energy lost by long strings goes into highly boosted smaller loops whose contributions to the GW spectrum are sub-dominant. Under those assumptions, the number of loops, contributing to the SGWB, produced per unit of time can be computed from the total energy flow into loops in Eq.~\eqref{eq:energyloss_loops}
\begin{equation}
\label{eq:loop-formation_rate}
\frac{dn}{dt_i} = \frac{\mathcal{F}_{\alpha}}{\gamma\,\mu\,\alpha\,t_i} \left.\frac{d\rho_{\infty}}{dt}\right|_\textrm{loop},
\end{equation}
with $\mathcal{F}_{\alpha} = 0.1$, $\gamma = \sqrt{2}$ and $\alpha = 0.1$.  In App.~\ref{sec:alpha_correlation_length}, we discuss the possibility to define the loop-size as a fixed fraction of the correlation length $L$ instead of a fixed fraction of the horizon size $t$. Especially, we show that the impact on the GW spectrum is negligible. The latter can be recast as a function of the loop-formation efficiency $\tilde{C}_{\rm eff}$ defined in Eq.~\eqref{eq:energyloss_loops}
\begin{equation}
\label{eq:LoopProductionFctBody2}
\frac{dn}{dt_i}=\mathcal{F}_{\alpha} \frac{\tilde{C}_{\rm eff}(t_i)}{\gamma \,\alpha \, t_i^4}.
\end{equation}
This is equivalent to choosing the following monochromatic horizon-sized loop-formation function
\begin{equation}
f(l,\,t_i) = \frac{\tilde{C}_{\rm eff}}{\alpha\,t_i^4} \delta(l-\alpha t_i).
\end{equation}
The assumptions leading to Eq.~\eqref{eq:LoopProductionFctBody2} are the ones we followed for our study and which are also followed by \cite{Cui:2017ufi, Cui:2018rwi}. Our results strongly depend on these assumptions and would be dramatically impacted if instead we consider the model discussed in the next paragraph.

\paragraph{A second population of smaller loops:}
The previous assumption - that the only loops relevant for the GW signal are the loops produced at horizon size - which is inspired from the Nambu-Goto numerical simulations of Blanco-Pillado et al. \cite{Blanco-Pillado:2013qja, Blanco-Pillado:2017oxo}, is in conflict with the results from Ringeval et al. \cite{Lorenz:2010sm, Ringeval:2017eww, Auclair:2019zoz}. In the latter works, the loop production function  is derived analytically starting from the correlator of tangent vectors on long strings, within the Polchinski-Rocha model \cite{Polchinski:2006ee, Polchinski:2007rg, Dubath:2007mf, Rocha:2007ni}. In the Polchinski-Rocha model, which has been tested in Abelian-Higgs simulations \cite{Hindmarsh:2008dw}, the gravitational back-reaction scale, i.e. the lower cut-off of the loop production function, is computed to be 
\begin{equation}
\Upsilon (G \mu)^{1+2\chi} \times t,
\end{equation} 
with $\Upsilon \simeq 10$ and $\chi \sim 0.25$.  Consequently, the gravitational back-reaction scale in the Polchinski-Rocha model is significantly smaller than the usual gravitational back-reaction scale, commonly assumed to match the gravitational radiation scale, $ \Gamma G \mu \, t$. Therefore, the model of Ringeval et al. predicts the existence of a second population of smaller loops which enhances the GW spectrum at high frequency by many orders of magnitude \cite{Ringeval:2017eww}. However, as raised by \cite{Blanco-Pillado:2019vcs}, the model of Ringeval et al. predicts the amount of long-string energy converted into loops, to be $\sim 200$ times larger than the one computed in the numerical simulations of Blanco-Pillado et al. \cite{Blanco-Pillado:2013qja}. These discrepancies between Polchinski-Rocha analytical modeling and Nambu-Goto numerical simulations remain to be understood. \\

\hspace{-1cm}
\begin{minipage}[c]{15cm}
\vspace{-0.5cm}
\subsection{The gravitational-wave spectrum}
\label{sec:SGWB}
For our study, we compute the GW spectrum observed today generated from CS as follows (see app.~\ref{app:derivationGWspectrum} for a derivation)
	\begin{equation}
	\Omega_{\rm{GW}}(f)\equiv\frac{f}{\rho_c}\left|\frac{d\rho_{\rm{GW}}}{df}\right|=\sum_k{\Omega^{(k)}_{\rm{GW}}(f)},
	\label{eq:SGWB_CS_Formula}
		\vspace{-0.25cm}
	\end{equation}
	where
	\begin{equation}
\Omega^{(k)}_{\rm{GW}}(f)=\frac{1}{\rho_c}\cdot\frac{2k}{f}\cdot\frac{\mathcal{F}_{\alpha}\,\Gamma^{(k)}G\mu^2}{\alpha(\alpha+\Gamma G \mu)}\int^{t_0}_{t_{\rm osc}}d\tilde{t}~ \frac{C_{\rm{eff}}(t_i)}{\,t_i^4}\left[\frac{a(\tilde{t})}{a(t_0)}\right]^5\left[\frac{a(t_i)}{a(\tilde{t})}\right]^3\Theta(t_i-t_{\rm osc})\Theta(t_i-\frac{l_*}{\alpha}),
		\vspace{-0.25cm}
	\label{kmode_omega}
	\end{equation}
	\vspace{-0.25cm}
	with
	\begin{eqnarray*}
	    \, \Theta &\equiv& \textrm{Heaviside function,}\\
		\mu, \, G, \, \rho_c &\equiv& \textrm{string tension, Newton constant, critical density,}\\
		 \, a &\equiv& \textrm{scale factor of the universe}\\
		 &&\textrm{(we solve the full Friedmann equation for a given energy density content)},\\
		 \, k &\equiv& \textrm{proper mode number of the loop (effect of high-k modes are discussed in App.~\ref{sec:study_impact_mode_nbr}.} \\
		 && \textrm{For technical reasons, in most of our plots, we restrict to $2\times 10^4$ modes),}\\
		\Gamma &\equiv& \textrm{gravitational loop-emission efficiency, }~(\Gamma \simeq 50 ~ \text{\cite{Blanco-Pillado:2017oxo}})\\
		\Gamma^{(k)} &\equiv& \textrm{Fourier modes of $\Gamma$, dependent on the loop small-scale structures,}\\
		&&(\Gamma^{(k)}\propto k^{-4/3}\textrm{ for cusps, e.g. \cite{Olmez:2010bi})} ,\\
		\mathcal{F}_{\alpha} &\equiv& \textrm{fraction of loops formed with size $\alpha$ ($\mathcal{F}_{\alpha}\simeq 0.1$), c.f. Sec.~\ref{sec:mainAssumptions}},\\ 
		C_\textrm{eff} &\equiv& \textrm{loop-production efficiency, defined in Eq.~\eqref{eq:loopEfficiency}, }\\
		&& \textrm{($C_\textrm{eff} $ is a function of the long-string mean velocity $\bar{v}$ and correlation length $\xi$,}\\
		&&  \textrm{both computed upon integrating the VOS equations, c.f. Sec.~\ref{sec:VOS})}\\
		\alpha &\equiv& \textrm{loop length at formation in unit of the cosmic time,}~(\alpha \simeq 0.1)\\
		&& \textrm{(we consider a monochromatic, horizon-sized loop-formation function, c.f. Sec.~\ref{sec:mainAssumptions}),}\\
		\tilde{t} &\equiv& \textrm{the time of GW emission},\\
		 f &\equiv& \textrm{observed frequency today}\\
		 &&\textrm{(related to frequency at emission $\tilde{f}$ through $f\,a(t_0)=\tilde{f}\,a(\tilde{t})$},\\
		 &&\textrm{related to loop length $l$ through $\tilde{f}=2k/l$},\\
		 &&\textrm{related to the time of loop production $t_i$ through $l = \alpha t_i -\Gamma G \mu(\tilde{t}-t_i)$)},\\
		 t_i &\equiv& \textrm{the time of loop production},\\
		 &&\textrm{(related to observed frequency and emission time $\tilde{t}$ through}\\
		 &&\textrm{$t_i (f, \, \tilde{t})= \frac{1}{\alpha+\Gamma G \mu} \left[ \frac{2k}{f}\frac{a(\tilde{t})}{a(t_0)} + \Gamma G \mu \, \tilde{t} \right]$\big),}\\
		t_0 &\equiv& \textrm{the time today},\\
		t_{\rm osc} &\equiv& \textrm{the time at which the long strings start oscillating, $t_{\rm osc} =\textrm{Max}[t_{\rm fric}, \, t_F]$,}\\
		&&\textrm{$t_F$ is the time of CS network formation, defined as $\sqrt{\rho_{\rm tot}(t_F)}\equiv\mu$ where $\rho_{\rm tot}$ is}\\
		&&\textrm{the universe total energy density. In presence of friction, at high temperature},\\
		&& \textrm{the string motion is damped until the time $t_{\rm fric}$, computed in app.~\ref{sec:thermal_friction},}\\
		l_* &\equiv&  l_{\rm c}\textrm{ for cusps and } l_{\rm k} \textrm{ for kinks in Eq.~\eqref{eq:length_cusps} and Eq.~\eqref{eq:length_kinks}} \\
		&&\textrm{(critical length below which the emission of massive radiation}  \\
		&&\textrm{is more efficient than the gravitational emission, c.f. Sec.~\ref{sec:massive_radiation})}.
	\end{eqnarray*}
\end{minipage}

\begin{figure}[]
\centering
\raisebox{0cm}{\makebox{\includegraphics[width=0.9\textwidth, scale=1]{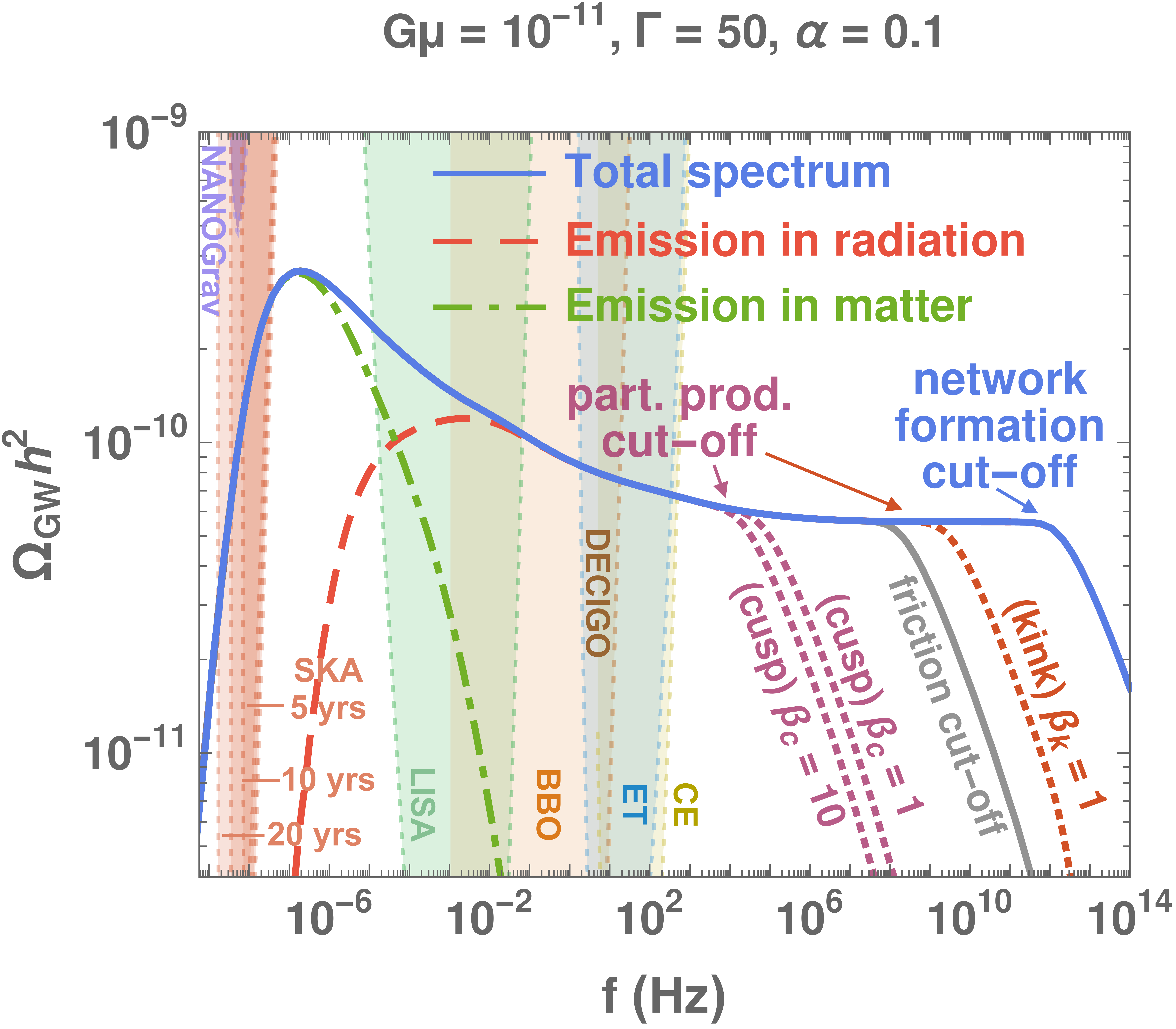}}}
\caption{\it \small GW spectrum from the scaling cosmic-string network evolving in a standard cosmology.
Contributions from GW emitted during radiation and matter eras are shown with red and green dashed lines respectively.
The high-frequency cut-offs correspond to either the time of formation of the network, c.f. Eq.~\eqref{eq:network_formation}, the time when friction-dominated dynamics become irrelevant, c.f.  App.~\ref{sec:thermal_friction}, or the time when gravitational emission dominates over massive particle production, for either kink or cusp-dominated small-scale structures, c.f. Sec.~\ref{sec:massive_radiation}. The cut-offs are described by Heaviside functions in the master formula in Eq.~\eqref{eq:SGWB_CS_Formula}.  In App.~\ref{sec:study_impact_mode_nbr}, we show that the slopes beyond the high-frequency cut-offs are given by $f^{-1/3}$. Colored regions indicate the integrated power-law sensitivity of future experiments, as described in app.~\ref{app:sensitivity_curves}.}
\label{sketch_scaling}
\end{figure}

\paragraph{A first look at the GW spectrum:}
Fig.~\ref{sketch_scaling} shows the GW spectrum computed with Eq.(\ref{eq:SGWB_CS_Formula}). The multiple frequency cut-offs visible on the figure, follow from the Heaviside functions in Eq.~\eqref{eq:SGWB_CS_Formula}, which subtract loops formed before network formation, c.f. Eq.~\eqref{eq:network_formation}, or when thermal friction freezes the network, c.f. App.~\ref{sec:thermal_friction}, or which subtract loops decaying via massive particle emission from cusps and kinks instead of GW, c.f. Sec.~\ref{sec:massive_radiation}. We indicate separately the contributions from the emission occurring before and after the matter-radiation equality. {One can see that loops emitting during the radiation era contribute to a flat spectrum whereas loops emitting during the matter era lead to a slope decreasing as $f^{-1/3}$. Similarly, the high-frequency cut-offs due to particle production, thermal friction, network formation, but also due to a second period of inflation (discussed in Sec.~\ref{sec:inflation}), give a slope $f^{-1/3}$.  In App.~\ref{sec:study_impact_mode_nbr}, we show that the presence of high-frequency modes are responsible for changing the slope $f^{-1}$, expected from the $(k=1)$-spectrum, to $f^{-1/3}$. }

\paragraph{Impact of the cosmology on the GW spectrum:} 
In a nutshell, the frequency dependence of the GW spectrum receives two contributions, a red-tilt coming from the redshift of the GW energy density and a blue-tilt coming from the loop-production rate $\propto t_i^{-4}$. On the one hand, the higher the frequency the earlier the GW emission, so the larger the redshift of the GW energy density and the more suppressed the spectrum. On the other hand, high frequencies correspond to loops formed earlier, those being more numerous, this increases the GW amplitude. Interestingly, during radiation-domination the two contributions exactly cancel such that the spectrum is flat.
As explained in more details in App.~\ref{sec:quadrupole_formula}, the flatness of the GW spectrum during radiation is intimately related to the independence of the GW emission power on the loop length. In the same appendix, we show that a change in the equation of state of the universe impacts the GW spectrum if it modifies at least one of the two following redshift factors: the redshift of the number of emitting loops and the redshift of the emitted GW.

For instance, when GW emission occurs during radiation but loop formation occurs during matter, the loop density redshifts faster. Then, the larger the frequency, the earlier the loop formation, and the more suppressed the GW spectrum (as $f^{-1}$ for $k=1$ and as $f^{-1/3}$ when taking into account high-k modes). Conversely, if loop formation occurs during kination, the loop density redshift slower and the GW gets enhanced at large frequency (as $f^1$).

\subsection{The frequency - temperature relation}
\paragraph{Relation between frequency of observation and temperature of loop formation:}
\label{sec:turning_point_scaling}
In app.~\ref{derive_turning_points}, we derive the relation between a detected frequency $f$ and the temperature of the universe when the loops, mostly responsible for $f$, are formed
\begin{align}
f=(6.7\times10^{-2}\textrm{ Hz})\left(\frac{T}{\textrm{GeV}}\right)\left(\frac{0.1\times 50 \times10^{-11}}{\alpha\,\Gamma G\mu}\right)^{1/2}\left(\frac{g_*(T)}{g_*(T_0)}\right)^{1/4}.
\label{turning_point_general_scaling}
\end{align}
We emphasize that Eq.~\eqref{turning_point_general_scaling} is very different from the relation obtained in the case of GW generated by a first-order cosmological phase transition. In the latter case, the emitted frequency corresponds to the Hubble scale at $T_p$ \cite{Caprini:2018mtu}
\begin{equation}
f = (19 \times 10^{-3}~\text{mHz}) \left( \frac{T_p}{100~\text{GeV}} \right)\left( \frac{g_*(T_p)}{100} \right)^{1/4}.
\end{equation}
In the case of cosmic strings, instead of being set by the Hubble scale at the loop-formation time $t_i$, the emitted frequency is further suppressed by a factor $(\Gamma G \mu)^{-1/2}$, which we now explain.
From the scaling law  $\propto t_i^{-4}$ of the loop-production function in Eq.\eqref{eq:LoopProductionFctBody2}, one can understand that the most numerous population of emitting loops at a given time $\tilde{t}$ is the population of loops created at the earliest epoch. They are the oldest loops\footnote{Note that they are also the smallest loops, with a length given by the gravitational radiation scale $\Gamma G \mu\,t$.}. Hence, a loop created at time $t_i$ contributes to the SGWB much later, at a time given by the loop half-lifetime $\tilde{t}_{\rm M}=\alpha\,t_i/2\Gamma G \mu$, c.f. Eq.~\eqref{eq:GWlifetime}. Therefore, the emitted frequency is dispensed from the redshift factor $a(\tilde{t}_{\rm M})/a(t_i)=(\tilde{t}_{\rm M}/t_i)^{1/2} \sim  (\Gamma G \mu)^{-1/2}$, and so, is higher. See app.~\ref{derive_turning_points} and its Fig.~\ref{cartoon_loop_maximal_decay} for more details.

\paragraph{The detection of a non-standard cosmology:}
During a change of cosmology, e.g a change from a matter to a radiation-dominated era, the long-string network evolves from one scaling regime to the other. The response of the network to the change of cosmology is quantified by the VOS equations, which are presented in Sec.~\ref{sec:VOS}. As a result of the transient evolution towards the new scaling regime, the turning-point frequency Eq.~\eqref{turning_point_general_scaling_app} associated to the change of cosmology  is lower in VOS than in the scaling network. The detection of a turning-point in a GW spectrum from CS by a future interferometer would be a smoking-gun signal for non-standard cosmology. Particularly, in Fig.~\ref{fig:turning_points_lines}, we show that LISA can probe a non-standard era ending around the QCD scale, ET/CE can probe a non-standard era ending around the TeV scale whereas DECIGO/BBO can probe the intermediate range. We show particular examples of long-lasting era in Sec.~\ref{sec:long_NS_era}. We focus on the particular case of a short matter era in Sec.~\ref{sec:interm_matter} and a short inflation era in Sec.~\ref{sec:inflation}, respectively. In the latter case, the turning-point frequency is even further decreased due to the string stretching which we explain in the next paragraph.

\paragraph{The detection of a non-standard cosmology (intermediate-inflation case):}
If the universe undergoes a period of inflation lasting $N_e$ e-folds, the correlation length of the network is stretched outside the horizon. After inflation, the network achieves a long transient regime lasting $\sim N_e$ other e-folds until the correlation length re-enters the horizon. Hence, the turning-point frequency in the GW spectrum, c.f. Eq.~\eqref{turning_point_inf}, receives a $\exp{N_e}$ suppression compared to Eq.~\eqref{turning_point_general_scaling} due to the duration of the transient. We give more details in Sec.~\ref{sec:inflation}.

		\paragraph{Cut-off frequency from particle production:}
\label{UVcutoff}
As discussed in the Sec.~\ref{sec:massive_radiation}, particle production is the main decay channel of loops shorter than
\begin{align}
l_*=\beta_m\frac{\mu^{-1/2}}{(\Gamma G\mu)^m},
\end{align}
where $m=1$ or $2$ for loops kink-dominated or cusp-dominated, respectively, and $\beta_m\sim\mathcal{O}(1)$.
The corresponding characteristic temperature above which loops, decaying preferentially into particles, are produced, is
\begin{equation}
T_* \simeq \beta_m^{-1/2}\, \Gamma^{m/2} \,\sqrt{\alpha}\,(G\mu)^{(2m+1)/4} \, M_{\rm pl} \simeq
\begin{cases}
\text{(0.2 EeV)} ~ \sqrt{\dfrac{\alpha}{0.1}} \,\sqrt{\dfrac{1}{\beta_c}} \left( \dfrac{G\mu}{10^{-15}}  \right)^{3/4}&\textrm{\hspace{0.5em}for kinks},\\[1em]
\text{(1 GeV)} ~ \sqrt{\dfrac{\alpha}{0.1}} \,\sqrt{\dfrac{1}{\beta_k}}\left( \dfrac{G\mu}{10^{-15}}  \right)^{5/4}&\textrm{\hspace{0.5em}for cusps}.\\
\end{cases}
\label{UVcutoff_T_app}
\end{equation}
We have used $l_*=\alpha \, t_i$, $H=1/(2t_i)$ and $\rho_{\rm rad}=3 M_{\rm pl}^2 H^2$. Upon using the frequency-temperature correspondence in Eq.~\eqref{turning_point_general_scaling}, we get the cut-off frequencies due to particle production 
\begin{equation}
f_* \simeq
\begin{cases}
\text{(1 GHz)} ~  \sqrt{\dfrac{1}{\beta_c}} \,\left( \dfrac{G\mu}{10^{-15}} \right)^{1/4}&\textrm{\hspace{0.5em}for kinks},\\[1em]
\text{(31 Hz)} ~ \sqrt{\dfrac{1}{\beta_k}} \left( \dfrac{G\mu}{10^{-15}}  \right)^{3/4}&\textrm{\hspace{0.5em}for cusps}.\\
\end{cases}
\label{UVcutoff_f_app}
\end{equation}
and which we show in most of our plots with dotted red and purple lines. Particularly, in Fig.~\ref{fig:turning_points_lines}, we see that particle production in the cusp-dominated case would start suppressing the GW signal in the ET/CE windows for string tension lower than $G\mu \lesssim 10^{-15}$. However, in the kink-dominated case, the spectrum is only impacted at frequencies much higher than the interferometer windows.  In App.~\ref{sec:study_impact_mode_nbr}, we show that the slope of the GW spectrum beyond the high-frequency cut-off $f_*$ is given by $f^{-1/3}$.

\subsection{The astrophysical foreground}

Crucial for our analysis is the assumption that the stochastic GW foreground of astrophysical origin can be substracted.  

LIGO/VIRGO has already observed three binary black hole (BH-BH) merging events \cite{TheLIGOScientific:2016qqj, Abbott:2016nmj, TheLIGOScientific:2016pea} during the first $4$-month observing run O1 in 2015, and seven additional BH-BH \cite{Abbott:2017vtc, Abbott:2017gyy, Abbott:2017oio, LIGOScientific:2018mvr} as well as one binary neutron star (NS-NS) \cite{TheLIGOScientific:2017qsa} merging events during the second $9$-month observing run O2 in 2017. And more events might still be discovered in the O2 data \cite{Venumadhav:2019lyq}.
According to the estimation of the NS merging rate following the detection of the first (and unique up to now) NS-NS merger event GW170817, NS-NS stochatisc background may be detectable after a $20$-month observing run with the expected LIGO/VIRGO design sensistivity in $2022+$ and in the most optimistic scenario, it might be detectable after $18$-month of the third observing run O3 who began in April, $1^{\rm st}$, 2019 \cite{ Abbott:2017xzg}.
Hence, one might worry about the possibility to distinguish the GW SGWB sourced by CS from the one generated by the astrophysical foreground.
However, in the BBO and ET/CE windows, the NS and BH foreground might be substracted with respective reached sensibilities $\Omega_{\rm GW} \simeq 10^{-15}$ \cite{Cutler:2005qq} and $\Omega_{\rm GW} \simeq 10^{-13}$ \cite{Regimbau:2016ike}.
In the LISA window, the binary white dwarf (WD-WD) foreground dominates over the NS-NS and BH-BH foregrounds \cite{Farmer:2003pa, Rosado:2011kv, Moore:2014lga}. The WD-WD galactic foreground, one order of magnitude higher than the WD-WD extragalactic \cite{Kosenko:1998mv}, might be substracted with reached sensibility $\Omega_{\rm GW} \simeq 10^{-13}$ at LISA \cite{Adams:2010vc, Adams:2013qma}. Hence, in the optimistic case where the foreground can be removed and the latter sensibility are reached one might be able to distinguished the signal sourced by CS from the one generated by the astrophysical foreground. Furthermore, the GW spectrum generated by the astrophysical foreground increased with frequency as $f^{2/3}$ \cite{Zhu:2012xw}, which is different from the GW spectrum generated by CS during radiation (flat), matter $f^{-1/3}$, inflation $f^{-1/3}$ or kination $f^1$.

\section{The Velocity-dependent One-Scale model}
\label{sec:VOS}

The master formula (\ref{eq:SGWB_CS_Formula}) crucially depends on the loop-production efficiency encoded in $C_{\rm eff}$. In this section, we discuss its derivation within the framework of the Velocity-dependent One-Scale (VOS) model.

\subsection{The loop-production efficiency}
\label{sec:loopProductionFct}

In a correlation volume $L^3$, a segment of length $L$ must travel a distance $L$ before encountering 
another segment. $L$ is the correlation length of the long-string network.  The collision rate, per unit of volume, is $\tfrac{\bar{v}}{L} \cdot \tfrac{1}{L^3} \sim \tfrac{\bar{v}}{L^4}$ where $\bar{v}$ is the long-string mean velocity. At each collision forming a loop, the network looses a loop energy $\mu \, L = \rho_{\infty} \, L^3$. Hence, the loop-production energy rate can be written as \cite{Kibble:1984hp}
\begin{equation}
\label{eq:energylossloop}
\left.\frac{d\rho_{\infty}}{dt}\right|_\textrm{loop}=\tilde{c}\, \bar{v}\frac{\rho_{\infty}}{L},
\end{equation}
where one can compute $\tilde{c}=0.23 \pm 0.04$ from Nambu-Goto simulations in expanding universe \cite{Martins:2000cs}. $\tilde{c}$ is the only free parameter of the VOS model. 
Hence, the loop-formation efficiency, defined in Eq.~\eqref{eq:energyloss_loops}, can be expressed as a function of the long-string parameters, $\bar{v}$ and $\xi \equiv L/t$,
\begin{equation}
\label{eq:loopEfficiency}
\tilde{C}_\textrm{eff} \equiv \sqrt{2}\,C_\textrm{eff}(t)=\frac{\tilde{c}\,\bar{v}(t)}{\xi^3(t)}.
\end{equation}
In app.~\ref{app:VOScalibration}, we discuss how our results are changed when considering a recent extension of the VOS model with more free parameters, fitted on Abelian-Higgs field theory numerical simulations \cite{Correia:2019bdl}, and taking into account the emission of massive radiation. Basically, the loop-formation efficiency $C_{\rm eff}$ is only decreased by a factor $\sim 2$.
In the following, we derive $\bar{v}$ and $\xi$ as solutions of the VOS equations.

\subsection{The VOS equations}

The VOS equations describe the evolution of a network of long strings in term of the mean velocity $\bar{v}$ and the correlation length $\xi = L/t$ \cite{Martins:1995tg, Martins:1996jp, Martins:2000cs, martins2016defect}. The latter is defined through the long string energy density $\rho_{\infty} \equiv \mu/L^2$.
Starting from the equations of motion of the Nambu-Goto string in a FRW universe, we can derive the so-called VOS equations (see app.~\ref{sec:VOS_proof} for a derivation)
\begin{align}
\label{eq:VOS_eq_body}
&\frac{dL}{dt}=HL \,( 1+ \bar{v}^2)+\frac{1}{2}\tilde{c}\,\bar{v}, \\
&\frac{d\bar{v}}{dt}=(1-\bar{v}^2)\left[\frac{k(\bar{v})}{L}-2H\bar{v}\right],
\end{align}
where
\begin{equation}
 k(\bar{v})=\frac{2\sqrt{2}}{\pi}(1-\bar{v}^2)(1+2\sqrt{2}\bar{v}^3)\frac{1-8\bar{v}^6}{1+8\bar{v}^6},
\end{equation}
is the so-called momentum parameter and is a measure of the deviation from the straight string, for which $ k(\bar{v})=1$ \cite{martins2016defect}.
The first VOS equation describes the evolution of the long string correlation length under the effect of Hubble expansion and loop chopping.
The second VOS equation is nothing more than a relativistic generalization of Newton's law where the string is accelerated by its curvature $1/L$ but is damped by the Hubble expansion after a typical length $H^{-1}$.

Numerical simulations \cite{Ringeval:2005kr, Vanchurin:2005pa, Martins:2005es, Olum:2006ix, BlancoPillado:2011dq} have shown that a network of long strings is first subject to a transient regime before reaching a scaling regime, in which the long string mean velocity $\bar{v}$ is constant and the correlation length grows linearly with the Hubble horizon $L = \xi \, t$. The values of the quantities $\bar{v}$ and $\xi $ depend on the cosmological background, namely the equation of state of the universe. Hence, when passing from a cosmological era 1 to era 2, the network accomplishes a transient evolution from the scaling regime 1 to the scaling regime 2. We  use the VOS equations to compute the time evolution of $\bar{v}$ and $\xi $ during the change of cosmology and then compute their impact on the CS SGWB.

\subsection{Scaling regime solution and beyond}
\label{sec:scalingVSvos}

\begin{figure}[h!]
\centering
\raisebox{0cm}{\makebox{\includegraphics[width=0.32\textwidth, scale=1]{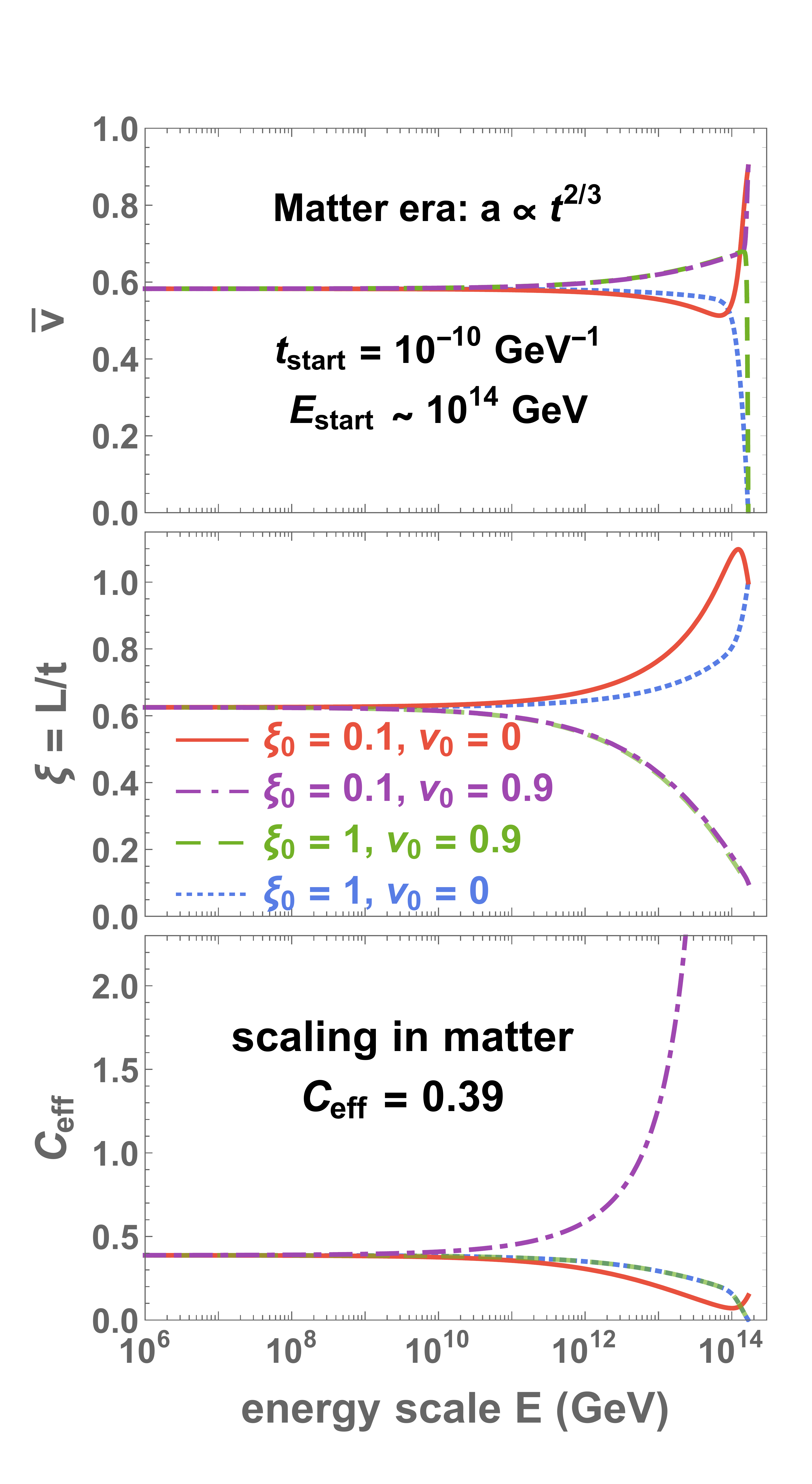}}}
\raisebox{0cm}{\makebox{\includegraphics[width=0.32\textwidth, scale=1]{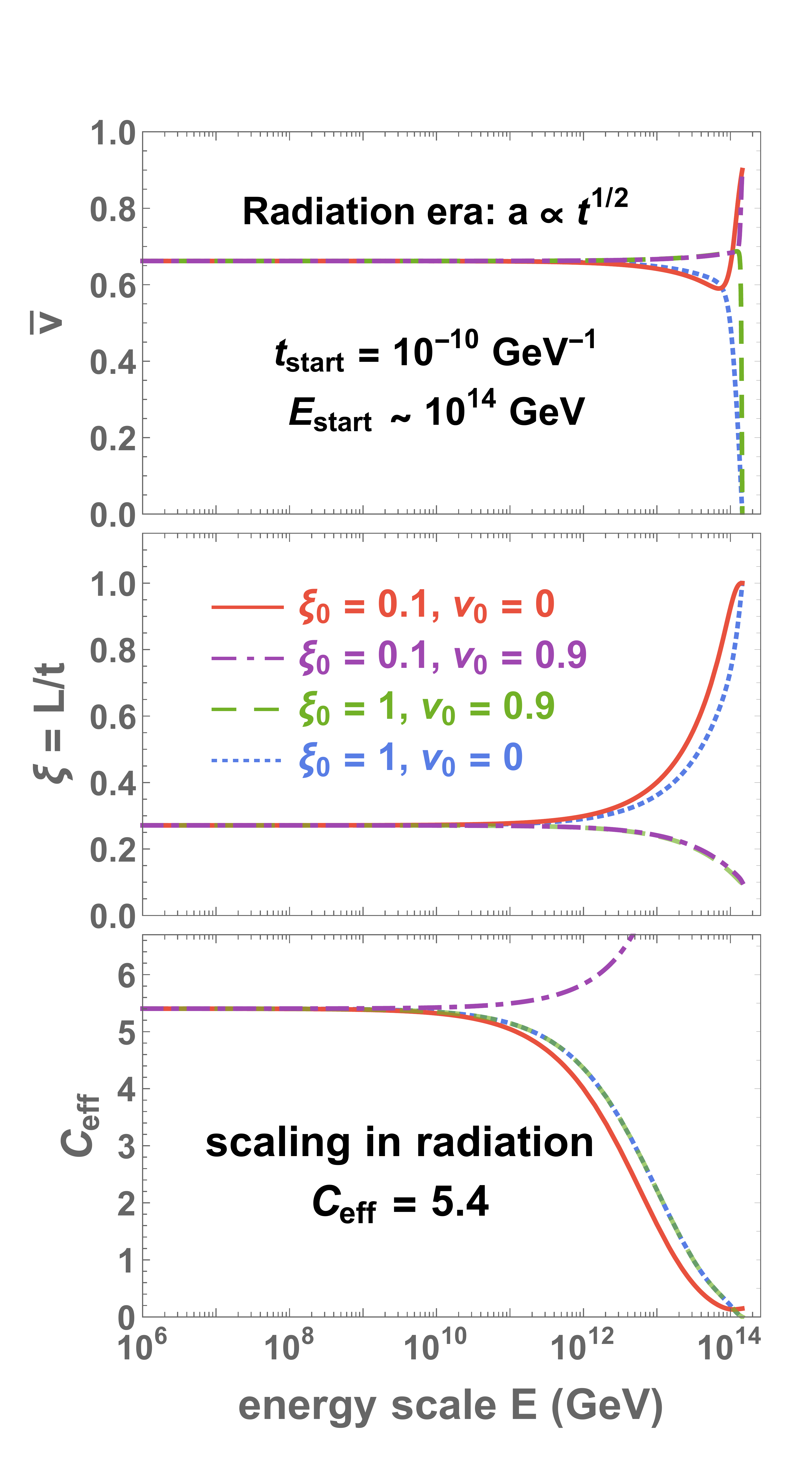}}}
\raisebox{0cm}{\makebox{\includegraphics[width=0.32\textwidth, scale=1]{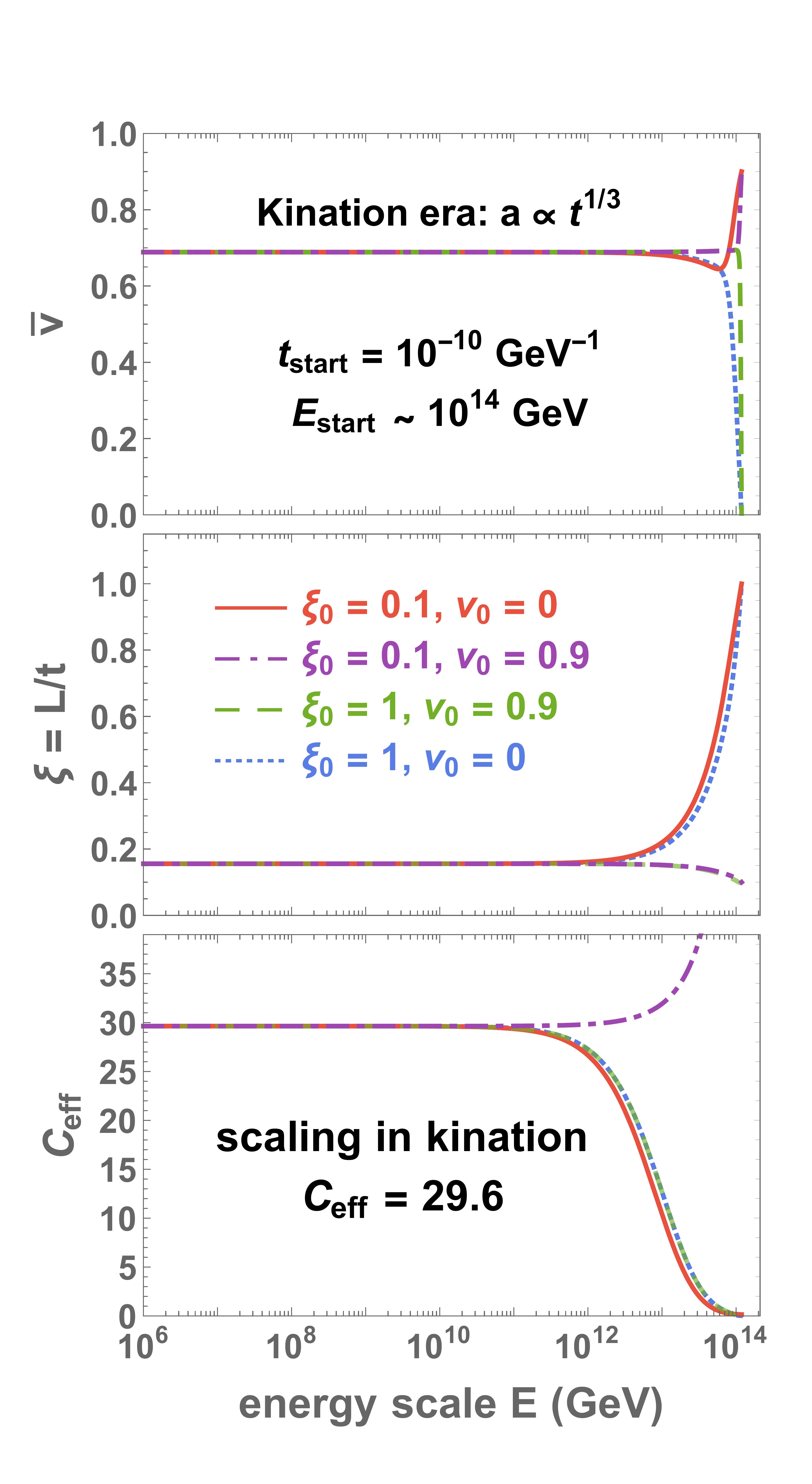}}}
\caption{\it \small Cosmic-string network evolving in the one-component universe with energy density $\rho\sim a^{-n}$ where $n=$ 3, 4 and 6 correspond to matter, radiation and kination, respectively. The long-string-network mean velocity $\bar{v}$, the correlation length $\xi$ and the corresponding loop-production efficiency $C_\textrm{eff}$ reach the scale-invariant solutions after the Hubble expansion rate has dropped by 2 orders of magnitude, independently of the initial conditions.}
\label{scaling_evolution}
\end{figure}

\paragraph{Scaling solution vs VOS solution:}
Fig.~\ref{scaling_evolution} shows the evolutions of $\xi,~\bar{v}$, and $C_\textrm{eff}$, from solving the VOS equations in Eq.~\eqref{eq:VOS_eq_body} with three equations of state, matter, radiation and kination. Regardless of the initial-condition choice, the network approaches a scaling solution where all parameters become constant. The energy scale of the universe has to decrease by some 4 orders of magnitude before reaching the scaling regime after the network formation. For a cosmological background evolving as $a\propto t^{2/n}$ with $n \geq 2$, the scaling regime solution is
\begin{equation}
\label{eq:VOS_scaling_solution}
\xi=\textrm{ constant} \textrm{\hspace{1.5em}and\hspace{1.5em}}\bar{v}=\textrm{ constant},
\end{equation}
with
\begin{equation}
\label{eq:VOS_scaling_solution2}
\textrm{with\hspace{1.5em}}\xi=\frac{n}{2}\sqrt{\frac{k(\bar{v})[k(\bar{v})+\tilde{c}]}{2(n-2)}}\quad\textrm{and\hspace{1em}}\bar{v}=\sqrt{\frac{n}{2}\frac{k(\bar{v})}{[k(\bar{v})+\tilde{c}]}\left(1-\frac{2}{n}\right)}.
\end{equation}
In order to fix the notation used in our plots, we define
\begin{itemize}
\item
\textbf{(Instantaneous) scaling network}: The loop-formation efficiency $C_{\rm eff}$, defined in Eq.~\eqref{eq:loopEfficiency}, is taken at its steady state value, given by Eq.~\eqref{eq:VOS_scaling_solution2}. In particular for matter, radiation and kination domination, one has 
\begin{equation}
\label{eq:Ceff_scaling}
C_\textrm{eff} \simeq 0.39,~5.4,~29.6 \qquad \text{for} \quad n=3,~4,~6.
\end{equation}
During a change of era $1\to 2$, $C_{\rm eff}$ is assumed to change instantaneously from the scaling regime of era $1$ to the scaling regime of era $2$. This is the assumption adopted in \cite{Cui:2017ufi, Cui:2018rwi}.
\item
\textbf{VOS network}: The loop-formation efficiency $C_{\rm eff}$, defined in Eq.~\eqref{eq:loopEfficiency}, is computed upon integrating the VOS equations in Eq.~\eqref{eq:VOS_eq_body}. During a change of cosmology, the long-string network experiences a transient regime.
\end{itemize}

\paragraph{Beyond the scaling regime in standard cosmology:}

In Fig.~\ref{ST_vos_scaling} and Fig.~\ref{ST_vos_scaling_ceff}, we compare the GW spectra and the $C_\textrm{eff}$ evolution, obtained with a scaling and VOS network. They are quite similar. The main difference arises from the change in relativistic degrees of freedom near the QCD confining temperature and from the matter-radiation transition. In contrast, predictions differ significantly when considering non-standard cosmology.

\paragraph{Beyond the scaling regime in non-standard cosmology:}

In Fig.~\ref{figure_preceding_NS_era}, in dashed vs solid, we compare the loop-production efficiency factor $C_\textrm{eff}$ and the corresponding GW spectra for  a scaling network and for  a VOS network. The VOS frequency of the turning point due to the change of cosmology is shifted to a lower frequency by a factor $\sim 22.5$ with respect to the corresponding scaling frequency.\footnote{ The turning-point frequency can even be smaller by ${\cal O}$(400) if in a far-future, a precision of the order of $1\%$ can be reached in the measurement of the SGWB, c.f. Eq.~\eqref{turning_point_general_scaling_app}.}. The shift results from the extra-time needed by the network to achieve its transient evolution to the new scaling regime. In the rest of this work, we go beyond the instantaneous scaling approximation used in \cite{Cui:2017ufi,Cui:2018rwi}.

\section{Standard cosmology}
\label{sec:standard}
\subsection{The cosmic expansion}
The SGWB from CS, c.f. master formula in Eq.~\eqref{eq:SGWB_CS_Formula}, depends on the cosmology through the scale factor $a$. We compute the later upon integrating the Friedmann equation
\begin{equation}
H^2=\frac{\rho}{3 M_{\rm pl}^2},\label{friedmann_eq}
\end{equation}
for a given energy density $\rho$. In the standard $\Lambda$CDM scenario, the universe is first dominated by radiation, then a matter era, and finally the cosmological constant so that we can write the energy density as 
\begin{equation}
\rho_\textrm{ST,0}(a)=\rho_{r,0}\,\Delta_R(T(a),T_0)\left(\frac{a}{a_0}\right)^4+\rho_{m,0}\left(\frac{a}{a_0}\right)^3+\rho_{k,0}\left(\frac{a}{a_0}\right)^2+\rho_{\Lambda,0},
\end{equation}
where $r,m,k$ and $\Lambda$ denote radiation, matter, curvature, and the cosmological constant, respectively. We take $\rho_i=\Omega_i h^2 \, 3 M_{\rm pl}^2 H_0^2$, where $H_0=100$~km/s/Mpc, $\Omega_{r}h^2 \simeq 4.2\times 10^{-5}$, $\Omega_{\rm m}h^2 \simeq 0.14$, $\Omega_k\simeq 0$, $\Omega_{\Lambda}h^2 \simeq 0.31$ \cite{Tanabashi:2018oca}. The presence of the function
\begin{equation}
\Delta_R=\left(\frac{g_*(T)}{g_*(T_0)}\right)\left(\frac{g_{*s}(T_0)}{g_{*s}(T)}\right)^{4/3},
\label{eq:DeltaR}
\end{equation}
comes from imposing the conservation of the comoving entropy $g_{*s}\,T^3\,a^3$, where the evolutions of $g_*$ and $g_{*,s}$ are taken from appendix C of \cite{Saikawa:2018rcs}.
We discuss the possibility of adding an extra source of energy density in the next sections, long matter/kination in Sec.~\ref{sec:long_NS_era}, intermediate matter in Sec.~\ref{sec:interm_matter} and intermediate inflation in Sec.~\ref{sec:inflation}.

\subsection{Gravitational wave spectrum}

Fig.~\ref{ST_vos_scaling} shows the dependence of the spectrum on the string tension.  The amplitude decreases with $G\mu$ due to the lower energy stored in the strings. Moreover, at lower $G\mu$, the loops decaying slower, the GW are emitted later, implying a lower redshift factor and a global shift of the spectrum to higher frequencies.
The figure also shows how the change in SM relativistic degrees of freedom introduces a small red-tilt which suppresses the spectrum by a factor $ \Delta_R^{-1} \sim 2.5$ at high frequencies.
We find that the amplitude of the GW spectrum at large frequency, assuming a standard cosmology, is given by 
\begin{equation}
\label{eq:lewicki_formula_GW_spectrum_radiation}
\Omega_{\rm GW}h^2 \simeq 15\pi\, \Delta_{R} \,\Omega_{r}h^2 \,C_{\rm eff}(n=4)\, \mathcal{F}_\alpha \,\left( \alpha\, G \mu /\Gamma\right)^{\! 1/2},
\end{equation}
where $\Omega_{r}h^2 \simeq 4.2\times 10^{-5}$ is the present radiation energy density of the universe \cite{Tanabashi:2018oca}.  
We provide an intuitive derivation based on the quadrupole formula in App.~\ref{sec:quadrupole_formula}.

\subsection{Deviation from the scaling regime}
Fig.~\ref{ST_vos_scaling_ceff} shows how the loop-formation efficiency $C_{\rm eff}$ varies during the change of SM relativistic degrees of freedom and the matter-radiation equality, upon solving the VOS equations, c.f. Sec.~\ref{sec:VOS}. We see the associated corrections to the spectrum in Fig.~\ref{ST_vos_scaling}, and which were already pointed out in \cite{Auclair:2019wcv}. The spectrum is enhanced at low frequencies because more loops are produced than when assuming that the matter era is reached instantaneously, c.f. Fig.~\ref{ST_vos_scaling_ceff}.

\subsection{Beyond the Nambu-Goto approximation}
\begin{figure}[h!]
\centering
\raisebox{0cm}{\makebox{\includegraphics[width=0.495\textwidth, scale=1]{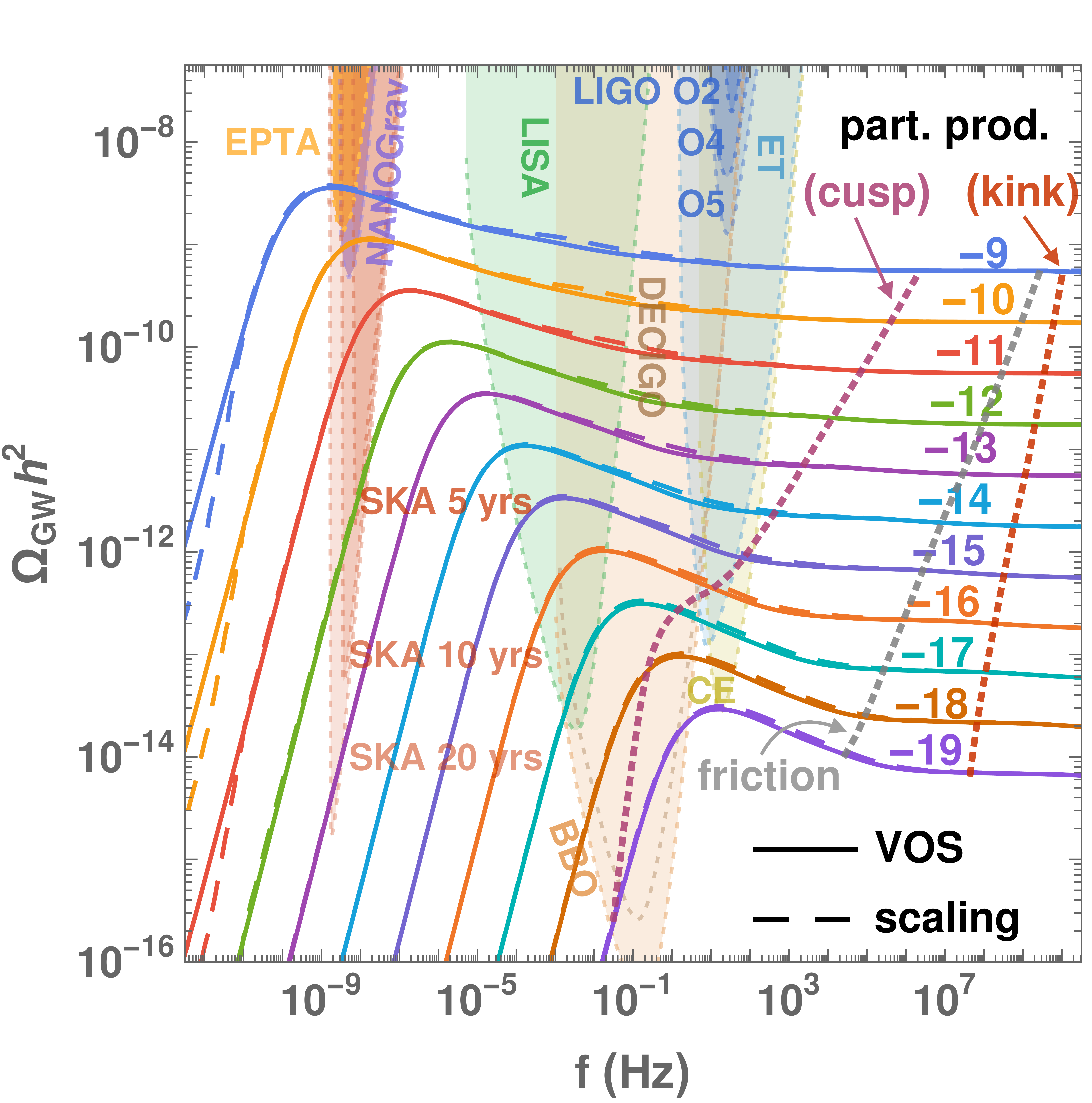}}}
\raisebox{0cm}{\makebox{\includegraphics[width=0.49\textwidth, scale=1]{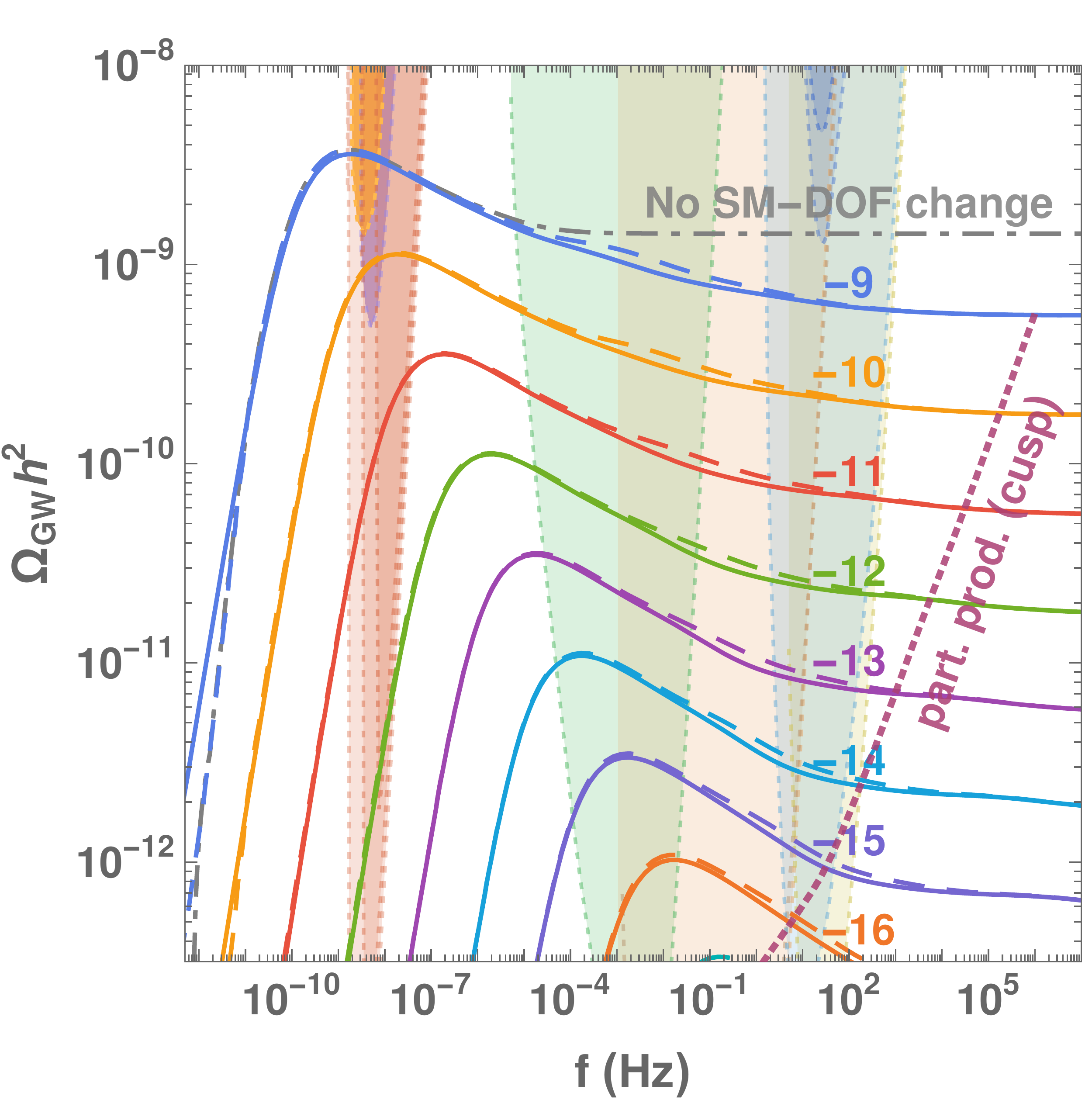}}}
\caption{\it \small \textbf{Left:} GW spectra from cosmic strings assuming either the scaling or VOS network, c.f. Sec.~\ref{sec:scalingVSvos}, evolving in the standard cosmological background.  Each line corresponds to string tension $G\mu = 10^{x}$, where $x$ is specified by a number on each line. Dotted lines show the spectral cut-offs expected due to particle production, c.f. Sec.~\ref{UVcutoff} and thermal friction, c.f. Sec.~\ref{sec:thermal_friction}, which depend on the nature of the loop small-scale structures: cusp or kink-dominated. \textbf{Right:} The zoom-in plot of the left panel shows the effects from the change of SM degrees of freedom on the scaling and VOS networks.}
\label{ST_vos_scaling}
\end{figure}

\begin{figure}[h!]
\centering
\raisebox{0cm}{\makebox{\includegraphics[width=0.49\textwidth, scale=1]{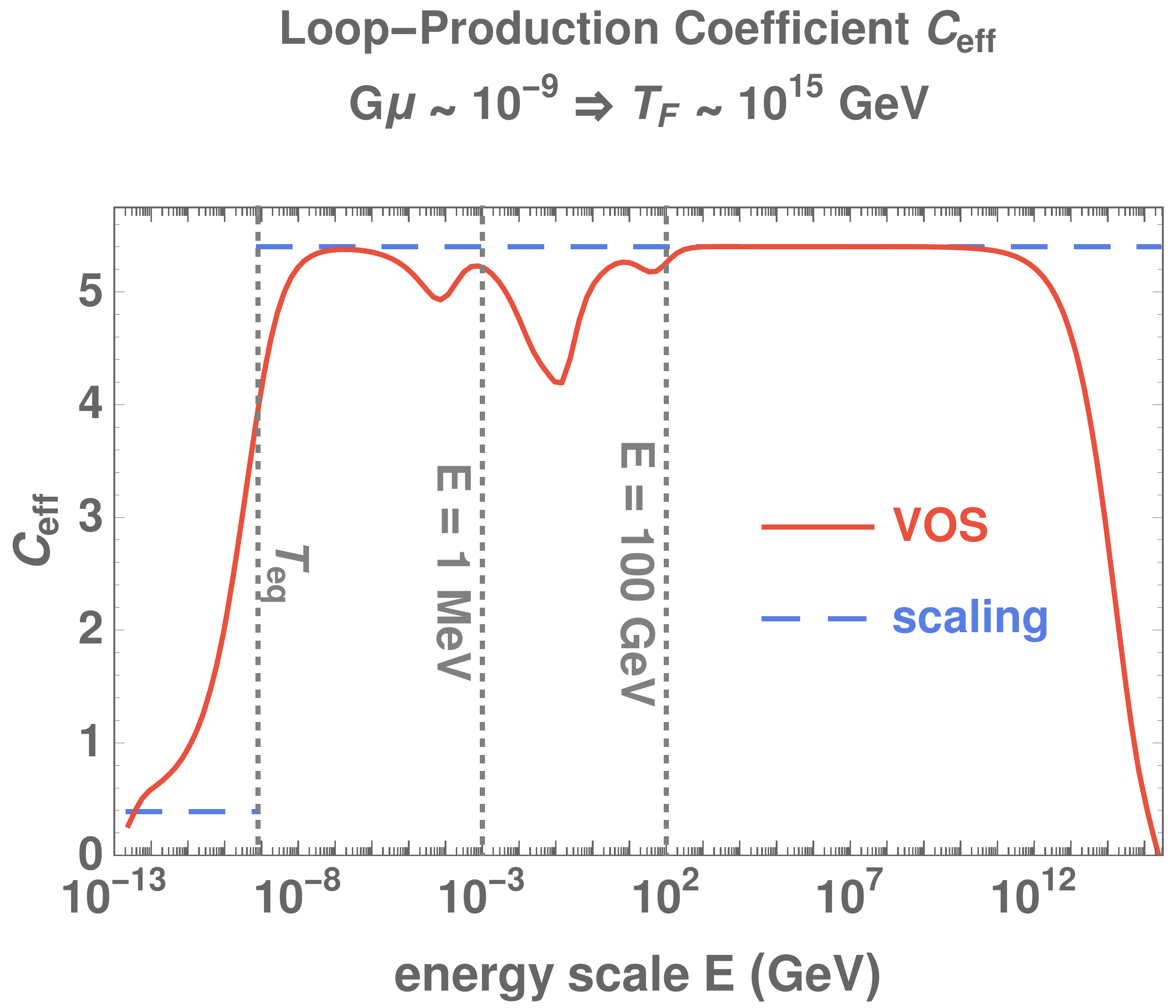}}}
\caption{\it \small  Comparison of the loop-production efficiency under the scaling assumption, where the attractor solution of the VOS equations is assumed to be reached instantaneously, and under the VOS assumptions, where one integrates the VOS equations. A standard cosmology is assumed.}
\label{ST_vos_scaling_ceff}
\end{figure}

Fig.~\ref{ST_vos_scaling} shows the possibility of a cut-off at high frequencies due to particle production, for two different assumptions regarding the loop small-scale structures: cusps or kinks domination, c.f. Sec.~\ref{sec:massive_radiation}. Above these frequencies, loops decays into massive radiation before they have time to emit GW. For kinky loops, the cut-off is outside any future-planned observational bands, while for cuspy loops, the cut-off might be in the observed windows for $G\mu \lesssim 10^{-15}$.


\begin{figure}[t!]
\centering
\raisebox{0cm}{\makebox{\includegraphics[width=0.49\textwidth, scale=1]{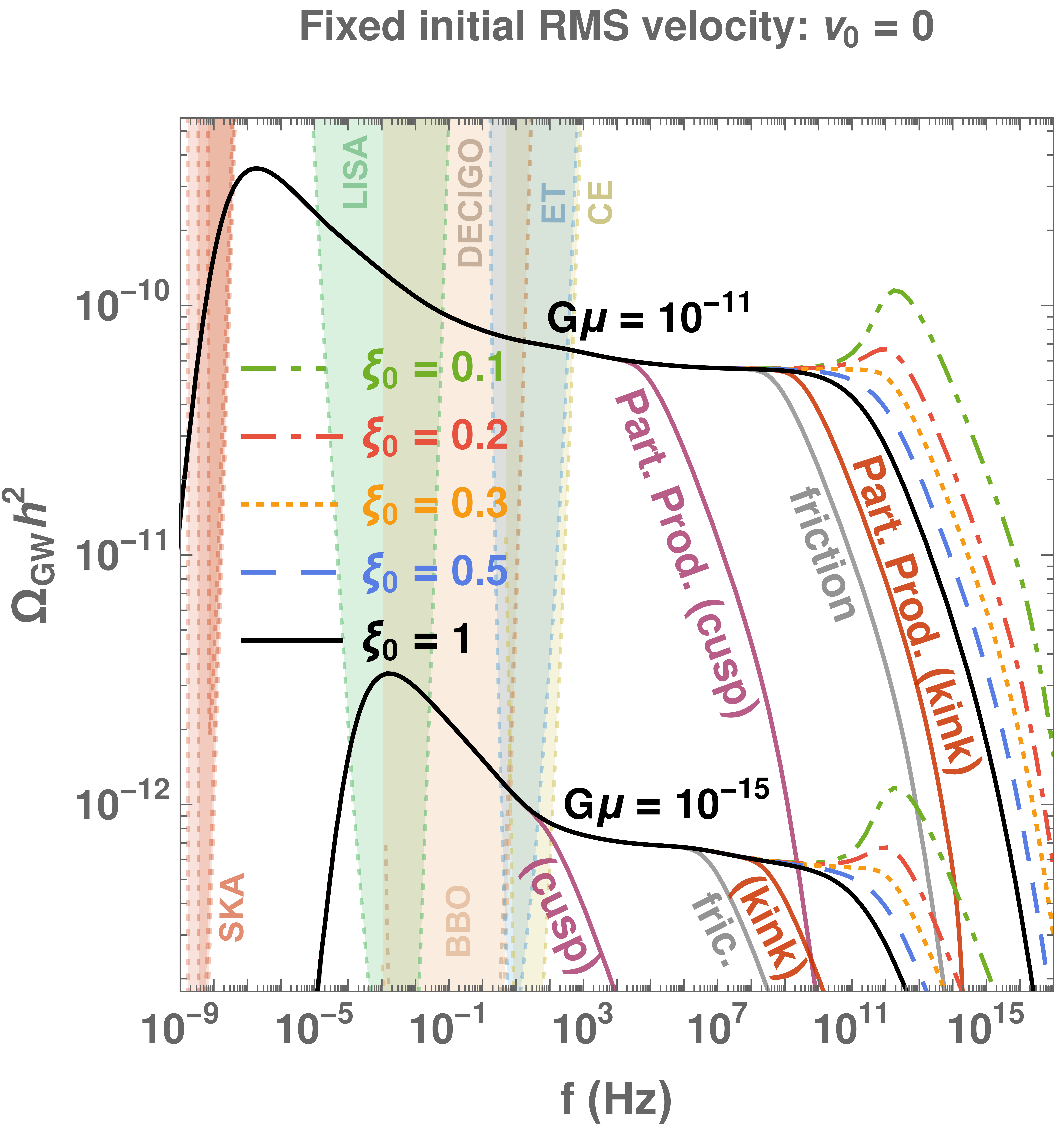}}}
\raisebox{0cm}{\makebox{\includegraphics[width=0.49\textwidth, scale=1]{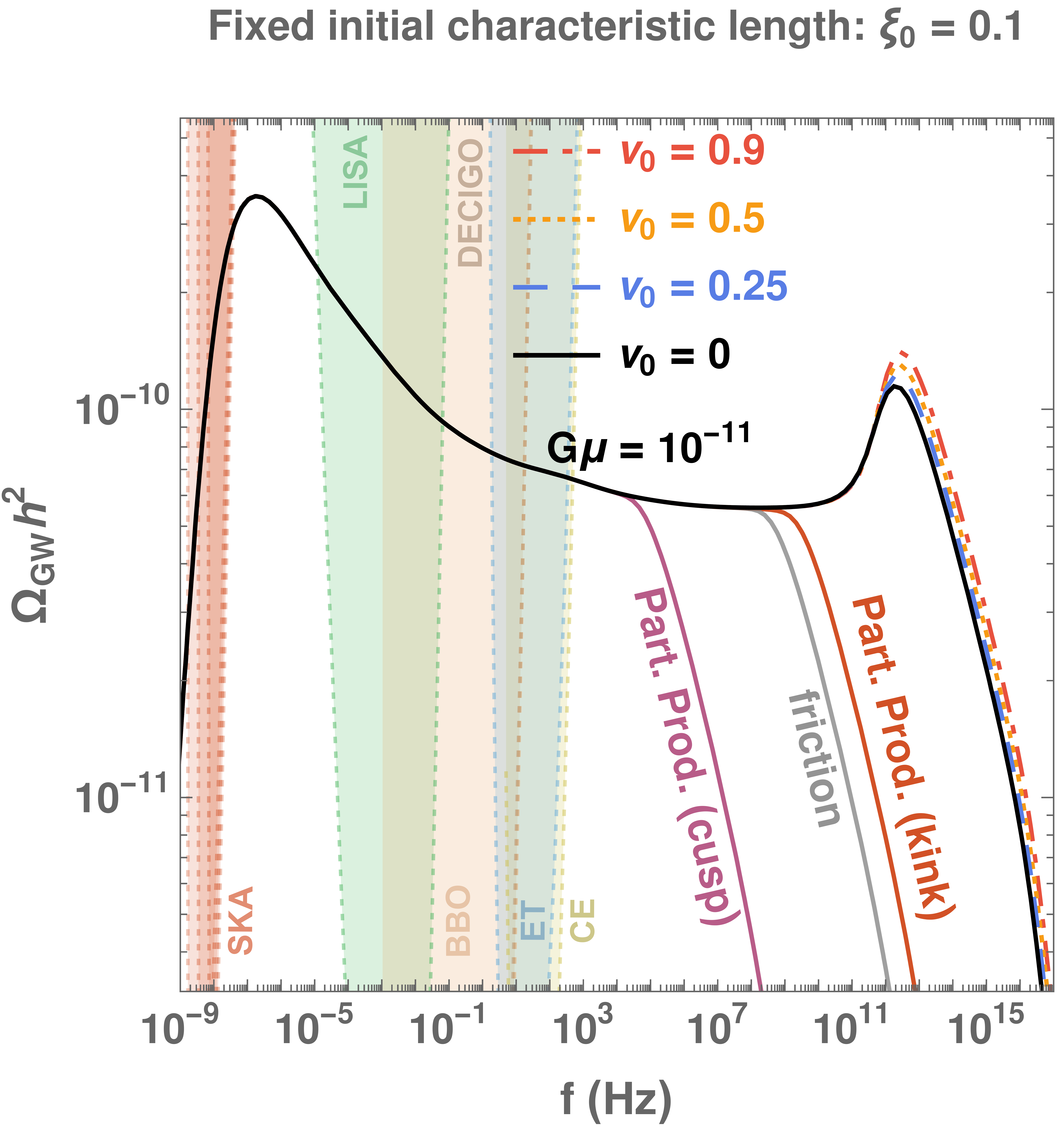}}}\\
\raisebox{0cm}{\makebox{\includegraphics[width=0.49\textwidth, scale=1]{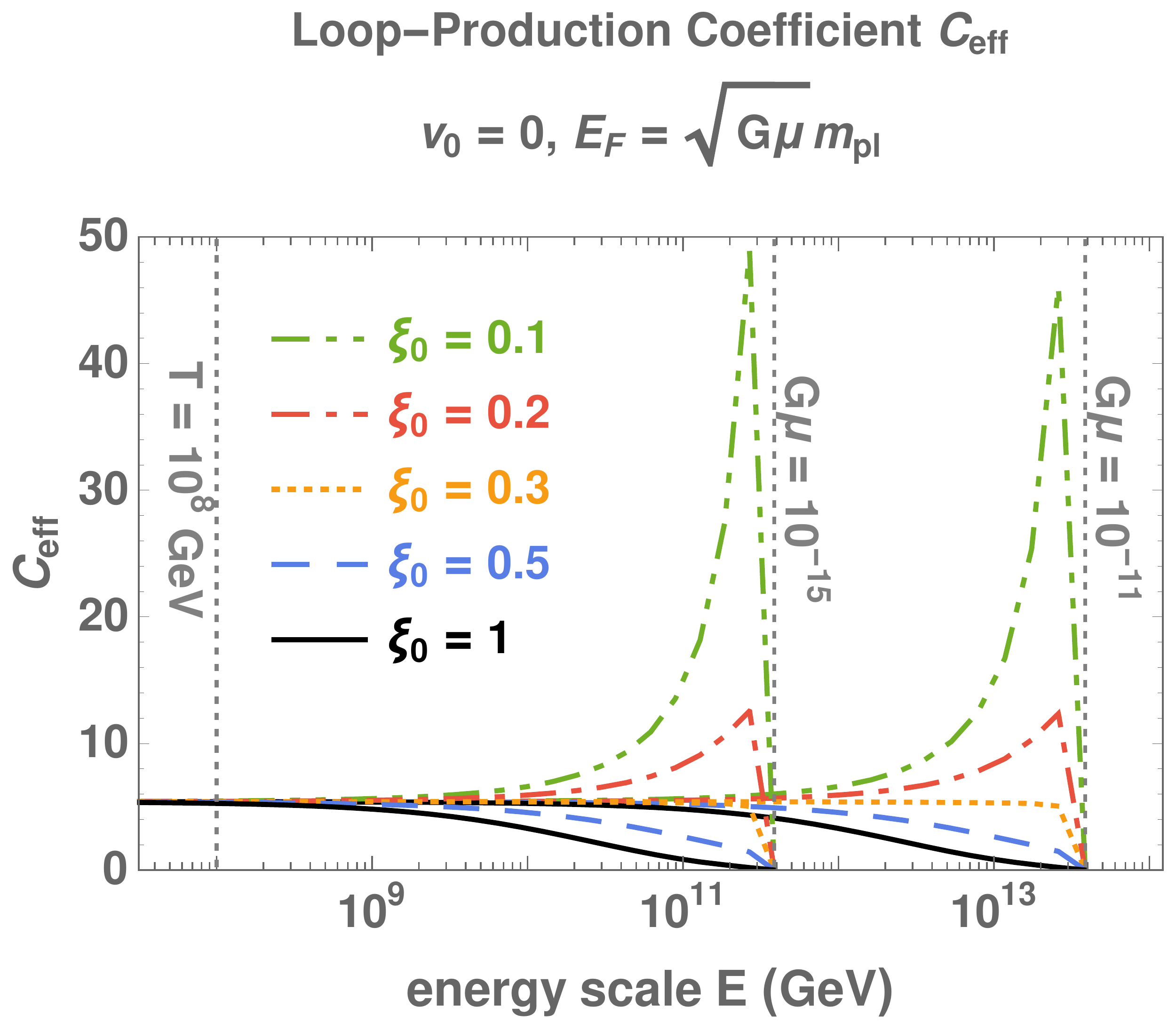}}}
\raisebox{0cm}{\makebox{\includegraphics[width=0.49\textwidth, scale=1]{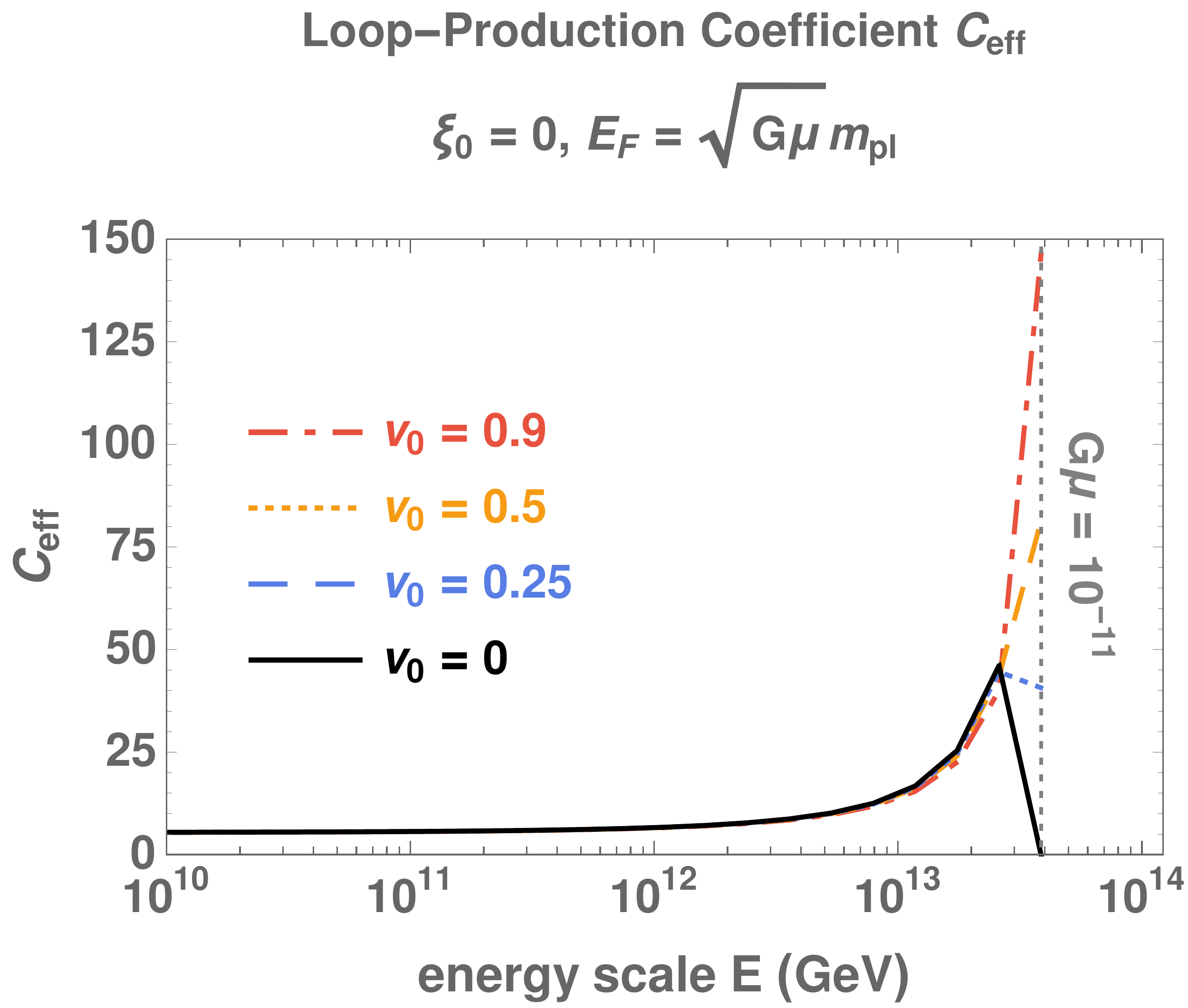}}}
\caption{\it \small \textbf{Top:} GW spectra assuming a standard cosmology. The network is formed at the temperature specified by the string tension: $T_F \sim m_{pl}\sqrt{G\mu}$. Initial conditions with a fixed initial mean velocity $\bar{v}_0$ (left) and a fixed initial correlation length scale $\xi_0$ (right) are applied. The cut-offs due to particle production, c.f. Sec.~\ref{UVcutoff}, and thermal friction, c.f.  App.~\ref{sec:thermal_friction}, are shown with purple, red and gray lines.  \textbf{Bottom:} The bumps at high frequencies come from the over-production of loops right after the network formation when $\xi/\bar{v}$ are taken smaller/larger than their scaling values.}
\label{ST_initial_conditions}
\end{figure}

\subsection{Initial network configuration}
Fig.~\ref{ST_initial_conditions} shows how the spectrum depends on the choice of initial conditions $\bar{v}_0$, $\xi_0$. As expected, only the region near the high-frequency cut-off, corresponding to loops created just after the network formation, is impacted. Such initial values lead to an overproduction of loops during the initial transient regime and to an enhancement of the spectrum. The impact of $\xi_0$ is stronger than the one of $\bar{v}_0$ because the loop-production efficiency scales as $C_{\rm eff} \propto \bar{v}/\xi^3$. We see that the smaller/larger $\xi_0$/$\bar{v}_0$, the higher the bump.

Note that the frequency of the bump is independent of $G\mu$. 
This can be understood upon plugging the temperature when the network is formed, $T_F\sim m_\textrm{pl}\sqrt{G\mu}$ into the $(f-T)$-correspondence  formula in Eq.~\eqref{turning_point_general_scaling}. Also note that at such a high temperature, the friction of the strings with the plasma might play a major role \cite{Vilenkin:1991zk}. 

The high-frequency bump could be a probe of the nature of the PT in the early universe, e.g. the initial correlation length, or a probe of the plasma-string interaction. This could be in principle a motivation for high-frequency GW experiments. However, the loops which would contribute to such high-frequency GW, might rather decay into particles, c.f. solid purple and red line in Fig.~\ref{ST_initial_conditions}.

In the next three sections, we will study the impact of different non-standard cosmologies on the SGWB from cosmic strings. Each cosmological history not only yields a distinct value for the scale factor of the universe today, $a_0$, thus a different amount of redshifting of gravitational waves in Eq.~(\ref{kmode_omega}), but also a distinct loop-production rate $\propto C_{\rm eff}/t_i^4$ due to a different formation time $t_i$ and a different loop-production efficiency $C_{\rm{eff}}$. 
In Sec.~\ref{sec:long_NS_era}, we assume that the radiation era was preceded by a long period of either matter domination or kination all the way after inflation.
In Sec.~\ref{sec:interm_matter} and Sec.~\ref{sec:inflation}, we assume instead some short eras of either matter domination or inflation, inside the radiation era.

\FloatBarrier
\section{Long-lasting matter or kination era}
\label{sec:long_NS_era}

\subsection{The non-standard scenario}
\begin{figure}[h!]
\centering
\raisebox{0cm}{\makebox{\includegraphics[width=0.6\textwidth, scale=1]{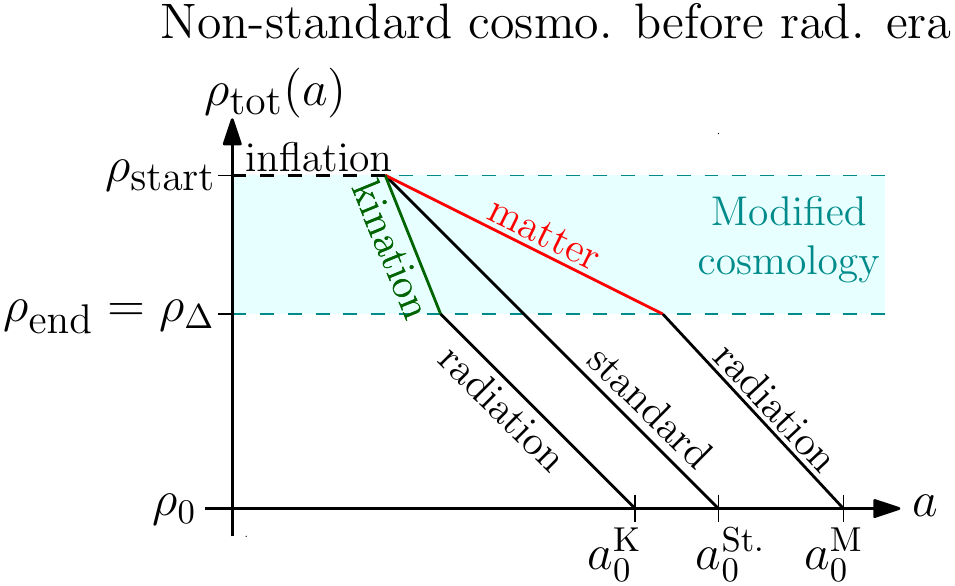}}}
\caption{\it \small Evolution of the energy density assuming a matter (M) and kination (K) era after inflation and before the radiation era. `St' refers to standard cosmology.  We suppose that the cosmic string network forms at the end of inflation with tension given by $G\mu\sim (\rho^{1/4}_\textrm{start}/m_\textrm{pl})^2$ (for instance the CS network can form through non-thermal dynamical symmetry breaking \cite{Shafi:1984tt,Vishniac:1986sk,Kofman:1986wm,Yokoyama:1988zza,Yokoyama:1989pa,Nagasawa:1991zr,Basu:1993rf,Freese:1995vp}).}
\label{figure_NS_era_long}
\end{figure}

In this section, we consider  the presence of a matter or kination-dominated era which  starts just after the end of  inflation, when the total energy density is $\rho_\textrm{start}=\rho_\textrm{inflation}$, and ends much later, at $\rho_{\rm end}$, when it becomes supplanted by the standard radiation-dominated era. 
At the end of the non-standard era, the temperature of the universe is $T_{\Delta}$. 
The energy density profile, sketched in Fig.~\ref{figure_NS_era_long}, is given  by
\begin{equation}
\rho_\textrm{tot}(a)=\begin{cases}
\rho_\textrm{start} \over{a_\textrm{start}}{a}^n+\rho_\textrm{late}(a) \hspace{2em}&\textrm{for }\rho_\textrm{start}>\rho>\rho_\textrm{end},\\[0.5em]
\rho_\textrm{end} \  \Delta_R(T_\textrm{end},T)\over{a_\textrm{end}}{a}^4+\rho_\textrm{late}(a)\hspace{2em}&\textrm{for }\rho<\rho_\textrm{end},
\end{cases}
\label{ns_cosmo_define_function}
\end{equation}
\begin{eqnarray*}
		\textrm{where \hspace{3em}}  
		\rho_\textrm{start}, \rho_\textrm{end} &\equiv& \textrm{ the starting and ending energy density of the non-standard cosmology},\\
		\rho_\textrm{late} &\equiv& \textrm{ the standard-cosmology energy density dominating at late times,}\\
		&&  \textrm{ e.g. the standard matter density, and cosmological constant}.\\
		\Delta_R &\mbox{is}& \mbox{given in Eq.}(\ref{eq:DeltaR}).
	\end{eqnarray*} 

%
%

\begin{figure}[h!]
\centering
\raisebox{0cm}{\makebox{\includegraphics[width=0.49\textwidth, scale=1]{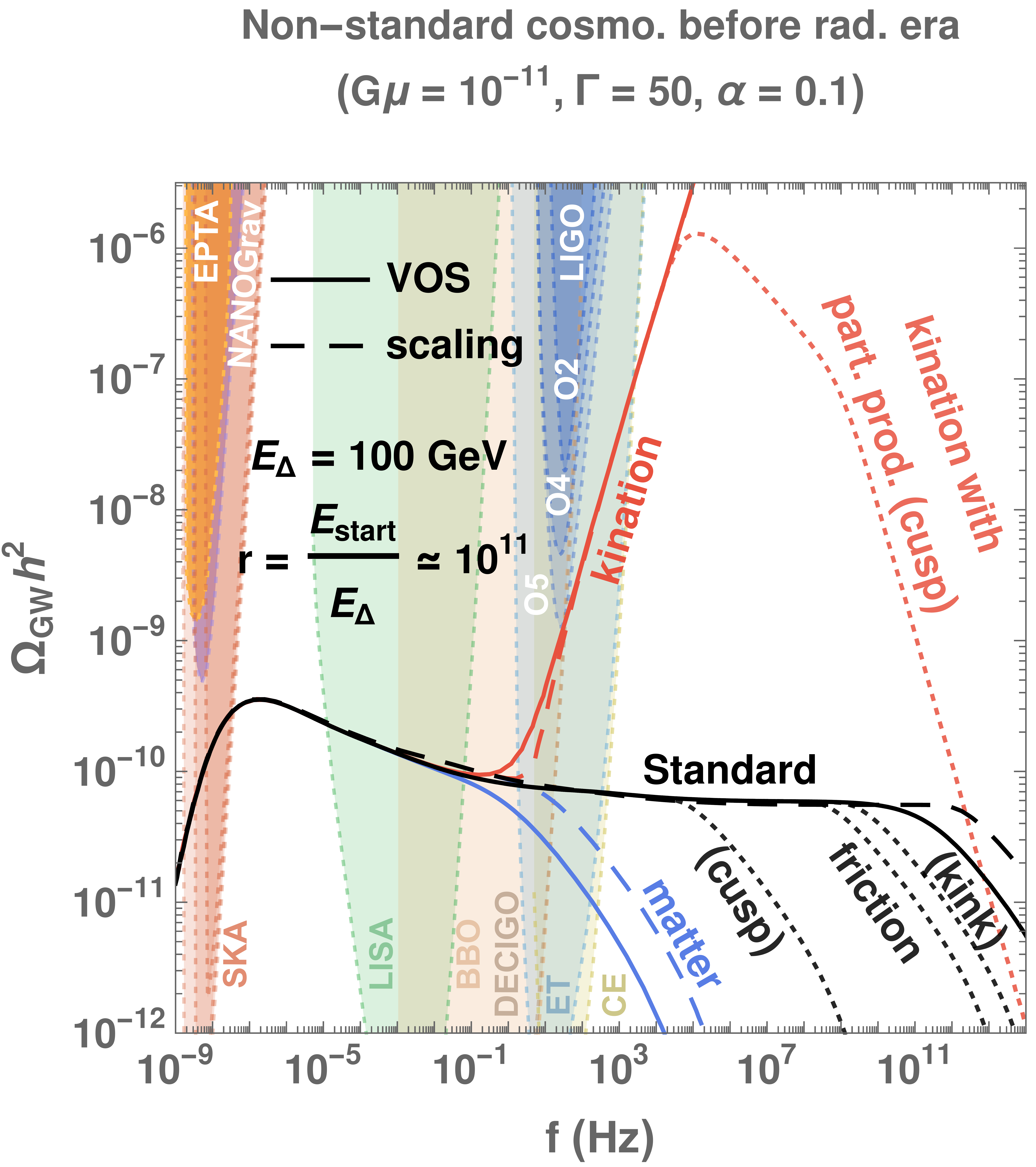}}}
\raisebox{1cm}{\makebox{\includegraphics[width=0.5\textwidth, scale=1]{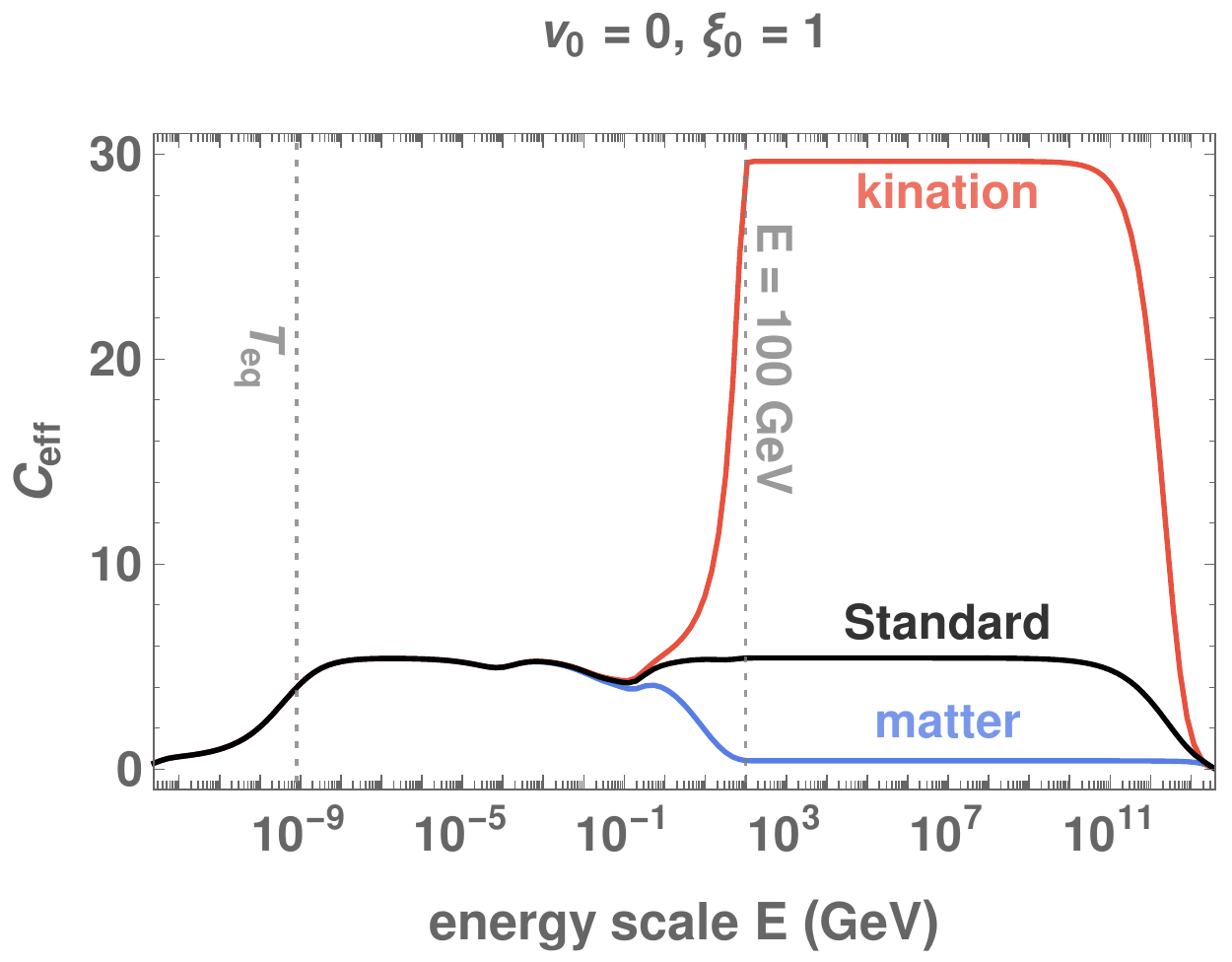}}}
\caption{\it \small \textbf{Left:} GW spectra from cosmic strings assuming either the scaling (dashed) or the VOS network (solid), c.f. Sec.~\ref{sec:scalingVSvos}, evolved in the presence of a non-standard era, either matter (blue) or kination-dominated (red), before the standard radiation era. The transient VOS evolution of the long-string network during the change of cosmology shifts the turning-point towards lower frequencies by $\mathcal{O}(25)$. The cut-offs due to particle production, c.f. Sec.~\ref{UVcutoff},  or thermal friction, c.f.  App.~\ref{sec:thermal_friction}, are shown with dotted lines. \textbf{Right:} The evolution of the loop-production efficiency for each cosmological background shows that the scaling solution is reached after a transient evolution corresponding to the Hubble rate dropping by on order of magnitude. The slower the expansion rate $a\propto t^{2/n}$, the slower the dilution of the long-string energy density $\rho_{\infty} \propto a^{-2}$ and the higher the needed loop-production efficiency $C_{\rm eff}$ in order to reach the scaling regime $\rho_{\infty} \propto t^{-2}$.}
\label{figure_preceding_NS_era}
\end{figure}

\subsection{Impact on the spectrum: a turning-point}
The resulting GW spectra are shown in Fig.~\ref{figure_preceding_NS_era} for long-lasting kination and matter eras starting at $E_\textrm{start}=m_{pl}\sqrt{G\mu}$ and ending at $E_\textrm{end}=E_\Delta= 100$ GeV with duration
		\begin{equation}
r\equiv\left(\frac{\rho_\textrm{start}}{\rho_\textrm{end}}\right)^{1/4}\equiv\left(\frac{E_\textrm{start}}{E_\Delta}\right)\simeq 10^{11}.
\end{equation}
For kination, the slower expansion of the universe means that loops are produced earlier when the loop-production is more efficient, c.f. Eq.~\eqref{eq:LoopProductionFctBody2}, which enhances the spectrum. 
For matter domination, we have the opposite behavior and the spectrum is suppressed. 
\paragraph{The turning-point frequency:}
\label{sec:turning_point_general}
A key observable is the frequency above which the GW spectrum differs from the one obtained in standard cosmology. This is the so-called \textit{turning-point} frequency $f_{\Delta}$.
It  corresponds to the redshifted-frequency emitted by the loops created during the change of cosmology at the temperature $T_{\Delta}$. In the instantaneous scaling approximation, c.f. dashed line in Fig.~\ref{figure_preceding_NS_era}, the turning-point frequency $f_{\Delta}$ is given by the $(T,\,f)$-correspondence relation 
\begin{align}
f_\Delta^{\textrm{scaling}}=(4.5\times10^{-2}\textrm{ Hz})\left(\frac{T_\Delta}{\textrm{GeV}}\right)\left(\frac{0.1\times 50 \times10^{-11}}{\alpha\,\Gamma G\mu}\right)^{1/2}\left(\frac{g_*(T_\Delta)}{g_*(T_0)}\right)^{1/4}.
\label{eq:turning_point_scaling}
\end{align}
However, the deviation from the scaling regime during the change of cosmology, c.f. Sec.~\ref{sec:scalingVSvos}, implies a shift to lower frequencies of the $(T,\,f)$-correspondence, by a factor $\sim 22.5$, c.f. solid vs dashed lines in Fig.~\ref{figure_preceding_NS_era}. The correct $(T,\,f)$-correspondence when applied to a change of cosmology is
\begin{align}
f_\Delta^{\textrm{VOS}}=(2\times10^{-3}\textrm{ Hz})\left(\frac{T_\Delta}{\textrm{GeV}}\right)\left(\frac{0.1\times 50 \times10^{-11}}{\alpha\,\Gamma G\mu}\right)^{1/2}\left(\frac{g_*(T_\Delta)}{g_*(T_0)}\right)^{1/4}.
\label{turning_point_general}
\end{align}
We fit the numerical factor in Eq.~\eqref{turning_point_general} (but also in Eq.~\eqref{turning_point_inf}) by imposing\footnote{The coefficient in Eq.~\eqref{turning_point_general} has been fitted upon considering the matter case $\Omega_\textrm{NS}= \Omega_\textrm{matter}$. Note that the turning-point in the kination case is slightly higher frequency by a factor of order 1, c.f. Fig.~\ref{figure_preceding_NS_era}.} the non-standard-cosmology spectrum $\Omega_\textrm{NS}$ to deviate from the standard-cosmology one $\Omega_\textrm{ST}$ by $10\%$ at the turning-point frequency,
\begin{align}
\left|\frac{\Omega_\textrm{NS}(f_\Delta)-\Omega_\textrm{ST}(f_\Delta)}{\Omega_\textrm{ST}(f_\Delta)}\right|\simeq 10\%.
\label{10per_criterion}
\end{align}
We are conservative here. Choosing $1\%$ instead of $10\%$ would lead to a frequency shift of the order of  ${\cal O}(400)$, c.f. Eq. \eqref{turning_point_general_scaling_app}.
Note that our Eq.~\eqref{turning_point_general} is numerically very similar to the one in \cite{Cui:2017ufi,Cui:2018rwi,Auclair:2019wcv} although an instantaneous change of the loop-production efficiency $C_{\rm eff}$ at $T_{\rm \Delta}$ is assumed in \cite{Cui:2017ufi,Cui:2018rwi,Auclair:2019wcv}. This can be explained if  in Ref.~\cite{Cui:2017ufi,Cui:2018rwi,Auclair:2019wcv},  the criterion in Eq.~\eqref{10per_criterion} is smaller than the percent level.

\begin{figure}[h!]
			\centering
			\raisebox{0cm}{\makebox{\includegraphics[height=0.48\textwidth, scale=1]{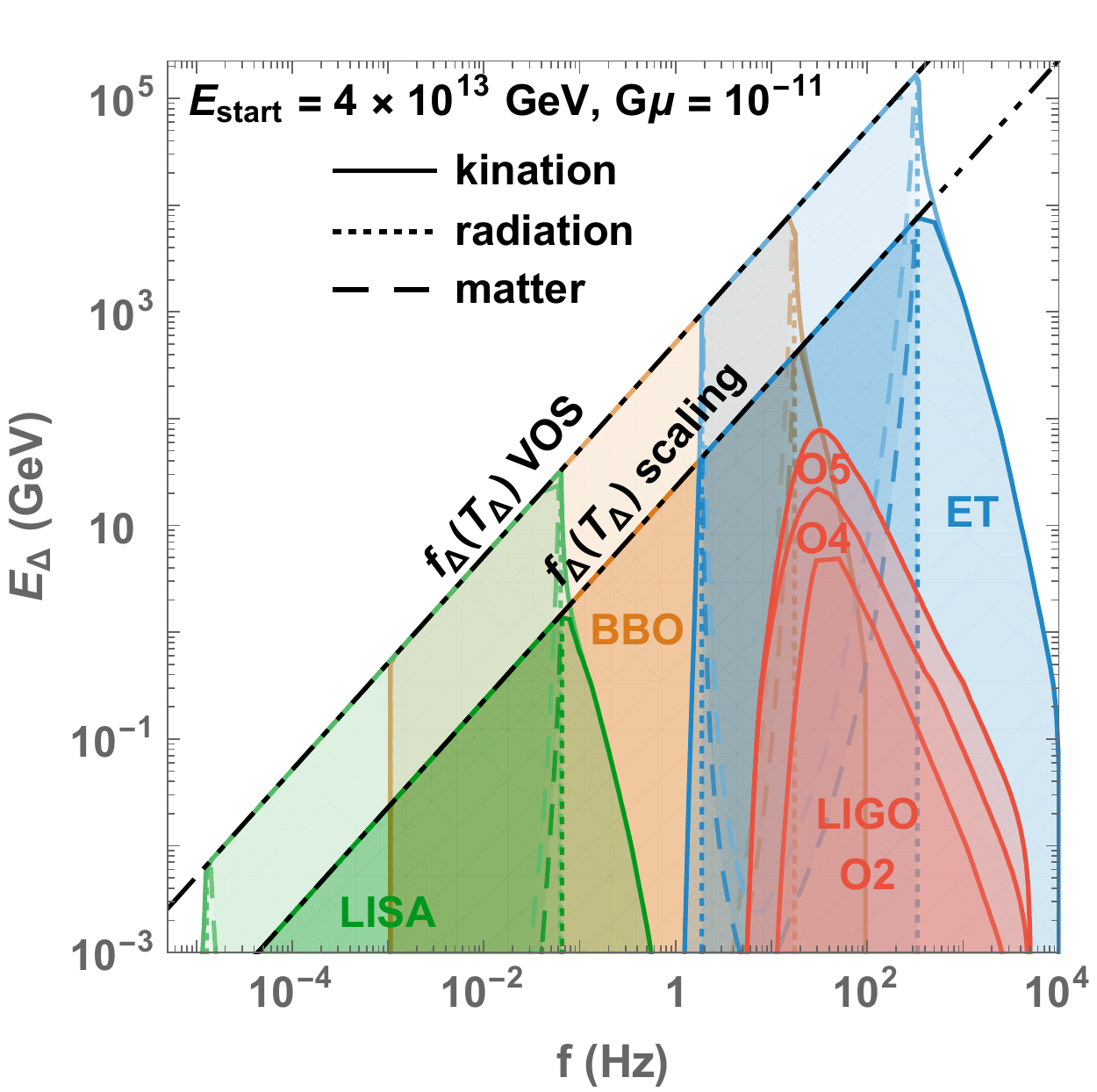}}}
			\raisebox{0cm}{\makebox{\includegraphics[height=0.48\textwidth, scale=1]{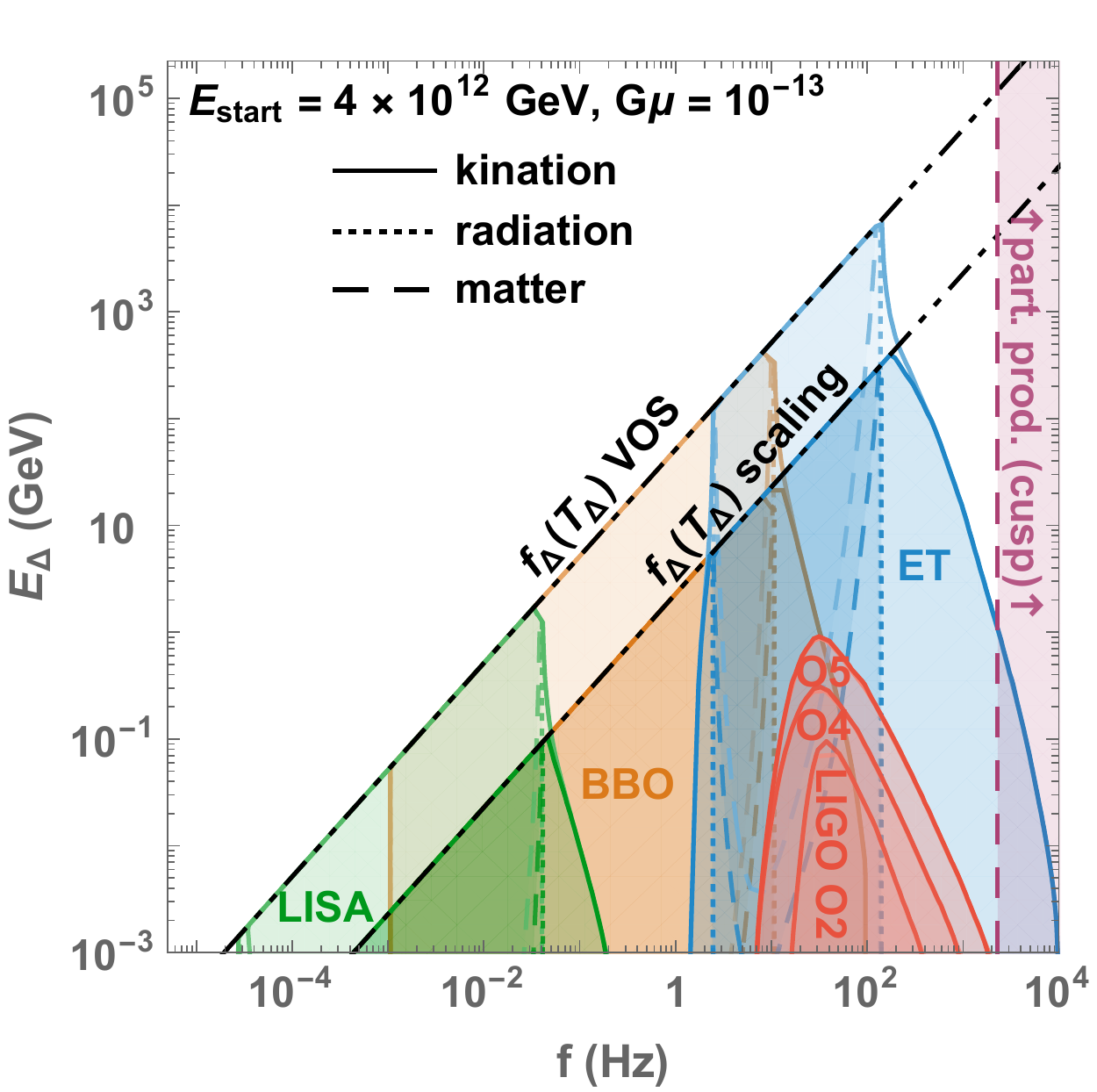}}}\\
			\raisebox{0cm}{\makebox{\includegraphics[height=0.48\textwidth, scale=1]{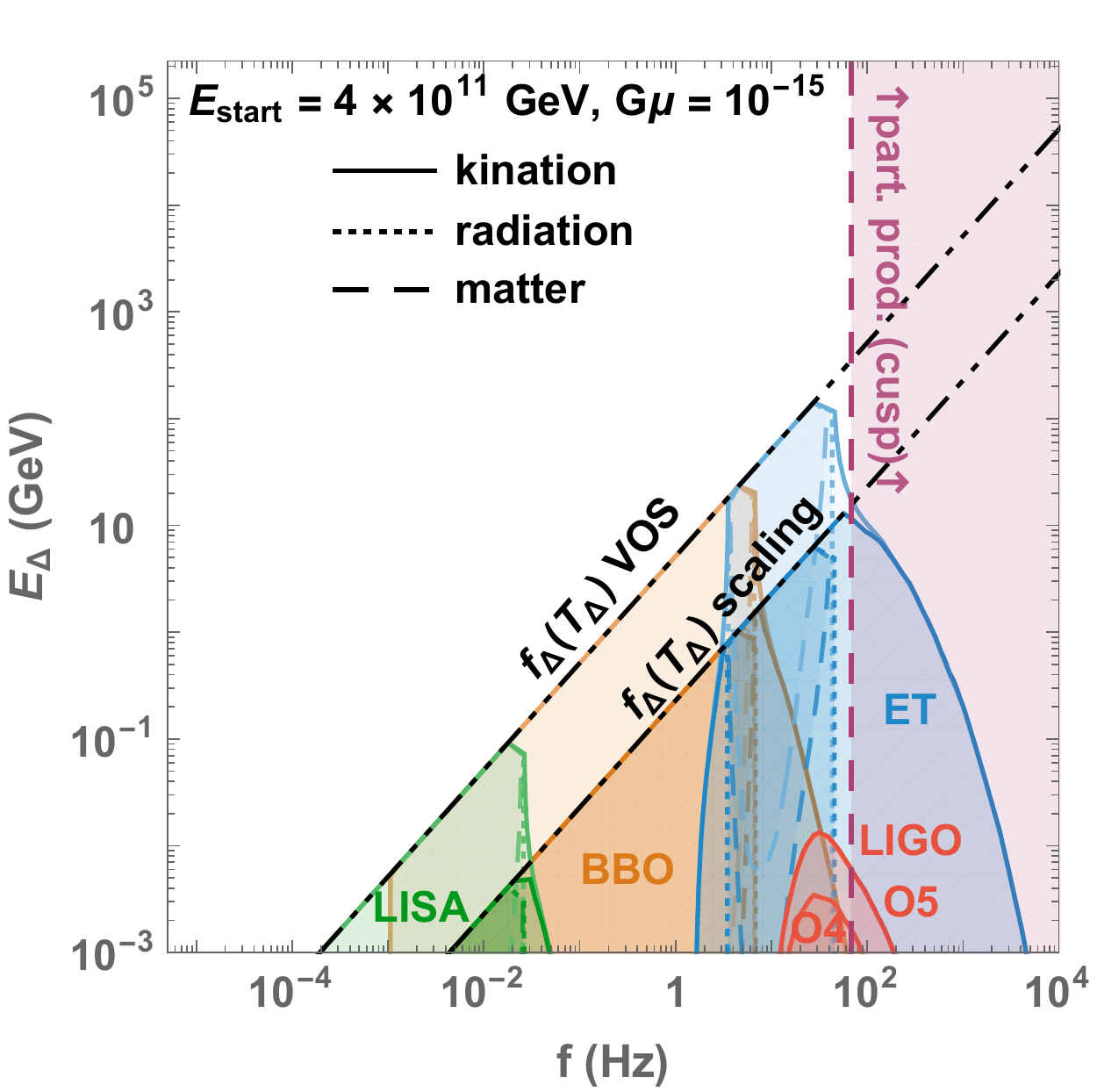}}}
			\hfill
			\caption{\it \small The colored regions show the detectability of a spectral suppression/enhancement, c.f. \textit{spectral-index prescription (Rx 2)} in Sec.~\ref{sec:Rx1VSRx2}, due to an early matter/kination era taking place before the standard radiation domination, assuming scaling and VOS networks, c.f. Sec.~\ref{sec:scalingVSvos}.  The limitation from particle production, c.f. Sec.~\ref{UVcutoff}, is considered in purple.}
				\label{fig:contour_all_f_n_sum}
		\end{figure}

\subsection{Constraints}

A long matter/kination era leads to a spectral suppression/enhancement which could lie within the observational windows of future GW observatories.
In Fig.~\ref{fig:contour_all_f_n_sum}, we show the constraints on an early long-lasting ($r\gtrsim10^{7}$) non-standard era ending at the temperature $T_{\Delta}$, for different values of $G\mu$. We assume that a non-standard era is detectable by a GW interferometer if the absolute value of the observed spectral index is larger than $0.15$. This corresponds to the \textit{spectral-index prescription (Rx 2)} discussed in Sec.~\ref{sec:Rx1VSRx2}.
We see that LISA, BBO/DECIGO and ET/CE can probe non-standard eras ending below $T_{\Delta} \simeq$ 10~GeV, $1$~TeV and $100$~TeV, respectively. LIGO can already constrain kination eras ending after $10$~GeV. The temperatures which we can probe are $\mathcal{O}(25)$ larger when assuming a VOS network (c.f. VOS dashed-dotted line in Fig.~\ref{fig:contour_all_f_n_sum}) compared to a scaling network (c.f. colored regions in Fig.~\ref{fig:contour_all_f_n_sum}), (see Sec.~\ref{sec:scalingVSvos} for the definitions of scaling and VOS networks).  Particle production starts to limit the observation for $G\mu \lesssim 10^{-15}$.

\subsection{A shorter period of kination}
\begin{figure}[h!]
\centering
\raisebox{0cm}{\makebox{\includegraphics[width=0.49\textwidth, scale=1]{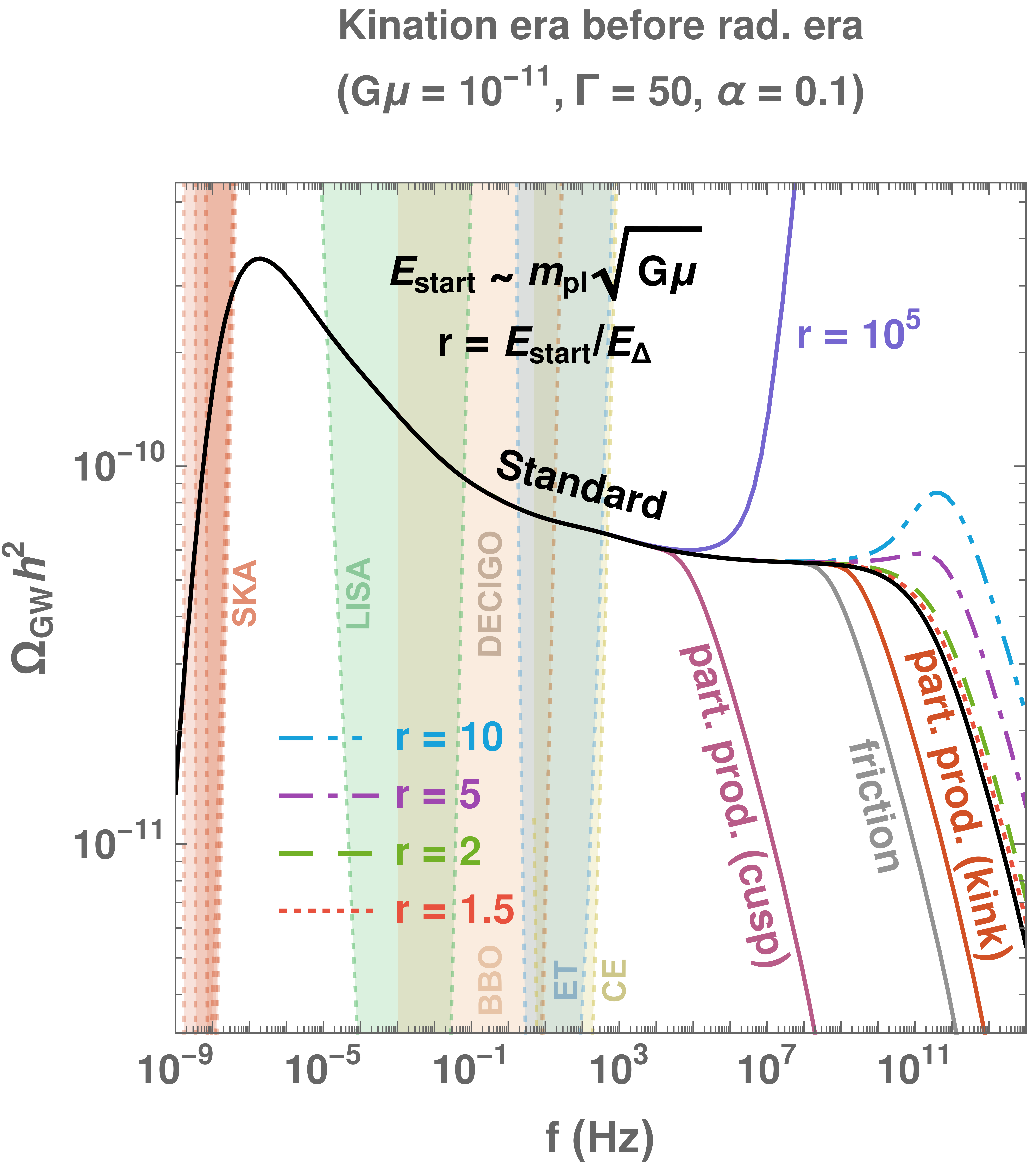}}}
\raisebox{0cm}{\makebox{\includegraphics[width=0.5\textwidth, scale=1]{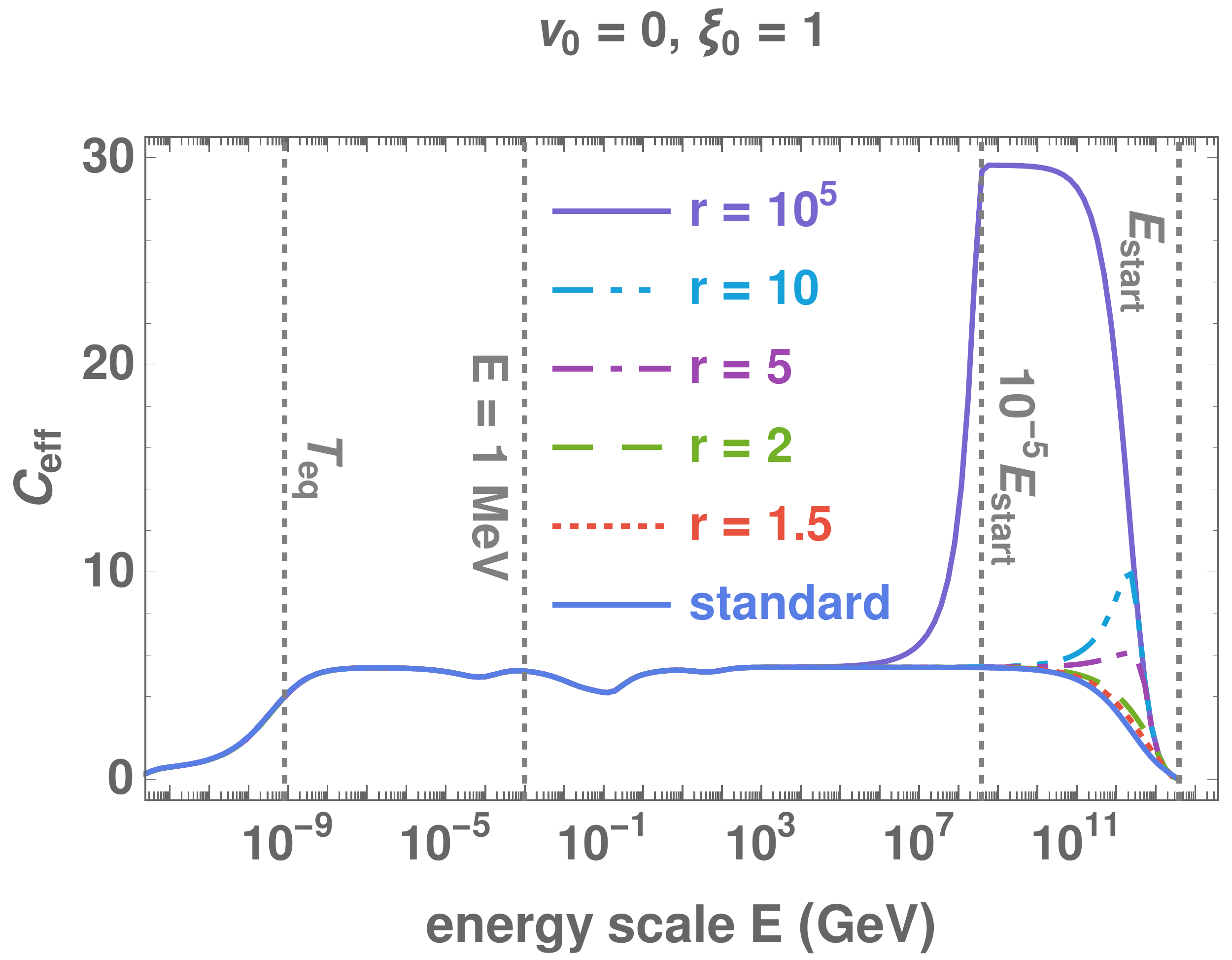}}}
\caption{\it \small \textbf{Left:} GW spectrum from the string network evolved in the presence of a short kination era after inflation.
The peak at high frequencies generated by loops created just after the network formation (see Fig.~\ref{ST_initial_conditions}) is enhanced by the effect of the kination era. The cut-offs due to particle production, c.f. Sec.~\ref{UVcutoff}, and thermal friction, c.f.  App.~\ref{sec:thermal_friction}, are shown with purple, red and gray lines. 
\textbf{Right:} The corresponding evolution of the loop-production efficiency shows that the scaling regime is never reached for short kination eras.}
\label{NS_KD_spectra}
\end{figure}

Interestingly, a short kination period can generate a bump in the spectrum. 
We show this  in the left panel of Fig.~\ref{NS_KD_spectra}. In fact, the network has no time to reach the scaling regime. Particularly, on the right panel of Fig.~\ref{NS_KD_spectra}, we show how the efficiency of the loop production grows with the duration of the kination era, without reaching its scaling value $C_{\rm eff} = 29.6$, c.f. Eq.~\eqref{eq:Ceff_scaling}. The bump gets higher for longer kination epoch since the network gets closer to its scaling solution. However, this high-frequency feature  may not be observable due to the high-frequency cutoff from particle production, c.f. solid purple and red lines in left panel of Fig.~\ref{NS_KD_spectra}

\FloatBarrier
\section{Intermediate matter era}
\label{sec:interm_matter}

\subsection{The non-standard scenario}
\begin{figure}[h!]
\centering
\raisebox{0cm}{\makebox{\includegraphics[width=0.49\textwidth, scale=1]{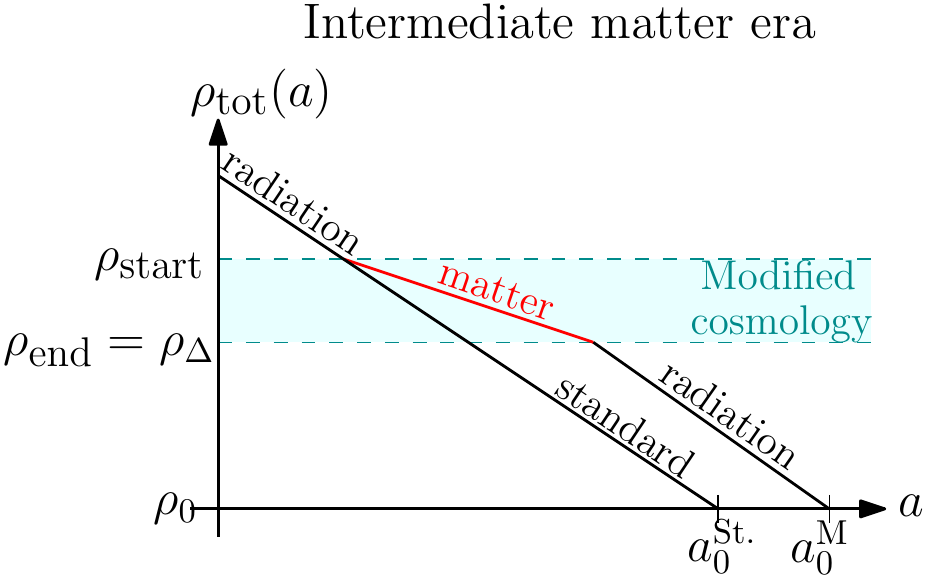}}}
\raisebox{0cm}{\makebox{\includegraphics[width=0.49\textwidth, scale=1]{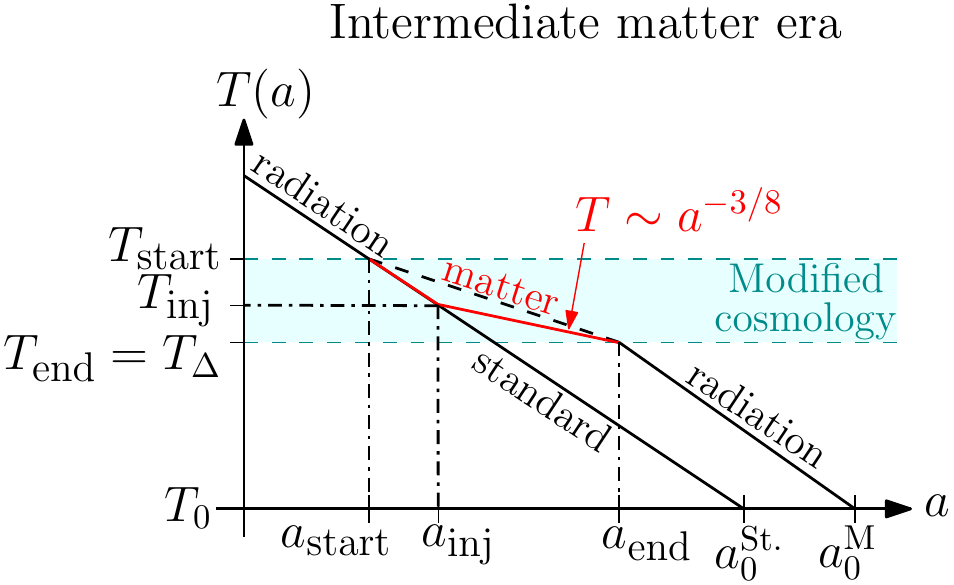}}}
\caption{\it \small Evolution of  the total energy density \textbf{(left)} and  the temperature \textbf{(right)} assuming the presence of an intermediate matter era. $T_{\rm inj}$ and $a_{\rm inj}$ are the temperature and scale factor at which the entropy injected by the decay of the matter content into radiation, starts to be effective, c.f. Fig.~2 in \cite{Cirelli:2018iax}. St: standard; M: matter.}
\label{fig:md_intermediate_diag}
\end{figure}

In this section, we consider the existence of an early-intermediate-matter-dominated era, following an earlier radiation era and preceeding the standard radiation era. The intermediate matter-dominated era starts when the matter energy density $\rho_{\rm matter} \propto a^{-3}$ takes over the radiation energy density $\rho_{\rm radiation} \propto a^{-4}$ and ends when the matter content decays into radiation, c.f. Fig.~\ref{fig:md_intermediate_diag}. The energy density profile is illustrated in Fig.~\ref{fig:md_intermediate_diag}
and can be written as
\begin{equation}
\rho_\textrm{tot}(a)=\begin{cases}
\rho^\textrm{st}_\textrm{rad}(a)+\rho_\textrm{late}(a)&\textrm{for }\rho>\rho_\textrm{start},\\[0.5em]
\rho_\textrm{start}\  \over{a_\textrm{start}}{a}^n+\rho_\textrm{late}(a)&\textrm{for }\rho_\textrm{start}>\rho>\rho_\textrm{end},\\[0.5em]
\rho_\textrm{end}   \Delta_R(T_\textrm{end},T) \ \over{a_\textrm{end}}{a}^4+\rho_\textrm{late}(a)\hspace{2em}&\textrm{for }\rho<\rho_\textrm{end}.
\end{cases}
\label{ns_cosmo_define_function}
\end{equation}
where
\begin{eqnarray*}
		\rho_\textrm{start}, \rho_\textrm{end} &\equiv& \textrm{ the starting and ending energy density of the non-standard cosmology},\\
		\rho_\textrm{late} &\equiv& \textrm{ the standard-cosmology energy density dominating at late times,}\\
		&&  \textrm{ e.g. the standard matter density, and cosmological constant}.\\
		\Delta_R &\mbox{is}& \mbox{given in Eq.}(\ref{eq:DeltaR}).
	\end{eqnarray*} 

\subsection{Impact on the spectrum: a low-pass filter}
\begin{figure}[h!]
\centering
\raisebox{0cm}{\makebox{\includegraphics[height=0.54\textwidth, scale=1]{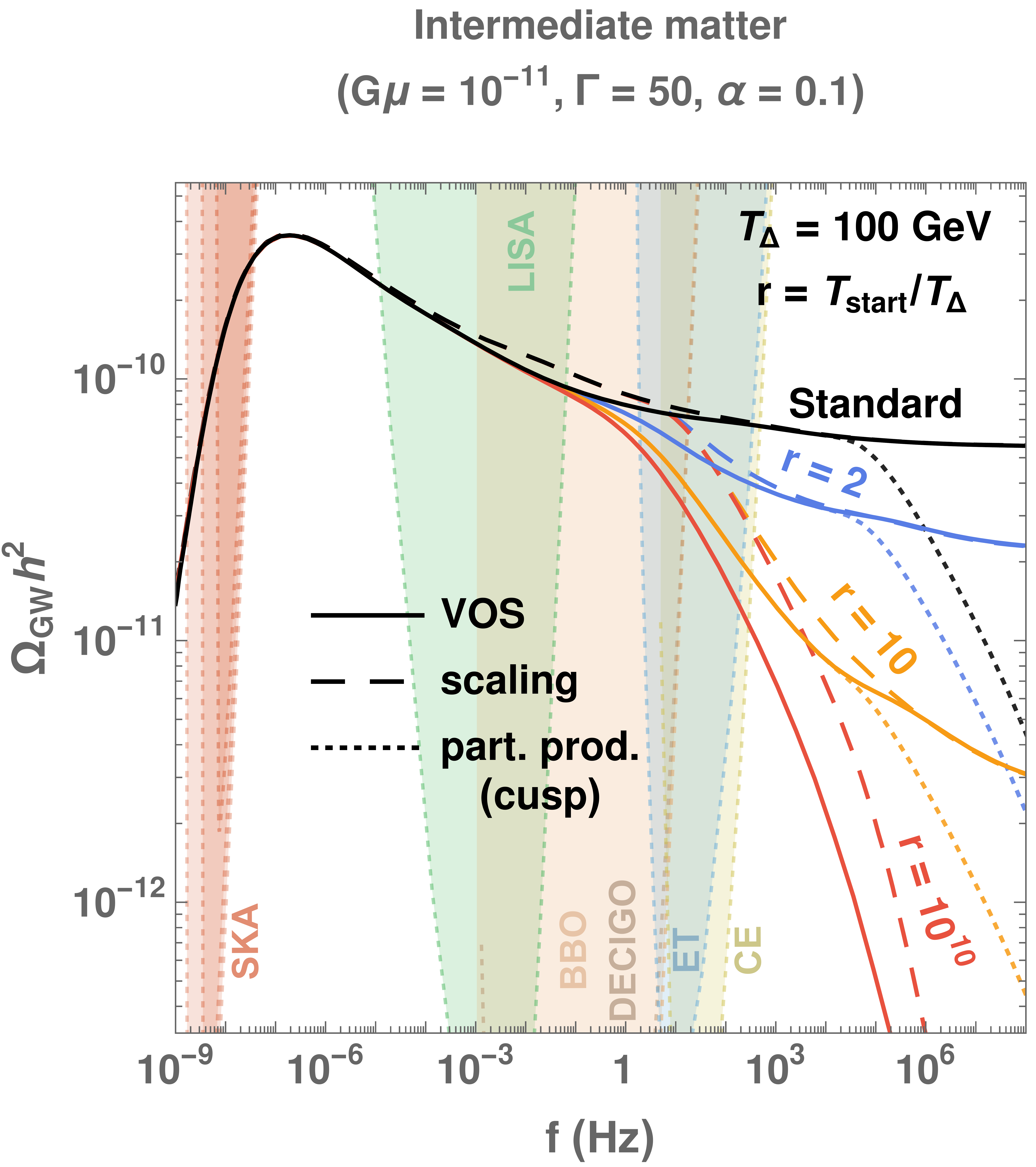}}}
\raisebox{0cm}{\makebox{\includegraphics[height=0.4\textwidth, scale=1]{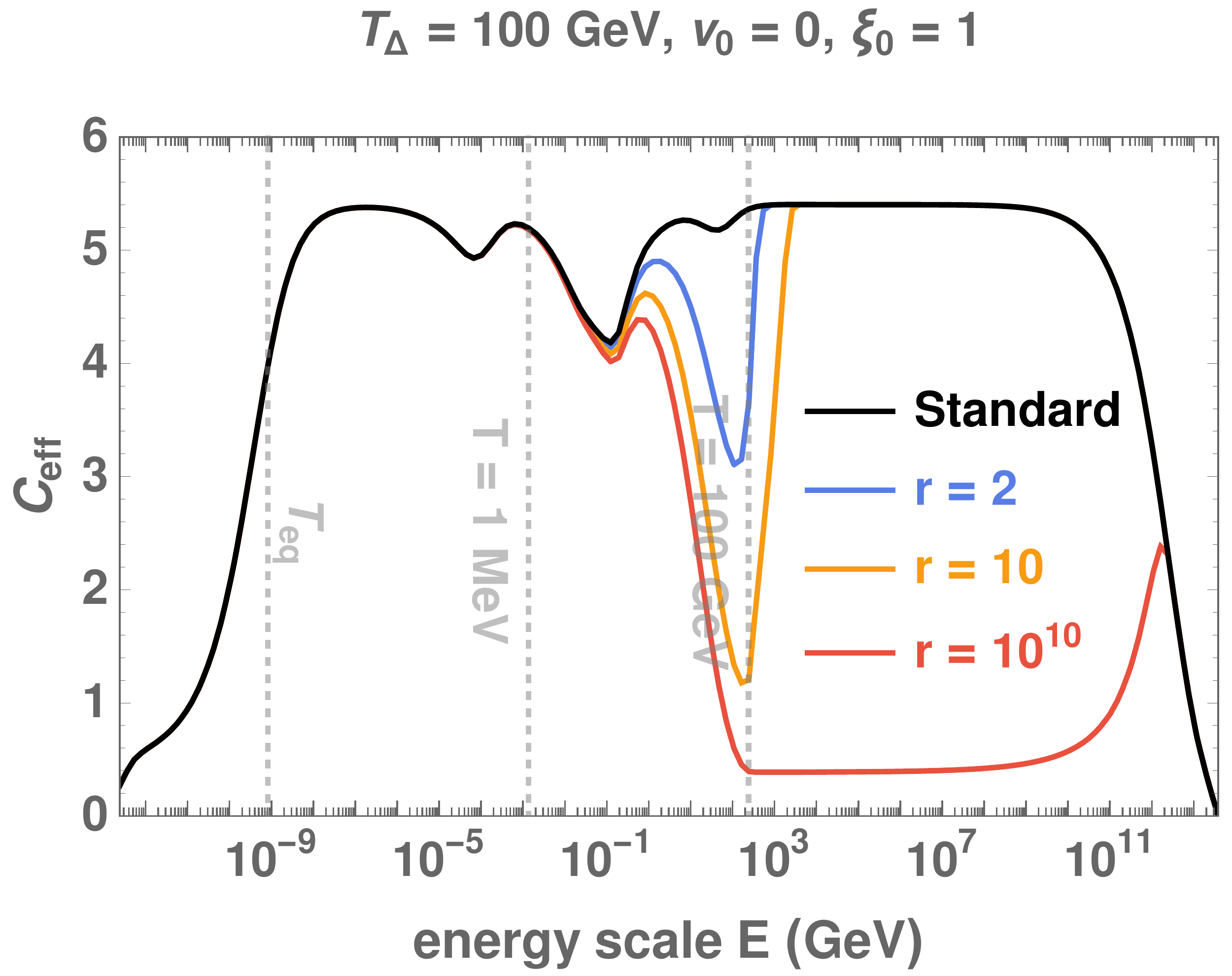}}}
\caption{\it \small \label{fig:VOSvsScaling_Ceff}  GW spectrum from an intermediate matter era starting at the temperature $T_{\rm start}$ and ending at $T_{\Delta}$. \textbf{Left:} The dashed-lines assume that the scaling regime in matter era switches instantaneously to the scaling regime in radiation era, meaning that $C_{\rm eff}$ varies discontinuously, whereas the plain lines incorporate the transient behavior solution of the VOS equations and shown on the right panel.  The cut-offs due to particle production, c.f. Sec.~\ref{UVcutoff}, are shown with dotted lines.   \textbf{Right:} Time evolution of the loop-production efficiency $C_{\rm eff}$ after solving the VOS equations, c.f. Sec.~\ref{sec:scalingVSvos}. }
\label{intermediate_matter_spectrum}
\end{figure}

In the left panel of Fig.~\ref{intermediate_matter_spectrum}, we show that an intermediate matter era blue-tilts the spectral index of the spectrum. Furthermore, at higher frequencies, corresponding to loops produced during the radiation era preceding the matter era, the spectrum recovers a flat scaling but is suppressed by the duration $r$ of the matter era
\begin{equation}
r=\frac{T_\textrm{start}}{T_{\rm \Delta}},
\end{equation}
where $T_{\rm \Delta} =T_{\rm end}$. By suppressing the high-frequency part of the spectrum, an early matter era acts on the CS spectrum as a low-pass filter.
The negative spectral index and the suppression can be understood from Fig.~\ref{fig:md_intermediate_diag}. Indeed, the universe, in the presence of an intermediate matter era, has expanded more than the standard universe. Hence at a fixed emitted frequency, loops are produced later and so are less numerous, implying less GW emission.
In the right panel of Fig.~\ref{intermediate_matter_spectrum}, we show that for short intermediate matter era, $r=2$ or $r=10$, the scaling regime in the matter era, which is characterized by $C_{\rm eff} = 0.39$, c.f. Eq.~\eqref{eq:Ceff_scaling}, is not reached. 

\subsection{Constraints}

\begin{figure}[h!]
			\centering
			\raisebox{0cm}{\makebox{\includegraphics[height=0.49\textwidth, scale=1]{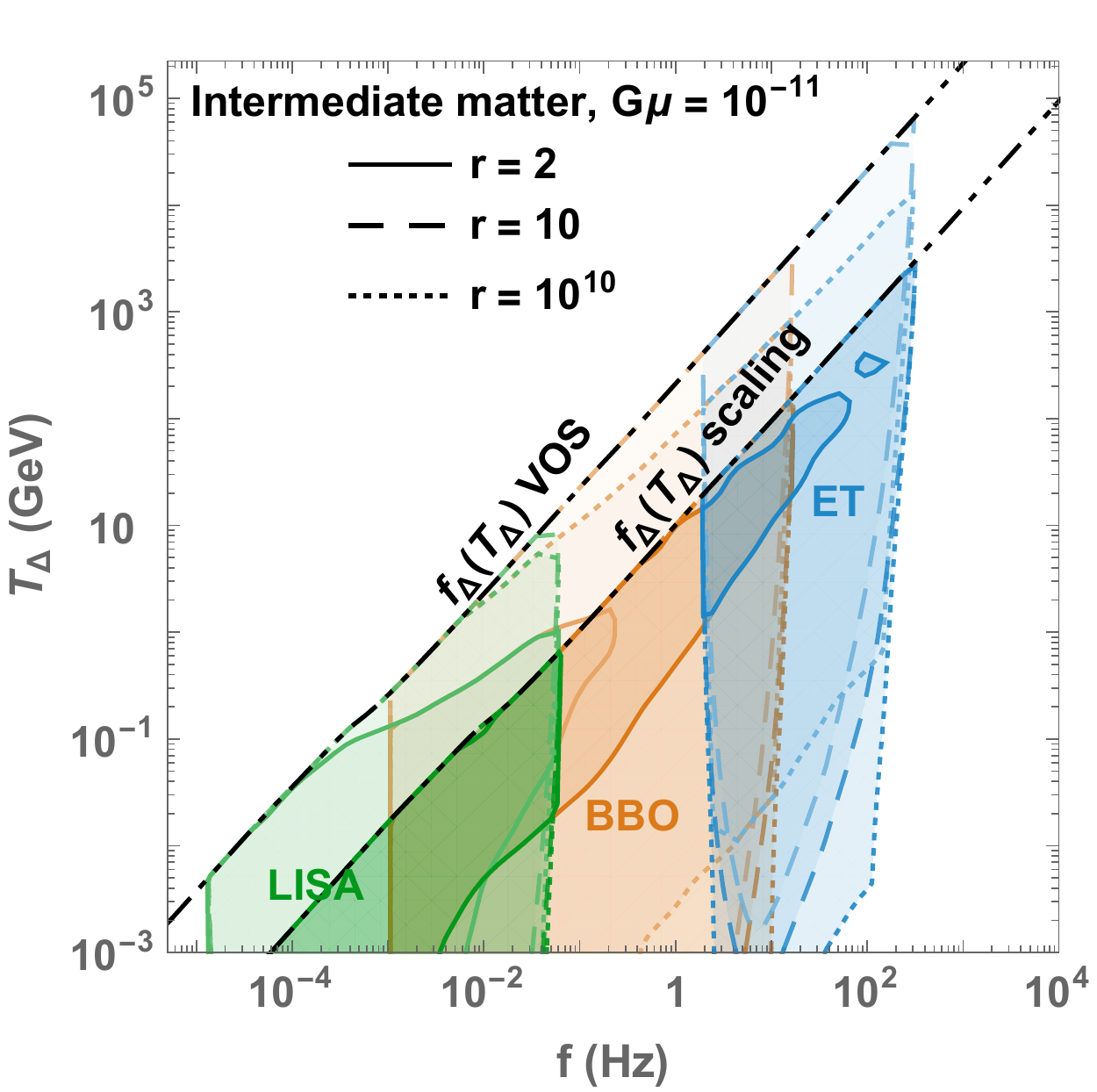}}}
			\raisebox{0cm}{\makebox{\includegraphics[height=0.49\textwidth, scale=1]{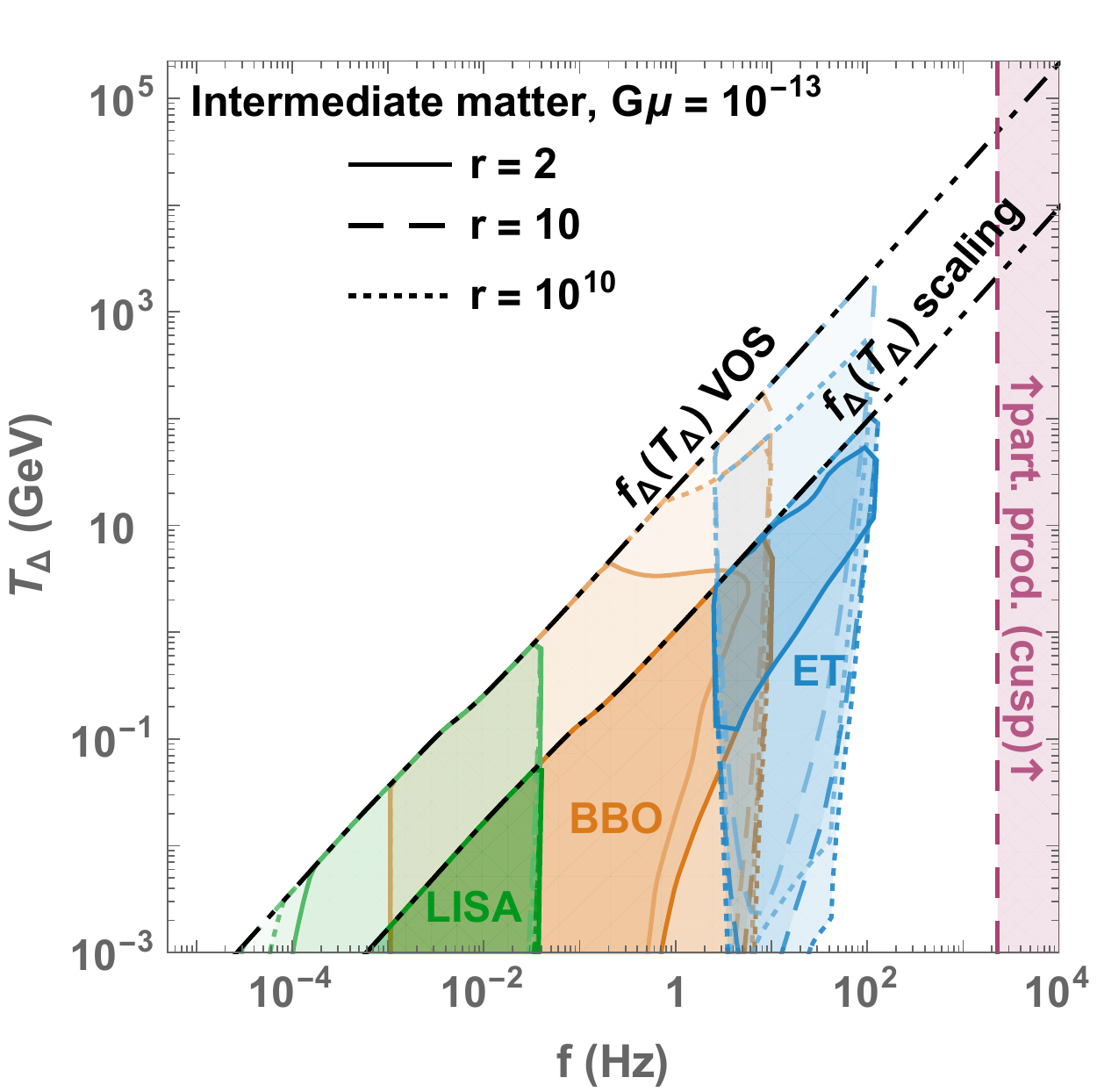}}}\\
			\raisebox{0cm}{\makebox{\includegraphics[height=0.49\textwidth, scale=1]{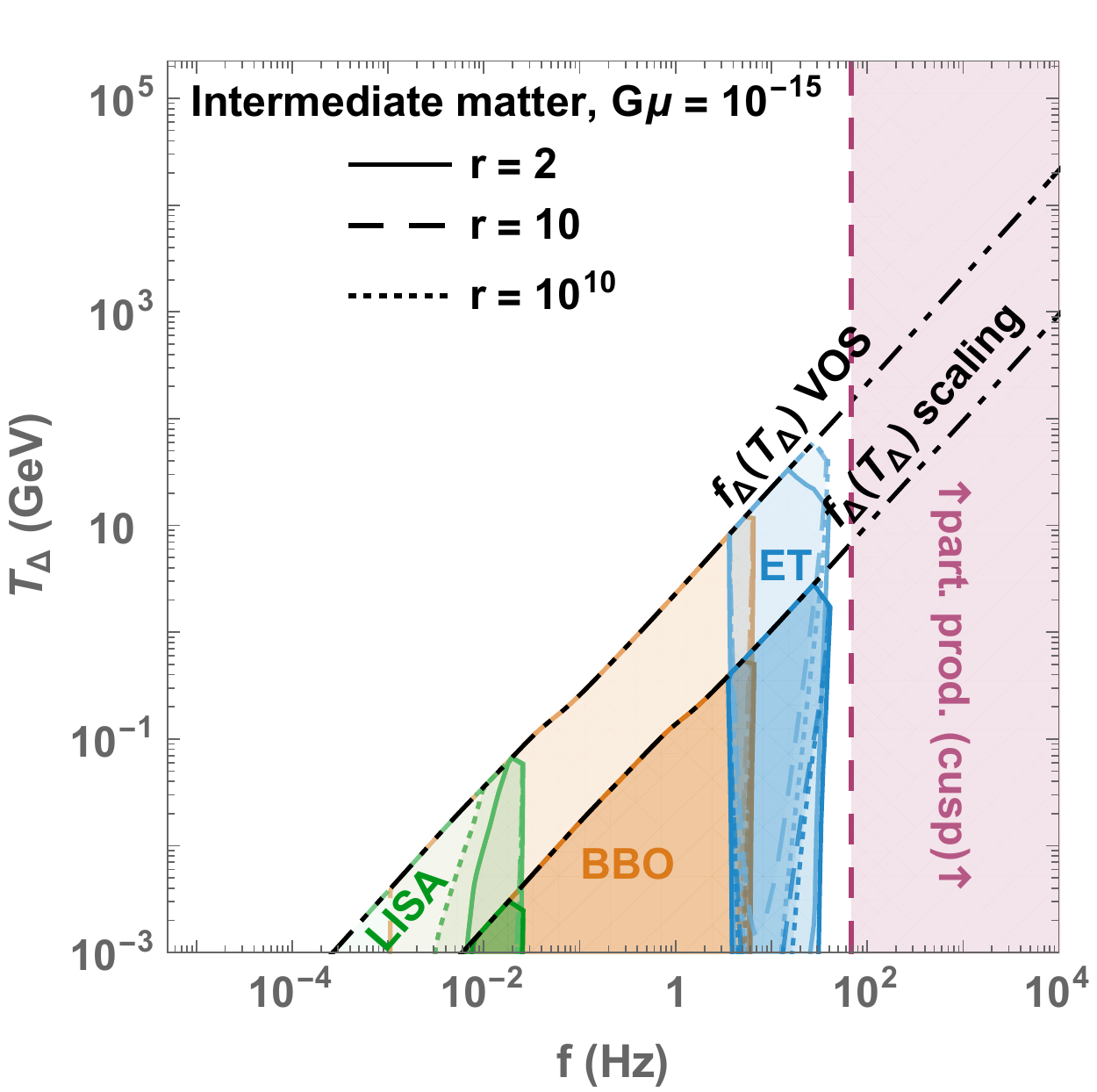}}}
			\hfill
			\caption{\it \small The colored regions show the detectability of the spectral suppression, c.f. \textit{spectral-index prescription (Rx 2)} in Sec.~\ref{sec:Rx1VSRx2}, due to a NS intermediate matter era with duration $r=T_\textrm{start}/T_\Delta$, assuming scaling and VOS networks, c.f. Sec.~\ref{sec:scalingVSvos}. Limitation from particle production, c.f. Sec.~\ref{UVcutoff}, is shown in purple.}
			\label{fig:contour_int}
		\end{figure}

In Fig.~\ref{fig:contour_int}, we show the constraints on the presence of an early-intermediate-non-standard-matter-dominated era starting at the temperature $r\,T_{\Delta}$ and ending at the temperature $T_{\Delta}$.
Matter eras as short as $r=2$ and ending at temperature as large as $100$~TeV could be probed by GW interferometers. We assume that an early-matter era is detectable if the spectral index is smaller than $-0.15$, c.f. \textit{spectral-index prescription (Rx 2)} in Sec.~\ref{sec:detectability}.
In a companion paper \cite{Gouttenoire:2019rtn}, we provide model-independent constraints on the abundance and lifetime of an unstable particles giving rise to such a non-standard intermediate matter era.

\section{Intermediate inflationary era}
\label{sec:inflation}

\subsection{The non-standard scenario}
Next, we consider the existence of a short inflationary period with a number of e-folds
\begin{align}
N_e \equiv \log\left(\frac{a_\textrm{start}}{a_\textrm{end}}\right),
\end{align}
smaller than $N_e \lesssim 20 \ll 60$, in order not to alter the predictions from the first inflation era regarding the CMB power spectrum. On the particle physics side, such a short inflationary period can be generated by a highly supercooled first-order phase transition.
It was stressed that nearly-conformal scalar potentials naturally lead to such short, with $N_e \sim 1-15$,  periods of inflation \cite{Konstandin:2011dr,vonHarling:2017yew,Bruggisser:2018mrt}. Those are well-motivated in new strongly interacting composite sectors arising at the TeV scale, as invoked to address the Higgs hierarchy problem and were first studied in a holographic approach \cite{Creminelli:2001th, Randall:2006py} (see also the review \cite{Caprini:2019egz}). As the results on the scaling of the bounce action for tunnelling and on the dynamics  of the phase transitions do essentially not depend on the absolute energy scale, but only on the shallow shape of the scalar potential describing the phase transition, those studies can thus be extended to a large class of confining phase transitions arising at any scale. In this section, we will take this inflationary scale as a free parameter.

\begin{figure}[h!]
\centering
\raisebox{0cm}{\makebox{\includegraphics[width=0.7\textwidth, scale=1]{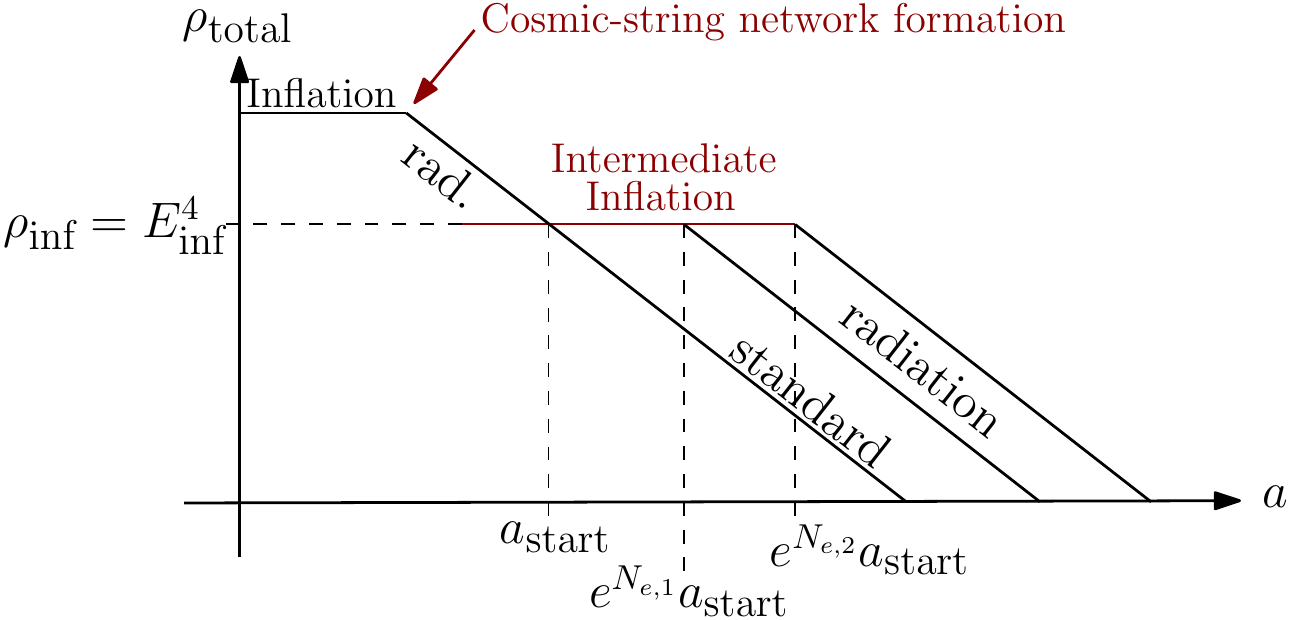}}}
\caption{\it\small Evolution of the total energy density assuming  the presence of an intermediate inflationary era characterised by the energy density $\rho_\textrm{inf}$, for two different durations (number of efolds), $N_{e,1}$ and $N_{e,2}$.}
\label{figure_inter_inflation}
\end{figure}

 We define the energy density profile as, c.f. Fig.~\ref{figure_inter_inflation}
\begin{equation}
\rho_\textrm{tot}(a)=\begin{cases}
\rho^\textrm{st}_\textrm{rad}(a)+\rho_\textrm{late}(a)&\textrm{for }\rho>\rho_\textrm{inf},\\[0.5em]
\rho_\textrm{inf}=E^4_\textrm{inf}\hspace{2em}&\textrm{for }\rho=\rho_\textrm{inf},\\[0.5em]
\rho_\textrm{inf}\Delta_R(T_\textrm{end},T)\over{a_\textrm{end}}{a}^4+\rho_\textrm{late}(a)\hspace{2em}&\textrm{for }\rho<\rho_\textrm{inf},
\end{cases}
\label{inf_define_function}
\end{equation}
where $\rho_\textrm{inf}$ is the total energy density of the universe during inflation and $E_\textrm{inf}\equiv \rho_\textrm{inf}^{1/4}$ is the corresponding energy scale.  The function $\Delta_R$ is defined in (\ref{eq:DeltaR}).
%

%
\begin{figure}[h!]
\centering
\raisebox{0cm}{\makebox{\includegraphics[width=0.9\textwidth, scale=1]{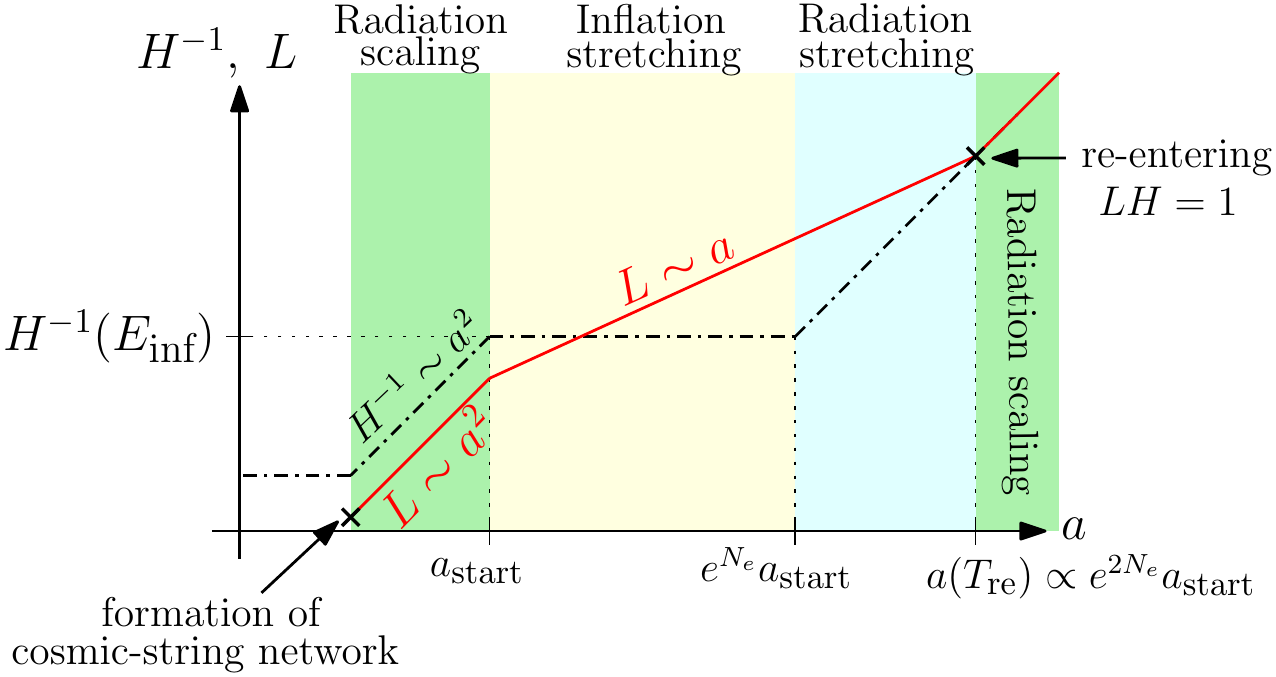}}}
\caption{\it\small After its formation, before inflation, the network enters the scaling regime with $L \sim a^2$ due to loop formation. During the $N_e$ e-folds of inflation, the network correlation length gets stretched out of the horizon by the rapid expansion and loop formation stops, thus $L\sim a$. After inflation, during radiation, the correlation length starts to re-enter the horizon and scales again as $L \sim a^2$.
}
\label{fig:stringsandinflation}
\end{figure}

\begin{figure}[h!]
\centering
\raisebox{0cm}{\makebox{\includegraphics[width=0.7\textwidth, scale=1]{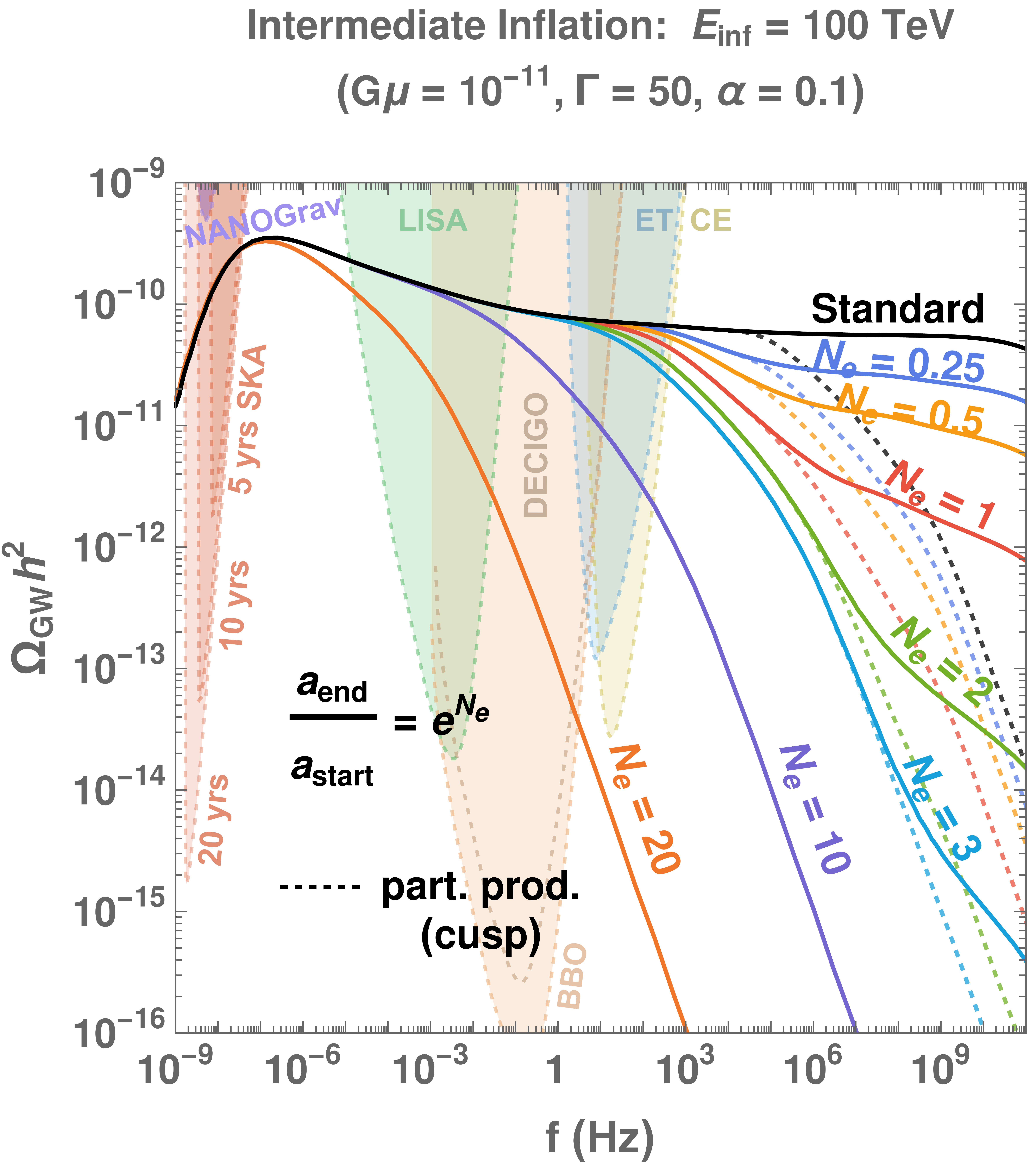}}}
\raisebox{0.65cm}{\makebox{\includegraphics[width=0.49\textwidth, scale=1]{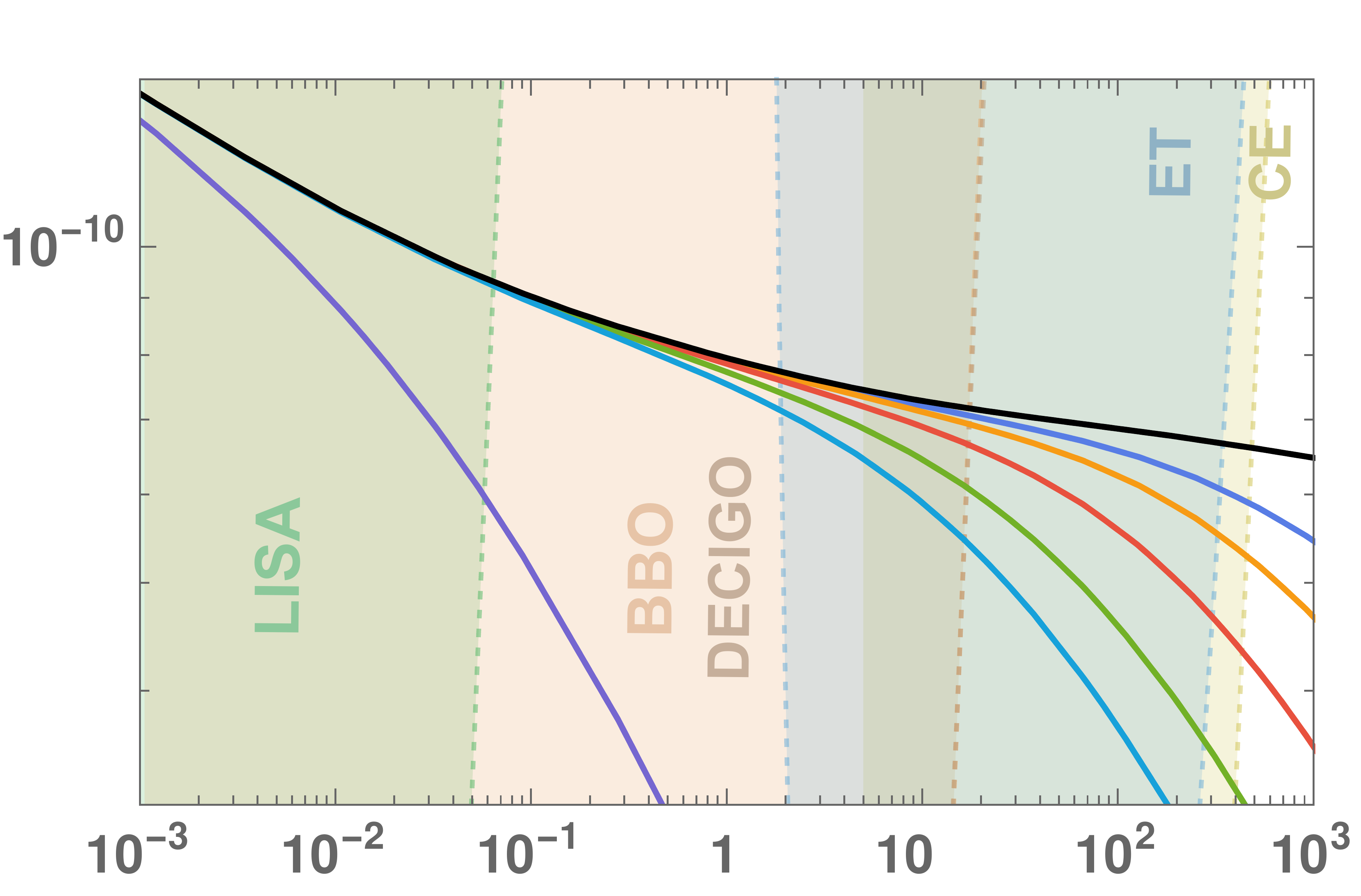}}}
\raisebox{0cm}{\makebox{\includegraphics[width=0.49\textwidth, scale=1]{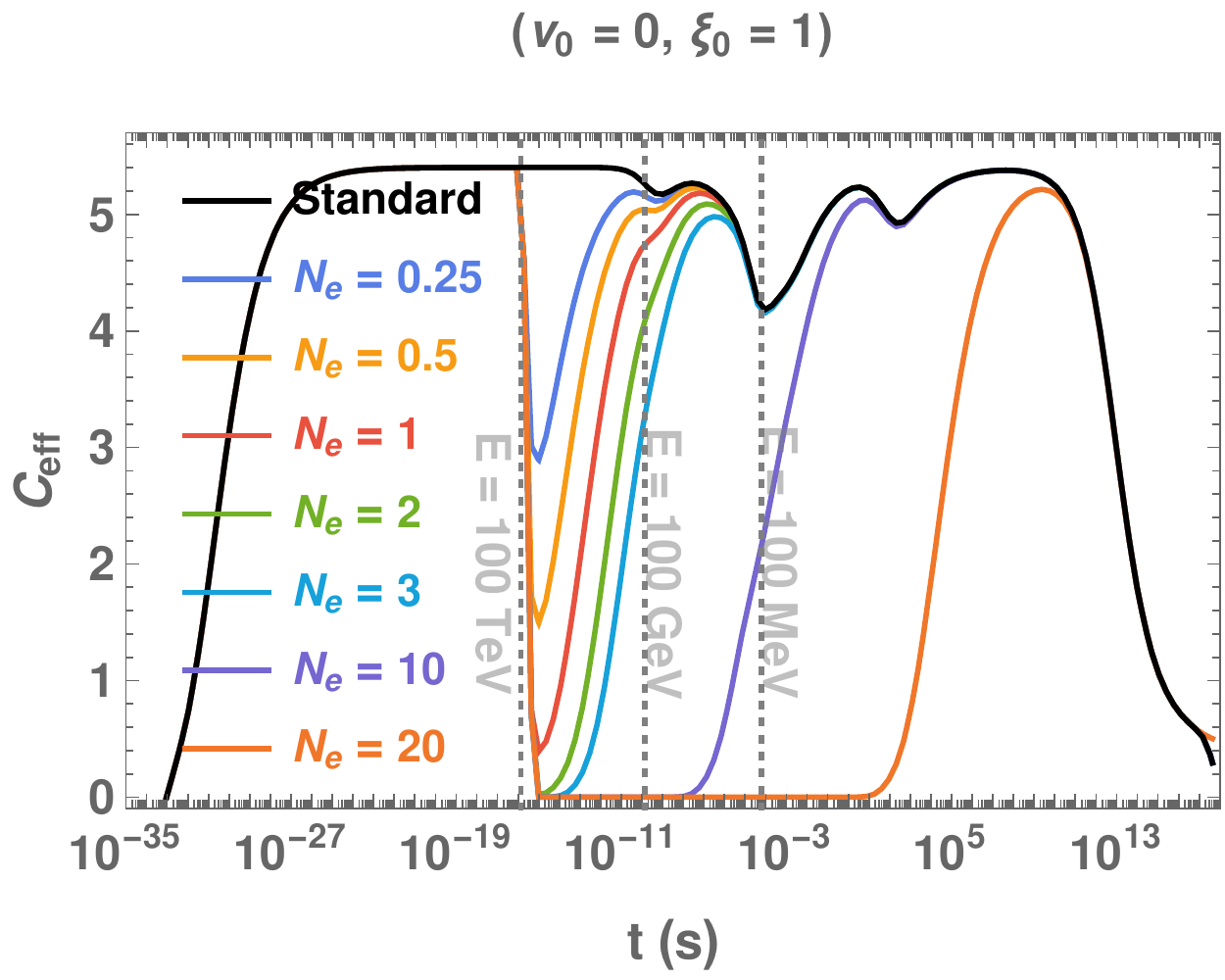}}}
\caption{\it \small \textbf{Top:} GW spectra from cosmic strings assuming either the scaling or the VOS network, evolved in the presence of a non-standard intermediate inflation era. Inflation directly affects the VOS parameters by stretching the strings beyond the horizon.  The transition between the $f^{-1/3}$ scaling after the turning point, to the $f^{-1}$ scaling at even larger frequencies, is an artefact due to total number of modes $k$ being fixed to $2\times 10^4$, see Fig.~\ref{fig:beautifulpeaks} for an extrapolation of the $f^{-1/3}$ behavior to arbitrary large frequencies and App.~\ref{sec:study_impact_mode_nbr} for more details. \textbf{Bottom:} The loop-production is suppressed and only becomes significant again when the correlation length re-enters the horizon. Limitations due to particle production, c.f. Sec.~\ref{UVcutoff}, are shown with dotted lines.}
\label{figure_spect_inflation}
\end{figure}

\subsection{The stretching regime and its impact on the spectrum}
\label{sec:turning_point_inf}
Fig.~\ref{figure_spect_inflation} shows how the fast expansion during inflation suppresses the GW spectrum for frequencies above a turning-point frequency $f_\Delta$ which depends on the number of e-folds. The larger the number of e-folds, the lower $f_\Delta$.  Indeed, during inflation, the loop-production efficiency $C_{\rm eff} \propto \xi^{-3}$ is severely suppressed, c.f. Fig.~\ref{figure_spect_inflation}, by the stretching of the correlation length $\xi$ beyond the Hubble horizon, and loop production freezes \cite{Guedes:2018afo}. 
After the end of inflation, one must wait for the correlation length to re-enter the horizon in order to reach the scaling regime again. The duration of the transient regime receives an enhancement factor $\exp{N_e}$. As a result, the turning-point frequency $f_\Delta$ receives a suppression factor $\exp{N_e}$ as derived below:
\begin{equation}
f_\Delta=(1.5\times10^{-4}\textrm{ Hz})\left(\frac{T_\textrm{re}}{\textrm{GeV}}\right)\left(\frac{0.1\times 50 \times 10^{-11}}{ \alpha \,\Gamma G\mu}\right)^{1/2}\left(\frac{g_*(T_\textrm{re})}{g_*(T_0)}\right)^{1/4},
\label{turning_point_inf}
\end{equation}
with $T_\textrm{re}$ the temperature at which the long-string network re-enters the Hubble horizon
\begin{equation}
T_\textrm{re}\simeq \frac{E_\textrm{inf}}{(0.1)\,g_*^{1/4}(T_\textrm{re})\,\exp(N_e)},
\label{eq:Tre-enter}
\end{equation}
where $(0.1)$ is the typical correlation length before the stretching starts. Note that the numerical factor in Eq.~\eqref{turning_point_inf} comes from the demanded precision of 10\% deviation, c.f. Eq.~\eqref{10per_criterion}. It can be lower by a factor $\sim 300$  if the 1\% precision is applied, as shown in Eq.~\eqref{turning_point_general_scaling_app_inf}.

\begin{figure}[h!]
\centering
\raisebox{0cm}{\makebox{\includegraphics[width=0.7\textwidth, scale=1]{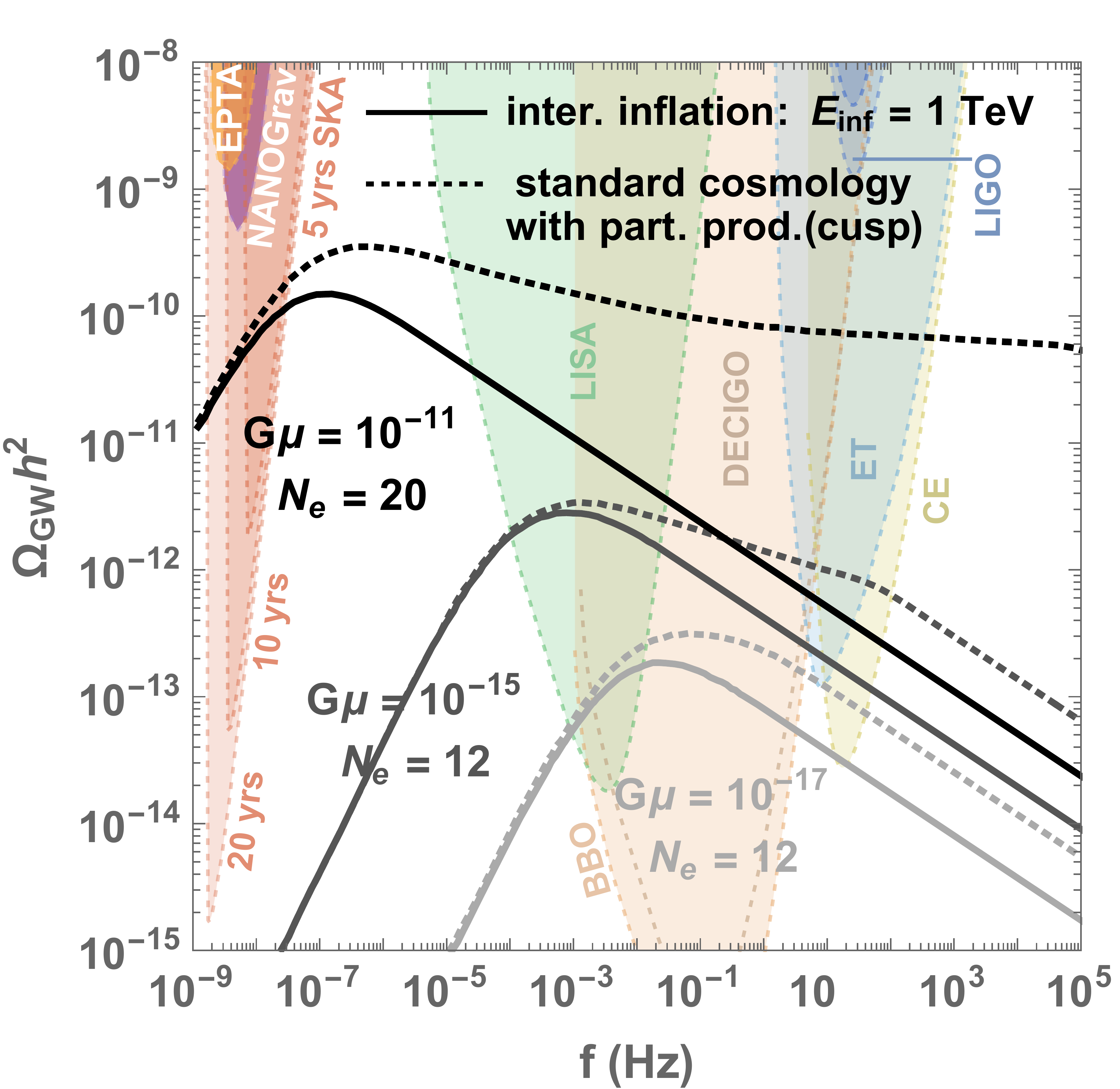}}}
\caption{\it\small In the case an intermediate inflationary era lasting for ${\cal O}(10)$ efolds, the SGWB from cosmic strings completely looses its scale invariant shape and has instead a peak structure. A TeV scale inflation era can lead to broad peaks either in the LISA or BBO band or even close to the SKA band, depending on the value of the string tensions $G\mu$, and the number of efolds $N_e$. At low $G\mu \lesssim 10^{-17},$ the spectrum manifests a peak structure even in standard cosmology because of the emission of massive particles at large frequencies, c.f. Sec.~\ref{sec:massive_radiation}. We introduce other realisations presenting a peak structure in Sec.~\ref{sec:peakspectrum}.  Here we extrapolate the $f^{-1/3}$ behavior to arbitrary large frequencies, which is equivalent to sum over an infinite number of proper modes $k$, see App.~\ref{sec:study_impact_mode_nbr}.  }
\label{fig:beautifulpeaks}
\end{figure}

Fig.~\ref{fig:beautifulpeaks} shows how a sufficiently long period of intermediate inflation can lead to SGWB with peak shapes in the future GW interferometer bands.
We emphasize that the change of the GW spectrum from CS in presence of a non-standard matter-dominated era, a short inflation, and particle production look similar. Therefore, the question of how disentangling each effect from one another deserves further studies.

Interestingly, in contrast with the SGWB which is dramatically impacted by an intermediate period of inflation, the short-lasting GW burst signals \cite{Damour:2000wa, Damour:2001bk, Siemens:2006yp, Olmez:2010bi, Ringeval:2017eww} remain preserved if the correlation length re-enters the horizon at a redshift higher than $\sim 5 \times 10^4$ \cite{Cui:2019kkd}. Indeed, the bursts being generated by the small scale structures, they have higher frequencies and then are emitted later than the SGWB, c.f. Fig.~2 in \cite{Ringeval:2017eww}.


\paragraph{Derivation of the turning-point formula (inflation case):}
		Let us review the chronology of the network in the presence of an intermediate-inflation period (see figure~\ref{fig:stringsandinflation}) in order to derive Eq.~\eqref{turning_point_inf}.
		In the early radiation era, the network has already been produced and reached the scaling regime before inflation starts. 
		The correlation length scale is of order $(0.1)t$ or equivalently
		\begin{align}
		L_\textrm{start} H_\textrm{start} \sim \mathcal{O}(0.1),
		\end{align}
		where $L$ is the correlation length of strings, and $H$ is the Hubble rate.
		When inflation begins, it stretches cosmic strings beyond the horizon with
		\begin{align}
		L\propto a \textrm{\hspace{1em} leading to \hspace{1em}} LH\gg1,
		\end{align} 
		within a few e-folds. 
		Later, the late-time energy density takes over inflation, but the network is still in the stretching regime $L\propto a$, i.e.
		\begin{align}
		\label{eq:network_radiation_length}
		LH\propto t^{(2-n)/n}\textrm{\hspace{1em} during the era with }\rho\propto a^{n}.
		\end{align} 		
		For $n>2$, the Hubble horizon will eventually catch up with the string length, allowing them to re-enter, and initiate the loop production.
		We consider the case where the universe is radiation-dominated after the inflation period and define the temperature $T_\textrm{re} $ of the universe when the long-string correlation length $L$ re-enters the horizon
		\begin{align}
		L_\textrm{re} H_\textrm{re} = 1,
		\end{align}
		where $L_\textrm{re}$ and $H_\textrm{re}$ are the correlation length and Hubble rate at the re-entering time. We can use Eq.~\eqref{eq:network_radiation_length} to evolve the correlation length, starting from the start of inflation up to the re-entering time
		\begin{align}
		1=L_\textrm{re} H_\textrm{re}&=\left(\frac{t_\textrm{re}}{t_\textrm{end}}\right)^{-1/2}L_\textrm{end}H_\textrm{end},\\
		&=\left(\frac{t_\textrm{re}}{t_\textrm{end}}\right)^{-1/2}\left(\frac{a_\textrm{end}}{a_\textrm{start}}\right)L_\textrm{start}H_\textrm{start},\\
		&\simeq \left(\frac{T_\textrm{re}}{T_\textrm{end}}\right)e^{N_e}(0.1)
		\end{align}
		We have used $t\propto T^{-2}$ during the radiation era and introduced the number $N_e$ of inflation e-folds.
		Finally, we obtain the re-entering temperature in terms of the number of e-folds $N_e$ and the inflationary energy scale $E_\textrm{inf}$ as
		\begin{align}
		\label{eq:Tre}
		T_\textrm{re}\simeq \frac{E_\textrm{inf}}{(0.1)\,g_*^{1/4}(T_\textrm{re})\,\exp(N_e)}.
		\end{align}
After plugging Eq.~\eqref{eq:Tre} into the VOS turning-point formula Eq.~\eqref{turning_point_general}, with $T_{\Delta}=T_{\rm re}$, and adjusting the numerical factor with the GW spectrum computed numerically, we obtain the relation in Eq.~\eqref{turning_point_inf} between the turning-point frequency and the inflation parameters $N_e$ and $E_{\rm inf}$.

\begin{figure}[h!]
			\centering
			\raisebox{0cm}{\makebox{\includegraphics[height=0.47\textwidth, scale=1]{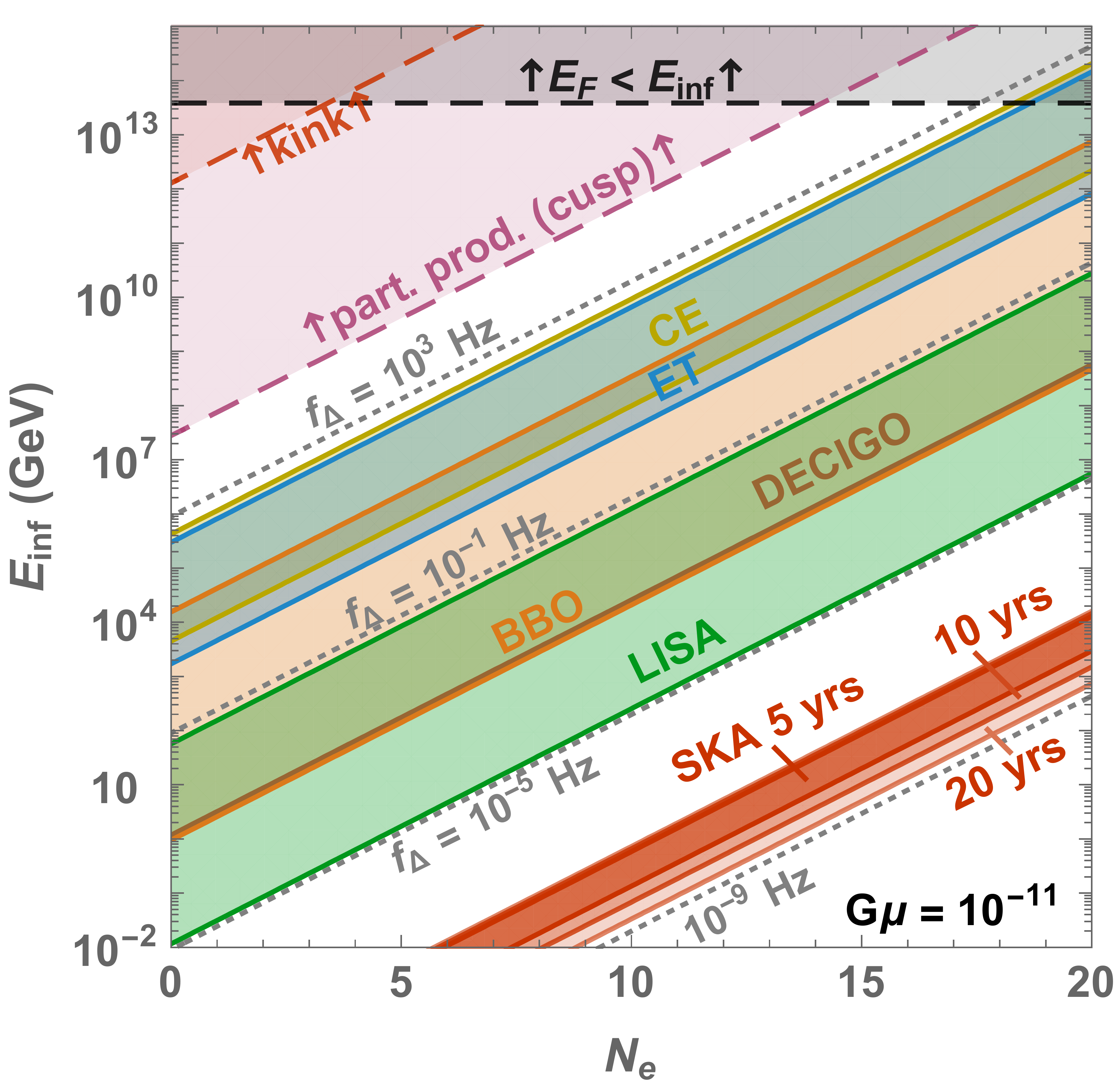}}}
			\raisebox{0cm}{\makebox{\includegraphics[height=0.47\textwidth, scale=1]{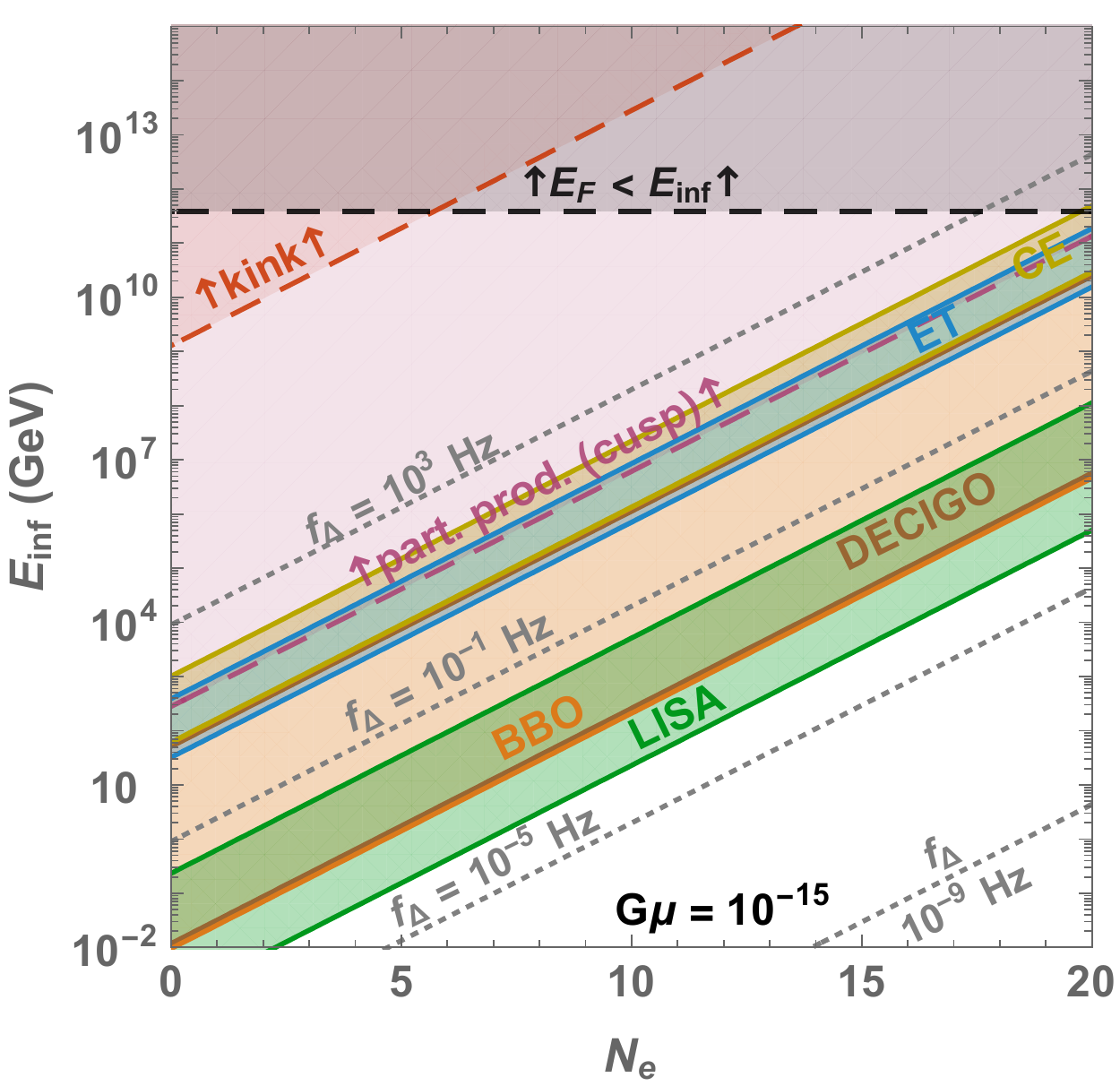}}}
			\hfill
			\caption{\it \small  Reach of future GW interferometers for probing an intermediate-inflation period with an energy scale $E_{\rm inf}$, lasting $N_e$ efolds. Colored regions correspond to the turning-points with amplitude higher than each power-law-sensitivity curve, c.f. \textit{turning-point prescription (Rx 1)} in Sec.~\ref{sec:Rx1VSRx2}. Gray dotted lines are turning-points, c.f. Eq.~\eqref{turning_point_inf}, for given frequencies. Red and purple dashed lines are limitations from particle production, c.f. Sec.~\ref{UVcutoff}.}
			\label{fig:contour_power_inflation1}
		\end{figure}
\begin{figure}[h!]
			\centering
			\raisebox{0cm}{\makebox{\includegraphics[height=0.495\textwidth, scale=1]{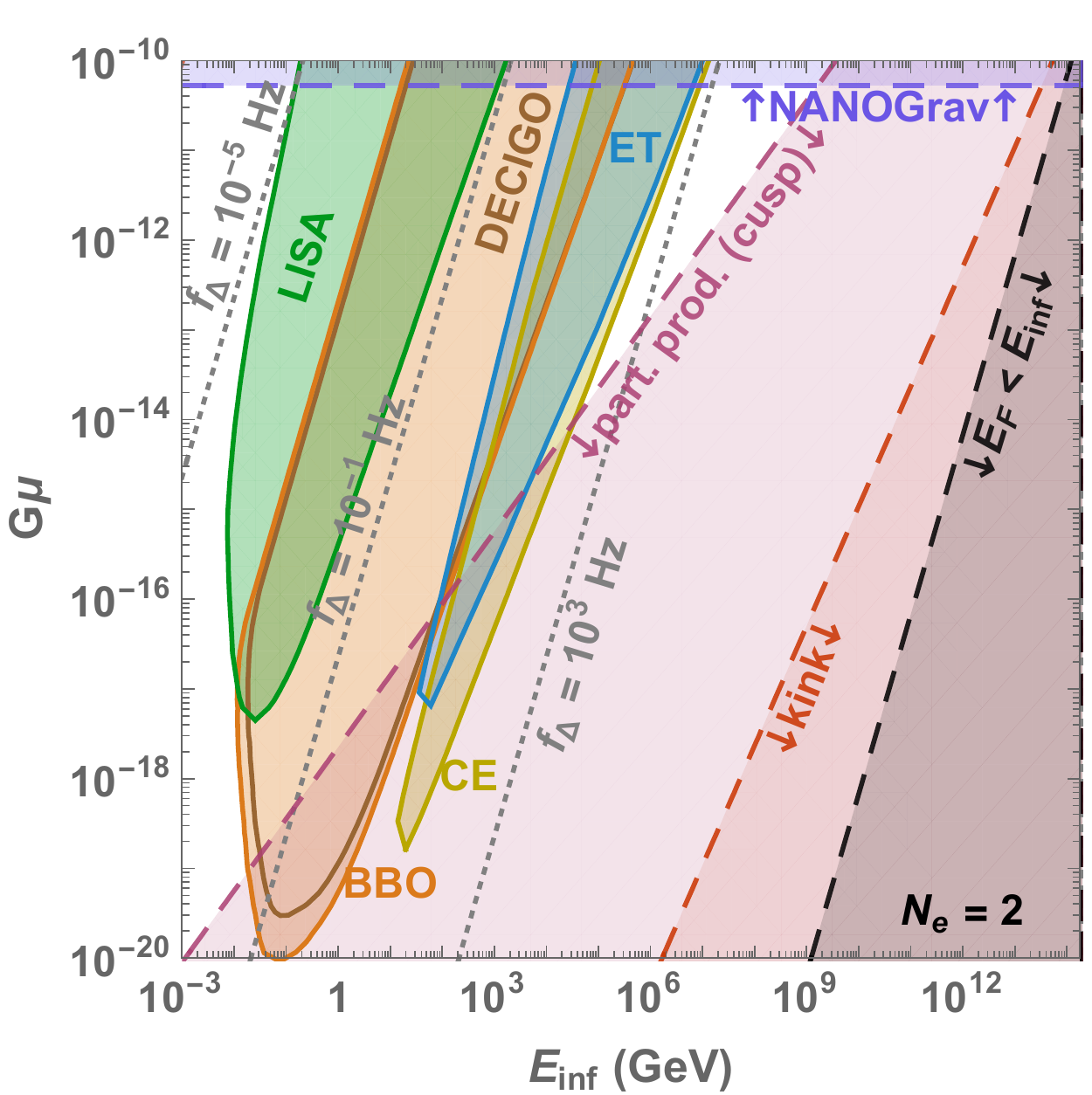}}}
			\raisebox{0cm}{\makebox{\includegraphics[height=0.495\textwidth, scale=1]{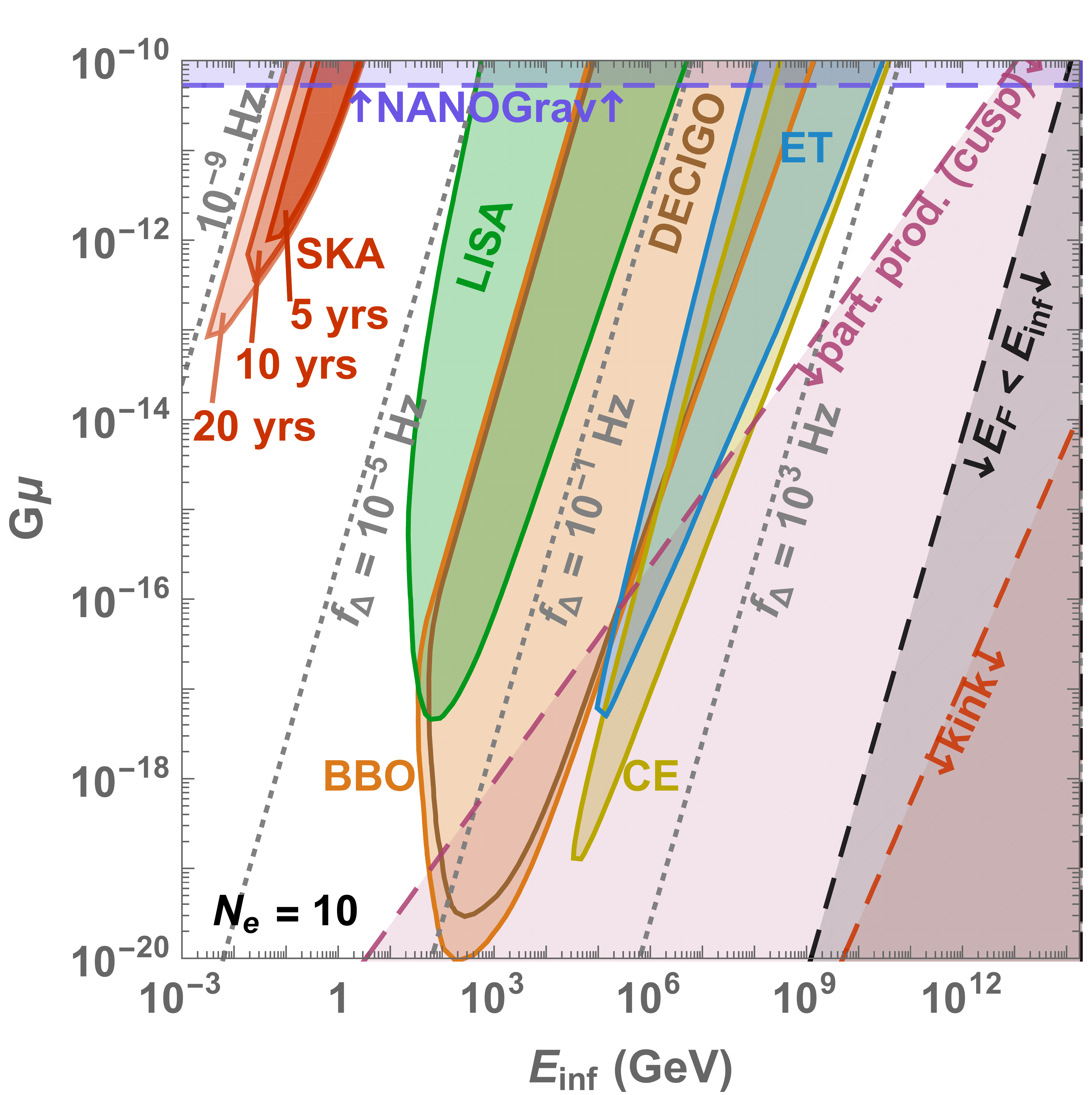}}}\\
			\raisebox{0cm}{\makebox{\includegraphics[height=0.495\textwidth, scale=1]{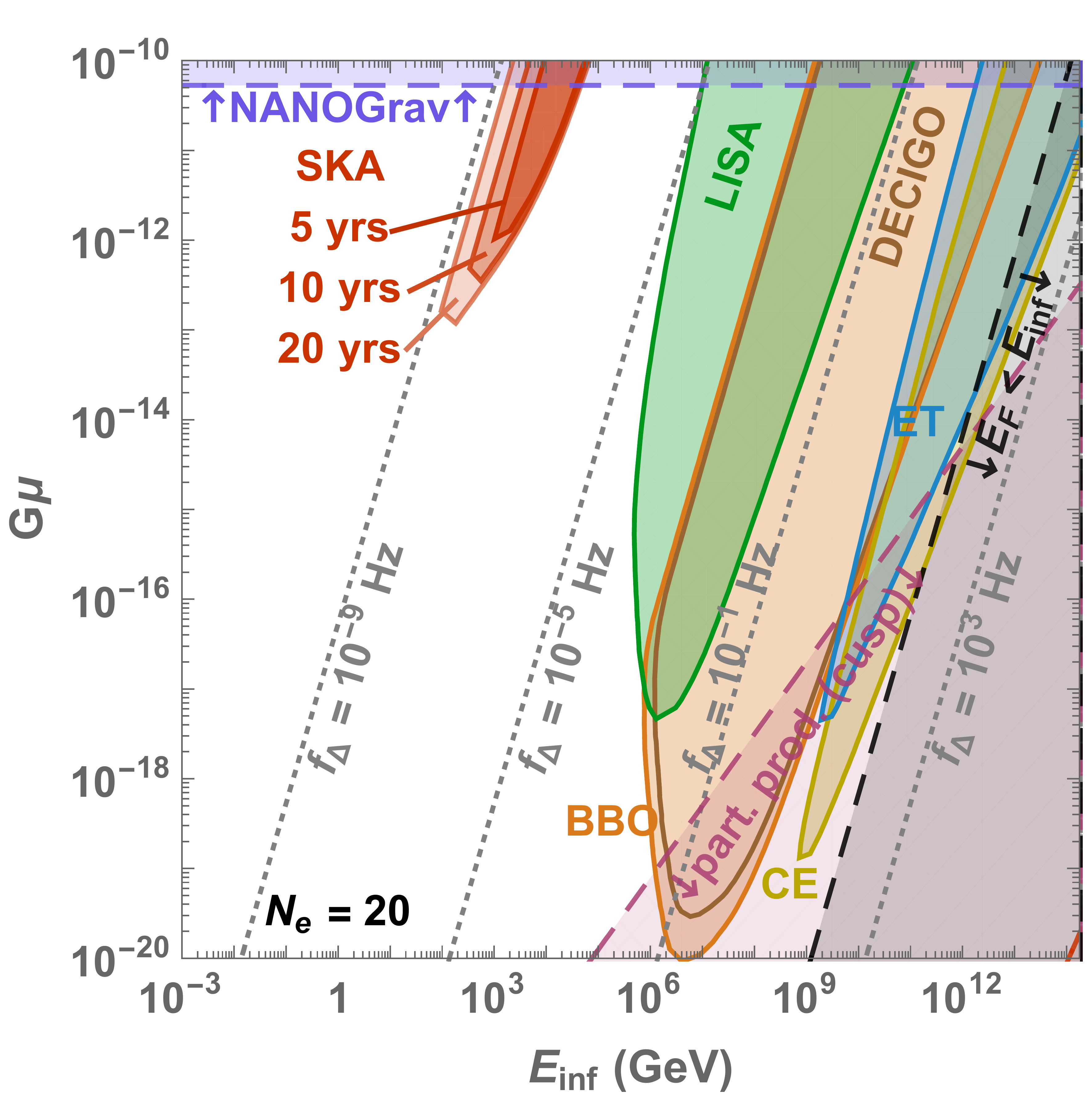}}}
			\hfill
			\caption{\it \small Constraints on intermediate inflation from CS detection by future GW observatories. The longer the intermediate inflation, the later the correlation length re-enters the horizon, the more shifted to lower frequencies the turning-point and the larger the inflation scale which we can probe.   Colored regions correspond to the turning-points with amplitude higher than each power-law-sensitivity curve, c.f. \textit{turning-point prescription (Rx 1)} in Sec.~\ref{sec:Rx1VSRx2}. The bound $E_F<E_\textrm{inf}$, where $E_F\sim m_{pl}\sqrt{G\mu}$ is the network-formation energy scale, guarantees that the CS network forms before the intermediate-inflation starts. Red and purple dashed lines are limitations from particle production, c.f. Sec.~\ref{UVcutoff}.}
			\label{fig:contour_power_inflation2}
		\end{figure}
%
\begin{figure}[h!]
			\centering
			\raisebox{0cm}{\makebox{\includegraphics[height=0.425\textwidth, scale=1]{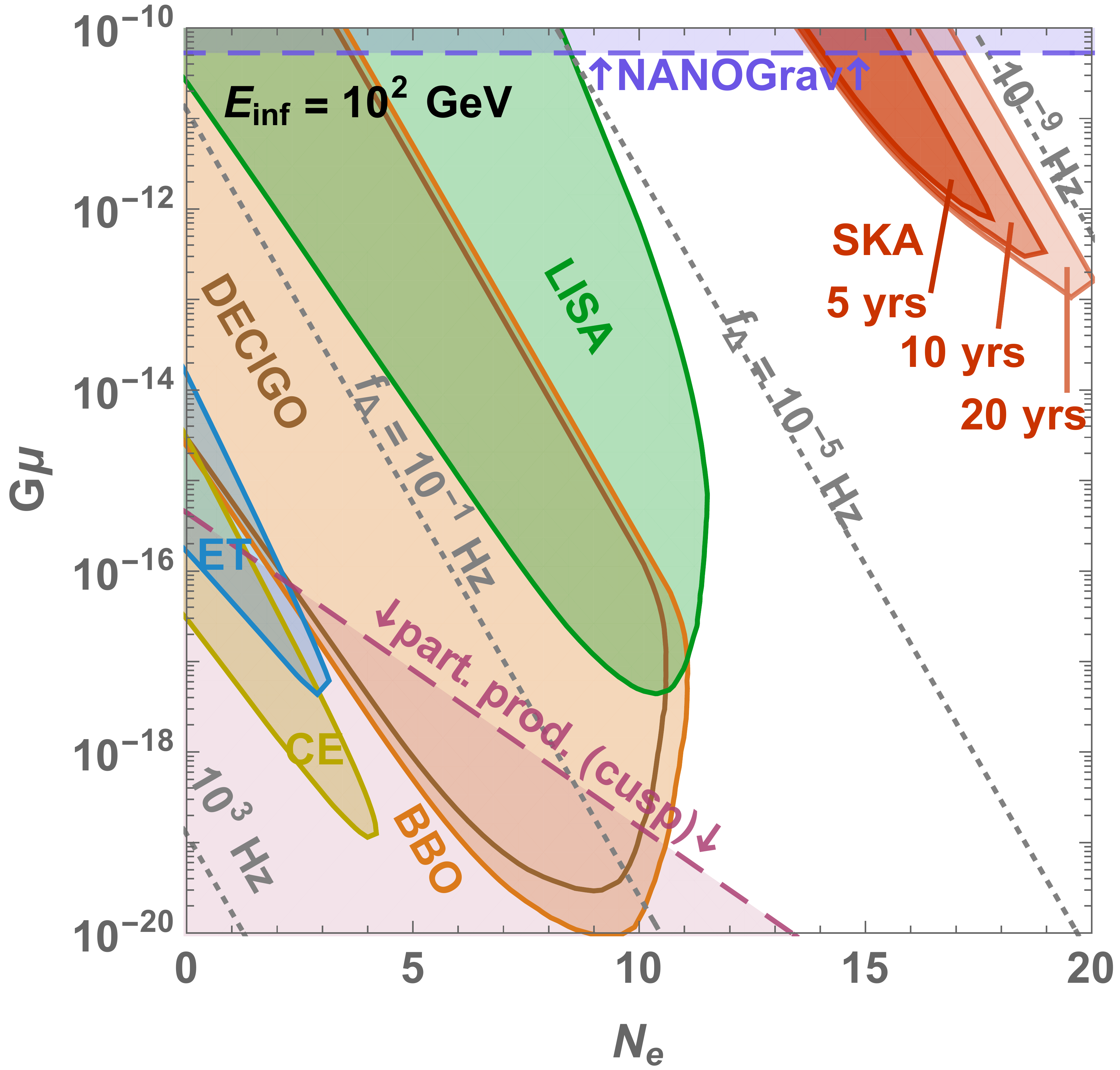}}}
			\raisebox{0cm}{\makebox{\includegraphics[height=0.425\textwidth, scale=1]{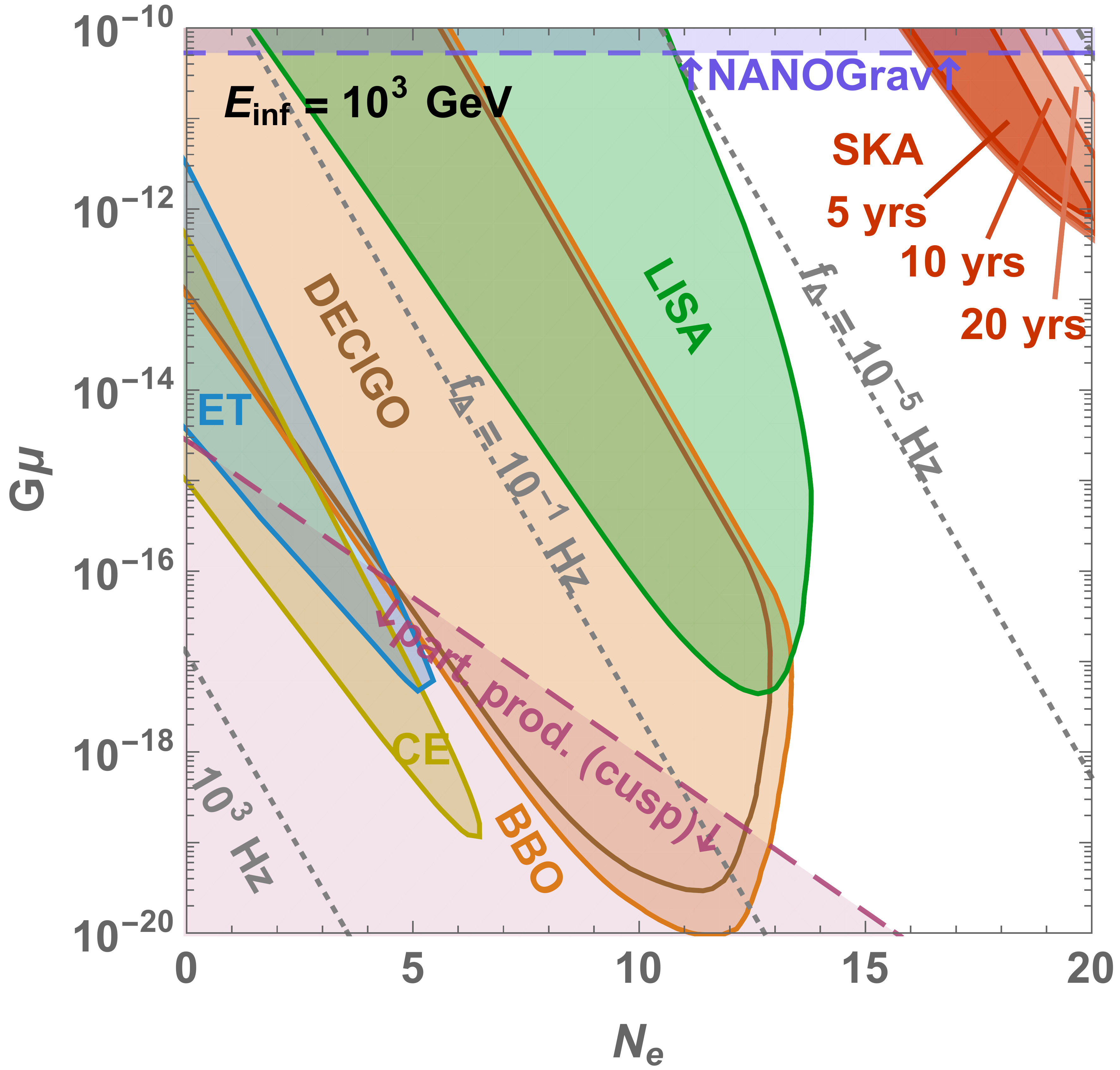}}}\\
			\raisebox{0cm}{\makebox{\includegraphics[height=0.425\textwidth, scale=1]{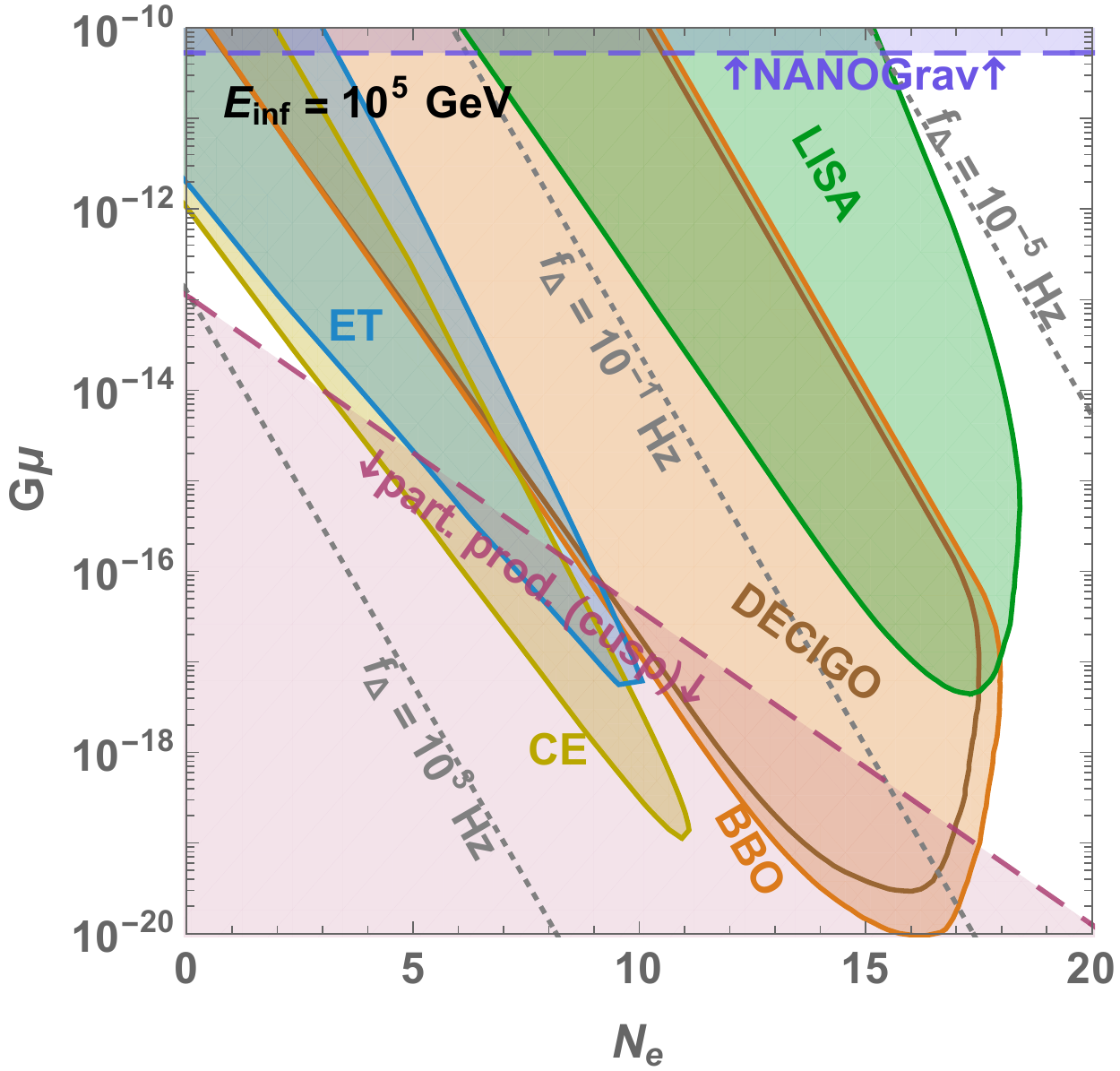}}}
			\raisebox{0cm}{\makebox{\includegraphics[height=0.425\textwidth, scale=1]{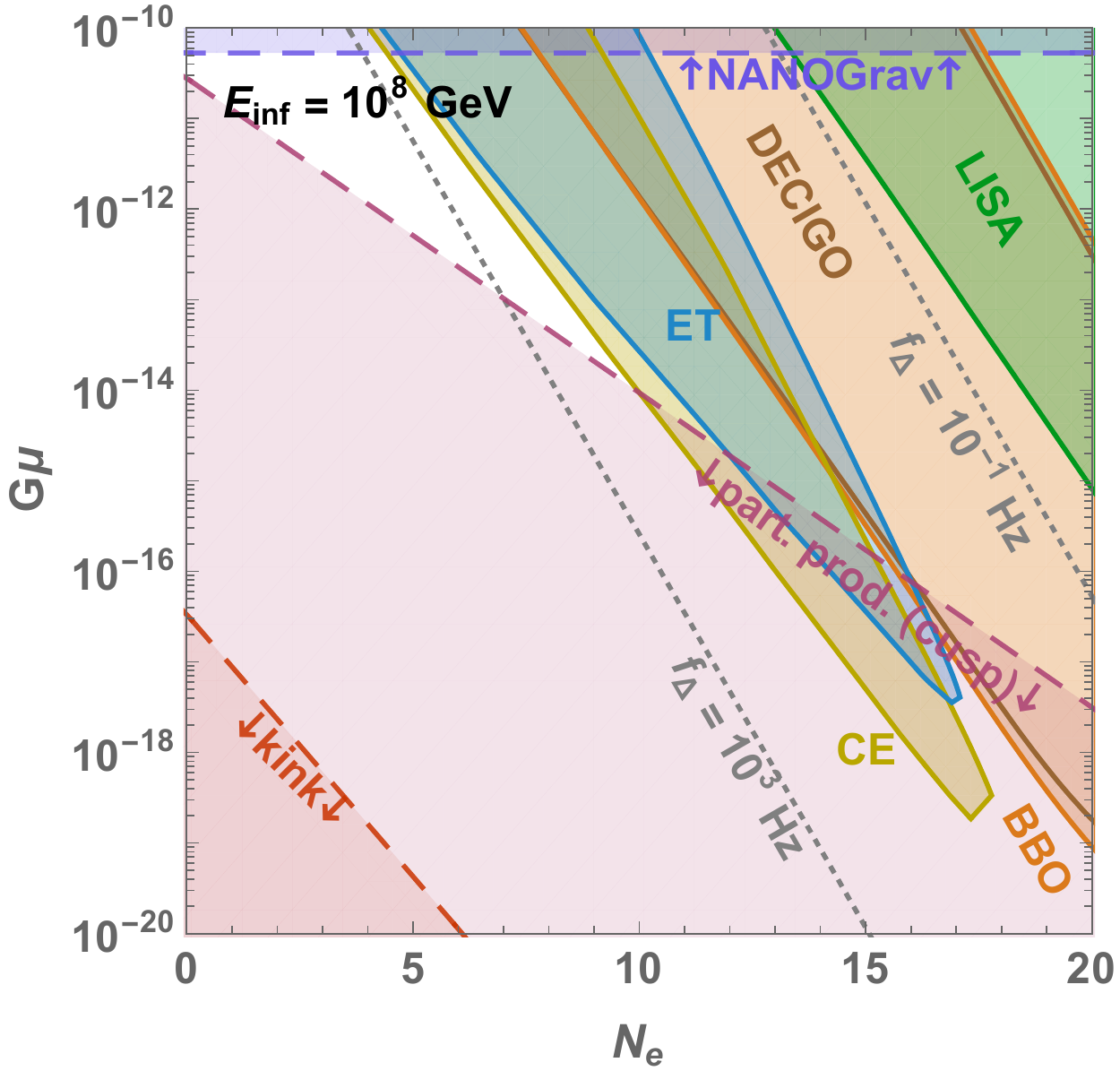}}}\\
			\raisebox{0cm}{\makebox{\includegraphics[height=0.425\textwidth, scale=1]{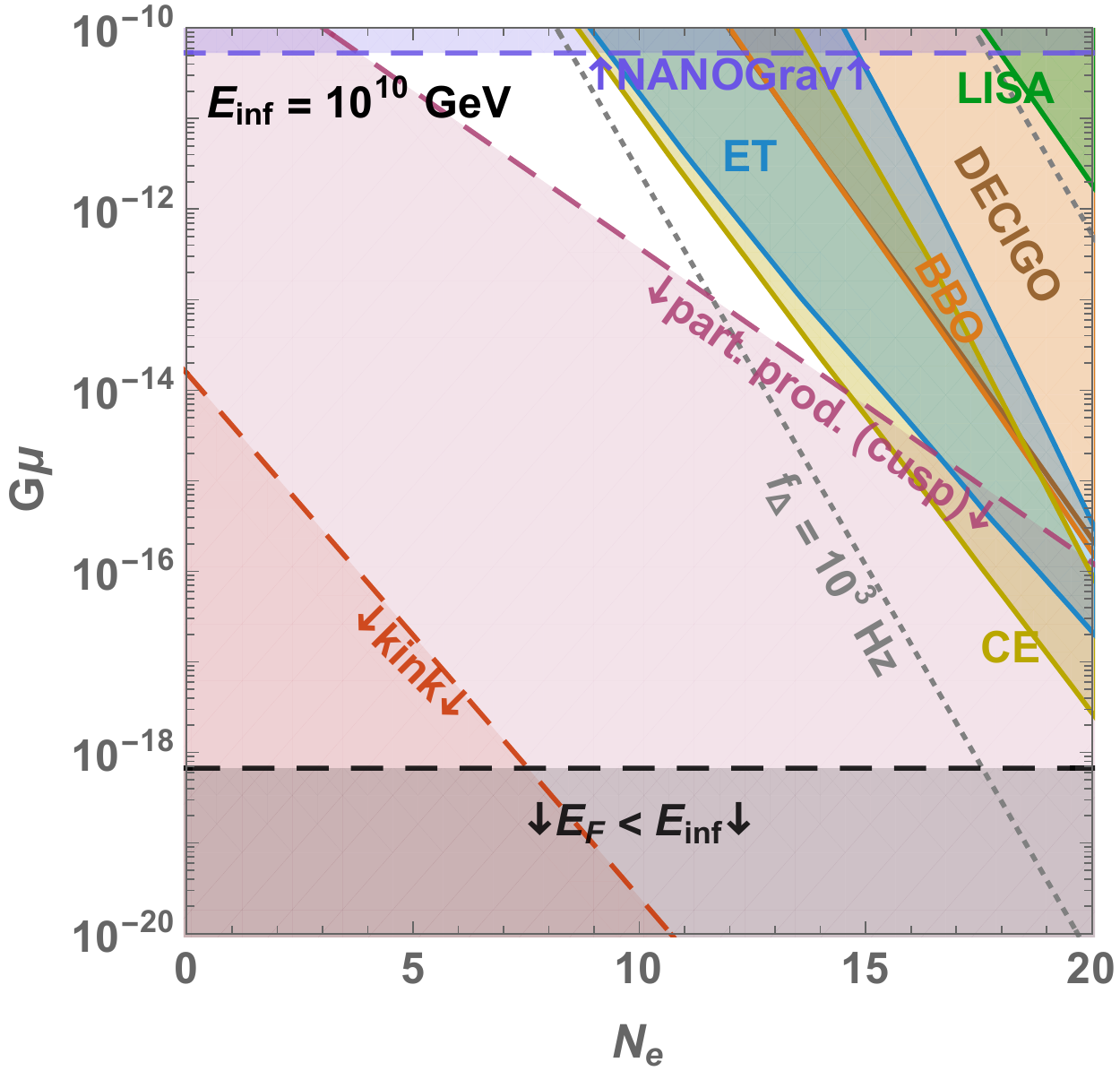}}}\\
			\hfill
		\caption{\it \small Prospect constraints on intermediate inflation if a GW interferometer detects a SGWB from CS with tension $G\mu$. The freezing of the long-string network due to the stretching of the correlation length outside the horizon allows to probe large inflationary scale $E_{\rm inf}$ for large number of efolds $N_e$.   Colored regions correspond to the turning-points with amplitude higher than each power-law-sensitivity curve, c.f. \textit{turning-point prescription (Rx 1)} in Sec.~\ref{sec:Rx1VSRx2}.  Red and purple dashed lines are limitations from particle production, c.f. Sec.~\ref{UVcutoff}.}
			\label{fig:contour_power_inflation3}
		\end{figure}
\FloatBarrier

\subsection{Constraints}

In Figs.~\ref{fig:contour_power_inflation1}, \ref{fig:contour_power_inflation2} and \ref{fig:contour_power_inflation3}, we show the constraints on an intermediate short inflation period in the planes $E_\textrm{inf}-N_e$, $G\mu-E_\textrm{inf}$, and $G\mu-N_e$, respectively. We follow the \textit{turning-point prescription (Rx 1)} defined in Sec.~\ref{sec:detectability}, which constrains a non-standard cosmology by using the detectability of the turning-point frequency defined by Eq.~\eqref{turning_point_inf}. 
The longer the intermediate inflation, the later the correlation length re-enters the horizon, the latter the long-string network goes back to the scaling regime, the lower the frequency of the turning-point and the larger the inflationary scale which can be probed.
The detection of a GW spectrum generated by CS by future GW observatories would allow to probe an inflationary energy scale $E_\textrm{inf}$ between $10^{-2}$~GeV and $10^{13}$~GeV assuming a number of e-folds $N_e \lesssim 20$.

\section{Peaked spectrum}
\label{sec:peakspectrum}

In this  section, we point out the possibility for the GW spectrum to exhibit a peak structure whenever a high and a low frequency cut-offs are close to each other.
As already discussed in the previous sections, a high-frequency cut-off can arise 
\begin{itemize}
\item [$\diamond$]
either in standard cosmology, mainly from particle production beyond the Nambu-Goto approximation, c.f. blue dotted line in Fig.~\ref{fig:peak_spectrum}, whose corresponding cut-off frequency is computed in Eq.~\eqref{UVcutoff_f_app}, but also possibly from friction or network formation, c.f. Fig.~\ref{sketch_scaling},
\item[$\diamond$]
or in non-standard cosmology, in the presence of an intermediate matter era, c.f. Sec.~\ref{sec:interm_matter}, or inflation era, c.f. Sec.~\ref{sec:inflation}. The associated cut-off frequencies are computed in Eq.~\eqref{turning_point_general} and Eq.~\eqref{eq:Tre}, respectively.
\end{itemize}
 Beyond the high-frequency cut-off, the GW spectrum is suppressed with a slope $f^{-1/3}$. In App.~\ref{sec:study_impact_mode_nbr}, we show that the $f^{-1/3}$ behavior, instead of $f^{-1}$ as expected from the $(k=1)$-spectrum, is due to the presence of the high-k modes.

Conversely, there are also phenomena which suppress the GW spectrum at \textit{low frequency}.
\begin{itemize}
\item[$\diamond$]
Simply because such low frequencies have not been emitted yet, c.f. blue line in Fig.~\ref{fig:peak_spectrum} below $0.1~$Hz, or Fig.~\ref{sketch_scaling} below $f\lesssim 10^{-7}~$Hz. The corresponding low-frequency cut-off is computed in Sec.~\ref{ref:low_cut_off_stable}.
\item[$\diamond$]
Or  the string network is metastable, as discussed in Sec.~\ref{ref:low_cut_off_metastable}
\end{itemize}
In Sec.~\ref{ref:GW_spectrum_peak_structure}, we show a variety of peak spectra, and compare them to peak spectra generated by bubble collision in first-order phase transitions.

\subsection{Low-energy cut-off of stable string network}
\label{ref:low_cut_off_stable}
The lowest frequency observed today is set by the inverse size of the main population of loops decaying today. This leads to the distinguished maximum of the standard GW spectrum around $0.1~$Hz of the blue line in Fig.~\ref{fig:peak_spectrum} or around $10^{-7}~$Hz in Fig.~\ref{sketch_scaling}.
As discussed in App.~\ref{derive_turning_points}, loops contributing to the frequency $f$ dominantly decay at $\tilde{t}_M$ defined by
\begin{equation}
\frac{2k}{f}\cdot\frac{a(\tilde{t}_{\rm M})}{a(t_0)}=\Gamma G\mu \times\tilde{t}_{\rm M}.
\end{equation}
Upon setting $\tilde{t}_M=t_0$, we deduce the frequency of the low-energy cut-off of any stable string network
\begin{equation}
f_\textrm{low}^{\rm stable}
= \frac{2}{\Gamma G\mu t_0}
\simeq (1.48\times 10^{-7} \textrm{ Hz})\left(\frac{50\times 10^{-11}}{\Gamma G\mu}\right),
\label{eq:peak_st_cutoff}
\end{equation}
where we have numerically adjusted the coefficient to fit with the GW spectrum.
This formula agrees with EPTA/NANOGrav constraints which bound $G\mu\lesssim 10^{-10}$ for $f\sim10^{-9}-10^{-8}$ Hz.

\subsection{Low-energy cut-off of metastable string network}
\label{ref:low_cut_off_metastable}

So far, we have only been considering a stable CS network.
However, there are mechanisms which can make the network decay, such as breaking into monopole ($M$) antimonopole ($\overline{M}$) pairs \cite{Vilenkin:1982hm,Martin:1996ea,Martin:1996cp,Leblond:2009fq}, Hubble-induced mass of flat direction in supersymmetric theories \cite{Kamada:2014qja,Kamada:2015iga},
symmetry restoration in runaway quintessential scenarios \cite{Bettoni:2018pbl}, or the formation of domain walls in the case of axionic string network \cite{Hiramatsu:2010yn,Hiramatsu:2012sc,Hiramatsu:2012gg,Ramberg:2019dgi}.
The decay of the string network can imprint a low-energy cut-off in the GW spectrum at a frequency much higher that the low-energy cut-off of stable string networks, c.f. Sec.~\ref{ref:low_cut_off_stable}.

In this work, we focus for illustration on the case of string breaking via nucleation of monopole-antimonopole pairs. 
Such a metastable string network can arise from a two-stage pattern of symmetry breaking
 \cite{Leblond:2009fq}
\begin{equation}
G \rightarrow H\times U(1) \rightarrow H,
\end{equation}
in which the first step generates monopoles, while the second one produces CS. If the overall vacuum manifold $G/H$ is simply connected, the CS ($S$) are topologically unstable \cite{Vilenkin:1982hm, Preskill:1992ck}. They can break under Schwinger production of monopole-antimonopole pairs ($M\bar{M}$), hence producing `dumbbells' $MS\bar{M}$, namely segments of string with monopoles attached at the two ends.\footnote{More complex hybrid topological objects, called $\mathcal{Z}_N$-string, can be generated from the breaking pattern $G \rightarrow H\times U(1) \rightarrow H\times \mathcal{Z}_N$ \cite{Vachaspati:1986cc}. They are monopoles connected to $N$ strings and are called `cosmic necklaces' for $N=2$ or `string web' for $N\geq 3$ \cite{Kibble:2015twa}. Their evolution is expected to be close to the scaling regime \cite{Vachaspati:1986cc, Siemens:2000ty,BlancoPillado:2007zr,Hindmarsh:2016dha,Hindmarsh:2018zch} if the energy loss due to the presence of monopoles is not too large \cite{Martins:2010ma}.} If the monopoles have unconfined flux which propagate outside the strings, their acceleration under the effect of the string tension up to ultra-relativistic velocities can lead to emission of ultra-high-energetic gauge radiation, possibly leading to observable ultra-high-energy cosmic rays \cite{Berezinsky:1997td,BlancoPillado:1999cy} or CMB distortion \cite{Kibble:2015twa}. If the monopoles do not carry unconfined flux, the only source of energy loss is through GW emission, whose emitted power is of the same order of magnitude as the one from CS loops \cite{Martin:1996cp} but with a spectrum extending to higher frequencies \cite{Martin:1996ea, Leblond:2009fq}. More precisely, the GW power radiated by a straight dumbbell is \cite{Martin:1996cp}
\begin{equation}
\label{eq:GW_power_dumbbells}
P_{\rm GW}^{\rm MS\bar{M}} \simeq \tilde{\Gamma}\,G\mu^2, \qquad\text{with}\quad \tilde{\Gamma}\equiv 8 \ln\left( \gamma_0 \right),
\end{equation}
where $\gamma_0$ is the maximal Lorentz factor reached by the monopoles. We follow \cite{Leblond:2009fq} and we set $\tilde{\Gamma} \sim 50$. We note the interesting possibility for dumbbells to explain Dark Matter if their lifetime is larger than the age of the universe \cite{Sanchez:2011mf,Terning:2019bhg}. 

The monopole-anti-monopole pair nucleation rate per unit length is \cite{Leblond:2009fq}
\begin{align}
\label{eq:monopole_instanton_rate}
\Gamma_d = \frac{\mu}{2\pi}\exp(-\pi\kappa),
\end{align}
where $\kappa\equiv m^2/\mu \gtrsim 1$ is the ratio of the monopole mass $m$ to the CS tension $\mu$.
As explained in Sec.~\ref{sec:mainAssumptions}, the main sources of SGWB generated by a stable network are the loops formed with a length $l=\alpha\,t_i$ where $t_i$ is the loop formation time and $\alpha \simeq 0.1$. The breaking rate growing linearly with the string length, the later the loops are formed, the more likely they break under $M\bar{M}$ nucleation. More precisely, the loops break when the age of the universe is equal to their lifetime upon breaking
\begin{equation}
\label{eq:breaking_time}
t_i \sim \left (\Gamma_d \times \alpha \,t_i \right)^{-1}.
\end{equation}
After $M\bar{M}$ nucleation, the loops become dumbbells $MS\bar{M}$ which shrinks under GW emission with power given by Eq.~\eqref{eq:GW_power_dumbbells}, until they totally disappear (or at least have their length divided by two) after a time
\begin{equation}
\label{eq:lifetime_dumbell}
\tilde{t}\simeq \frac{\alpha\,t_i}{2\tilde{\Gamma} G\mu}.
\end{equation}
Upon plugging the large-loop-breaking time in Eq.~\eqref{eq:breaking_time}, in the dumbbell-lifetime under GW emission in Eq.~\eqref{eq:lifetime_dumbell}, we obtain the age of the universe after which no loops remain\footnote{Note that we have only considered broken loops and we have neglected the additional GW emission coming from the broken long strings of the network.} and after which GW emission stops
\begin{equation}
\label{eq:stop_GW_emission_time}
\tilde{t}_\textrm{stop}\sim\sqrt{\frac{1}{\Gamma_d\times\tilde{\Gamma}  G\mu}},
\end{equation}
which agrees with the estimation in \cite{Leblond:2009fq}.
The frequency $f_\textrm{break}$ emitted by this population of broken large-loops just before they disappear, at $\tilde{t}_\textrm{stop}$, corresponds to the lowest frequency of the GW spectrum, and it obeys, c.f. Eq.~\eqref{eq:lifetime_dumbell} and Eq.~\eqref{eq:tdelta_fdelta_eq_line1},
\begin{equation}
\frac{2k}{f_\textrm{break}}\cdot\frac{a(\tilde{t}_\textrm{stop})}{a(t_0)}= \Gamma G\mu \times \tilde{t}_\textrm{stop}.
\end{equation}
For string breaking during a radiation-dominated era, $\frac{a(\tilde{t}_\textrm{stop})}{a(t_0)}=\left(\frac{\tilde{t}_\textrm{stop}}{t_\textrm{eq}}\right)^{1/2}\left(\frac{\tilde{t}_\textrm{eq}}{t_0}\right)^{2/3}$, and we get
\begin{align}
f_\textrm{break}\simeq 2\left(2\pi z_\textrm{eq} G \,\tilde{\Gamma}^3\right)^{-1/4}\left(G\mu \cdot t_0\right)^{-1/2}\exp(-\frac{\pi \kappa}{4}).
\end{align}
Taking into account the summation of higher frequency modes and the more accurate cosmological history, we give the numerically-fitted version of the cut-off frequency due to string breaking through monopole-anti-monopole nucleation
\begin{align}
f_\textrm{break}
&\simeq (1.82\times 10^{19} \textrm{ Hz})\left(\frac{50}{\tilde{\Gamma}}\right)^{3/4}\left(\frac{10^{-11}}{G\mu}\right)^{1/2}\exp(-\frac{\pi \kappa}{4}).
\label{eq:break_cutoff}
\end{align}
where we have used $z_\textrm{eq}=3360$.
The smaller the separation between the monopole mass and the string tension $\kappa = m^2/\mu >1$, the faster the string breaking, and the higher the cut-off frequency $f_\textrm{break}$.

\subsection{GW spectrum}
\label{ref:GW_spectrum_peak_structure}
\begin{figure}[h!]
			\centering
			\raisebox{0cm}{\makebox{\includegraphics[height=0.5\textwidth, scale=1]{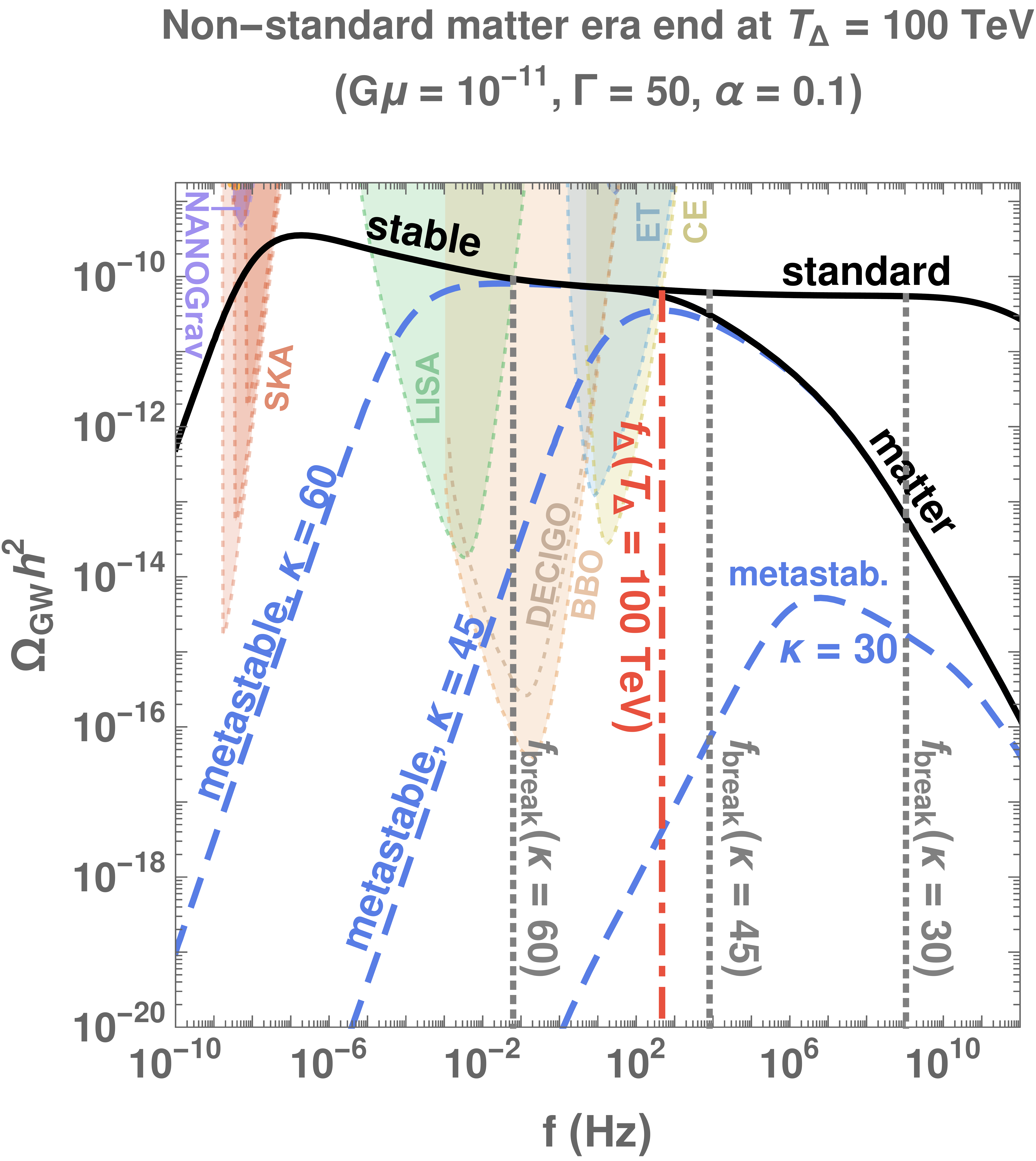}}}
			\raisebox{0cm}{\makebox{\includegraphics[height=0.5\textwidth, scale=1]{Figures/metastable_MD}}}\\
			\hfill
		\caption{\it \small { GW spectra from metastable string networks for different ratios of monopole mass to CS tension $\kappa\equiv m^2/\mu$, leading to different low-frequency cut-offs $f_\textrm{break}$ (vertical gray dotted). By decreasing $\kappa$, the different GW constraints can be relaxed. \textbf{Left:} Upon introducing a matter era with high-frequency cut-off $f_\Delta$ (vertical red dot-dashed), we can get a peak shape when $f_\textrm{break}\gtrsim f_\Delta$. \textbf{Right:} A rich variety of spectral shapes can be obtained by combining cut-off from metastability to long-lasting matter era (blue), short-lasting matter era (red) with duration $r = T_\textrm{start}/T_{\Delta}= 100$, or particle production (dotted).  } }
			\label{fig:peak_spectrum_formation}
		\end{figure}

The GW spectrum from metastable strings can be visualized in blue lines in the left panel of Fig.~\ref{fig:peak_spectrum_formation}, with different monopole-mass to string-tension ratios $\kappa\equiv m^2/\mu$, leading to different low-frequency cut-offs $f_\textrm{Low}=f_\textrm{break}$. 
By decreasing $\kappa \lesssim 65$, we can evade the current GW constraints from EPTA/NANOGrav on the string tension, but also from the CMB, such that the current constraint comes from LIGO O2, $G\mu \lesssim 5\times 10^{-6}$ \cite{Buchmuller:2019gfy}. The latter can also be relaxed if $\kappa \lesssim 40$ \cite{Buchmuller:2019gfy}. Hence, Schwinger production of monopole-anti-monopole pairs constitutes an interesting proposal to revive GUT strings following the symmetry breaking pattern 
$SO(10) \rightarrow G_{\rm SM}\times U(1)_{\rm B-L} \rightarrow G_{\rm SM}$ \cite{Buchmuller:2019gfy}.
The black lines represent the GW spectrum from stable strings in the presence of an early long-lasting matter era, leading to a high-frequency cut-off $f_\textrm{High}=f_\Delta$. 
Eventually, a peaked spectrum can be generated when 
\begin{align}
f_\textrm{Low}\geq f_\textrm{High}.
\end{align}
The higher the ratio $f_\textrm{Low}/f_\textrm{High}$, the more suppressed the peak amplitude relative to the spectrum without peak.

In the right plot in Fig.~\ref{fig:peak_spectrum_formation}, we lower the temperature $T_{\Delta}$ at which the matter era ends, in order to bring the peak spectrum within the GW interferometers windows, and show in red the case of a short-matter era $T_{\rm start}/T_{\Delta}=100$. A rich variety of spectral shapes can be obtained by combining different cut-off effects.

\subsection{Comparison of peaked GW spectra of different physical origins}
\begin{figure}[h!]
			\centering
			\raisebox{0cm}{\makebox{\includegraphics[height=0.5\textwidth, scale=1]{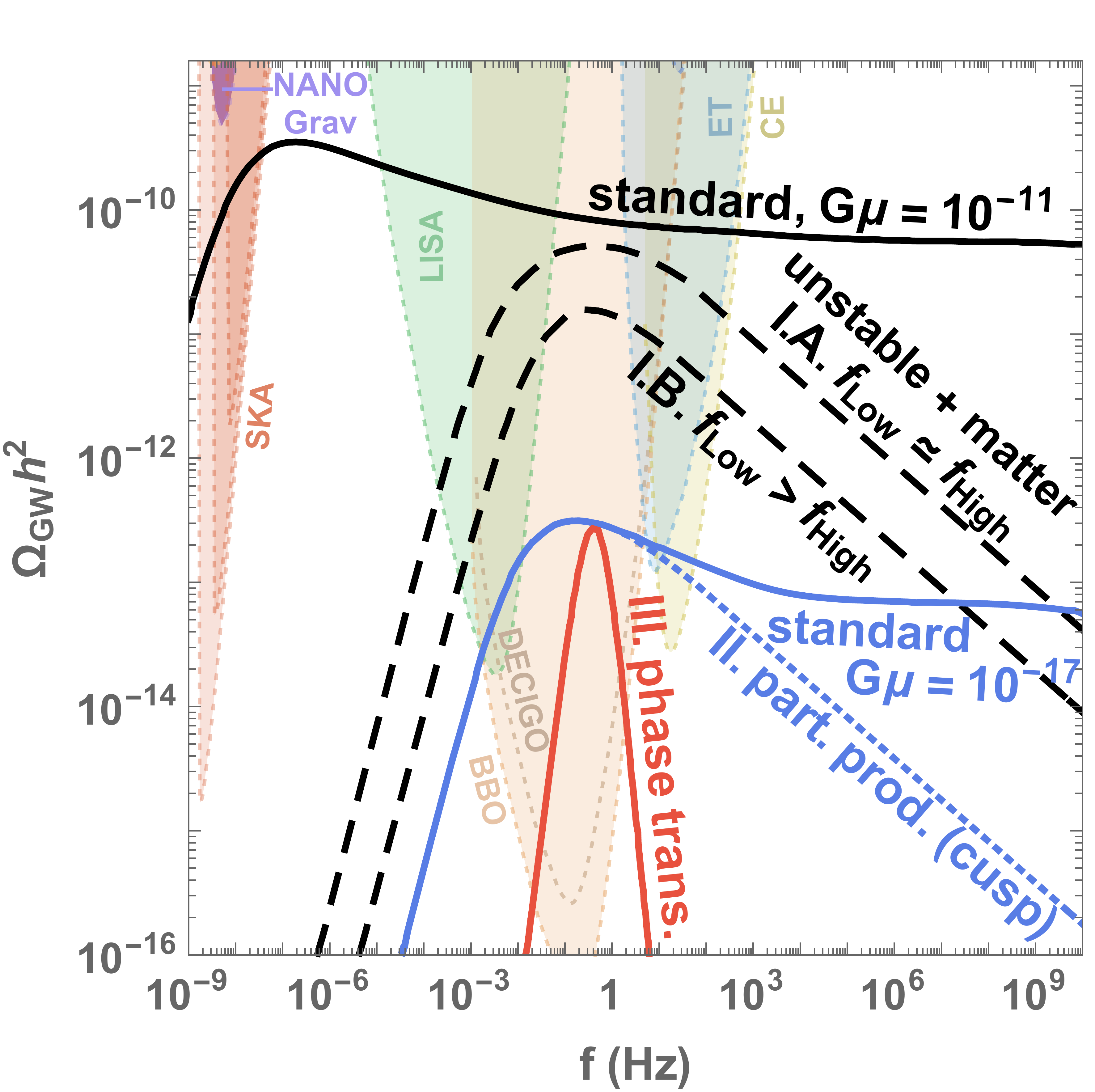}}}\\
			\hfill
		\caption{\it \small { Peaked spectra with same peak frequencies, corresponding to the benchmark scenarios I, II and III, described in the text and in Table~\ref{tab:metastable}.} {Here we extrapolate the $f^{-1/3}$ behavior to arbitrary large frequencies, which is equivalent to sum over an infinite number of proper modes $k$, see App.~\ref{sec:study_impact_mode_nbr}.  }} 
			\label{fig:peak_spectrum}
		\end{figure}

In Fig.~\ref{fig:peak_spectrum}, we show three types of peaks, whose precise parameter choices are detailed in Table~\ref{tab:metastable}.
\begin{itemize}
\item [I.]
In black dashed lines, we show GW spectra from metastable string networks, c.f. Sec.~\ref{ref:low_cut_off_metastable}, in the presence of an early long-lasting matter-dominated era, c.f. Sec.~\ref{sec:interm_matter}, for two different metastable-to-matter cut-off-frequency ratios. The same high-frequency cut-off can also be produced from an intermediate inflation era, c.f. Sec.~\ref{sec:inflation}. The slopes are $f^{3/2}$ and low frequencies and $f^{-1/3}$ are high frequencies, c.f. App.~\ref{sec:study_impact_mode_nbr}. 
\item [II.]
In blue lines, we show a GW spectrum from a stable network, which low-frequency cut-off is discussed in Sec.~\ref{ref:low_cut_off_stable}. We assume the presence of cusps responsible for particle production, leading to the high-frequency cut-off in dotted line, c.f. discussion in Sec.~\ref{UVcutoff}. The slopes are $f^{3/2}$ at low frequencies and $f^{-1/3}$ are high frequencies, c.f. App.~\ref{sec:study_impact_mode_nbr}. 
\item [III.]
In red line, we show a GW spectrum from a first-order phase transition assuming non-runaway bubble-walls, generated by sound waves \cite{Caprini:2015zlo}. It should be distinguishable from the peak spectrum from CS since the peak is thinner and the slopes are steeper: $f^3$ at low frequencies and $f^{-4}$ at high frequencies ($f^{-5/3}$ for turbulence).  Also, the slopes of a GW spectrum from first-order phase transitions assuming run-away bubble-walls, generated by scalar gradient, are also steeper than the CS case \cite{Caprini:2015zlo}: $f^3$ at low frequencies (or $f^1$ \cite{Jinno:2017fby}) and $f^{-1}$ at high frequencies (or $f^{-1.5}$ \cite{Cutting:2018tjt}).

\end{itemize}

\begin{table}[h!]
		\centering
		\label{tab:metastable}
\begin{tabular}{|c|c|c|}
\hline 
scenario                                                    & lower cut-off                   & higher cut-off                     \\ \hline
\begin{tabular}[c]{@{}c@{}}\\[-0.75em]I.A. metastable strings with long-lasting matter era:\\ $G\mu = 10^{-11}, ~ T_\Delta = 200$ GeV, $\kappa = 55$ \\[0.25em]\end{tabular}  
                               & $f_\textrm{break}\simeq 3.2$ Hz                     &  $f_\Delta\simeq 1$ Hz                          \\ \hline     
\begin{tabular}[c]{@{}c@{}}\\[-0.75em]I.B. metastable strings with long-lasting matter era:\\ $G\mu = 10^{-11}, ~ T_\Delta = 10$ GeV, $\kappa = 52.5$\\[0.25em] \end{tabular}  
                               & $f_\textrm{break}\simeq 23$ Hz                         & $f_\Delta\simeq 0.04$ Hz                      \\ \hline                                                                                              
\begin{tabular}[c]{@{}c@{}}\\[-0.75em]II. stable strings with standard cosmology and\\particle production (cusps): $G\mu = 10^{-17}, ~ \beta_c  = 1$  \\[0.25em]\end{tabular}                    & $f_\textrm{low}^{\rm stable}\simeq 0.15$ Hz                         & $f_\textrm{cusp}\simeq 2.9$ Hz                      \\ \hline
\begin{tabular}[c]{@{}c@{}}\\[-0.75em]III. first-order phase transition generated from\\ acoustic waves in standard cosmology \cite{Caprini:2015zlo}:\\$T_p = 5$ TeV, $\alpha = 0.1,~ \beta = 10^3,~ v_w \simeq 1$\\[0.25em]  \end{tabular}                                      &                 \multicolumn{2}{l|}{   \hspace{1.5 cm}   $f_\textrm{peak} \simeq 0.41$ Hz  }                     \\ \hline 
\end{tabular}
\caption{\it \small { Benchmark scenarios I, II and III, described in the text. The corresponding GW spectra are shown in Fig.~\ref{fig:peak_spectrum}.}}
\end{table}

\FloatBarrier

\section{Detectability of spectral features}
\label{sec:detectability}

\begin{figure}[h!]
			\centering
			\raisebox{0cm}{\makebox{\includegraphics[width=0.49\textwidth, scale=1]{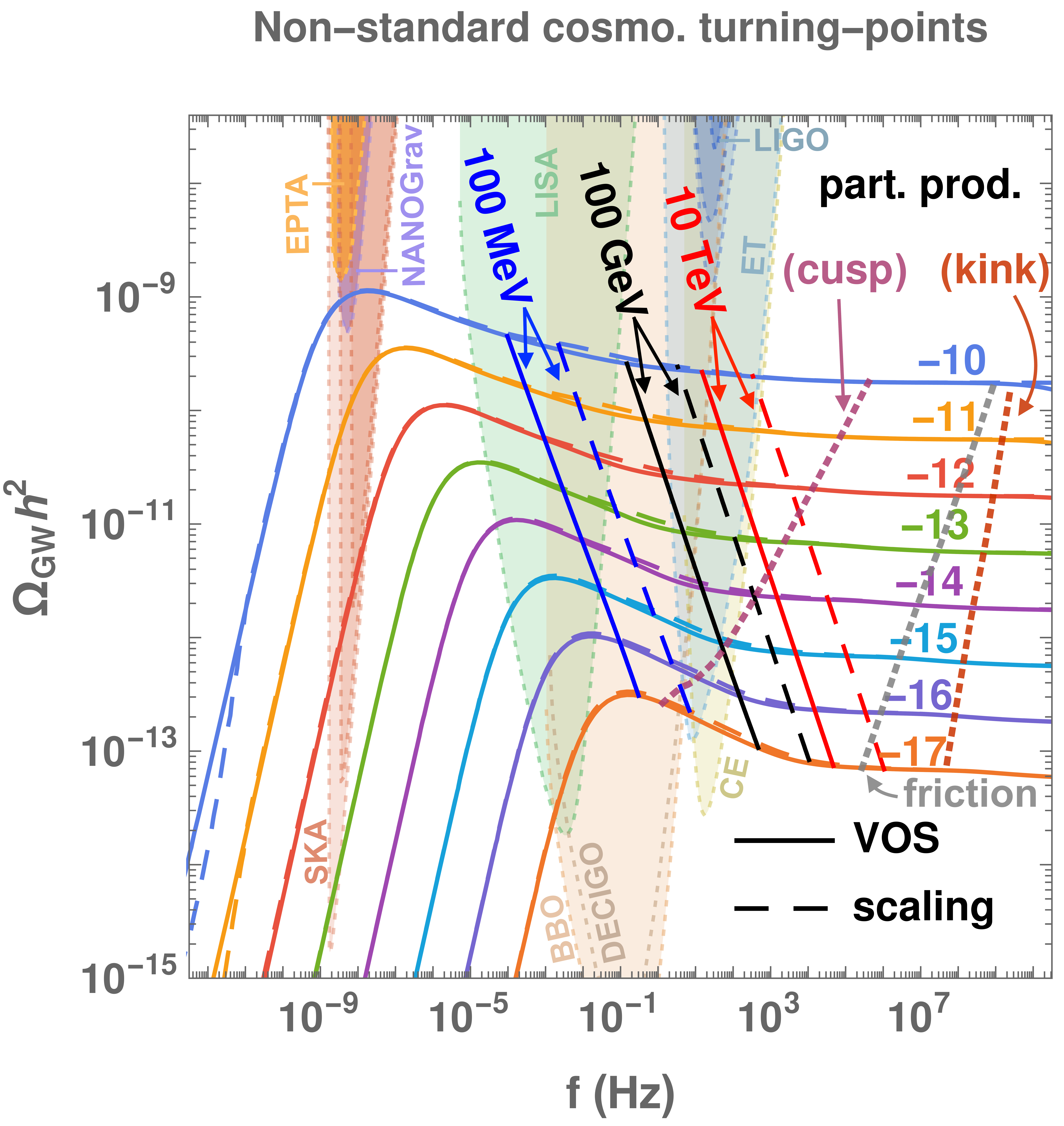}}}
			\raisebox{0cm}{\makebox{\includegraphics[width=0.49\textwidth, scale=1]{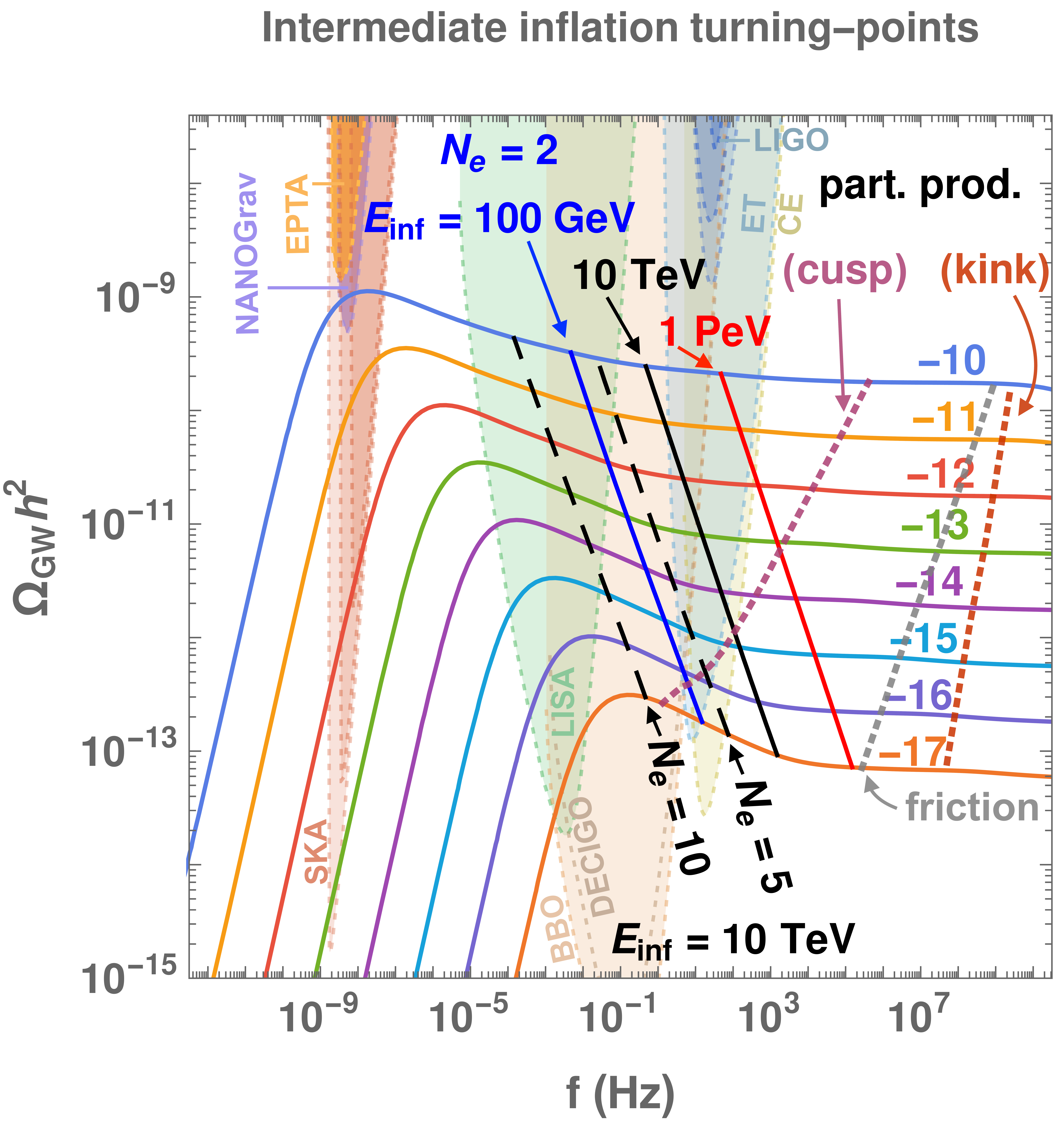}}}
			\caption{\it \small \textbf{Left:} Straight solid and dashed lines are a collection of VOS  and scaling turning points, given by Eq.~\eqref{turning_point_general} and Eq.~\eqref{turning_point_general_scaling}  respectively,  for general non-standard cosmologies ending at temperature $T_\Delta$. The displayed spectra assumes a standard cosmology.   Each spectrum corresponds to string tension $G\mu = 10^{x}$, where $x$ is specified by a number on each line.\textbf{Right:}  We show the turning-points, given by Eq.~\eqref{turning_point_inf}, for intermediate inflation lasting for $N_e$ e-folds and taking place at the energy scale $E_{\rm inf}$. 
The dotted lines in the two panels show the cut-off frequencies due to particle productions, c.f. Sec.~\ref{UVcutoff}, and thermal friction, c.f. Sec.~\ref{sec:thermal_friction}, for each value of $G\mu$.}
				\label{fig:turning_points_lines}
		\end{figure}

\subsection{Two prescriptions}
\label{sec:Rx1VSRx2}
We aim at using the would-be detection of a SGWB spectrum generated by CS  to constrain an early non-standard era. We assume the detection of a SGWB from CS by a detector $(i)$ with sensitivity $\Omega_{\rm sens}^{(i)}(f)$ 
\begin{equation}
\Omega_{\rm sens}^{(i)}(f)  \gtrsim \Omega_{\rm GW} (f).
\end{equation}
The power-law integrated sensitivity (PLS) curves of the different experiments are computed in app.~\ref{app:sensitivity_curves}.
We propose two prescriptions for detecting a non-standard era $(T_{\Delta}, \, r)$:
\begin{itemize}
\item
\textbf{Rx 1} \textit{(turning-point prescription)}: The frequency $f_{\Delta}$ of the turning-point where the standard and non-standard spectra meet must be inside the interferometer window. GW detected at frequency $f_{\Delta}$ have been emitted by loops formed during the change of cosmology at the temperature $T_\Delta$.   The relation between $f_{\Delta}$ and $T_{\Delta}$ is given in Eqs.~\eqref{eq:turning_point_scaling}, \eqref{turning_point_general} for a non-inflationary non-standard era, or Eq.~\eqref{turning_point_inf} for an intermediate inflation era, using the $10\%$ prescription  (see Eqs.~\eqref{turning_point_general_scaling_app} and \eqref{turning_point_general_scaling_app_inf} for other cases).
\item
\textbf{Rx 2} \textit{(spectral-index prescription)}: The absolute value of the observed spectral index $\beta$, which is defined as $ \Omega_{\rm GW}(f) \propto f^{\beta}$, is larger than $0.15$.
\end{itemize}
The second prescription assumes that the non-standard era tilts the spectral index by more than a benchmark value. We checked that the choice of the precise benchmark value, e.g. $0.15$, has very little impact on the results.

\begin{figure}[h!]
			\centering
			\raisebox{0cm}{\makebox{\includegraphics[height=0.49\textwidth, scale=1]{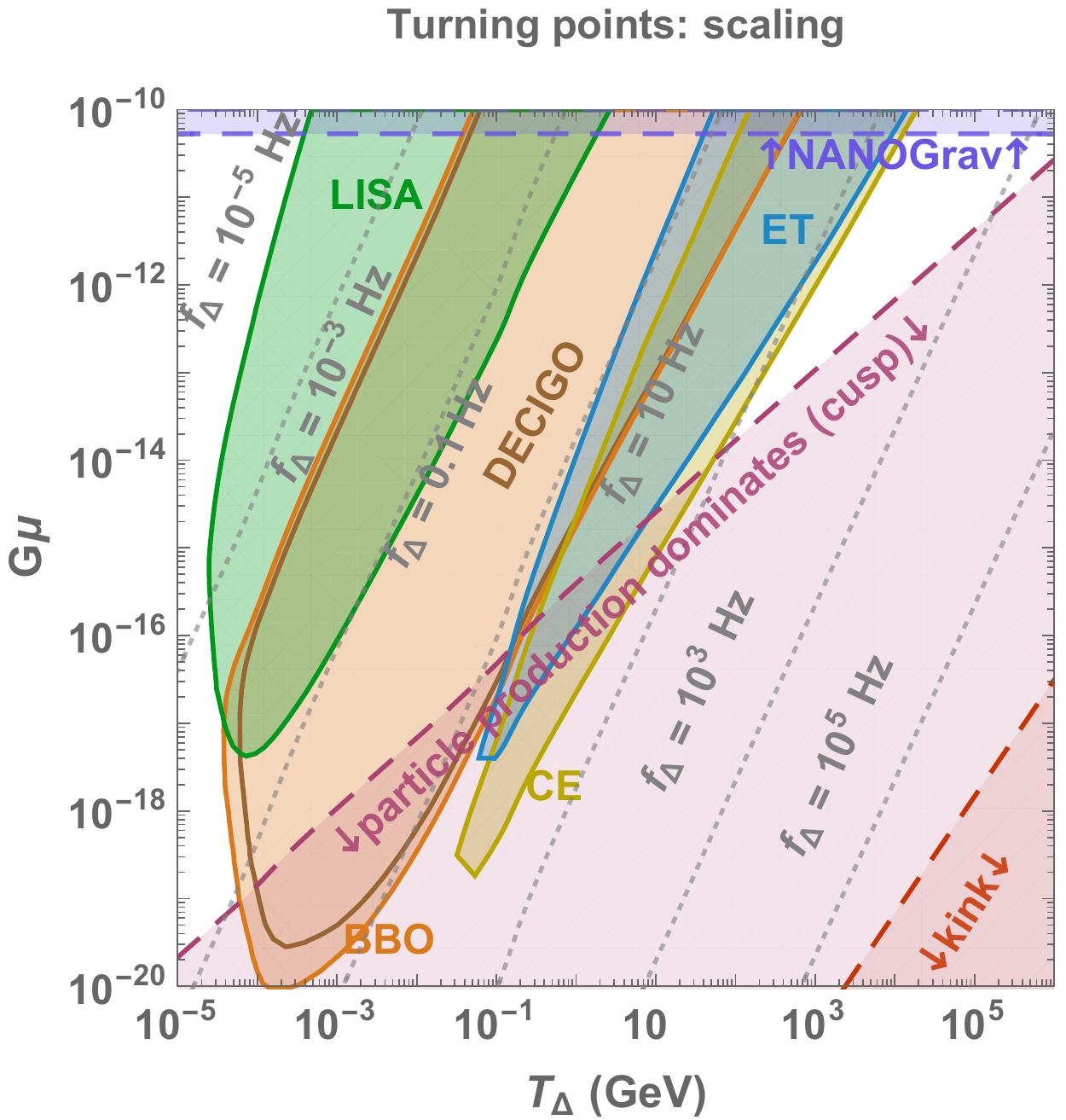}}}
			\raisebox{0cm}{\makebox{\includegraphics[height=0.49\textwidth, scale=1]{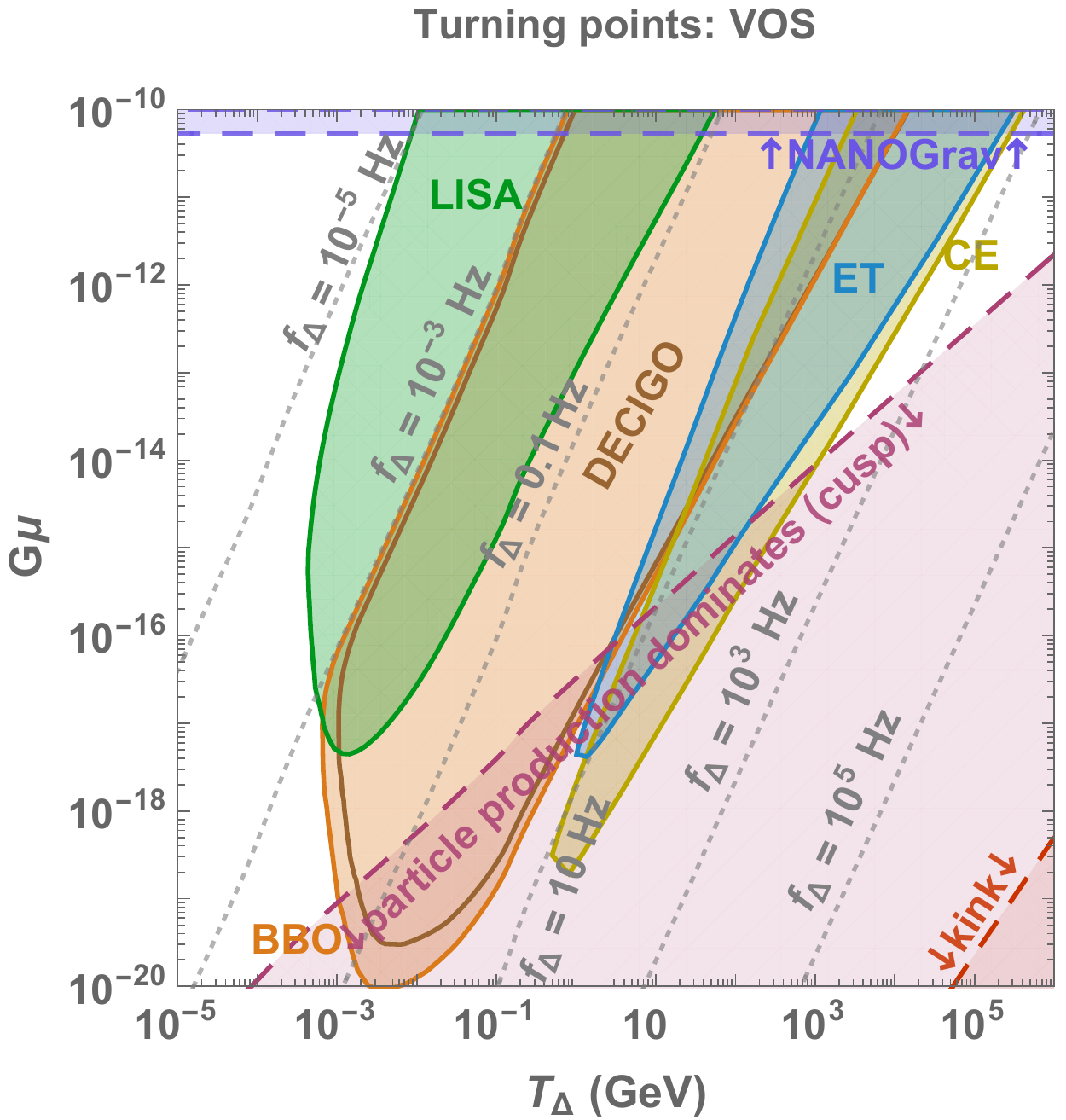}}}\\
			\raisebox{0cm}{\makebox{\includegraphics[height=0.49\textwidth, scale=1]{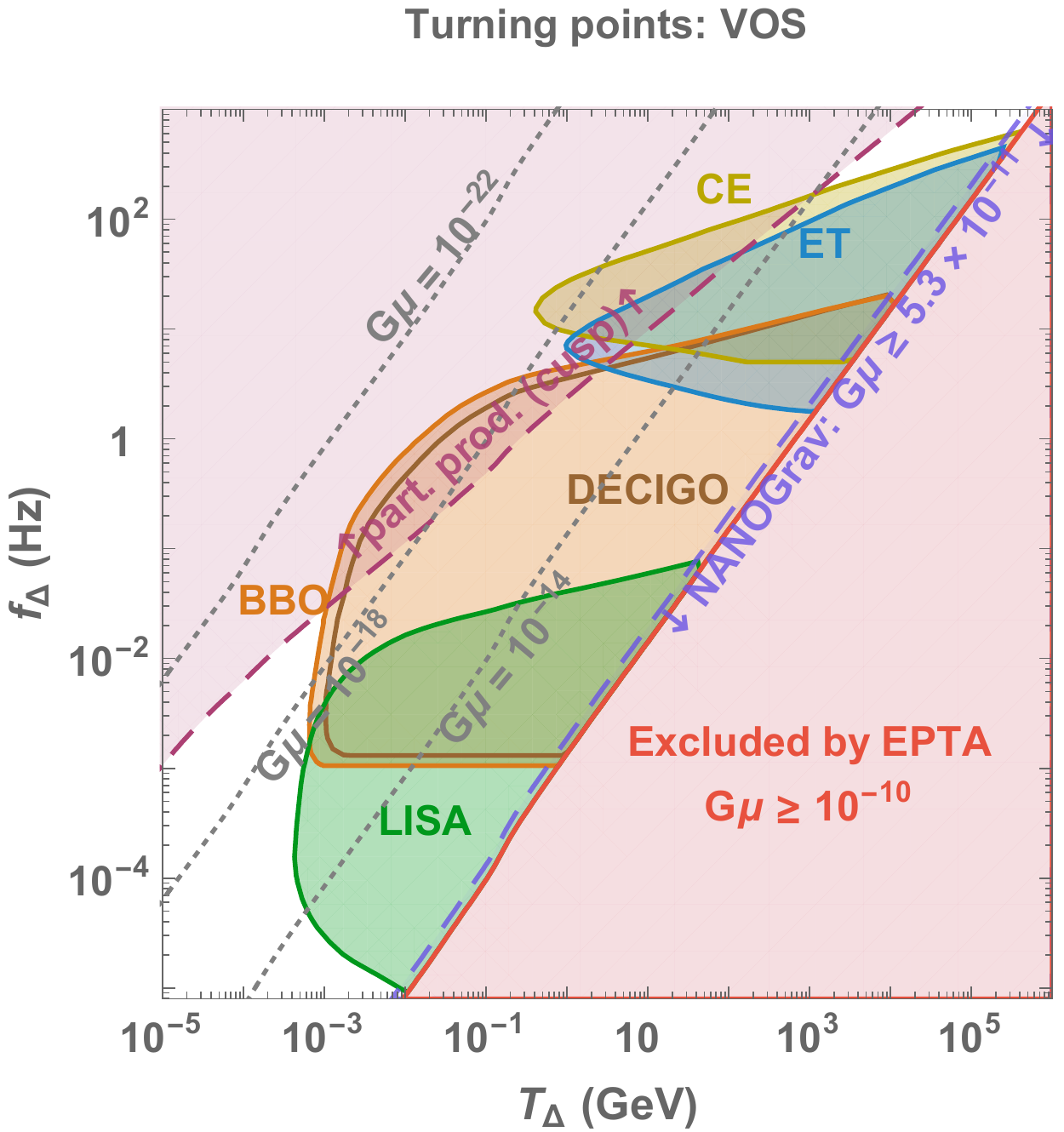}}}
			\raisebox{0cm}{\makebox{\includegraphics[height=0.49\textwidth, scale=1]{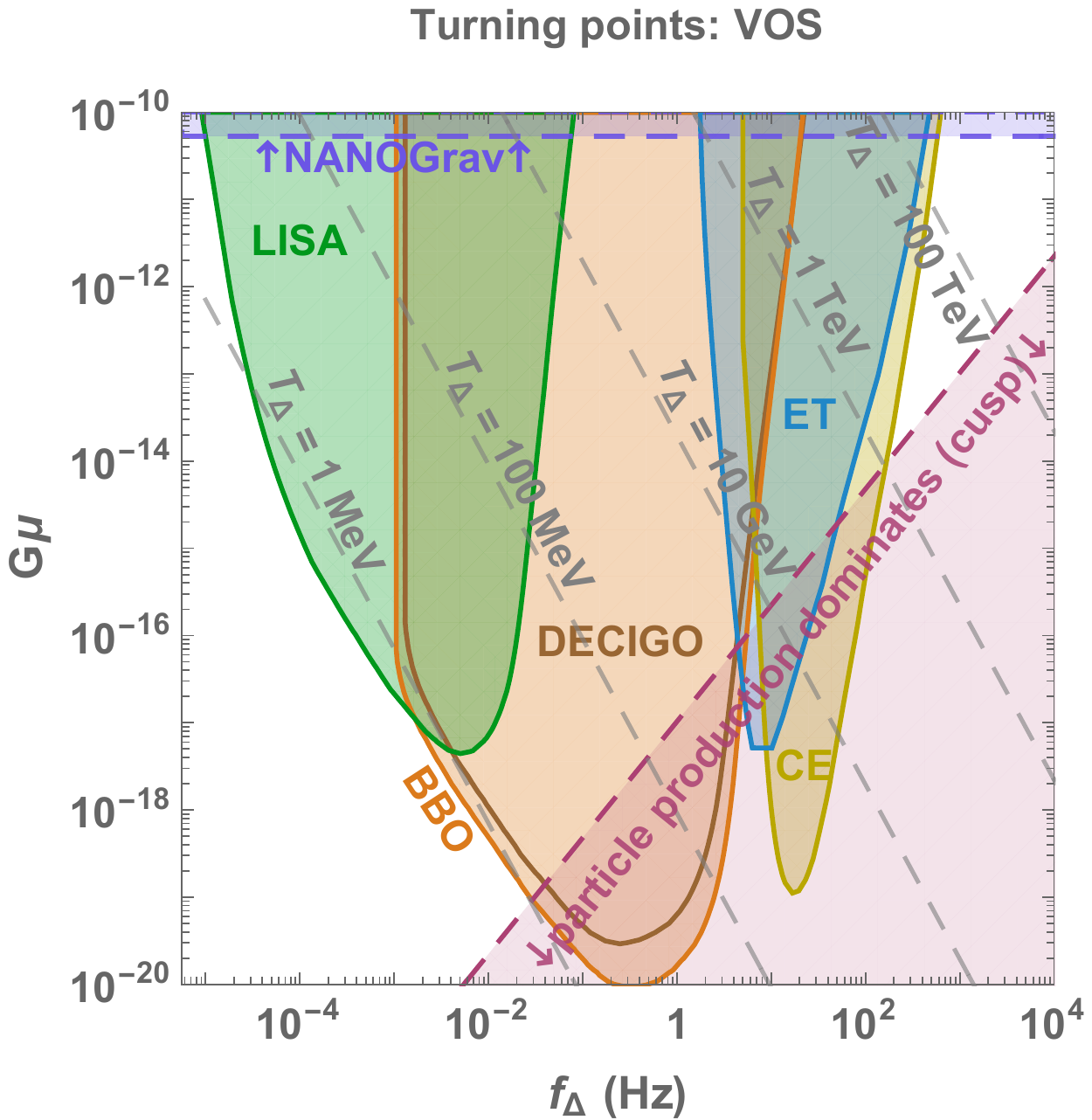}}}\\
			\hfill
			\caption{\it \small \textbf{Top:} Comparison between the detectability of the turning-points in $G\mu-T_\Delta$ planes assuming a scaling network (left) with the one assuming the full VOS evolution (right), c.f. Sec.~\ref{sec:scalingVSvos}, evolved in non-inflationary-non-standard-eras. Gray dotted lines are turning-points for given frequencies, c.f. Eq.~\eqref{turning_point_general_scaling} for scaling netwok and Eq.~\eqref{turning_point_general} for VOS network. \textbf{Bottom:}  Detectability of turning points in the planes $f_\Delta-T_\Delta$ and $G\mu-f_\Delta$ assuming a VOS network. Limitations from particle production (see Sec.~\ref{UVcutoff}) and bounds from EPTA are also included. }
		\label{fig:contour_power}
		\end{figure}

\subsection{More details on the turning-point prescription}
Turning points of non-inflationary-non-standard-eras, c.f. Eq.~\eqref{turning_point_general}, are plotted in the left panel of figure~\ref{fig:turning_points_lines}, for different values of $G\mu$ and temperatures $T_{\rm \Delta}$ at the end of the non-standard era. We show the shift to lower frequencies by a factor $\sim 22.5$ due to the deviation from the scaling regime during the change of cosmology. 

Turning points in the special intermediate-inflation-era scenario, c.f. Eq.~\eqref{turning_point_inf}, are plotted in the right panel of figure~\ref{fig:turning_points_lines}, for different inflation scales $E_{\rm inf}$ and e-fold numbers $N_{\rm e}$. Due to the stretching of the correlation length outside the horizon and the necessity to wait that it re-enters in order to reach the scaling regime, the longer the inflation the lower the turning-point frequency.

With the solid purple and red lines, we show the expected cut-off frequencies above which the GW spectrum is expected to be suppressed due to the domination of massive particle production over gravitational emission, in the benchmark cases where the loop small-scale structures are dominated either by cusps or kinks. Hereby, we show the possibility of losing the information about the cosmological evolution when the turning-points are at higher frequencies than the particle-production cut-off.

In figure~\ref{fig:contour_power}, we show the detectability of a turning point at frequency $f_{\Delta}$, corresponding to a change of cosmology taking place at the temperature $T_{\Delta}$, in the plane $G\mu-T_\Delta$, $f_\Delta-T_\Delta$, and $G\mu-f_\Delta$. We compare the turning-point formula, defined in Eq.~\eqref{turning_point_general} in the VOS regime with the one defined in eq~\eqref{turning_point_general_scaling} in the scaling regime.

Some of these plots were already presented in \cite{Cui:2017ufi,Cui:2018rwi} (for long matter and kination era), assuming that the scaling regime holds during the change of cosmology. Our plots turn out to be similar due  to their different choice of precision in the determination of the turning point frequency, see criterion in Eq.~\eqref{10per_criterion}.
	
\subsection{Comparative reach of different observatories at a glance}

\begin{figure}[t!]
\centering
\raisebox{0cm}{\makebox{\includegraphics[width=0.565\textwidth, scale=1]{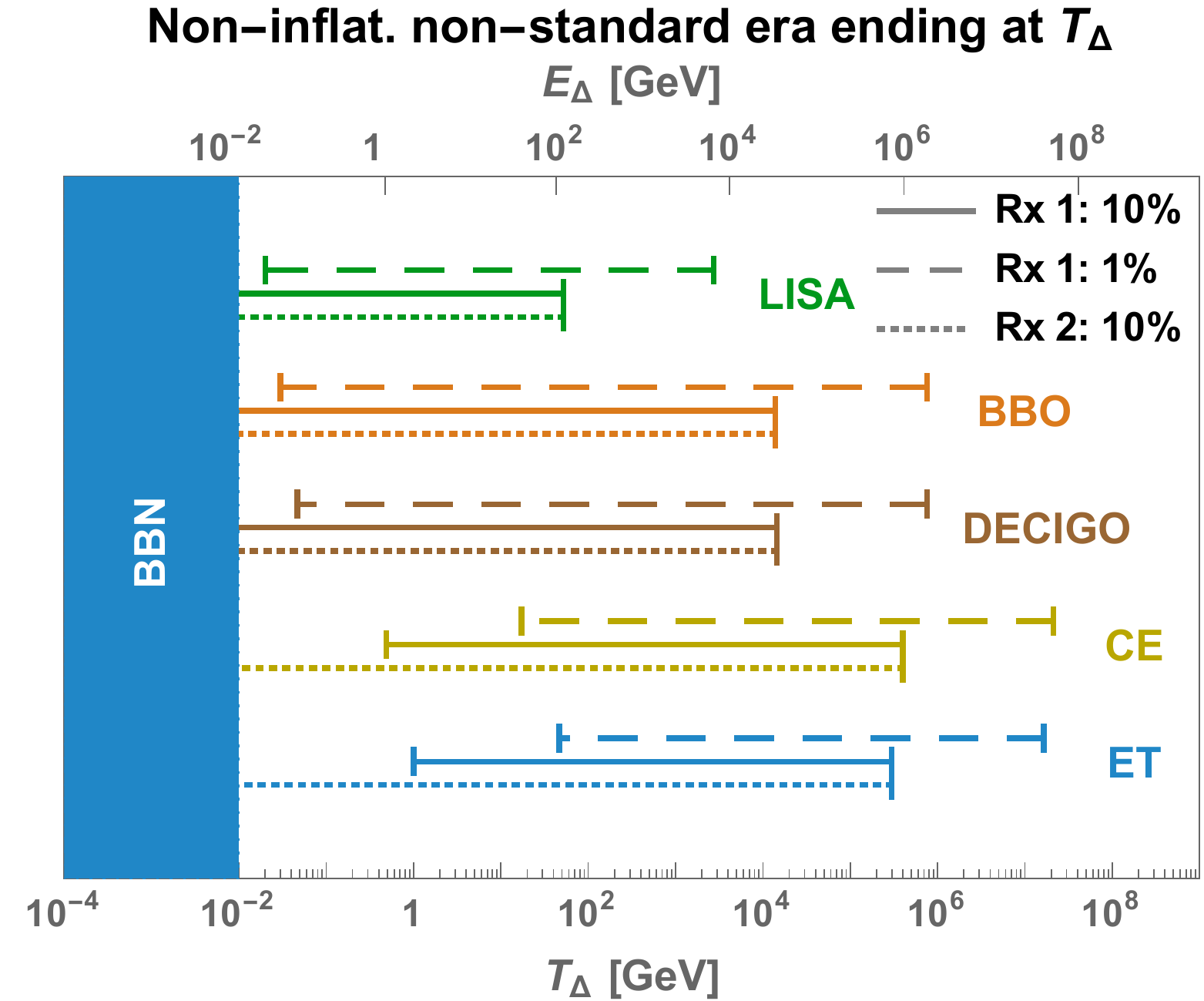}}}\\[0.5em]
\raisebox{0cm}{\makebox{\includegraphics[width=0.55\textwidth, scale=1]{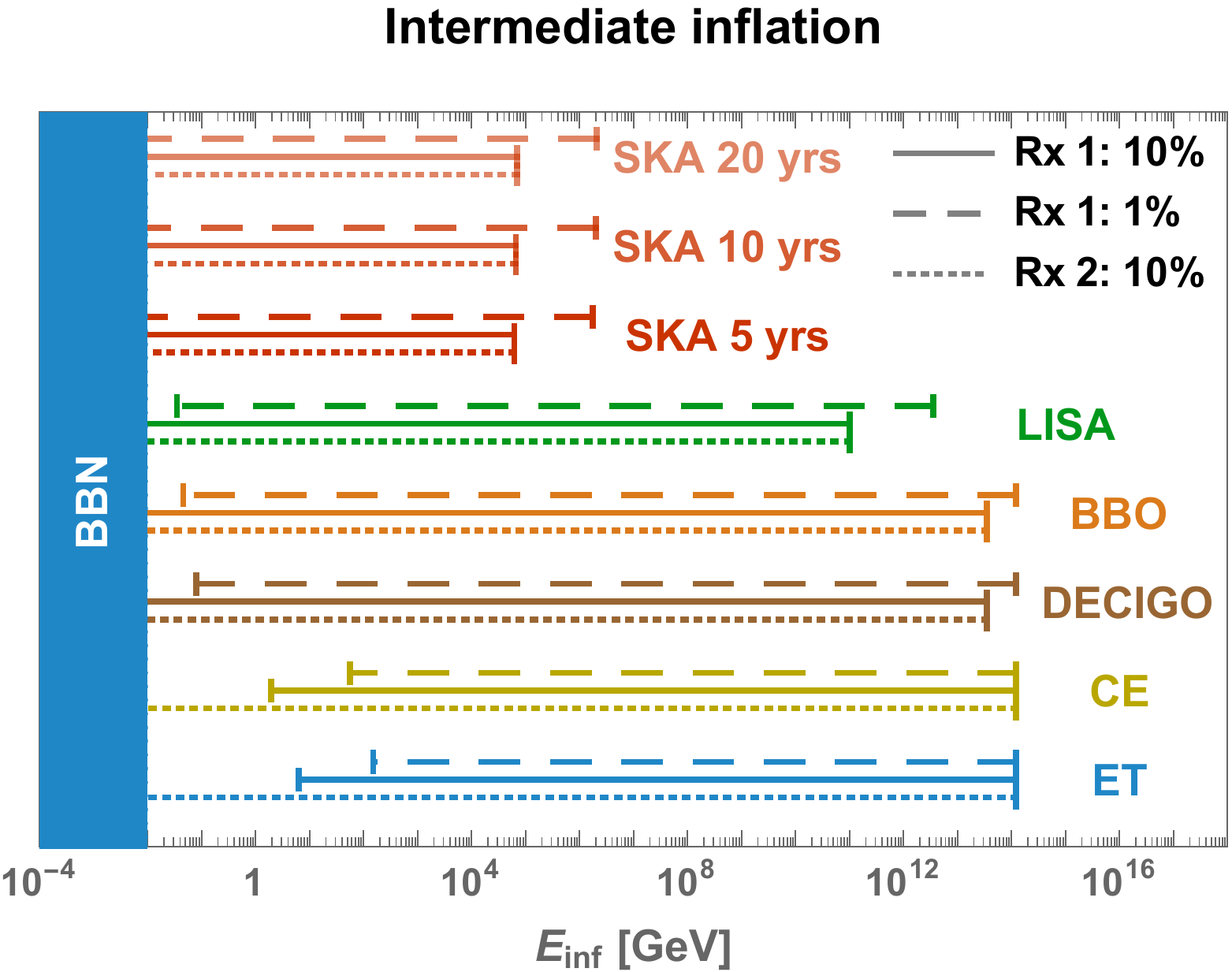}}}
\caption{\it \small \textbf{Top:}  Sensitivity to the energy scale $E_\Delta$ of the universe at the end of any non-inflationary non-standard era  for each future GW interferometer. The connection to $E_\Delta$ is given by the observation of the turning-point frequency defined in Eq.~(\ref{turning_point_general_scaling_app}). The width of the bands includes varying the string tension for $G\mu<10^{-10}$. The dotted, dashed and solid lines correspond to different observational prescriptions defined in Sec.~\ref{sec:Rx1VSRx2}. \textbf{Bottom:} Sensitivity to the energy scale $E_{\rm inf}$ of an intermediate inflationary era  for each future GW interferometer as well as for future radio telescope SKA. The connection to $E_{\rm inf}$ is given by the observation of the turning-point frequency defined in Eq.~(\ref{turning_point_general_scaling_app_inf}). The width of the band also includes varying the number of efolds of inflation $N_e$ up to 20.}
\label{fig:summary1}
\end{figure}

Our analysis shows the complementarity between distinct experiments to probe different energy scales and durations of the non-standard era, as well as different string tensions. 
 As our contour plots in Fig.~\ref{fig:contour_all_f_n_sum}, ~\ref{fig:contour_int}, ~\ref{fig:contour_power_inflation1}, ~\ref{fig:contour_power_inflation2}, ~\ref{fig:contour_power_inflation3}, ~\ref{fig:contour_power} show, it is not really possible to associate a given new physics energy scale to a given frequency band of observation. In fact, a given frequency band can probe different energy scales, depending on the nature and duration of the non-standard era, and the value of the string tension. Still, we provide some overall comparison in Fig.~\ref{fig:summary1} of the reach of each experiment on the energy scale of the non-standard era. The precise numbers depend on the definition of the observable used to probe the non-standard physics. For any non-inflationary non-standard era (upper plot), we use the turning point frequency defined in Eq.(\ref{turning_point_general_scaling_app}). It depends sensitively on the precision of the measurement. A realistic value
 is $\sim 10\%$. For comparison, we also show results for the idealistic case of 1$\%$. These plots include variation of $G\mu$. In the bottom plot of Fig.~\ref{fig:summary1},  we show the case of an intermediate inflationary era, which depends very sensitively on the number of efolds, while the turning point frequency does not depend on the duration of the non-standard era for other non-standard cosmologies. The upper plot applies to any non-standard era with equation of state $\rho\sim a^{-n}$ with any $n$ between 3 and 6, independently of the duration of this non-standard era.
 Interestingly, radio telescope SKA can be sensitive to a low-scale inflationary era. 
 Note that the bands in  Fig.~\ref{fig:summary1}  are calculated by neglecting particle production, which will affect mostly ET and CE, c.f. Figs.~\ref{fig:contour_power} and \ref{fig:contour_power_inflation2}. The lower bound on  $T_\Delta$ and $E_\textrm{inf}$ should be weaker by 1 order of magnitude for ET and CE if there is a cutoff from particle production.

\section{Summary and conclusion}
\label{sec:conclusion}

In standard cosmology, the GW spectrum generated by a network of Nambu-Goto cosmic strings (and mainly due to emission by loops)  is nearly scale-invariant. 
Its potential observation by third-generation interferometers would be a unique probe of new effects beyond the standard models of particle physics and cosmology. 
Such opportunity  was pointed out in \cite{Cui:2017ufi, Cui:2018rwi,Auclair:2019wcv,Guedes:2018afo,Chang:2019mza}. 

Deriving firm conclusions is still premature as
theoretical predictions of the GW spectrum from CS are subject to a number of large uncertainties. 
Still, we feel that the extraordinary potential offered by future GW observatories to probe 
high energy physics has not yet been explored, and in a series of papers, we are starting to scrutinise 
 how much can be learnt, even if only in the far-future, after those planned GW observatories will have reached their expected long-term sensitivity and the astrophysical foreground will have been fully understood.

Deviations in the cosmological history  with respect to standard cosmology not only change the redshifting factor of  GW but also modify the time of loop formation and the loop-production efficiency.
We presented  predictions for the resulting GW spectra under a number of assumptions which we have comprehensively  reviewed. 

We extend previous works in several directions, as listed in the introduction.

A particular feature of gravitational waves from cosmic strings is the relation between the observed frequency and the GW production mechanism. In contrast with short-lasting cosmological sources of  gravitational waves, such as phase transitions, where the frequency is simply related to the Hubble radius at the time of GW emission, for cosmic strings the time of the dominant GW emission is much later than the time of  loop production, by a factor $\sim 1/(G\mu)$, such that the observed frequency is higher due to a smaller redshift.
We stressed that a given interferometer may be sensitive to very different energy scales, depending on the nature and duration of the non-standard era, and the value of the string tension. 
This goes  against usual paradigms. For instance, 
it is customary to talk about LISA as a window on the EW scale \cite{Grojean:2006bp,Caprini:2019egz}. This does not apply for GW from cosmic strings, as LISA could either be a window on a non-standard matter era at the  QCD scale or on a 10 TeV inflationary  era, meaning that the GW observed in the LISA band have been emitted by loops that were created at the QCD epoch, or at a 10 TeV epoch depending on the nature of the new physics responsible for the non-standard era.
Interestingly, the Einstein Telescope and Cosmic Explorer offer a window of observation on the highest scales, up to $10^{14}$ GeV inflationary eras. They can also be windows on the EW and TeV scales, as will be discussed in more details in \cite{Gouttenoire:2019rtn}.
BBO/DECIGO could probe new physics in an intermediate range.  Finally, radio telescope SKA may be sensitive to a TeV scale inflationary era. 
The goal of this paper was to stress 
the very rich variety of spectral shapes that can be obtained by combining different physical effects. In particular, we showed how peaked shapes can arise naturally in a large variety of models.

We apply these findings to probe well-motivated particle physics scenarios in \cite{Gouttenoire:2019rtn}. Particularly generic are intermediate matter eras triggered by cold  heavy particles arising in UV completions of the Standard Model. In  \cite{Gouttenoire:2019rtn}, we show on specific models how a new  uncharted particle theory space can be probed from analysis of SGWB from CS.

Finally, one important  question  will be to work out how to distinguish a stage of matter or inflationary expansion, which both lead to a suppression of the GW spectrum, from the cutoff induced by particle production from small loops. Both predict a cutoff at high frequencies and lead to similar spectra.
Interestingly, particle production by cosmic string networks can be probed through cosmic rays and bring complementary non-gravitational information on the SGWB.
Besides, the complementarity between different GW instruments will be crucial here as the detection of the low-frequency peak of the spectrum (due to the transition from the standard radiation to the standard matter era) can enable to probe the string tension and to break the degeneracy between different spectral predictions.
The possibility to reconstruct the spectral shape of a SGWB was  analysed in \cite{Caprini:2019pxz} using LISA data only.
In the case of a SGWB generated by CS, which can span more than twenty decades in frequency, it will be crucial to use data from different interferometers  (and even from radio telescopes) to probe the full spectrum.

\medskip

\section*{Acknowledgements}

We thank  Marek Lewicki, Dani\`{e}le Steer, and Lara Sousa for useful discussions. We also thank Valerie Domcke and the DESY summer students 2018 Anna Kormu and Sam Wikeley for their participation at an early stage of the project. 
This work is supported by the Deutsche Forschungsgemeinschaft under Germany's Excellence Strategy - EXC 2121 ,,Quantum Universe'' - 390833306. 
The work of Y.G. is partly supported by a PIER Seed Project funding `Dark Matter at 10 TeV and beyond, a new goal for cosmic-ray experiments' (Project ID PIF-2017-72). 
P.S. acknowledges his master-degree scholarship from the Development and Promotion of Science and Technology Talents project (DPST), Thailand.

\appendix

\section{Constraints on cosmic strings from BBN, gravitational lensing, CMB and cosmic rays}
\label{app:phenoCS}

By confronting our theoretical predictions for the GW spectrum from CS with the sensitivity curves of EPTA \cite{Lentati:2015qwp} and NANOGrav  \cite{Arzoumanian:2018saf} (which we take from \cite{Breitbach:2018ddu}), we derived the respective bounds
$G \mu \lesssim 2 \times 10^{-10}$ (EPTA) and $G \mu \lesssim 5 \times10^{-11}$ (NANOGrav),
as discussed in Sec.~\ref{subsec:Gmuconstraints}.
For this reason, we only considered  in our analysis $G \mu$ values smaller than $5\times 10^{-11}$.
In Sec.~\ref{sec:GW_BBN}, we give the constraints on the string tension from not changing the expansion rate of the universe at BBN. They are much weaker than the ones from Pulsar Timing Arrays but they can become relevant in presence of kination.

In Sec.~\ref{sec:gravitational_lensing} and Sec.~\ref{sec:temperature_anisotropy}, we give bounds from gravitational lensing and CMB observables. They are also much weaker than the ones from Pulsar Timing Arrays but they have the strong advantage to be independent of our assumptions for the theoretical prediction of the GW background. 
Finally, in Sec.~\ref{sec:particle_prod_pheno}, we discuss the possibility of probing CS from the massive particle production in the presence of kinks and cusps.

\subsection{GW constraints from BBN}

\label{sec:GW_BBN}
As a sub-component of the total energy density of the universe, the amount of GW can impact the expansion rate of the universe which is strongly constrained by BBN and CMB. More precisely, any non-standard energy density can act as an effective number of neutrino relics
\begin{equation}
N_{\rm eff} = \frac{8}{7}\left( \frac{\rho_{\rm tot}-\rho_{\gamma}}{\rho_\gamma} \right)\left( \frac{11}{4} \right)^{4/3},
\end{equation}
which is constrained by CMB measurements \cite{Aghanim:2018eyx} to $N_{\mathsmaller{\rm eff}} = 2.99_{-0.33}^{+0.34}$ and by BBN predictions \cite{Mangano:2011ar, Peimbert:2016bdg} to $N_{\mathsmaller{\rm eff}} = 2.90_{-0.22}^{+0.22}$ whereas the SM prediction \cite{Mangano:2005cc, deSalas:2016ztq, Escudero:2020dfa} is $N_{\mathsmaller{\rm eff}}  \simeq 3.045$.
Using $\Omega_{\gamma} h^2 \simeq 2.47\times 10^{-5}$ \cite{Tanabashi:2018oca}, we obtain the following bound on the GW spectrum 
\begin{equation}
\int^{f_{\rm high}}_{f_\textrm{BBN}}\frac{df}{f}h^2 \Omega_\textrm{GW}(f) \leq 5.6 \times 10^{-6} ~ \Delta N_\nu,
\end{equation}
where $f_{\rm high}$ is the frequency today of the first GW produced,  $f_\textrm{BBN}$ is the frequency today of the GW produced at BBN, and we set $\Delta N_\nu \leq 0.2$. The value of $f_\textrm{BBN}$ depends on the source of GW.
For CS, the temperature at BBN, $T_\textrm{CMB}\simeq 1$ MeV, translates via Eq.~\eqref{fdeltaApp} to the frequency 
\begin{equation}
f_\textrm{BBN} \simeq 8.9 \times 10^{-5}~\textrm{Hz}~\left(\frac{0.1\times 50 \times 10^{-11}}{\alpha \Gamma G \mu}\right)^{1/2}.
\end{equation}
\begin{figure}[h!]
\centering
\raisebox{0cm}{\makebox{\includegraphics[width=0.5\textwidth, scale=1]{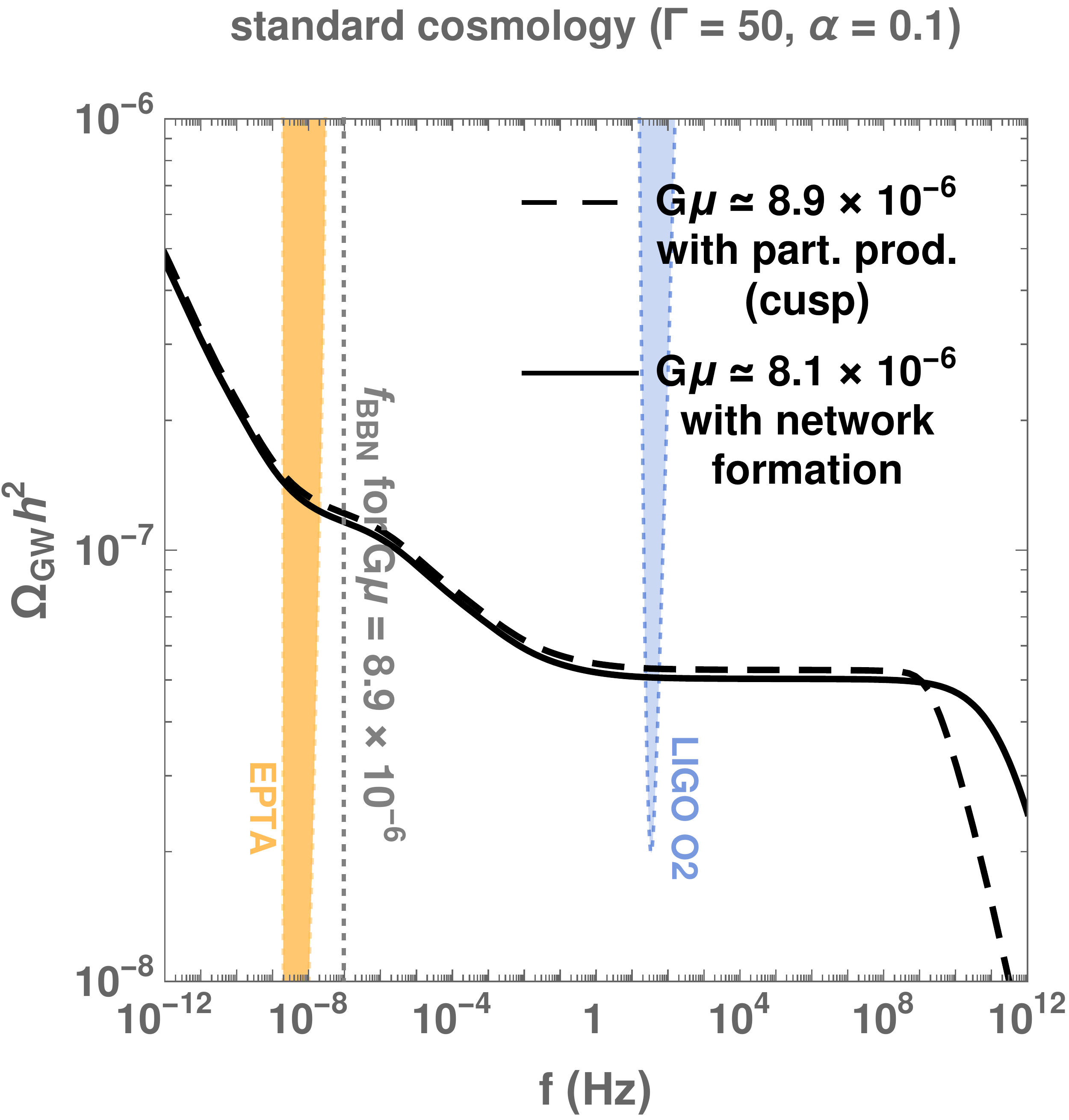}}}
\caption{\it \small 
Two GW spectra which saturate the BBN bounds, assuming a VOS string network, c.f. Sec.~\ref{sec:scalingVSvos}, evolving in standard cosmology. The solid line assumes a cut-off due to network formation whereas the dashed line assumes a cut-off due to particle production from cusps. The dotted vertical line is the frequency emitted when BBN starts. We compare the BBN bounds to the bounds from EPTA and LIGO O2. }
\label{excessive_GW}
\end{figure}
In Fig.~\ref{excessive_GW}, we show the GW spectra which saturate the BBN bound for two different high-frequency cut-offs.
We can see that the lower the cut-off, the higher the upper bound on $G\mu$ due to less GW present at the time of BBN. Assuming the presence of the cut-off due to particle production from cusps, we obtain 
\begin{equation}
\text{BBN:}  \qquad h^2\Omega_\textrm{GW}(f)\lesssim 8.9\times 10^{-6}.
\end{equation}
We expect the BBN bounds to become softer in presence of non-standard matter or inflation era but tighter in presence of an early kination era. For instance, scenarios of inflation followed by a stiff equation of state (e.g. quintessential inflation \cite{Peebles:1998qn}) are dramatically jeopardized by the BBN bounds \cite{Figueroa:2019paj}.
Similarly, in the case of CS, we find that the maximally allowed string tension is $G\mu \simeq 3.9 \times 10^{-15},~3.8 \times 10^{-17},~\textrm{and }2.9 \times 10^{-20}$ for long-lasting kination era ending at temperature $T_\Delta = 100$ TeV, $1$ TeV, and $1$ GeV, respectively.

\subsection{Gravitational lensing}
\label{sec:gravitational_lensing}
The presence of energy confined within the core of CS affects the spacetime around them. The metric near a CS is locally flat but globally conical \cite{Vilenkin:1981zs}. Photons from a distant celestial object travelling in the vicinity of a CS are subject to gravitational lensing effects. The corresponding constraint $G\mu \lesssim 3 \times 10^{-7}$ has been derived from the search of gravitational lensing signatures of CS in the high-resolution wide-field astronomical surveys GOODS \cite{Christiansen:2008vi} and COSMOS \cite{Christiansen:2010zi}. It has been claimed that constraints from gravitational lensing surveys at radio frequencies like LOFAR and SKA could reach $G\mu \lesssim 10^{-9}$ \cite{Mack:2007ae}. 

\subsection{Temperature anisotropies in the CMB}
\label{sec:temperature_anisotropy}
There are two possible effects from CS on  temperature fluctuations in the CMB:
\begin{enumerate}
\item 
CS moving through the line-of-sight can induce Doppler shifts on the photons coming from the last scattering surface, known as the Kaiser-Stebbins-Gott effect \cite{Gott:1984ef, Kaiser:1984iv,Bouchet:1988hh}, potentially leaving line-like discontinuities in the CMB.
\item 
A CS moving in the primordial plasma leaves surdensity perturbations, the so-called wakes \cite{Silk:1984xk}, possibly imprinted in the CMB temperature anisotropy.
Due to the stochastic behavior of the Kibble mechanism, these perturbations are decoherent and give rise to a CMB spectrum without acoustic peaks \cite{Pogosian:1999np}.
\end{enumerate}
Lattice numerical computation of the temperature anisotropy in Abelian-Higgs \cite{Lizarraga:2014xza, Lizarraga:2016onn}, Nambu-Goto \cite{Pogosian:1999np, Battye:2010xz, Charnock:2016nzm} or global strings \cite{Lopez-Eiguren:2017dmc} have constrained the string tension to $G\mu \lesssim \, \text{few} \times 10^{-7}$ \cite{Ade:2013xla}. Constraints of the same magnitude can be found from non-gaussianities \cite{Ringeval:2012tk, Ade:2013xla,Ciuca:2019oka}. Also, the same signatures as in the CMB can be imprinted in the $21$~cm power spectrum, and an experiment with a collecting area of $10^4-10^6$~km$^2$ might constrain $G\mu \lesssim 10^{-10}-10^{-12}$ \cite{Khatri:2008zw}.


\subsection{Non-gravitational radiation}

\label{sec:particle_prod_pheno}

As discussed in Sec.~\ref{sec:massive_radiation}, the presence of small-scale structures on local strings, cusps and kinks, invalidates the Nambu-Goto approximation and implies the radiation of massive particles. Therefore, CS have been proposed as a possible mechanism for generating non-thermal Dark Matter  \cite{Jeannerot:1999yn, Matsuda:2005fb, Cui:2008bd, Long:2019lwl}.

At a cusp, the string can reach ultrarelativistic velocities. Therefore, CS have been pointed \cite{MacGibbon:1989kk, Bhattacharjee:1998qc, Berezinsky:1998ft, Berezinsky:2011cp} as a possible candidate for the detection of ultra-high energy cosmic rays \cite{bird1994cosmic} above the Greisen-Zatsepin-Kuzmin (GZK) cut-off, around $10^{20}$~eV \cite{PhysRevLett.16.748, zatsepin1966upper,Abbasi:2007sv}, even though the expected flux at earth is generally too small to be detected \cite{Bhattacharjee:1989vu,Srednicki:1986xg,Gill:1994ic}.

More precisely, upon introducing a coupling between the SM and the dark $U(1)'$ from which the CS result, e.g. a Higgs portal or a kinetic mixing, an effective interaction between SM particles and the CS arises \cite{Hyde:2013fia}. In that case, the formation of cusps and kinks on the string radiate SM particles \cite{Long:2014mxa}. The expected gamma-ray flux at the earth is too low to be observed by Fermi-Lat \cite{Long:2014lxa} , also if we assume that all the massive particles radiated by the CS subsequently decay into gamma-ray \cite{Auclair:2019jip}. However, for cusp domination and for large coupling between the SM and $U(1)'$, the flux of high-energy neutrino might be measured by the future experiments SKA and LOFAR for $G\mu \sim [10^{-14},\, 10^{-16}]$ \cite{Long:2014lxa}. Also, the distortions in the CMB may be detected by the future telescope PIXIE for $G\mu \sim [10^{-12},\, 10^{-14}]$ \cite{Long:2014lxa}. Finally, depending on the magnitude of the SM-$U(1)'$ coupling, the  BBN constraints can already exclude values of string tensions between $10^{-8} \gtrsim G\mu \gtrsim 10^{-14}$ \cite{Long:2014lxa}. 

Constraints from particle emission apply on an interval of values for $G\mu$, and not as upper bound like for gravitational emission ~\cite{Bhattacharjee:1989vu}.
For longer lifetimes $\propto (\Gamma G \mu)^{-1}$, there are more loops and we expect a larger flux of emitted particles while  gravitational emission grows with $G\mu$. At small $G\mu$, loops decay preferentially into particles, c.f. sec.~\ref{sec:massive_radiation}. In that case, the expected flux of emitted particles increases with the string tension which controls the power of the particle emission. Therefore, there exists a value of $G\mu$ for which the expected flux of emitted particles is maximal. This is the value of $G\mu$ when particle production is as efficient as gravitational production. For example for loops created at the recombination time, the value of $G\mu$ maximizing the cosmic ray production is $10^{-18}$ \cite{Laliberte:2019kpb}.

\paragraph{Superconducting Cosmic Strings:} an other possibility for generating large particle production is to couple the CS with electromagnetic charge carriers and to spontaneously break electromagnetic gauge invariance inside the vortex \cite{Witten:1984eb}. Upon moving through cosmic magnetic fields, Superconducting Cosmic Strings (SCS) are able to develop a large electric current $\mathcal{I}$. The formation of cusps on SCS is expected to emit bursts of electromagnetic radiation \cite{Vilenkin:1986zz, Spergel:1986uu, Copeland:1987yv, BlancoPillado:2000xy}, up to very high energies, set by the string tension $ \sqrt{\mu}\sim 10^{13}~\textrm{GeV}\,\sqrt{G\mu/10^{-15}}$, hence leading to high-energy gamma-rays \cite{Babul:1987lza, Berezinsky:2001cp, Cheng:2010ae}. 
Hence, SCS could be an explanation for the observed gamma-ray bursts at high redshifts, which depart from the predictions from star-formation-history \cite{Cheng:2010ae}.
However, the expected photon flux at earth is larger in the radio band than in the gamma-ray band 
\cite{Vachaspati:2008su, Zadorozhna:2009zza, Cai:2011bi} (but also mostly generated by kinks instead of cusps \cite{Cai:2012zd}). Thus, it has been proposed that SCS could be 
an explanation for the Fast-Radio-Burst events \cite{Yu:2014gea, Ye:2017lqn, Brandenberger:2017uwo} for string tensions in the range $G\mu \sim [10^{-12},\, 10^{-14}]$ and string currents $\mathcal{I}\sim [10^{-1}, \, 10^2]~\text{GeV}$ \cite{Ye:2017lqn}.
Electromagnetic emission from SCS lead to CMB distortions \cite{Sanchez:1988ek, Sanchez:1990kj, Tashiro:2012nb}. A next-generation telescope like PIXIE \cite{Kogut:2011xw} would exclude string tensions $G\mu \sim 10^{-18}$, for string currents as low as $\mathcal{I} \sim 10^{-8}~$GeV \cite{Tashiro:2012nb}.  Also, electromagnetic radiation, by increasing the ionization fraction of neutral hydrogen, can affect the CMB temperature and polarization correlation functions at large angular scales, leading to the contraint $\mathcal{I}\lesssim 10^{7}$~GeV \cite{Tashiro:2012nv}. Note that ionization of neutral hydrogen has been studied in \cite{Laliberte:2019kpb} in the case of non-superconducting strings. Additionally, the radio emission from SCS can increase the depth of the 21 cm absorption signal, and EDGES data excludes the SCS tension $G\mu \sim 10^{-13}$ for string currents as low at $\mathcal{I}\sim 10$~GeV. Finally, emission of boosted charge carriers from SCS cusps moving in a cosmic magnetic field $B$, has been studied in \cite{Berezinsky:2009xf}, and provide a possible explanation for high-energy neutrino above $10^{20}$~eV, for $G\mu \sim [10^{-14},\,10^{-20}]$.

\section{Derivation of the GW spectrum from CS}
\label{app:derivationGWspectrum}

In this appendix we provide the steps leading to Eq.~(\ref{kmode_omega}).

\subsection{From GW emission to detection}
The GW energy density spectrum today is defined as 
\begin{equation}
\Omega_{\rm GW}(f) = \frac{f}{\rho_{c}}\, \left| \frac{d \rho_{\rm GW}(f, \, t_0)}{df} \right|.
\end{equation}
After emission, the GW energy density redshifts as radiation,  $\rho_{\rm GW}\propto a^{-4}$, so the GW energy density per unit of frequency redshifts as
\begin{equation}
\frac{d \rho_{\rm GW}(f, \, t_0)}{df} =\frac{d \rho_{\rm GW}(\tilde{f}, \, \tilde{t})}{d\tilde{f}}  \, \left(\frac{a(\tilde{t})}{a(t_0)}\right)^3
\end{equation}
where the frequency at emission $\tilde{f}$ is related to the frequency today $f$ through 
$$\tilde{f}=\frac{a(t_0)}{a(\tilde{t})} f.$$  
\subsection{From loop production to GW emission}
After its formation at $t_{i}$, a loop shrinks through emission of GW with a rate $\Gamma G \mu$ so that its length evolves as, c.f.~Sec.~\ref{sec:masslessradiation}
\begin{equation}
\label{eq:CSlength_app}
l(t) = \alpha t_i -\Gamma G \mu(t-t_i),
\end{equation}
where $\alpha$ is the length at formation in units of the horizon size.
The resulting GW are emitted at a frequency $\tilde{f}$ corresponding to one of the proper modes of the loop, i.e. 
\begin{equation}
\label{eq:GWspecfreq_app}
\tilde{f}=\frac{2k}{l}, \qquad  k\in\mathbb{Z}^{+}.
\end{equation}  
The GW energy rate emitted by one loop through the mode $k$ is, c.f. Sec.~\ref{sec:masslessradiation}
\begin{equation}
\frac{dE_{\rm GW}^{(k)}}{dt}=\Gamma^{(k)}\, G \mu^2, \qquad \text{with} \quad\sum_k\Gamma^{(k)} = \Gamma,
\end{equation}
where 
\begin{equation}
\label{eq:Gamma_k}
\Gamma^{(k)}= \frac{\Gamma \, k^{-4/3} }{ \sum_{p=1}^{\infty}p^{-4/3}} \simeq \frac{\Gamma \, k^{-4/3} }{ 3.60},
\end{equation}
which assumes that the GW emission is dominated by cusps.
The GW energy density spectrum resulting from the emission of all the decaying loops until today is
\begin{equation}
\frac{d\rho_{\rm GW}(\tilde{f}, \, \tilde{t})}{d\tilde{f}} = \int_{t_F}^{t_0} d\tilde{t} \, \frac{dE_{\rm GW}}{d\tilde{t}} \, \frac{d n(\tilde{f}, \, \tilde{t})}{d\tilde{f}},
\end{equation}
where $d n(\tilde{f}, \, \tilde{t})/d\tilde{f}$ is the number density of loops emitting GW at frequency $\tilde{f}$ at time $\tilde{t}$ and $t_0$ is the age of the universe today. Loops start being created at time of CS network formation $t_F$,  after the damped evolution has stopped, c.f. Sec.~\ref{sec:VOS_proof}.  \\
\subsection{The loop production}
In Sec.~\ref{sec:mainAssumptions}, we assume the loop-formation rate to be
\begin{equation}
\label{eq:LoopProductionFctBody_app}
\frac{dn}{dt_i}=(0.1)\frac{C_{\rm eff}(t_i)}{\alpha \, t_i^4},
\end{equation}
where $C_{\rm eff}(t_i)$ is the loop-formation efficiency.
We deduce the loop number density per unit of frequency
\begin{align}
\label{eq:loop_density_Jac}
\frac{d n(\tilde{f}, \, \tilde{t})}{d\tilde{f}} &=\left[\frac{a(t_i)}{a(\tilde{t})}\right]^3  \, \frac{dn}{dt_i} \cdot \frac{dt_i}{dl} \cdot \frac{dl}{d\tilde{f}} \\
&=\left[\frac{a(t_i)}{a(\tilde{t})}\right]^3  \,  \sum_{k} (0.1)\frac{C_\textrm{eff}(t_i)}{t_i^4} \cdot \frac{1}{\alpha \, (\alpha+\Gamma G\mu)} \cdot \frac{2k}{f^2} \left[\frac{a(\tilde{t})}{a(t_o)}\right]^2.
\end{align}
\subsection{The master equation}
Finally, we get the GW energy density spectrum
\begin{align}
\label{eq:master_eq_app}
\Omega_{\rm GW}(f) &=\sum_{k} \Omega_{\rm GW}^{(k)}(f) \notag \\ 
&= \sum_{k}\frac{1}{\rho_{c}}\, \frac{2k}{f} \, \frac{\mathcal{F}_{\alpha}\, \Gamma^{(k)} \, G\mu^2}{\alpha \, (\alpha+\Gamma\,G\mu)} \int_{t_{\rm osc}}^{t_{0}} d\tilde{t} \; \frac{C_\textrm{eff}(t_{i})}{t_{i}^4}\, \left[\frac{a(\tilde{t})}{a(t_0)}\right]^5 \, \left[\frac{a(t_i)}{a(\tilde{t})}\right]^3 \, \theta(t_i-t_{\rm osc})\,\theta(t_i-\frac{l_*}{\alpha}).
\end{align}
The first Heaviside function stands for the time $t_{\rm osc}$ at which long-strings start oscillating, either just after formation of the long-string network or after that friction becomes negligible, c.f. Sec.~\ref{sec:thermal_friction}. The second Heaviside function stands for the energy loss into particle production which is more efficient than GW emission for loops of length smaller than a characteristic length $l_*$, which depends on the string small-scale structure, c.f. sec~\ref{sec:massive_radiation}.
The time $t_i$ of formation of the loops, which emit at time $\tilde{t}$ and which give the detected frequency $f$, can be determined from Eq.~\eqref{eq:CSlength_app} and Eq.~\eqref{eq:GWspecfreq_app}
\begin{equation}
\label{eq:def_t_i}
t_i (f, \, \tilde{t})= \frac{1}{\alpha+\Gamma G \mu} \left[ \frac{2k}{f}\frac{a(\tilde{t})}{a(t_0)} + \Gamma G \mu \, \tilde{t} \right].
\end{equation}
Note that the contribution coming from the higher modes are related to the contribution of the first mode by
\begin{equation}
\label{eq:mode_k_vs_mode_1}
 \Omega_{\rm GW}^{(k)}(f) = k^{-4/3} \,  \Omega_{\rm GW}^{(1)}(f/k).
\end{equation}

\subsection{The GW spectrum from the quadrupole formula}
\label{sec:quadrupole_formula}

\paragraph{In standard cosmology:}
The scaling behavior $\Omega_{\rm GW} \propto  \sqrt{G \mu } \times f^0$, e.g. discussed along Eq.~\eqref{eq:lewicki_formula_GW_spectrum_radiation}, can be understood qualitatively from the quadrupole formula for the power emission of GW \cite{maggiore2008gravitational,Vachaspati:1984gt} 
\begin{equation}
\label{eq:GW_power_quadrupole}
P_{\rm GW}\sim N_{\rm loop} \, \frac{G}{5} \left(Q^{'''}_{\rm loop}\right)^2,
\end{equation}
where the triple derivative of the quadrupole of a loop is simply the string tension
\begin{equation}
\label{eq:GW_quadrupole}
Q^{'''}_{\rm loop} \sim \textrm{mass}\times \textrm{length}^2/\textrm{time}^{3} \sim \mu.
\end{equation}
 During the scaling regime, the number of loops formed at time $t_i$ scales as $t_i^{-3}$. Hence, the number of loops formed at time $t_i$, evaluated at a later time $\tilde{t}$ is 
\begin{equation}
\label{eq:loop_number_naive}
N_{\rm loop} \sim  \left(\frac{\tilde{t}}{t_i} \right)^3 \left(\frac{t_i }{ \tilde{t} }\right)^{\! 3/2},
\end{equation}
where the second factor accounts for the redshift as $a^{-3}$ of the loops between $t_i$ and $\tilde{t}$ during radiation.
Since GW redshift as radiation, their energy density today is
\begin{equation}
\label{eq:GW_amplitude_naive}
\Omega_{\rm GW} \sim \Omega_{\rm rad}~ \frac{\rho_{\rm GW}(\tilde{t})}{\rho_{\rm rad}(\tilde{t})} \sim  \Omega_{\rm rad}  \left( G \mu \right)^2 \left(\frac{ \tilde{t} }{t_i }\right)^{\! 3/2},
\end{equation}
where we assumed radiation-domination at $\tilde{t}$
\begin{equation}
\label{eq:rho_rad_radiation}
\rho_{\rm rad}(\tilde{t}) \sim G^{-1} \,\tilde{H}^2 ~\frac{\rho_{\rm rad}(\tilde{t})} {\rho_{\rm tot}(\tilde{t}) } ~\sim
G^{-1} ~\tilde{t}^{-2},
\end{equation}
and where we used that the energy density of GW at $\tilde{t}$ is
\begin{equation}
\rho_{\rm GW}(\tilde{t}) \sim \left( P_{\rm GW} ~\tilde{t} \right) / \,\tilde{t}^{\,3}.
\end{equation}
with the GW power $ P_{\rm GW}$ defined in Eq.~\eqref{eq:GW_power_quadrupole}.
From Eq.~\eqref{eq:loop_number_naive}, one can see that, at a fixed formation time $t_i$, the later the GW emission, the more numerous the loops. Hence, the dominant contribution to the SGWB from a given population of loops formed at $t_i$ occurs after one loop-lifetime, c.f. Eq.~\eqref{eq:GWlifetime}, at 
\begin{equation}
\label{eq:main_emission_naive}
\tilde{t}_{\rm M} \sim  \frac{\alpha \, t_i }{\Gamma G\mu}.
\end{equation}
Upon plugging Eq.~\eqref{eq:main_emission_naive} into Eq.~\eqref{eq:GW_amplitude_naive}, one gets
\begin{equation}
\label{eq:GW_amplitude_final_naive}
\Omega_{\rm GW} \propto  \sqrt{G\mu}\times f^0.
\end{equation}
From Eq.~\eqref{eq:GW_amplitude_naive}, we can see that the GW spectrum during radiation is set by a combination of the strength of the GW emission from loops, $(G\mu)^2$, and the loop-lifetime $\tilde{t}_{\rm M}/t_i$, c.f. Eq.~\eqref{eq:main_emission_naive}. Both are set by the triple derivative of the loop-quadrupole $Q^{'''}_{\rm loop}\sim \mu$. Hence we understand that the flatness in frequency during radiation is closely linked to the independence of the triple derivative of the loop-quadrupole \footnote{{ In contrast, the GW spectrum generated by domain walls during radiation is not flat since in that case the triple derivative of the wall-quadrupole depends on the emission time: $Q^{'''}_{\rm DW}\sim \sigma\,\tilde{t}$, where $\sigma$ is the wall energy per unit of area. Hence, the energy density fraction in GW before wall annihilation \cite{Saikawa:2017hiv} increases with time $\Omega_{\rm GW}^{\rm DW} \sim \left( G\, \sigma \,\tilde{t} \right)^2$.}}, c.f. Eq.~\eqref{eq:GW_quadrupole}, on the loop length, and therefore on the frequency. \\

\paragraph{In non-standard cosmology:}
\label{sec:quadrupole_formula_NS}
In presence of non-standard cosmology, possibly being different during loop formation $a(t_i)\propto t_i^{2/n}$ and GW emission $a(\tilde{t})\propto \tilde{t}^{2/m}$, one must modify how the number of emitting loops, c.f. Eq.~\eqref{eq:loop_number_naive}, and the GW energy density, c.f. Eq.~\eqref{eq:rho_rad_radiation}, get redshifted. Namely, Eq.~\eqref{eq:loop_number_naive} and Eq.~\eqref{eq:rho_rad_radiation}, here denoted by $\rm Std.$, become
\begin{equation}
\label{eq:loop_number_naive_NS}
N_{\rm loop} \sim \left(\frac{\tilde{t}}{t_i} \right)^3\left(\frac{a(t_i)}{a(\tilde{t})}\right)^3\propto N_{\rm loop}^{\rm Std.} \left(\frac{t_i}{\tilde{t}}\right)^{\frac{6}{m}-\frac{3}{2}},
\end{equation}
and
\begin{equation}
\label{eq:rho_rad_radiation_NS}
\frac{\rho_{\rm rad}(\tilde{t})} {\rho_{\rm tot}(\tilde{t}) }\sim \left( \frac{a(\tilde{t})}{a(t_{\rm end})} \right)^{n-4} \propto ~\tilde{t}^{\,\frac{2(4-n)}{n}},
\end{equation}
where $t_{\rm end}$ is the ending time of the non-standard cosmology. 
The GW spectrum, c.f. Eq.~\eqref{eq:GW_amplitude_naive}, depends on the combination of the loop- and GW-redshift factors in Eq.~\eqref{eq:loop_number_naive_NS} and Eq.~\eqref{eq:rho_rad_radiation_NS}. Upon plugging the scaling $t_i \propto f^{-2}$ and $\tilde{t}_M \propto f^{-2}$ (which are themselves deduced from $t_i \propto a(\tilde{t})/f$, c.f. Eq.~\eqref{eq:GWspecfreq_app}, and Eq.~\eqref{eq:main_emission_naive}), we obtain
\begin{equation}
\label{eq:spectral_index_omega_GW}
\Omega_{\rm GW} \propto f^{4(1 - \frac{3}{m} - \frac{1}{n})}.
\end{equation}
When loop formation at $t_i$ occurs during matter/kination but GW emission at $\tilde{t}_{\rm M}$ occurs during radiation, $(m,\, n) = (3, \, 4)/(6, \, 4)$, we find that the GW spectrum scales like $f^{-1}$/$f^1$. { In Sec.~\ref{sec:study_impact_mode_nbr}, we show that the presence of high-frequency modes $k\gg 1$ turns the $f^{-1}$ behavior to $f^{-1/3}$. }

\subsection{Impact of the high-frequency proper modes of the loop}
\label{sec:study_impact_mode_nbr}
	
			  \begin{figure}[h!]
				\centering
				\centering
			\includegraphics[width=0.45\textwidth]{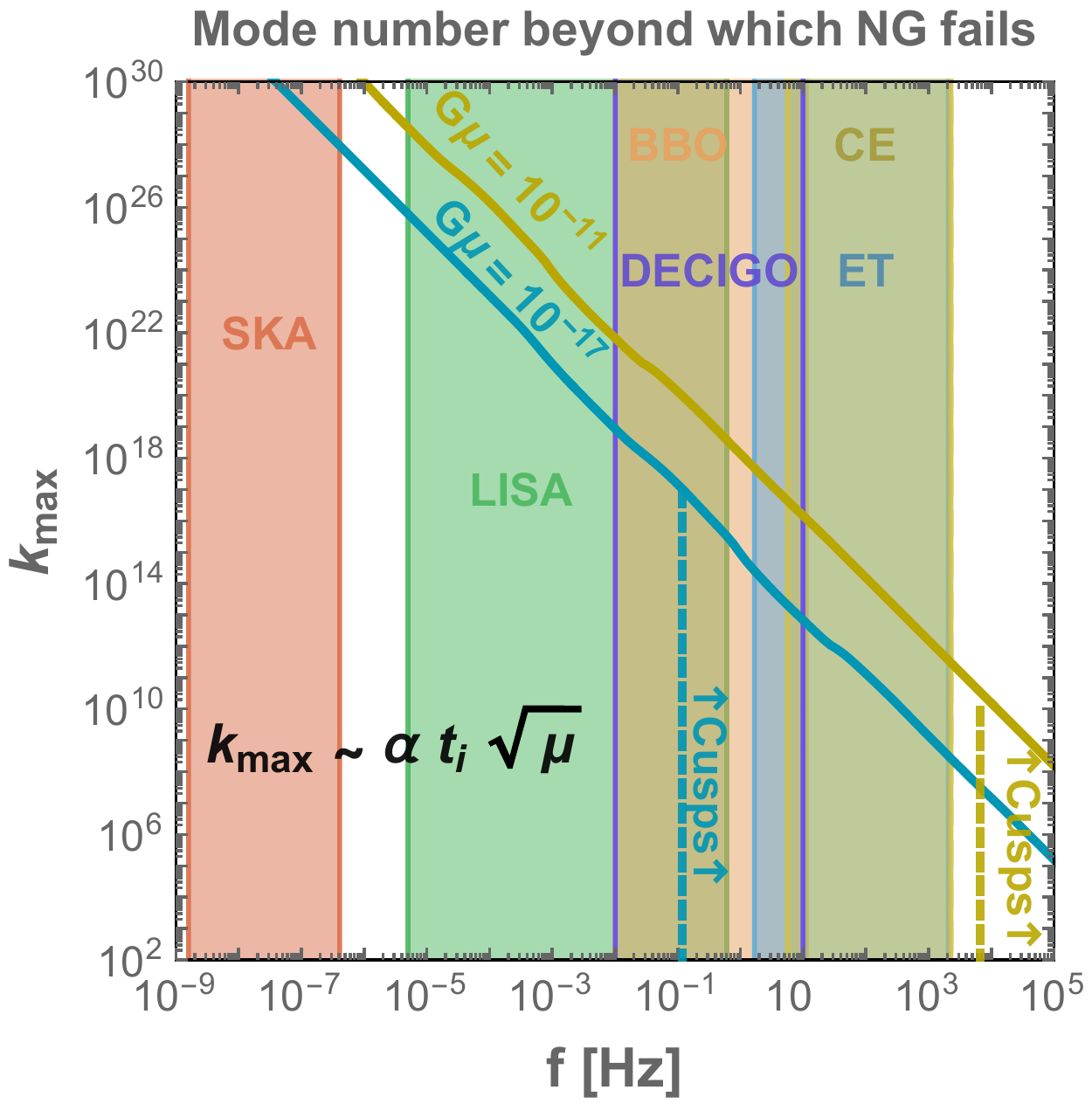}
			\includegraphics[width=0.45\textwidth]{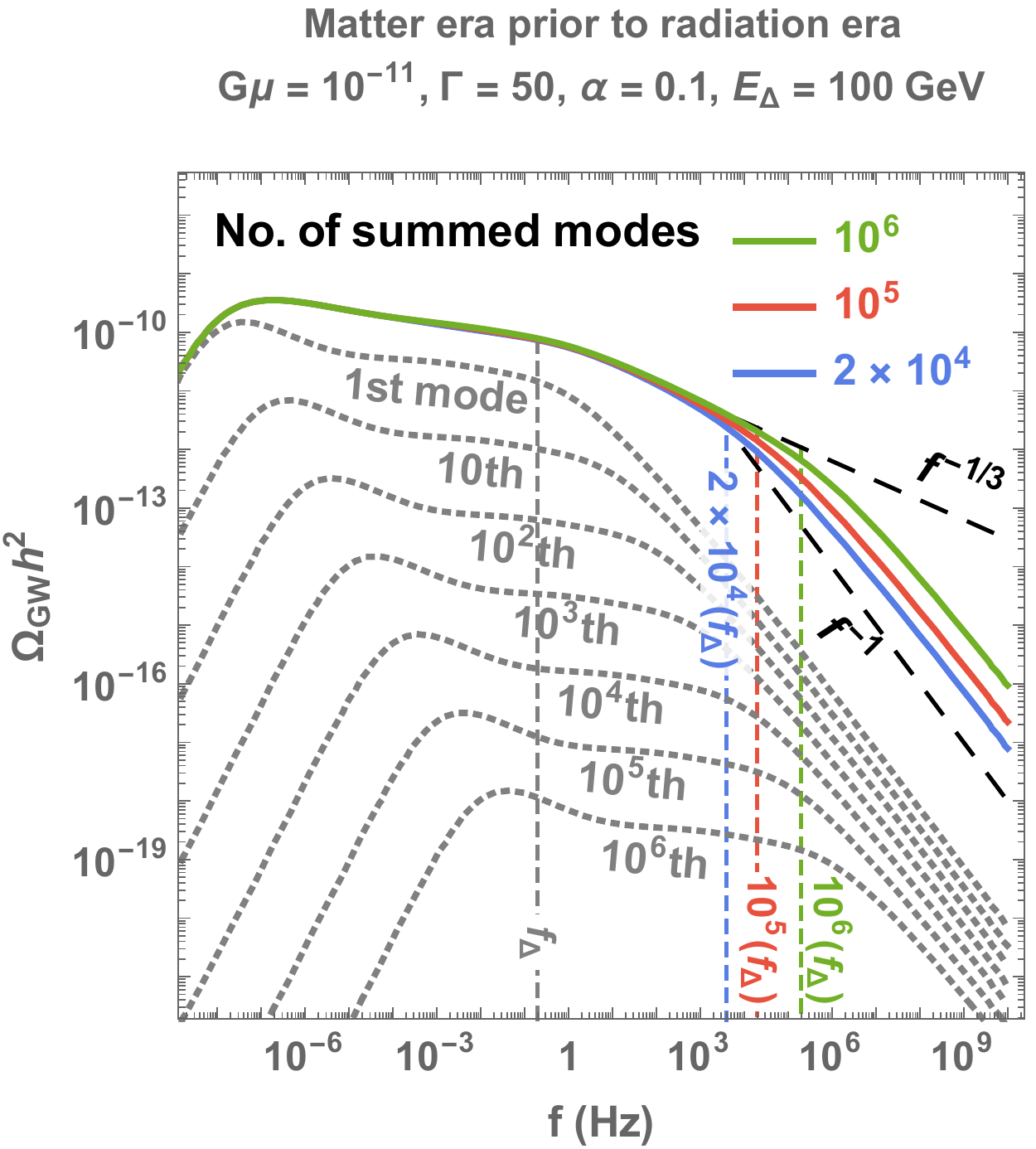}
				\caption{\it \small \textbf{Left:} Maximal mode number $k_{\rm max}$ beyond which we can not trust the Nambu-Goto approximation anymore. It occurs when the wavelength of the oscillation, given by $\alpha \, t_i/k$ where $t_i$ is the Hubble horizon when the loop forms, becomes of the order of the loop thickness $\mu^{-1/2}$. We can see that in the different interferometer windows, $k_{\rm max}$ is extremely large, often much larger than the maximal mode number tractable numerically $\sim 10^6$. \textbf{Right:} Decomposition of a GW spectrum under the contributions coming from the different proper modes of the loop. We can see that high-k modes are responsible for the change of slope $f^{-1/3} \, \rightarrow \,f^{-1}$ between the physical turning point frequency $f_{\rm \Delta}$ and a second, artificial, turning-point $f_{\rm max}$, given by $f_{\rm max} = k_{\rm max}\,f_{\rm \Delta}$, c.f. Eq.~\eqref{eq:second_turning_point}, where $k_{\rm max}$ is the total number of modes chosen for doing the computation, here $2\times 10^4$, $10^5$ and $10^6$. Except when explicitly specified, for technical reasons we fix $k_{\rm max}=2 \times 10^4$ modes in all the plots of our study. }
				\label{fig:max_mode_number}
			\end{figure}

\paragraph{The motivation:}
When computing the GW spectrum from cosmic strings, given by Eq.~\eqref{eq:master_eq_app}, we are confronted with an infinite sum over the proper modes $k$ of the loop. The number of modes before we violate the Nambu-Goto approximation is very large, as shown in the left panel of Fig.~\ref{fig:max_mode_number}. 
In what follows, we study the impact of the high-frequency modes on the GW spectrum.
From Eq.~\eqref{eq:GWspecfreq_app}, Eq.~\eqref{eq:Gamma_k} and Eq.~\eqref{eq:master_eq_app}, we can see that the spectrum for the $k^{\rm th}$ mode is related to the fundamental spectrum $k=1$ through Eq.~\eqref{eq:mode_k_vs_mode_1}, which we rewrite here
\begin{equation}
\label{eq:mode_k_vs_mode_1_(2)}
\Omega_{\rm GW}^{(k)}(f) = k^{-\delta} \,  \Omega_{\rm GW}^{(1)}(f/k),
\end{equation}
In this study, we fix $\delta = 4/3$ since we assume that the small-scale structure is dominated by cusps. However, the results of the present section apply to any small-scale structure described by $\delta$.

\paragraph{Case of a fundamental spectrum with a flat slope:}
At first, if we assume that the one-mode spectrum is flat, $\Omega_{\rm GW}^{(1)}(f) \propto f^0$, then the total spectrum is a simple rescaling of the fundamental spectrum by the Riemann zeta function
\begin{equation}
\Omega_{\rm GW}(f) =\zeta\left(\delta\right) \,\Omega_{\rm GW}^{(1)}(f),
\end{equation}
where in particular, $\zeta(4/3)=\sum_k k^{-4/3} \simeq 3.60$.

\paragraph{Case of a fundamental spectrum with a slope $f^{-1}$:}
Now, we consider the case where the fundamental spectrum has a slope $f^{-1}$, as expected in thepresence of an early matter era, c.f. Eq.~\eqref{eq:spectral_index_omega_GW}, but also, in the presence of high-frequency cut-offs. The high-frequency cut-offs in the spectrum are described by Heaviside functions in the master formula in Eq.~\eqref{eq:master_eq_app}, of the type $\Theta\left( t_i \, - \, t_{\rm \Delta}  \right)$, where $t_{\rm \Delta}$ is the cosmic time when loop formation starts, assuming it is suppressed before on. The time $t_{\rm \Delta}$ can correspond to either the time of formation of the network, c.f. Eq.~\eqref{eq:network_formation}, the time when friction-dominated dynamics become irrelevant, c.f.  App.~\ref{sec:thermal_friction}, the time when gravitational emission dominates over massive particle production, c.f. Sec.~\ref{sec:massive_radiation}, or the time when the string correlation length re-enters the Hubble horizon after a short period of second inflation, c.f. Sec.~\ref{sec:inflation}.  The slope of the $k=1$ spectrum beyond the cut-off frequency can be read from Eq.~\eqref{eq:master_eq_app} after injecting Eq.~\eqref{eq:def_t_i} and $t_i=t_{\rm \Delta}$, where we find
\begin{equation}
\label{eq:first_mode_spectrum}
\Omega_{\rm GW}^{(1)}(f) =  \Omega_\Delta \Theta\left( -f + f_\Delta  \right) \,+\, \Omega_\Delta \,  \frac{f_\Delta }{f} \,\Theta\left( f - f_\Delta  \right).
\end{equation}
The fundamental spectrum is flat until $f_\Delta $ and then shows a slope $f^{-1}$ beyond.
The total spectrum, summed over all the proper modes, can be obtained from Eq.~\eqref{eq:mode_k_vs_mode_1_(2)} and Eq.~\eqref{eq:first_mode_spectrum}
\begin{equation}
\label{eq:spectrum_slope_m1}
\Omega_{\rm GW}(f)  = \sum_{k=1}^{\rm k_{\Delta }} \,\frac{\Omega_\Delta }{k^{\delta}}\, k \,\frac{f_\Delta }{f}  \,+\, \sum_{k=k_\Delta }^{\rm k_{\rm max}} \frac{\Omega_\Delta  }{k^{\delta}},
\end{equation}
where $k_{\rm max}$ is the maximal mode, chosen arbitrarily, and $k_\Delta $ is the critical mode defined such that modes with $k<k_\Delta $ have a slope $f^{-1}$ whereas modes with $k>k_\Delta $ have a flat slope. For a given frequency $f$, the critical mode number $k_\Delta$ obeys
\begin{equation}
\label{eq:critical_mode}
k_\Delta   \simeq  \frac{f}{f_\Delta }.
\end{equation}
We now evaluate Eq.~\eqref{eq:spectrum_slope_m1} in the large $k_\Delta $ limit, while still keeping $k_\Delta < k_{\rm max}$
\begin{equation}
\label{eq:spectrum_slope_m1_(2)}
\Omega_{\rm GW}^{1\ll k_\Delta <k_{\rm max}}(f)  \simeq  \Omega_\Delta \frac{f_\Delta }{f} k_\Delta ^{2-\delta}  \,+\, \frac{1}{\delta-1} \frac{\Omega_\Delta }{k_\Delta ^{\delta-1}} ,
\end{equation}
where we have used the asymptotic expansion of the Euler-Maclaurin formula for the first term and the asymptotic expansion of the Hurwitz zeta function for the second term. Finally, after injecting Eq.~\eqref{eq:critical_mode}, we get
\begin{equation}
\label{eq:spectrum_slope_m1_(3)}
\Omega^{1\ll k_\Delta <k_{\rm max}}_{\rm GW}(f)  \simeq \frac{\delta}{\delta-1}\Omega_\Delta  \left( \frac{f_\Delta }{f} \right)^{\delta - 1} \propto \begin{cases}
f^{-1/3} & \mbox{for cusps ($\delta=4/3$) } \\
f^{-2/3} & \mbox{kinks ($\delta = 5/3$)} \\
f^{-1}  &  \mbox{kink-kink collisions ($\delta = 2$) }\\
\end{cases}
\end{equation}
We conclude that the spectral index beyond a high-frequency turning point $f_\Delta $ due to an early matter era, a second inflation era, particle production, thermal friction domination, or the formation of the network, is modified by the presence of the high-k modes in a way that depends on the small-scale structure. Particularly, if the small-scale structure is dominated by cusps, we find a slope $-1/3$. We comment on the possibility to get information about the nature of the small-structure from detecting a GW spectrum from CS with a decreasing slope. The study \cite{Blasi:2020wpy} was the first one to point out the impact of the high-frequency modes on the value of a decreasing slope.

\paragraph{Impact of fixing the total number of proper modes:}
For technical reasons we are unavoidably forced to choose a maximal number of modes $k_{\rm max}$. We now study the dependence of the GW spectrum on the choice of $k_{\rm max}$. The evaluation of Eq.~\eqref{eq:spectrum_slope_m1} for $k_\Delta  > k_{\rm max}$ leads to
\begin{equation}
\label{eq:spectrum_slope_m1_(3)}
\Omega^{1\ll k_{\rm max} <k_\Delta }_{\rm GW}(f) = \zeta\left(\delta-1\right) \,\Omega_\Delta \,  \frac{f_\Delta }{f}.
\end{equation}
Hence, in addition to the initial physical turning point $f_\Delta $, where the slope changes from flat to $f^{-1/3}$, there is a second artificial turning point $f_{\rm max}$ given by
\begin{equation}
\label{eq:second_turning_point}
f_{\rm max} = k_{\rm max}\,f_{\Delta },
\end{equation}
where the slope changes from $f^{-1/3}$ to $f^{-1}$.  We show the different behaviors in the right panel of Fig.~\ref{fig:max_mode_number}.

\paragraph{Case of a fundamental spectrum with a slope $f^{+1}$:}
As last, we comment on the case where the fundamental spectrum has a slope $f^1$, as in the case of an early kination era, c.f. Eq.~\eqref{eq:spectral_index_omega_GW}. Repeating the same steps as in Eq.~\eqref{eq:spectrum_slope_m1}, we obtain
\begin{equation}
\label{eq:spectrum_slope_p1}
\Omega_{\rm GW}(f)  = \zeta\left(\delta+1\right)\,\Omega_\Delta \,\frac{f}{f_\Delta } ,
\end{equation}
hence the slope of the full spectrum is the same as the slope of the fundamental spectrum.

\section{Derivation of the frequency - temperature relation}
\label{derive_turning_points}
In this appendix, we compute the correspondence between an observed frequency $f$ and the temperature $T$ of the universe when the loops responsible for that frequency have been formed. 

		\begin{figure}[h!]
				\centering
				\centering
			\includegraphics[width=0.6\textwidth]{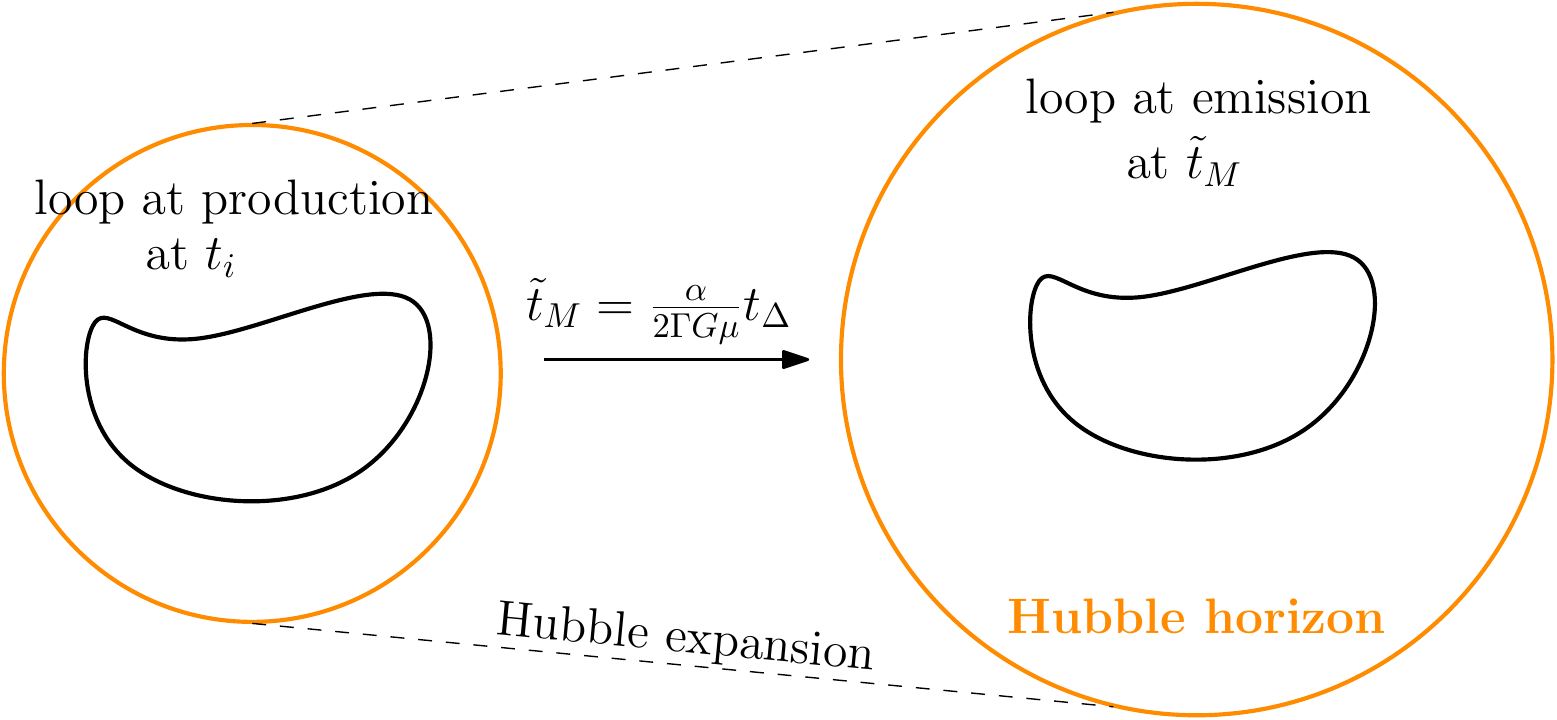}
				\caption{\it \small Loops produced at time $t_i$ contribute to the GW spectrum much later, when they have accomplished half of their lifetime, at $\tilde{t}_{\rm M}\simeq\alpha\, t_i/  (2\Gamma G \mu)$. Hence GW emitted from cosmic-string loops are exempt from a redshift factor $a(\tilde{t}_{\rm M})/a(t_i)$ so have much higher frequency than GW produced from other sources at the same energy scale.}
				\label{cartoon_loop_maximal_decay}
			\end{figure}
\subsection{In standard cosmology}
According to the scaling of the loop-formation rate ${dn}/{dt_i}\propto t_i^{-4}$, the main contribution to the GW emission at time $\tilde{t}$ comes from the loops created at the earliest epoch. Correspondingly, loops created at $t_i$ contribute to the spectrum as late as possible, at the \textit{main emission time} $\tilde{t}_{\rm M}$. The latest emission time is set by the loop lifetime $\alpha \, t_i/ \Gamma G \mu$, where $\alpha$ is the loop-length at formation in horizon unit, c.f. Eq.~\eqref{eq:GWlifetime}. Hence, a loop produced at time $t_i$ mainly contributes to the spectrum, much later c.f. figure~\ref{cartoon_loop_maximal_decay}, at a time
\begin{equation}
\label{eq:t_M}
\tilde{t}_{\rm M} \simeq \frac{\alpha \, t_i}{ 2 \Gamma G \mu},
\end{equation}
where the factor $1/2$ is found upon maximizing the loop-formation rate ${dn}/{dt_i}\propto t_i^{-4}$ and upon assuming $\alpha\gg \Gamma G\mu$.
The loop length after half the loop lifetime, in Eq.~\eqref{eq:t_M}, is equal to half the length at formation $\alpha\, t_i/2$, c.f. Eq.~\eqref{eq:CSlength_app}. Hence the emitted frequency is set by
\begin{eqnarray}
\label{eq:tdelta_fdelta_eq_line1}
\alpha \, t_i &\simeq &\frac{4}{f}\frac{a(\tilde{t}_M)}{a(t_0)},\\
&\simeq & \frac{4}{f}\frac{a(\tilde{t}_M)}{a(t_\textrm{eq})}\frac{a(t_\textrm{eq})}{a(t_0)} \\
&\simeq& \frac{4}{f}\left(\frac{\tilde{t}_M}{t_\textrm{eq}}\right)^{1/2}\left(\frac{\,t_\textrm{eq}}{t_0}\right)^{2/3}
\label{tdelta_fdelta_eq}
\end{eqnarray}
where we used $f \,a(t_0)/a(\tilde{t}) = 2k/l$ and only considered the first Fourier mode $k=1$, c.f. Eq.~\eqref{eq:GWspecfreq_0}. By merging Eq.~\eqref{eq:t_M} and Eq.~\eqref{tdelta_fdelta_eq}, we obtain the relation between an observed frequency $f$ and the time $t_i$ of loop formation
\begin{equation}
f\simeq \sqrt{\frac{8z_\textrm{eq}}{\alpha\Gamma G\mu}}\left(\frac{t_\textrm{eq}}{t_i}\right)^{1/2}t_0^{-1},
\end{equation}
where the redshift at matter-radiation equality is  $z_{\rm eq} = \Omega_{\rm C}/\Omega_{\gamma} \simeq 3360$, and  $t_{\rm eq} \simeq 51.8$~kyrs (from integrating Eq.~\eqref{friedmann_eq}) and $t_0\simeq 13.8~$Gyrs \cite{Tanabashi:2018oca}.
Finally, using entropy conservation, we obtain the relation between the frequency $f$ at observation and the temperature $T$ of the universe when the corresponding loops are formed
		 \begin{align}
		 \nonumber
		f &\simeq \sqrt{\frac{8}{z_\textrm{eq}\alpha\Gamma G\mu}}\left(\frac{g_*(T)}{g_*(T_0)}\right)^{1/4}\left(\frac{T}{T_0}\right) \, t_0^{-1}\\
		&\simeq (6.7\times10^{-2}\textrm{ Hz})\left(\frac{T}{\textrm{GeV}}\right)\left(\frac{0.1\times50\times10^{-11}}{\alpha\Gamma G\mu}\right)^{1/2}\left(\frac{g_*(T)}{g_*(T_0)}\right)^{1/4}.
		\label{fdeltaApp}
		\end{align}
\subsection{During a change of cosmology}

The derivation of (\ref{fdeltaApp})  does not take into account the time-variation of  $C_{\rm eff}$.
It assumes that loops are produced and decayed during the scaling regime in the radiation era.
An observable to test the non-standard cosmology is 
 the frequency $ f_{\Delta}$ of the \textit{turning-point} defined as the frequency at which the GW spectrum starts to deviate from the standard-cosmology behavior and the spectral index changes.
We obtain different fitted values for this turning point frequency depending on the prescription. We quote below different expressions, depending whether we assume that the spectrum can be measured with a $10\%$ precision, and $1\%$ respectively. We compare the predictions obtained using a scaling and VOS network:
\begin{align}
f_{\Delta}\simeq \textrm{ Hz} \left(\frac{T_{\Delta}}{\textrm{GeV}}\right)\left(\frac{0.1\times50\times10^{-11}}{\alpha\Gamma G\mu}\right)^{1/2}\left(\frac{g_*(T_{\Delta})}{g_*(T_0)}\right)^{1/4} \times 
\begin{cases}
2 \times 10^{-3} &\textrm{for VOS}, 10 \% \\
45 \times 10^{-3} &\textrm{for scaling}, 10 \% \\
0.04 \times 10^{-3} &\textrm{for VOS}, 1 \% \\
15 \times 10^{-3} &\textrm{for scaling}, 1 \% \\
\end{cases}
\label{turning_point_general_scaling_app}
\end{align}
 Therefore, the turning point frequency is lower in VOS than in scaling  by a factor $\sim 22.5$ if we define the turning-point frequency by  an amplitude deviation of $10\%$ with respect to standard cosmology, and by a factor $\sim 375$ for a deviation of $1\%$. The loops contributing to this part of the spectrum have been formed at the time of the change of cosmology. When the cosmology changes, the network achieves a transient evolution in order to reach the new scaling regime. The long-string network needs extra time to transit from one scaling regime to the other, hence the shift in the relation between observed frequency and temperature of loop formation at the turning-point, c.f. Sec.~\ref{sec:turning_point_general}.

\subsection{In the presence of an intermediate inflation period}
		The above derivation of the relation between the observed frequency and the time of loop production assumes that cosmic-string loops are constantly being produced throughout the cosmic history.
		It does not apply if the network experiences an intermediate era of inflation.
		This case is discussed in Sec.~\ref{sec:turning_point_inf} and the turning-point formulae are, for a given precision
		\begin{align}
f_{\Delta}\simeq \textrm{ Hz} \left(\frac{T_{\Delta}}{\textrm{GeV}}\right)\left(\frac{0.1\times50\times10^{-11}}{\alpha\Gamma G\mu}\right)^{1/2}\left(\frac{g_*(T_{\Delta})}{g_*(T_0)}\right)^{1/4} \times 
\begin{cases}
1.5 \times 10^{-4} &\textrm{for } 10 \% \\
5 \times 10^{-6} &\textrm{for } 1 \% \\
\end{cases}
\label{turning_point_general_scaling_app_inf}
\end{align}

\subsection{Cut-off from particle production}
The cutoff frequency due to particle production is given in Sec.~\ref{UVcutoff}.

\section{Derivation of the VOS equations}
\label{sec:VOS_proof}

\subsection{The Nambu-Goto  string in an expanding universe}
The Velocity-dependent One-Scale equations (VOS) in Eq.~\eqref{eq:VOS_eq_body}, describe the evolution of a network of long strings in term of the mean velocity $\bar{v}$ and the correlation length $\xi = L/t$, see the original papers \cite{Martins:1995tg, Martins:1996jp, Martins:2000cs} or the recent review \cite{martins2016defect}. The set of points visited by the Nambu-Goto  string during its time evolution form a 2D manifold, called the world-sheet, described by time-like and space-like coordinates $\mathtt{t}$ and $\sigma$. The embedding of the 2D world-sheet in the 4D space-time is described by $x^\mu(\mathtt{t},\, \sigma)$ where $\mu =1,\,2,\,3,\,4$. The choice of the word-sheet coordinates being arbitrary, we have two gauge degrees of freedom which we can fix by imposing $\mathbf{\dot{x}}\cdot \mathbf{x'} = 0$ and $\mathtt{t}=\tau$ where $\tau$ is the conformal time of the expanding universe. The dot and prime denote the derivatives with respect to the time-like and space-like world-sheet coordinates, $ \mathbf{\dot{x}}\equiv d\mathbf{x}/d\mathtt{t} $ and  $ \mathbf{x'}\equiv d\mathbf{x}/d\sigma $. Then, the equations of motion of the Nambu-Goto string in a FRW universe are \cite{Turok:1984db}
\begin{eqnarray}
\mathbf{\ddot{x}}+2\mathcal{H}(1-\mathbf{\dot{x}}^2)\mathbf{\dot{x}}&=&\frac{1}{\epsilon}\left(\frac{\mathbf{x'}}{\epsilon}\right)',\\
\dot{\epsilon}+2\mathcal{H}\,\mathbf{\dot{x}}^2\,\epsilon&=&0,
\end{eqnarray}
where $\mathcal{H}\equiv \dot{a}/a=Ha$ and $\epsilon\equiv \sqrt{\mathbf{x'}^2/(1-\mathbf{\dot{x}}^2})$ is the coordinate energy per unit of length.
\subsection{The long-string network}
The macroscopic evolution of the long string network can be described by the energy density 
\begin{equation}
\label{vos_long_E}
\rho_{\infty}=\frac{E}{a^3}=\frac{\mu}{a^2(\tau)}\int \epsilon \,d\sigma\equiv\frac{\mu}{L^2},
\end{equation}
 and the root-mean-square averaged velocity
\begin{equation}
\bar{v}^2\equiv\langle\mathbf{\dot{x}}^2\rangle=\frac{\int\mathbf{\dot{x}}^2\epsilon \,d\sigma}{\int\epsilon \,d\sigma},
\label{vos_avg_v}
\end{equation}
where we recall that $\mu$ is the CS linear mass density.
\subsection{VOS 1: the correlation length}
Differentiating Eq.~\eqref{vos_long_E} gives the evolution of the energy density in an expanding universe
\begin{eqnarray}
\frac{d\rho_{\infty}}{dt}&=&\frac{d\rho_{\infty}}{d\tau}\cdot\frac{d\tau}{dt}=\frac{1}{a}\cdot\frac{d\rho_{\infty}}{d\tau},\\
&=&\frac{\mu}{a}\left[\frac{d}{d\tau}\left(\frac{1}{a^2}\right)\int\epsilon \,d\sigma + \frac{1}{a^2}\int\frac{d\epsilon}{d\tau}\,d\sigma\right],\\
&=&-2\frac{\mu}{a^3}\mathcal{H}\left[\int\epsilon \,d\sigma + \int\mathbf{\dot{x}}^2\epsilon \,d\sigma\right],\\
&=&-2H\rho_{\infty} \,( 1+ \bar{v}^2).
\label{E_network}
\end{eqnarray}
Moreover, after each string crossing, the network transfers energy into loops with a rate given by Eq.~\eqref{eq:energylossloop} and we get
\begin{equation}
\frac{d\rho_{\infty}}{dt}=-2H\rho_{\infty} \, ( 1+ \bar{v}^2)-\tilde{c} \, \bar{v}\frac{\rho_{\infty}}{L},
\end{equation}
which after using Eq.~\eqref{vos_long_E}, leads to the first VOS equation
\begin{equation}
\label{eq:VOS_L}
\text{VOS 1:} \qquad \frac{dL}{dt}=HL \,( 1+ \bar{v}^2)+\frac{1}{2}\tilde{c}\,\bar{v}.
\end{equation}
We neglect the back-reaction on long strings from gravitational emission which is suppressed with respect to the loop-chopping loss term by $O(G\mu)$. The case of global strings, for which however, the back-reaction due to particle production may play a role, is considered in App.~\ref{sec:VOS_global}.

\subsection{Thermal friction}
\label{sec:thermal_friction}

In addition to the Hubble friction, there can be friction due to the interactions of the strings with particles of the plasma, leading to the retarding force \cite{Vilenkin:1991zk}
\begin{equation}
F = \rho \, \sigma \, \bar{v} = \beta \, T^3\, \bar{v},
\end{equation}
where $\rho \sim T^4$ is the plasma energy density and $\sigma \sim T^{-1}$ is the cross-section per unit of length. The effect of friction is to damp the string motion and to suppress the GW spectrum \cite{Garriga:1993gj}. For gauge strings, a well-known realisation of friction is the interaction of charged particles with the pure gauge fields existing outside the string, the so-called Aharonov-Bohm effect \cite{Aharonov:1959fk}. In such a case, the friction coefficient $\beta$ is given by \cite{Vilenkin:1991zk}
\begin{equation}
\beta=2\pi^{-2}\zeta (3)\sum_i g_i \sin^2(\pi \nu_i),
\end{equation}
with
\begin{eqnarray*}
i &\equiv&\textrm{\hspace{1em} relativistic particle species ($m_i \ll T$),}\\
 g_i &\equiv& \textrm{\hspace{1em} number of relativistic degrees of freedom of $i$} \times \begin{cases}3/4 \textrm{ (fermion),}\\1 \textrm{ (boson),} \end{cases} \\
2\pi\,\nu_i \equiv e_i\,\Phi &\equiv& \textrm{\hspace{1em}  phase-shift of the wave-function of particle $i$ when transported on a}\\
&& \textrm{\hspace{1em}  close path around the string. $e_i$ being its charge under the associated }\\
&& \textrm{\hspace{1em}  gauge group and $\Phi$ the magnetic field flux along the string.}
\end{eqnarray*}
The friction term in the first VOS equation, Eq.~\eqref{eq:VOS_L}, becomes
\begin{equation}
2H\bar{v}^2\longrightarrow\frac{\bar{v}^2}{l_d} \equiv 2H\bar{v}^2+\frac{\bar{v}^2}{l_f},
\label{VOS_frictionlength_intro}
\end{equation}
where we introduced a friction length due to particle scattering $l_f \equiv \mu/(\sigma\, \rho) = \mu/(\beta T^3)$ \cite{Martins:1995tg, Martins:1996jp}, and the associated effective friction length $l_d$. At large temperature, the large damping due to the frictional force prevents the CS network to reach the scaling regime until it becomes sub-dominant when $2H \lesssim 1/l_f$, so after the time
\begin{align}
\label{eq:eq_friction}
t_\textrm{fric} &\simeq (2.5\times 10^{-5})~\left(\frac{106.75}{g_*(t_\textrm{fric})} \right)^{3/2}~\beta^2 ~(G\mu)^{-2}~ t_{\rm pl},\\
&\simeq (1.4\times10^{-4})~\left(\frac{g_*(t_\textrm{F})}{106.65} \right)^{1/2}\left(\frac{106.75}{g_*(t_\textrm{fric})} \right)^{3/2}~\beta^2 ~(G\mu)^{-1} ~ t_{F},
\end{align}
where $t_{\rm pl} \equiv \sqrt{G}$ and where the network formation time $t_F$ is the cosmic time when the energy scale of the universe is equal to the string tension $\rho_{\rm tot}^{1/2}(t_F) \equiv \mu$. 
For friction coefficient $\beta = 1$, the friction becomes negligible at the temperatures $T_* \simeq 4$~TeV for $G\mu=10^{-17}$, $T_* \simeq 400$~TeV for $G\mu=10^{-15}$, $T_* \simeq 40$~PeV for $G\mu=10^{-13}$, hence respectively impacting the SGWB only above the frequencies $20$~kHz, $200$~kHz, $2$~MHz, c.f. Eq.~\eqref{turning_point_general}, which are outside the GW interferometer windows, c.f. Fig.~\ref{ST_vos_scaling}. 

\begin{figure}[h!]
\centering
\raisebox{0cm}{\makebox{\includegraphics[width=0.49\textwidth, scale=1]{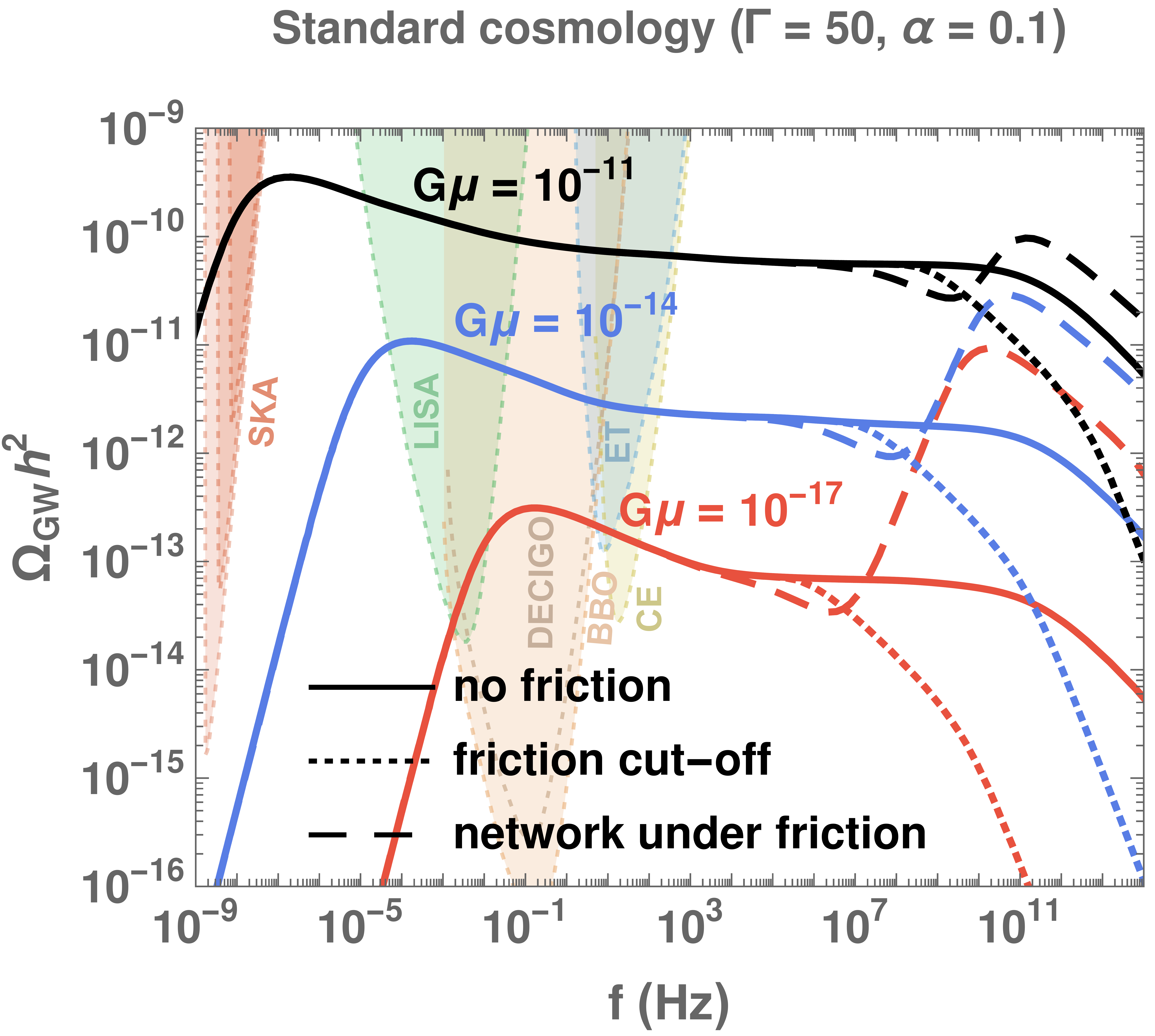}}}
\caption{\it \small GW spectrum from CS assuming no thermal friction (solid lines), thermal friction only at the level of the long-string network,  i.e. upon including eq. \eqref{VOS_frictionlength_intro} in the VOS equations (dashed lines) or thermal friction taken at the loop level, i.e. by removing GW emissions anterior to $t_{\rm fric}$ defined in Eq.~\eqref{eq:eq_friction} (dotted lines). See text for more details. A standard cosmology is assumed. } 
\label{fig:friction_spectrum}
\end{figure}

In Fig.~\ref{fig:friction_spectrum}, we show the impact of thermal friction on the GW spectrum from CS in two different ways.
\begin{itemize}
\item [$\diamond$]
\textbf{Network under friction}\textit{ (dashed lines in Fig.~\ref{fig:friction_spectrum}) :} the thermal friction is only taken into account at the level of the long-string network. Concretely, by simply including the friction term in Eq.~\eqref{VOS_frictionlength_intro} in the VOS equations. The GW peak at high frequency is due to the loop over-production by the frozen network, followed by a fast relaxation (with a little oscillatory behavior) to the scaling regime once friction becomes negligible with respect to Hubble expansion. This approach is insufficient since it assumes that the GW power emitted by loops is still given by $\Gamma G\mu^2$ with $\Gamma \simeq 50$ and therefore it does not take into account the damping of the oscillations of the loops under which we expect $\Gamma \rightarrow 0$.
\item [$\diamond$]
\textbf{GW emission cut-off}\textit{ (dotted lines in Fig.~\ref{fig:friction_spectrum}) :}  The damping of the loop oscillations is now taken into account by discarding all GW emissions happening earlier than $t_{\rm fric}$ in Eq.~\eqref{eq:eq_friction}, when thermal friction is larger than Hubble friction. Technically, the time $t_{\rm osc}$ of first loop oscillations in Eq.~\eqref{kmode_omega} is set equal to $t_\textrm{fric}$ in Eq.~\eqref{eq:eq_friction}. 
\end{itemize}
In many of our plots, e.g. Fig.~\ref{sketch_scaling} or Fig.~\ref{ST_vos_scaling}, we show the GW spectrum in presence of thermal friction with a gray line, computed according to the second prescription above, entitled `GW emission cut-off'.  Note that in most cases, the effect of friction manifests itself at very high frequencies, outside the observability band of planned interferometers. It could however become relevant if those high frequencies could be probed in future experiments.

\subsection{VOS 2: the mean velocity}
Differentiating Eq.~\eqref{vos_avg_v} gives the evolution of the averaged velocity, which constitutes the second VOS equation
\begin{equation}
\label{eq:VOS_v}
\text{VOS 2:}  \qquad \qquad \frac{d\bar{v}}{dt}=(1-\bar{v}^2)\left[\frac{k(\bar{v})}{L}-\frac{\bar{v}}{l_d}\right],
\end{equation}
with
\begin{equation}
k(\bar{v}) \equiv \frac{\left<(1-\mathbf{\dot{x}}^2)\,(\mathbf{\dot{x}} \cdot \mathbf{u}) \right>}{\bar{v}\,(1-\bar{v}^2)},
\end{equation}
where $\mathbf{u}$ is the unit vector aligned with the radius of curvature $\propto {d^2\mathbf{x}}/{d\sigma^2}$. $k(\bar{v})$ indicates the degree of wiggliness of the string. More precisely, $k(\bar{v})=1$ for a straight string and $k(\bar{v})\lesssim 1$ once we add small-scale structures.
We use the results from numerical simulations \cite{Martins:2000cs}
\begin{equation}
\label{eq:momentum_operator}
 k(\bar{v})=\frac{2\sqrt{2}}{\pi}(1-\bar{v}^2)(1+2\sqrt{2}\bar{v}^3)\frac{1-8\bar{v}^6}{1+8\bar{v}^6}.
\end{equation}
Eq.~\eqref{eq:VOS_v} is a relativistic generalization of Newton's law where the string is accelerated by its curvature $1/L$ but is damped by the Hubble expansion and plasma friction after a typical length $1/l_d$.
Equation \eqref{eq:VOS_v} neglects the change in long string velocity $\bar{v}$ due to loop formation as proposed in \cite{Avelino:2020ubr}.

\section{Extension of the original VOS model}
\label{app:VOScalibration}

\subsection{VOS model from Nambu-Goto  simulations}

In our study, we describe the evolution of the long-string network through the VOS model, defined by the equations in Eq. \eqref{eq:VOS_eq_body}. The only free parameter of the model is the loop-chopping efficiency $\tilde{c}$, which is computed to be 
\begin{equation}
\textbf{NG:}\qquad \tilde{c}=0.23\pm 0.04
\end{equation}
from Nambu-Goto  network simulations in an expanding universe \cite{Martins:2000cs}.
\subsection{VOS model from Abelian-Higgs simulations}
Abelian-Higgs (AH) field theory simulations in both expanding and flat spacetime suggest a larger value \cite{Moore:2001px, Martins:2003vd}
\begin{equation}
\label{eq:AH_original}
\textbf{AH:}\qquad \tilde{c}=0.57\pm 0.04.
\end{equation}
Indeed, in Abelian-Higgs simulations, no loops are produced below the string core size so the energy loss into loop formation is lower. Consequently, the loop-chopping efficiency must be increased to maintain scaling.
\FloatBarrier	
\begin{figure}[h!]
			\centering
			\raisebox{0cm}{\makebox{\includegraphics[height=0.53\textwidth, scale=1]{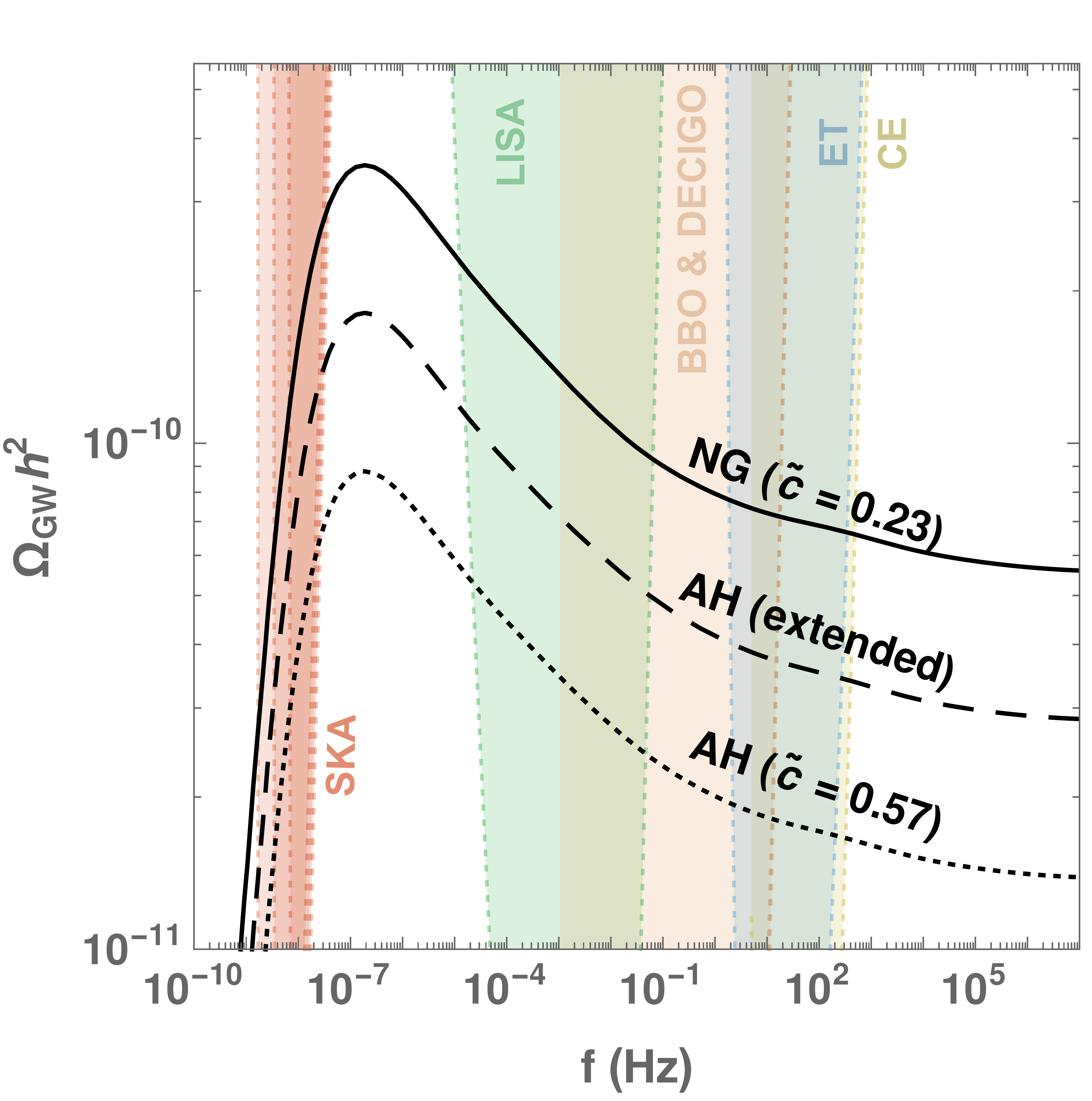}}}
			\raisebox{0cm}{\makebox{\includegraphics[height=0.4\textwidth, scale=1]{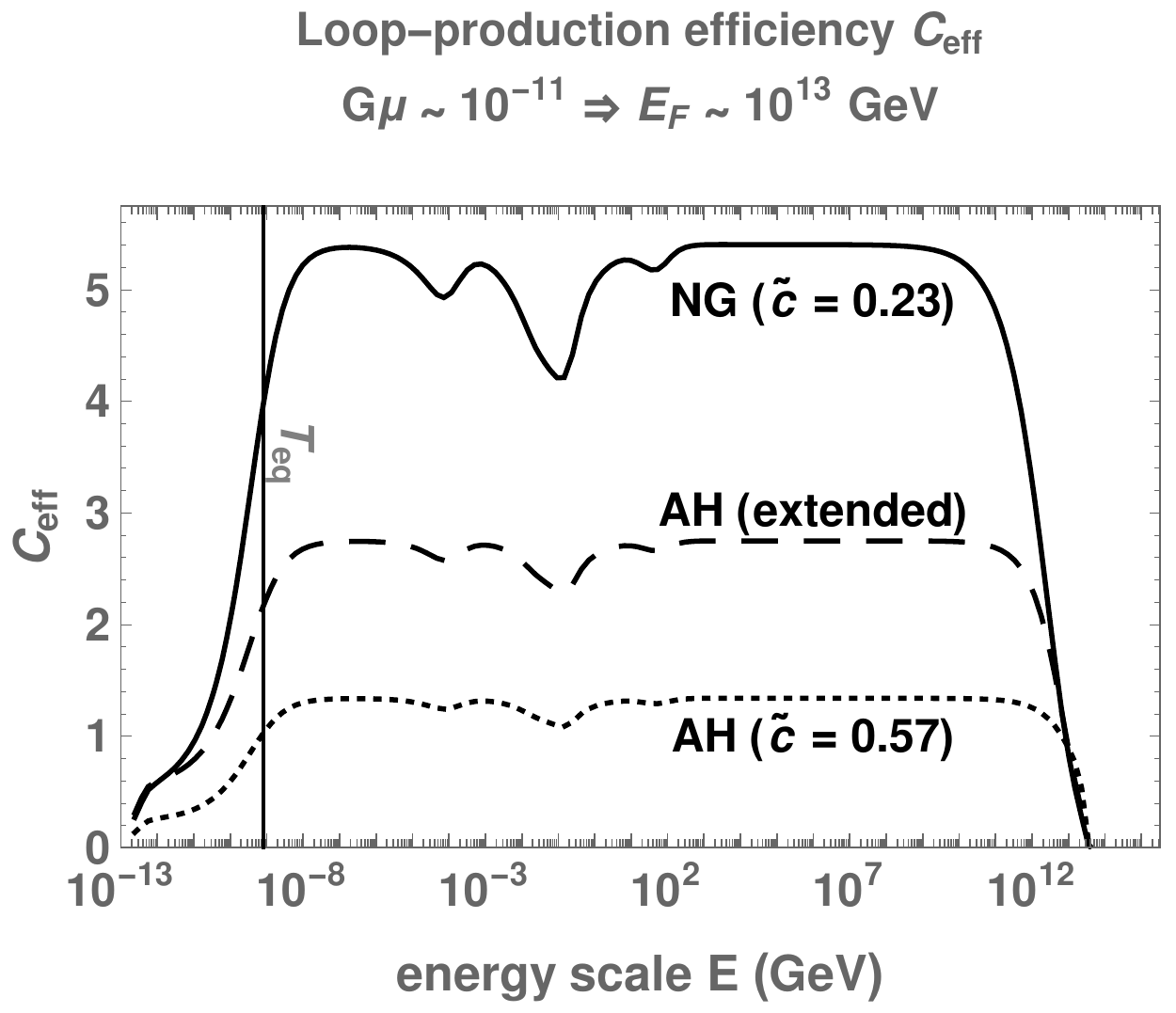}}}
			\hfill
			\caption{\it \small \textbf{Left:} GW spectra with different VOS modellings of the long-string network evolution. The VOS models are either based on Nambu-Goto simulations (solid line - $\tilde{c}=0.23$) \cite{Martins:2000cs} or abelian-Higgs (AH) field theory simulations (dashed line - $\tilde{c}=0.57$) \cite{Moore:2001px, Martins:2003vd}, possibly extended to include particle production \cite{Correia:2019bdl} (dotted line). \textbf{Right:} The corresponding loop-production efficiency for each VOS model.}
			\label{fig:VOSmodelS}
		\end{figure}
		\FloatBarrier

%
\subsection{VOS model from Abelian-Higgs  simulations with particle production}
\label{app:AH_extended_VOS}
In Abelian-Higgs simulations, the loops produced at the string core scale are non-linear lumps of field, called ``proto-loops", which decay fast into massive radiation. Therefore, a recent work \cite{Correia:2019bdl} extends the VOS model by including a term in Eq. \eqref{eq:VOS_L} to account for the emission of massive radiation at the string core scale. The energy-loss function $F(v)$ is modified as
\begin{align}
\label{eq:Fv_energy_loss_function}
\left.F(\bar{v})\right|_\textrm{original}=\frac{\tilde{c}\bar{v}}{2}\textrm{\hspace{1em}}\Rightarrow\textrm{\hspace{1em}}\left.F(\bar{v})\right|_\textrm{extended}=\frac{\tilde{c}\bar{v}+d[k_0-k(\bar{v})]^r}{2},
\end{align}
and the momentum operator $k(v)$, c.f. Eq.~\eqref{eq:momentum_operator}, accounting for the amount of small-scale structures in the string, is modified to 
\begin{align}
\label{eq:kv_AH_extended}
 k(\bar{v})=\frac{2\sqrt{2}}{\pi}\frac{1-8\bar{v}^6}{1+8\bar{v}^6} \textrm{\hspace{1em}}\Rightarrow\textrm{\hspace{1em}} \left.k(\bar{v})\right|_\textrm{extended}=k_0\frac{1-(q \bar{v}^2)^\beta}{1+(q \bar{v}^2)^\beta},
\end{align}
where more free parameters have been introduced. With Abelian-Higgs simulations, one finds \cite{Correia:2019bdl}
\begin{equation}
\label{eq:AH_extended}
\textbf{AH extended:}\qquad \tilde{c}=0.31,
\end{equation}
as well as $d=0.26$, $k_0=1.27$, $r = 1.66$, $q = 2.27$, and $\beta = 1.54$. In Abelian-Higgs extended, the loop-chopping efficiency, c.f. Eq.~\eqref{eq:AH_extended}, is smaller than the one in the original Abelian-Higgs model, c.f. Eq.~\eqref{eq:AH_original}. Indeed, because of the additional energy loss through massive-radiation, less energy loss via loop-chopping is needed to maintain scaling.

In figure~\ref{fig:VOSmodelS}, we compare the GW spectra in the different VOS models.  The difference in amplitude comes from the difference in the number of loops, set by $C_{\rm eff}$. The larger the loop-chopping efficiency $\tilde{c}$, the smaller the loop-formation efficiency $C_{\rm eff}$. This counter-intuitive result can be better understood by looking at table.~\ref{tab:tableVOS}. A larger loop-chopping efficiency $\tilde{c}$ implies a larger loop formation rate only during the transient regime. In the scaling regime, a larger loop-chopping efficiency $\tilde{c}$ implies a more depleted long-string network and then a larger correlation length $\xi$. Hence, the long-string network is more sparse and so the rate of loop formation via string crossing is lower.
\begin{table}[]
\centering
\begin{tabular}{c|ccc}
 \hline \hline\\[-0.75em]
\begin{tabular}[c]{@{}c@{}}scaling in radiation\\ dominated universe\end{tabular}& \begin{tabular}[c]{@{}c@{}}NG\\ $\tilde{c}=0.23$\end{tabular} & \begin{tabular}[c]{@{}c@{}}AH \\ $\tilde{c}=0.57$\end{tabular} & \begin{tabular}[c]{@{}c@{}}AH extended\\ $\tilde{c}=0.31$\\ ($d,~k_0,~r,~q,~\beta$)\end{tabular} \\ \\[-0.75em]\hline\\[-0.75em]
$\bar{v}$                   & 0.66                                           & 0.62                                           & 0.59                                                    \\
$\xi$                  & 0.27                                           & 0.57                                           & 0.36                                                    \\
$C_\textrm{eff}$                & 5.4                                            & 1.3                                            & 2.8                                                     \\[0.25em] \hline \hline
\end{tabular}
\caption{\it \small Values of mean velocity $\bar{v}$, correlation length $\xi$, and loop-production efficiency $C_\textrm{eff}$ in radiation scaling regime with different VOS calibrations.}
\label{tab:tableVOS}
\end{table}

\section{GW spectrum from global strings}
\label{app:global_strings}

 The main distinction with loops from global strings is that they are short-lived whereas loops from local strings are long-lived. This results in different GW spectra in both frequency and amplitude as  we discuss in detail below.

\subsection{The presence of a massless mode}

For global string, the absence of gauge field implies the existence of a massless Goldstone mode, with logarithmically-divergent gradient energy, hence leading to the tension $\mu_{\rm g}$, c.f. Eq.~\eqref{tension_string_exp}
\begin{equation}
\label{eq:string_tension_global_app}
\mu_{\rm g}  \equiv \mu_{\rm l} \, \ln\left( \frac{H^{-1}}{\delta} \right) \simeq \mu_{\rm l} \, \ln\left( \eta\,t \right)  , \quad \text{with} \quad \mu_{\rm l} \equiv 2 \pi \eta^2,
\end{equation}
where $\eta$ is the scalar field VEV, $\mu_{\rm l}$ is the tension of the would-be local string (when the gauge coupling is switched on) and $\delta \sim \eta^{-1}$ is the string thickness.
Goldstones are efficiently produced by loop dynamics with the power
\begin{equation}
P_{\rm Gold}=\Gamma_{\rm Gold}\, \eta^2,
\label{eq:power_goldstone_app}
\end{equation}
where $\Gamma_{\rm Gold} \approx 65$ \cite{Vilenkin:1986ku, Chang:2019mza}, causing loops to decay with a rate
\begin{equation}
\frac{d l_{\rm g}}{dt}= \frac{dE}{dt}\frac{dl}{dE} \equiv \kappa \equiv \frac{\Gamma_{\rm Gold}}{2 \pi \, \ln\left( \eta\,t \right) }.
\end{equation}
Therefore, the string length evolving upon both GW and Goldstone bosons emission reads
\begin{equation}
\label{eq:length_string_app}
l_{\rm g}(t) = \alpha t_i - \Gamma G \mu_{\rm g}  (t - t_i) - \kappa (t - t_i ).
\end{equation}

\subsection{Evolution of the global network}
\label{sec:VOS_global}
The Velocity-One-Scale equations, presented in App.~\ref{sec:VOS_proof},
\begin{align}
\label{eq:VOS_global_1}
&\frac{dL}{dt}=HL \,( 1+ \bar{v}^2)+\left.F(\bar{v})\right|_\textrm{global}, \\
\label{eq:VOS_global_2}
&\frac{d\bar{v}}{dt}=(1-\bar{v}^2)\left[\frac{k(\bar{v})}{L}-\frac{\bar{v}}{l_d}\right],
\end{align}
are modified to include the additional energy-loss due to Goldstone production. Namely, the energy-loss coefficient $F(\bar{v})$, c.f. Eq.~\eqref{eq:Fv_energy_loss_function}, becomes
\begin{equation}
\label{eq:Fv_global}
\left.F(\bar{v})\right|_\textrm{local} =\frac{\tilde{c}\bar{v}+d[k_0-k(\bar{v})]^r}{2}  \textrm{\hspace{1em}} \Rightarrow\textrm{\hspace{1em}} \left.F(\bar{v})\right|_\textrm{global}=\left.F(\bar{v})\right|_\textrm{local} + \frac{s \, v^6}{2\ln\left( \eta\,t\right)},
\end{equation}
where the constant $s$ controlling the efficiency of the Goldstone production, is inferred from lattice simulations \cite{Klaer:2017qhr}, to be $s \simeq 70$ \cite{Martins:2018dqg}.
However, the momentum operator $k(v)$ in Eq.~\eqref{eq:VOS_global_2}, is unchanged with respect to the local case
\begin{align}
\label{eq:kv_AH_extended_2}
 k(\bar{v})=k_0\frac{1-(q \bar{v}^2)^\beta}{1+(q \bar{v}^2)^\beta}.
\end{align}
with $k_0=1.37$, $q=2.3$, $\beta = 1.5$, $\tilde{c} = 0.34$, $d=0.22$, $r=1.8$ \cite{Correia:2019bdl}.
Here through Eq.~\eqref{eq:Fv_global} and Eq.~\eqref{eq:kv_AH_extended_2}, we follow \cite{Klaer:2019fxc} and consider the extended VOS model based on Abelian-Higgs simulations, proposed in \cite{Correia:2019bdl} and already discussed in app.~\ref{app:AH_extended_VOS}, in which we have simply added the backreaction of Goldstone production on long strings in Eq.~\eqref{eq:Fv_global}.
We have checked that we can neglect the thermal friction due to the contact interaction of the particles in the plasma with the string, c.f. Eq.~\eqref{VOS_frictionlength_intro} for which the interaction cross-section is given by the Everett formula in \cite{Vilenkin:2000jqa}. 

In order to later compute the GW spectrum, we defined the loop-formation efficiency, analog of the local case in Eq.~\eqref{eq:loopEfficiency}
\begin{equation}
C_{\rm eff}^\textrm{g} = \tilde{c}\, \bar{v}/\xi^3,
\end{equation}
with $\bar{v}$ and $\xi \equiv L/t$ obeying the VOS equations in Eq.~\eqref{eq:VOS_global_1} and Eq.~\eqref{eq:VOS_global_2}. Due to the logarithmic dependence of the string tension on the cosmic time, the scaling regime is slightly violated. Consequently, the loop-formation efficiency plotted in right panel of Fig.~\ref{fig:global_strings_nonst}, never reaches a constant value. Hence, in this study we model the network based on VOS evolution, rather than using the scaling solutions. Only for enormous value of $ \ln\left( \eta\,t \right) $ corresponding to cosmic times much larger than the age of the Universe today, we find that the solutions to the modified VOS equations in Eq.~\eqref{eq:VOS_global_1} and Eq.~\eqref{eq:VOS_global_2} reach a scaling regime $C_\textrm{eff}=$ 0.46, 2.24, 6.70 for matter-, radiation-, and kination-dominated universe, respectively. By comparing to the values found in \cite{Chang:2019mza}, our results agree only for the radiation case. Note that the dependence of the string network parameters $\bar{v}$ and $\xi \equiv L/t$ on the logarithmically-time-dependent string tension arises only through the term of Goldstone production energy loss in Eq.~\eqref{eq:Fv_global}.

\subsection{The GW spectrum}
Finally, the GW spectrum generated by loops of global strings is given by a similar expression as for the local case, c.f. the mattress equation in Eq.~\eqref{eq:SGWB_CS_Formula}, 
	\begin{equation}
	\Omega^{\rm g}_{\rm{GW}}(f)\equiv\frac{f}{\rho_c}\left|\frac{d\rho_{\rm{GW}}^{\rm g}}{df}\right|=\sum_k{\Omega^{(k),\, \rm g}_{\rm{GW}}(f)},
	\label{eq:SGWB_CS_Formula_global}
	\end{equation}
	where
	\begin{equation}
\Omega^{(k), \, \rm g}_{\rm{GW}}(f)=\frac{1}{\rho_c}\cdot\frac{2k}{f}\cdot\frac{\mathcal{F}_{\alpha}\,\Gamma^{(k)}G\mu_{\rm g} ^2}{\alpha(\alpha+\Gamma G \mu_{\rm g} + \kappa)}\int^{t_0}_{t_F}d\tilde{t}~ \frac{C_{\rm{eff}}^{\rm g}(t_i^{\rm g})}{\,(t_i^{\rm g})^4}\left[\frac{a(\tilde{t})}{a(t_0)}\right]^5\left[\frac{a(t_i^{\rm g})}{a(\tilde{t})}\right]^3\Theta(t_i^{\rm g}-t_F).
	\label{kmode_omega}
	\end{equation}
 We have checked that we can safely neglect massive radiation. The loop formation time $t_i^{\rm g}$ is related to the emission time $\tilde{t}$ after using $l_{\rm g}(\tilde{t}) =0$ in Eq.~\eqref{eq:length_string_app}
\begin{equation}
t_i^{\rm g} = \frac{\Gamma G\mu_{\rm g} + \kappa} {\alpha + \Gamma G\mu_{\rm g} + \kappa}\tilde{t}.
\end{equation}

\begin{figure}[h!]
			\centering
			\raisebox{0cm}{\makebox{\includegraphics[width=0.65\textwidth, scale=1]{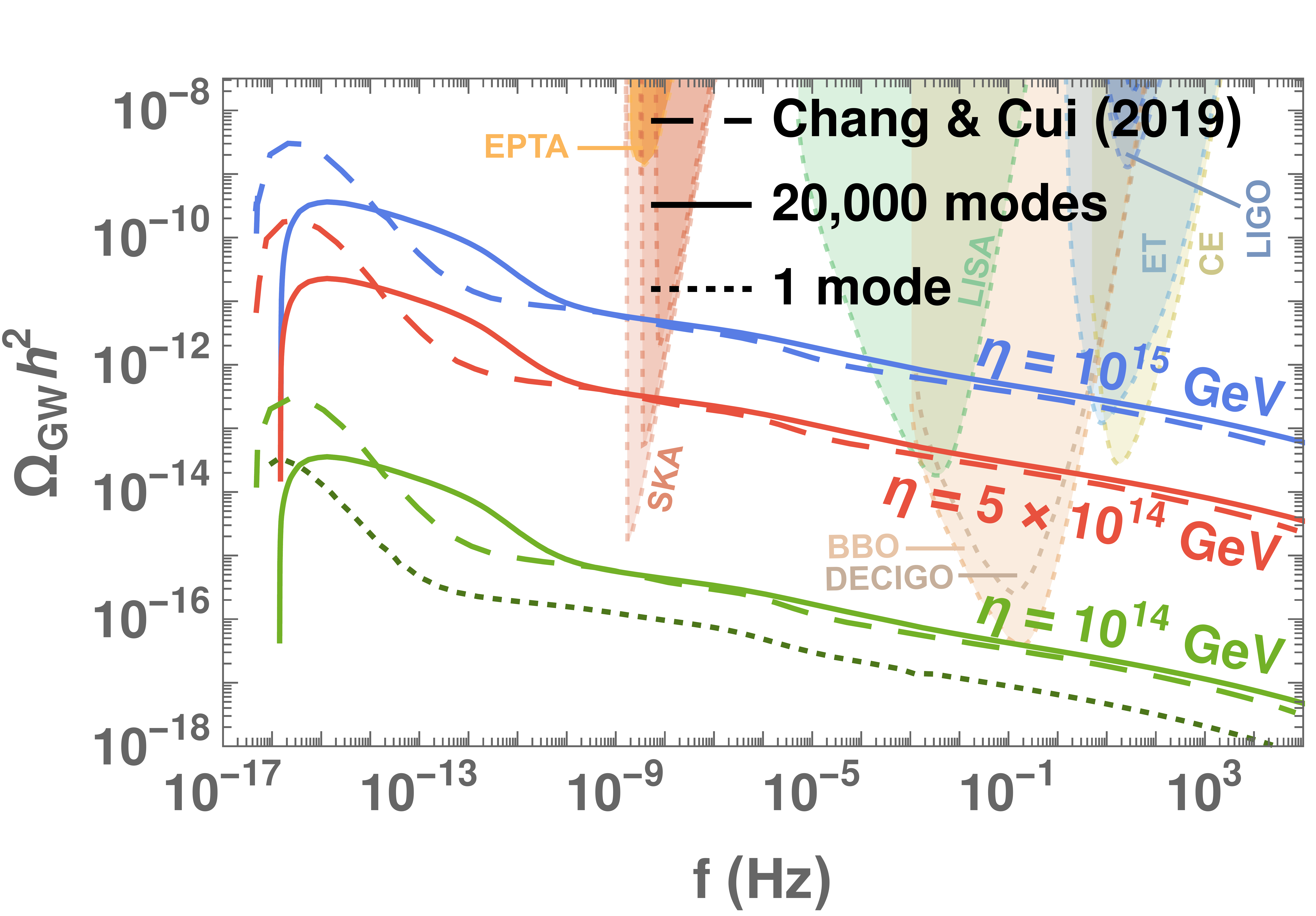}}}\\
			\hfill
		\caption{\it \small   GW spectrum from global cosmic strings assuming VOS network formed at energy scale $\eta$, evolved in the standard cosmology. The shape of the GW spectrum computed in \cite{Chang:2019mza} resembles the first mode $k=1$ of our spectrum.}
			\label{fig:global_strings_st}
		\end{figure}

We plot the GW spectrum from local strings in Fig.~\ref{fig:global_strings_st} and compare to the spectrum computed in \cite{Chang:2019mza}.
With respect to \cite{Chang:2019mza}, we find a lower value for $C_\textrm{eff}$ during the late matter-dominated universe ($0.46$ instead of $1.32$) which implies a smaller spectral bump, while the radiation contributions are considerably the same. Moreover, the shapes of the spectra are different. An explanation could be the summation over the high-frequency modes (up to $k=2\times 10^4$ in our case) which smoothens the spectrum. The spectrum from \cite{Chang:2019mza} resembles to the first-mode of our spectrum. 

The constraints on the inflation scale, $6 \times 10^{13}$~GeV, from the non-detection of the fundamental B-mode polarization patterns in the CMB \cite{Ade:2018gkx, Akrami:2018odb}, implies the upper bound $T_{\rm reh} \lesssim 5 \times 10^{16}$~GeV on the reheating temperature, assuming instantaneous reheating. Hence, assuming that the network is generated from a thermal phase transition, we are invited to impose $\eta \lesssim 5 \times 10^{16}$. 
However, a stronger restriction arises because of the CMB constraint on strings tensions
\begin{equation}
\left.G\mu_\textrm{g}\right|_{\textrm{CMB}}= \left.2\pi \left(\frac{\eta}{m_\textrm{pl}}\right)^2\log(\eta\, t_\textrm{CMB})\right.\lesssim 10^{-7} \quad \rightarrow \quad \eta \lesssim 1.4 \times 10^{15},
\end{equation}
where we use $t_\textrm{CMB}\simeq 374$ kyr. Hence, we restrict to $\eta \lesssim 10^{15}$~GeV as in \cite{Chang:2019mza}.

\subsection{Global versus local strings}
The local spectrum is recovered upon setting $ \ln\left( \eta\,t \right) =1$, $\kappa =0$ and $s=0$ in Eq.~\eqref{eq:string_tension_global_app}, Eq.~\eqref{eq:length_string_app} and Eq.~\eqref{eq:Fv_global}.
Because of the faster decay, in the global case, the main emission time $\tilde{t}_{\rm M}^{\rm g}$ is shorter than its local counterpart $\tilde{t}_{\rm M}^{\rm \,l}$, c.f. Eq.~\eqref{eq:t_M},
\begin{align}
\label{eq:t_M_global}
\tilde{t}_{\rm M}^{\rm g} &\simeq \frac{\alpha + \Gamma G\mu_{\rm g} + \kappa }{ \Gamma G \mu_{\rm g} + \kappa}\frac{ t_i}{2}, \\ 
 & \sim \frac{\Gamma G \mu_{\rm l}}{ \alpha} ~\tilde{t}_{\rm M}^{\rm \,l},
\end{align}
where in the last line, we have assumed $\kappa \gg \alpha, \tilde{t}_{\rm M}^{\rm \,l}$.
Therefore, the frequency $f$ of a GW emitted by a loop produced at the temperature $T$, is lowered  compared to the local case computed in Eq.~\eqref{fdeltaApp}, by a factor
\begin{align}
\label{eq:global_vs_local_freq}
f\Big|_{\rm global} &\simeq (4.7\times10^{-6}\textrm{ Hz})\left(\frac{T}{\textrm{GeV}}\right)\left(\frac{0.1}{\alpha }\right)\left(\frac{g_*(T_i)}{g_*(T_0)}\right)^{1/4} ,\\
&\sim \left(\frac{\Gamma G \mu_{\rm l}}{\alpha}\right)^{1/2} f\Big|_{\rm local}.
\label{eq:global_vs_local_freq_2}
\end{align}
In contrast to the local case, the frequency is independent of the string scale and we explain its origin at the end of the section.
We can rewrite the GW spectrum in Eq.~\eqref{eq:SGWB_CS_Formula_global} as
\begin{equation}
\label{eq:Omega_global}
\Omega^{\rm g}_{\rm{GW}}(f) \simeq \frac{1}{\rho_c}\cdot\frac{2}{f} \cdot\frac{\mathcal{F}_\alpha\,\Gamma G\mu_{\rm g}^2}{\alpha(\alpha+\Gamma G \mu_{\rm g} + \kappa)}\cdot C_{\rm{eff}}^{\rm g}\cdot t_0^{-5/2}\cdot(\tilde{t}_{\rm M}^{\rm g})^{1+\frac{5}{2}-\frac{3}{2}}\cdot t_i^{-4 + \frac{3}{2}} .
\end{equation}
Then, from Eq.~\eqref{eq:t_M_global} and $\alpha \,t_{i}^{\rm g} \simeq 4\,a(\tilde{t}_{\rm M}^{\rm g})/a(t_0) f^{-1}$ in Eq.~\eqref{eq:tdelta_fdelta_eq_line1}, one obtains
\begin{align*}
&\tilde{t}_{M}^{\rm g} \simeq \frac{1}{t_0}\frac{4}{\alpha^2} \left( \frac{1}{f} \right)^2 \left( \frac{\alpha+\Gamma G \mu_{\rm g} + \kappa}{\Gamma G \mu_{\rm g} + \kappa}\right)^2, \\
&t_{i}^{\rm g} \simeq \frac{1}{t_0}\frac{8}{\alpha^2} \left( \frac{1}{f} \right)^2 \left( \frac{\alpha+\Gamma G \mu_{\rm g} + \kappa}{\Gamma G \mu_{\rm g} + \kappa}\right).
\end{align*}
From comparing the global GW spectrum $\Omega_{\rm GW}^{\rm g}$ to the local GW spectrum $\Omega_{\rm GW}^{\rm l} = \Omega_{\rm GW}^{\rm g}( \ln\left( \eta\,t \right) =1,\, \kappa=0, \, s=0)$, one obtains
\begin{equation}
\label{eq:ratio_global_to_local_GW}
\frac{\Omega_{\rm GW}^{\rm g}}{\Omega_{\rm GW}^{\rm l}} \sim \frac{\alpha}{\kappa}\left( \frac{\Gamma G \mu_{\rm l}}{\alpha} \right)^{3/2} \left( \frac{\mu_{\rm g}}{\mu_{\rm l}}\right)^2 \frac{C_{\rm eff}^{\rm g}}{C_{\rm eff}^{\rm l}},
\end{equation}
where we have assumed $\kappa \gg \alpha,\, \Gamma G \mu_{\rm g} $. 
Upon using Eq.~\eqref{eq:lewicki_formula_GW_spectrum_radiation} and Eq.~\eqref{eq:ratio_global_to_local_GW}, we get\footnote{Upon restoring the dependence on the different parameters appearing in Eq.~\eqref{eq:Omega_global}, we obtain
\begin{equation}
\label{eq:Omega_global_2}
\Omega_{\rm GW}^{\rm g} \simeq  25 \Delta_{R} \,\Omega_{r}h^2 \,C^{\rm g}_{\rm eff}(n=4) \, \mathcal{F}_\alpha \, \frac{\Gamma}{\Gamma_{\rm Gold}}  \log^3\left(\eta \,\tilde{t}_{M}^{\rm g}\right) \frac{\eta^4}{M_{\rm pl}^4},
\end{equation}
which can be compared with its local equivalent in Eq.~\eqref{eq:lewicki_formula_GW_spectrum_radiation}.}

\begin{equation}
\label{eq:ratio_global_to_local_GW_2}
\Omega_{\rm GW}^{\rm l} \simeq \Omega_{r}h^2\frac{\eta}{M_{\rm pl}}, \qquad \text{and}\qquad \Omega_{\rm GW}^{\rm g} \simeq \Omega_{r}h^2 \log^3\left(\eta \,\tilde{t}_{M}^{\rm g}\right) \frac{\eta^4}{M_{\rm pl}^4},
\end{equation}
where $\Omega_{r}h^2 \simeq 4.2\times 10^{-5}$ is the present radiation energy density of the universe \cite{Tanabashi:2018oca}.

As shown in Fig.~\ref{fig:global_strings_st}, in consequence of the strong dependence of the GW amplitude on the string scale $\eta$, only global networks above $\eta \gtrsim 5 \times 10^{14}~$GeV can be detected by LISA or CE whereas EPTA or BBO/DECIGO can probe $\eta \gtrsim  10^{14}~$GeV. Also note the logarithmic spectral tilt of the GW spectrum due to the logarithmic dependence of the global string tension on the cosmic time.

In summary, GW spectra from local and global strings manifest differences in frequency, Eq.~\eqref{eq:global_vs_local_freq_2}, and amplitude, Eq.~\eqref{eq:ratio_global_to_local_GW_2}, because local loops are long-lived whereas global loops are short-lived. In the local case, the dominant GW emission at $\tilde{t}_{\rm M}^{\rm l} \sim t_i^{\rm l}/G\mu_{\rm l}$ occurs much after the loop formation time $t_i^{\rm l}$, after one loop lifetime. However, the emitted frequency is fixed by $~(t_i^{\rm l})^{-1}$. Hence, as discussed in Sec.~\ref{sec:turning_point_scaling}, the observed frequency is exempted from a redshift factor given by $\sqrt{G\mu_{\rm l}}$. In the global case, the loops are short-lived and the time of dominant GW emission coincides with the time of loop formation. Hence, the emitted frequency redshifts more and the spectrum is shifted to the left. The GW spectrum is also reduced by the redshift factor $(G\mu)^{3/2}$.

\begin{figure}[h!]
			\centering
			\raisebox{0cm}{\makebox{\includegraphics[width=0.495\textwidth, scale=1]{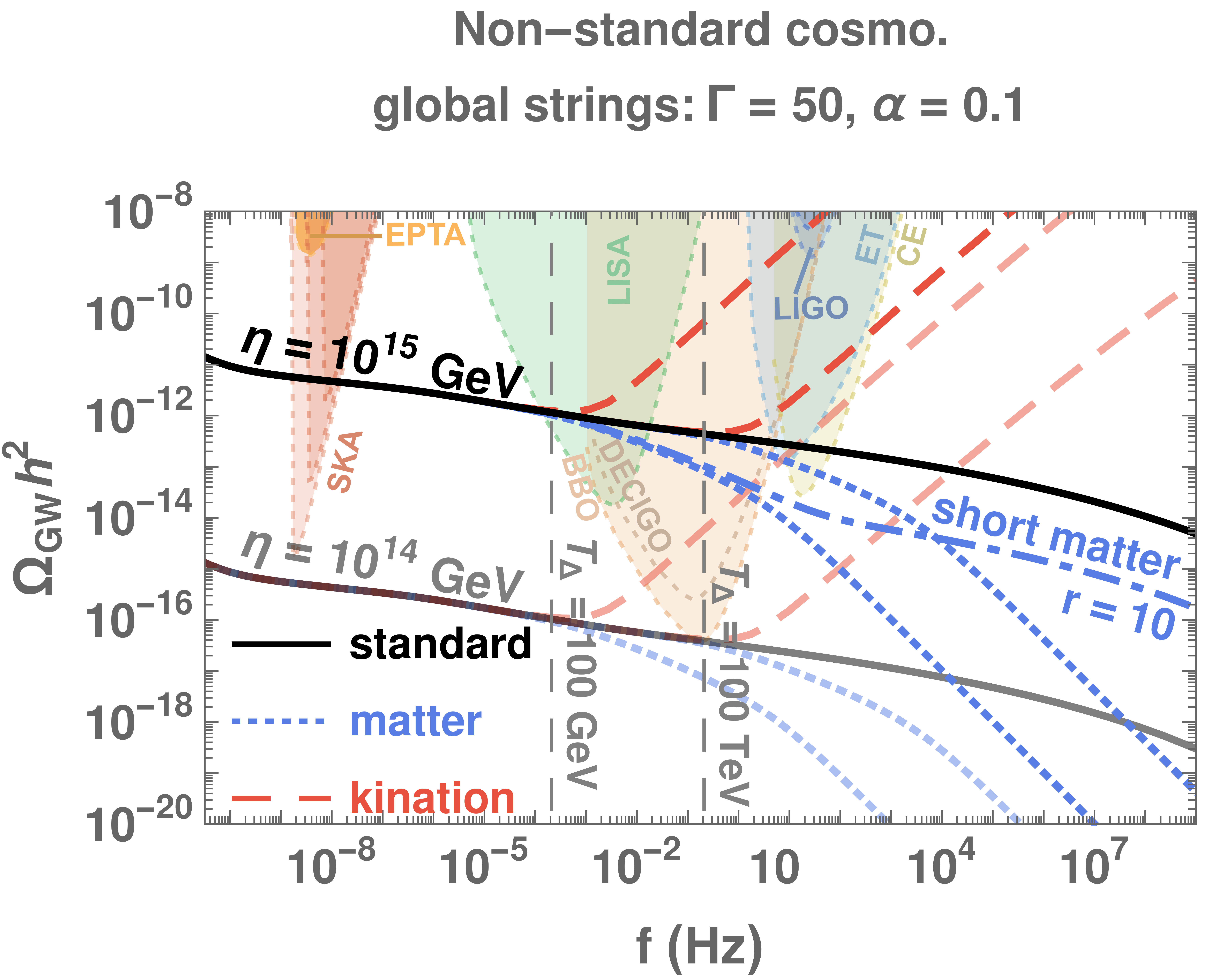}}}
			\raisebox{0cm}{\makebox{\includegraphics[width=0.475\textwidth, scale=1]{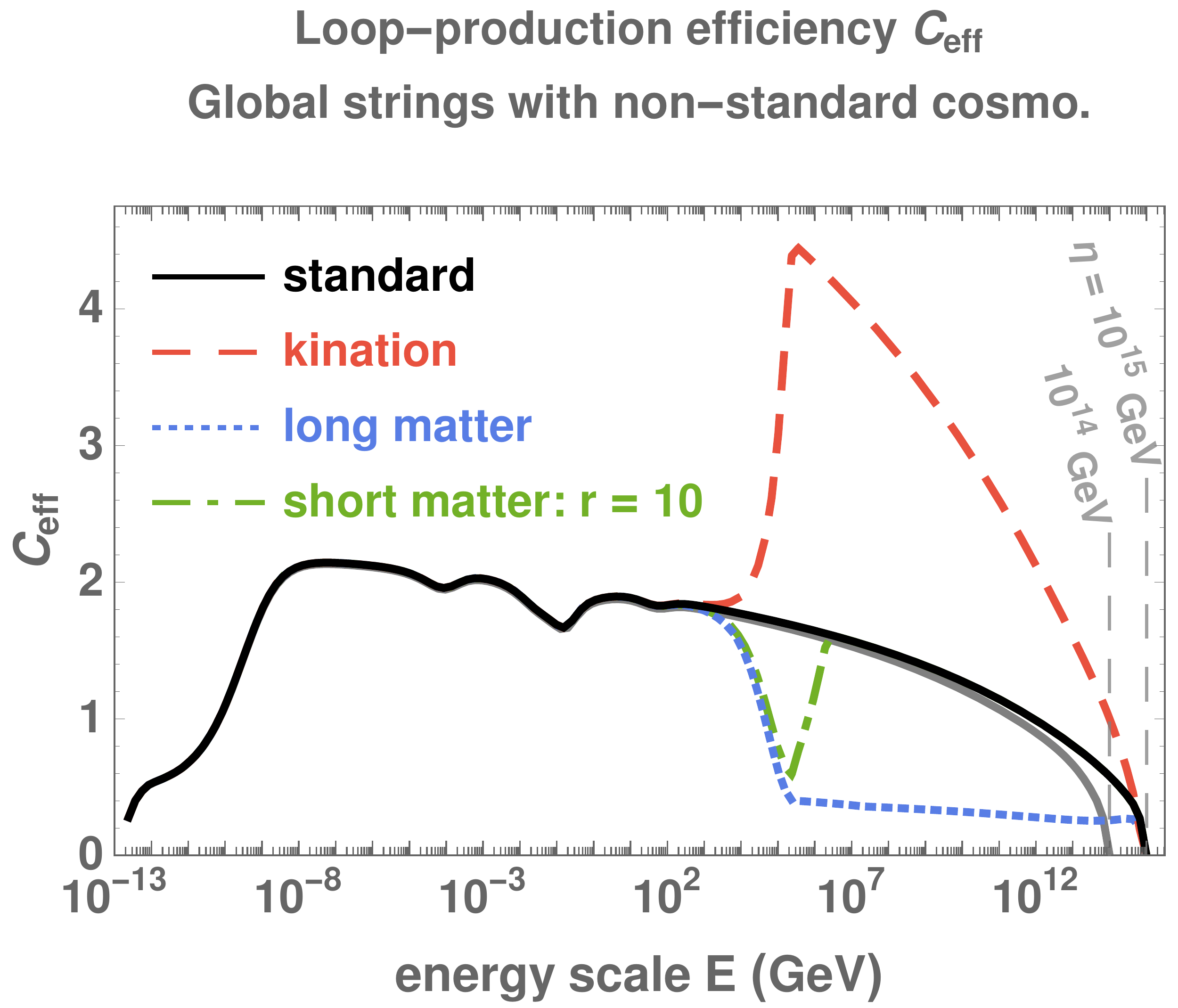}}}\\
			\hfill
		\caption{\it \small   \textbf{Left:} GW spectrum from the global cosmic strings assuming VOS network, evolving in the presence of a non-standard era, either long-lasting matter (dotted), intermediate matter (dot-dashed), or kination (dashed), ending at the temperature $T_\Delta=100$ GeV or $100$ TeV. The turning-point frequency is independent of the string tension. \textbf{Right:} The evolution of the loop-production efficiency for each cosmological background never reaches a plateau, in contrast to local strings in which case the scaling regime is an attactor solution, c.f. right panel of Fig.~\ref{figure_preceding_NS_era}. Indeed, for global strings the scaling behavior is logarithmically violated due to the inclusion of energy loss through Goldstone production in the VOS equations, c.f. Sec.~\ref{sec:VOS_global}.}
			\label{fig:global_strings_nonst}
		\end{figure}
		
		\begin{figure}[h!]
			\centering
			\raisebox{0cm}{\makebox{\includegraphics[width=0.5\textwidth, scale=1]{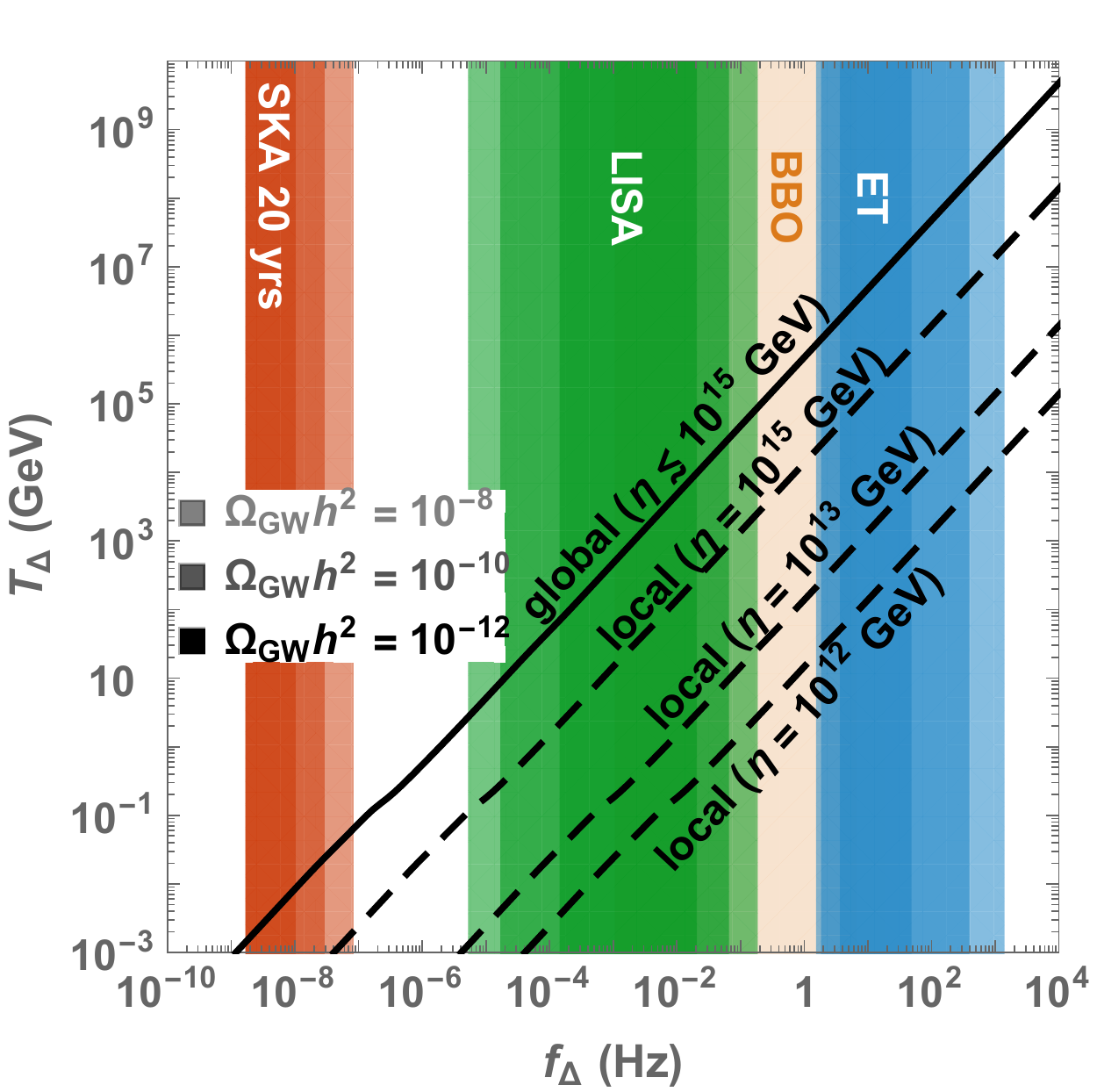}}}\\
			\hfill
		\caption{\it \small Detectability of the turning-points in the $T_\Delta - f_\Delta$ plane. The solid line represents turning-points for global strings formed at any energy scale $\eta$, while dashed lines are turning points from local strings, similar to lines in the bottom-left panel of Fig.~\ref{fig:contour_power}. Shaded areas correspond to the frequencies probed by each observatory assuming SGWB of amplitudes $\Omega_{\textrm{GW}} h^2 = 10^{-8}$, $10^{-10}$ and $10^{-12}$. The plot is totally inspired from \cite{Chang:2019mza}.} 
			\label{fig:global_strings_contour}
		\end{figure}
		
		\subsection{As a probe of non-standard cosmology}

The impact of non-standard cosmology on the GW spectra of global strings is shown in Fig.~\ref{fig:global_strings_nonst}.
The frequency of the turning point corresponding to a change of cosmology at a temperature $T_\Delta$ is given by Eq.~\eqref{eq:global_vs_local_freq}. We report here a numerically-fitted version 
\begin{align}
f^{\textrm{glob}}_{\Delta}\simeq \textrm{ Hz} \left(\frac{T}{\textrm{GeV}}\right)\left(\frac{0.1}{\alpha}\right)\left(\frac{g_*(T)}{g_*(T_0)}\right)^{1/4} \times 
\begin{cases}
8.9 \times 10^{-7} &\textrm{for } 10 \% \\
7.0 \times 10^{-8} &\textrm{for } 1 \% \\
\end{cases},
\label{turning_point_globalstring}
\end{align}
where the detection criterion is defined as in Eq.~\eqref{10per_criterion}. In contrast to local strings, the turning-point is independent of the string tension. 

We now consider the reach of global strings for probing a non-standard cosmology. Fig.~\ref{fig:global_strings_contour} shows the detectability of the turning-points by future GW experiments. Due to the string-scale independence, the global-string detectability collapses onto a line. Because of the shift of the spectrum to lower frequencies by a factor $\sim \sqrt{G\mu}\sim \eta/M_{\rm pl}$, c.f. Eq.~\eqref{eq:global_vs_local_freq_2}, GW from global string networks can probe earlier non-standard-cosmology and larger energy scales with respect to GW from local strings.

\section{Impact of the cosmology on the loop size at formation}\label{varying_alpha_spectrum}
\label{app:impactcosmo-loop-size}
\subsection{Loop size as a fraction of the Hubble horizon}

In this section, we discuss  the validity of defining the loop-size at formation, $\l_i$, as a constant fraction $\alpha$ of the Hubble horizon size $t_i$
\begin{align}
\label{eq:loop_size_horizon}
\text{Terminology I :} \quad l_i = \alpha \, t_i.
\end{align}
Nambu-Goto simulations \cite{Blanco-Pillado:2013qja} suggest $\alpha \simeq 0.1$. In this first prescription, we neglect the effects of a change of cosmology on $\alpha$. The advantage of the second prescription presented in Sec.~\ref{sec:alpha_correlation_length} is to account analytically for these effects. 

The time derivative of the length of the loop at its formation is simply 
\begin{align}
\frac{dl}{dt}=\alpha+\Gamma G\mu,
\end{align}
which leads to the GW spectrum in Eq. \eqref{eq:SGWB_CS_Formula}. Outside the redshift factors, in this first prescription the loop-production efficiency $C_\textrm{eff}$ is the only parameter depending on the cosmology in the GW spectrum formula.

\subsection{Loop size as a fraction of long-string correlation length}
\label{sec:alpha_correlation_length}

Loops being formed by inter-commutation of long strings, the appropriate length scale setting their size at formation should be the correlation length $L$ of the long-string network, and not the Hubble horizon size $t_i$. Therefore, the loop-size at formation $\l_i$ in Eq.~\eqref{eq:loop_size_horizon} should be replaced by the more natural definition, pointed out first in \cite{Sousa:2013aaa}
\begin{align}
\text{Terminology II :} \quad l_i\equiv\alpha_L \, L_i = \alpha_L \, \xi \, t_i,
\label{eq:loop_size_xi}
\end{align}
where $\alpha_L$ is the new constant loop-size parameter. The two definitions of the loop-size parameter in Eq.~\eqref{eq:loop_size_horizon} and Eq.~\eqref{eq:loop_size_xi} are related through
\begin{equation}
\alpha = \alpha_L \, \xi.
\end{equation}
We assume that $\alpha_L$ is a constant independent of the equation of state of the universe \cite{Auclair:2019wcv}. We set its value to $\alpha_{\rm L} = 0.37$ in order to match $\alpha = 0.1$ during radiation-domination. Therefore, the dependence of $l_i$ on the change of cosmology is directly tied to $\xi$. 

The GW spectrum formula, c.f. Eq.~\eqref{eq:master_eq_app}, which depends on the time derivative of the loop length at the production, c.f. Eq.~\eqref{eq:loop_density_Jac}
\begin{align}
\frac{dl}{dt}=\alpha_L\frac{d}{dt}(\xi t)+\Gamma G\mu,
\end{align}
becomes
\begin{align}
\Omega^{(k)}_{\rm{GW}}(f)=&\frac{1}{\rho_c}\frac{2k}{f}(0.1)\,\Gamma^{(k)}G\mu^2 
\int^{t_0}_{t_{\rm osc}}d\tilde{t}~ 
\frac{\tilde{c} \bar{v}(t_i)}{\gamma \alpha_L}
\left[\frac{1}{\xi(t_i) t_i}\right]^{4}
\left[\frac{1}{\alpha_L\frac{d}{dt}(\xi t)+\Gamma G\mu}\right] \nonumber \\
&\times\left[\frac{a(\tilde{t})}{a(t_0)}\right]^5\left[\frac{a(t_i)}{a(\tilde{t})}\right]^3\Theta(t_i-t_{\rm osc})\Theta(t_i-\frac{l_*}{\alpha}),
\label{kmode_omega_alphachange}
\end{align}
where $t_i$ is the root solution of
\begin{equation}
l(t)=\alpha_L \xi(t_i)t_i - \Gamma G\mu (t-t_i).
\label{ti_alphachange}
\end{equation}

\begin{figure}[h!]
			\centering
			\raisebox{0cm}{\makebox{\includegraphics[height=0.365\textwidth, scale=1]{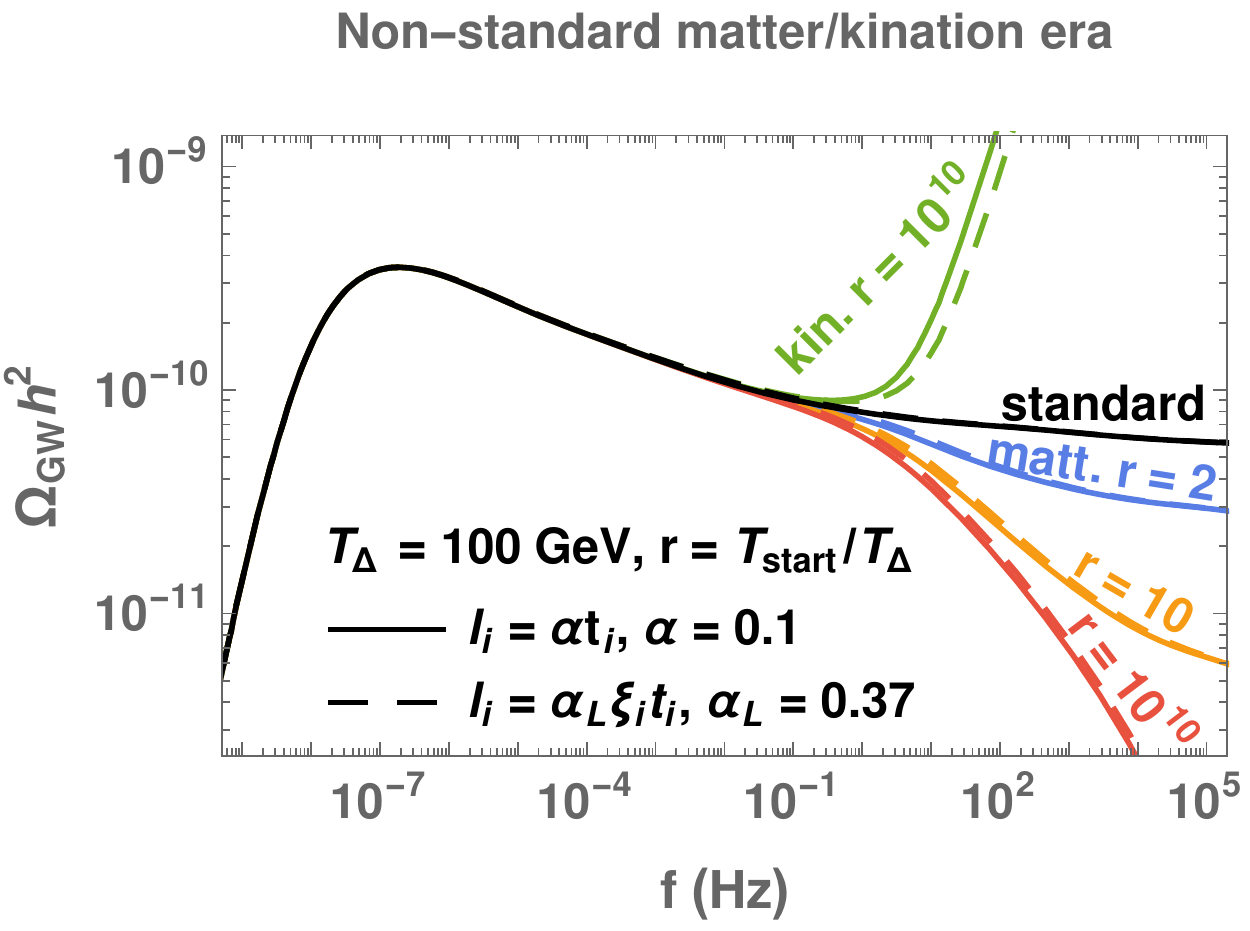}}}
			\raisebox{0cm}{\makebox{\includegraphics[height=0.365\textwidth, scale=1]{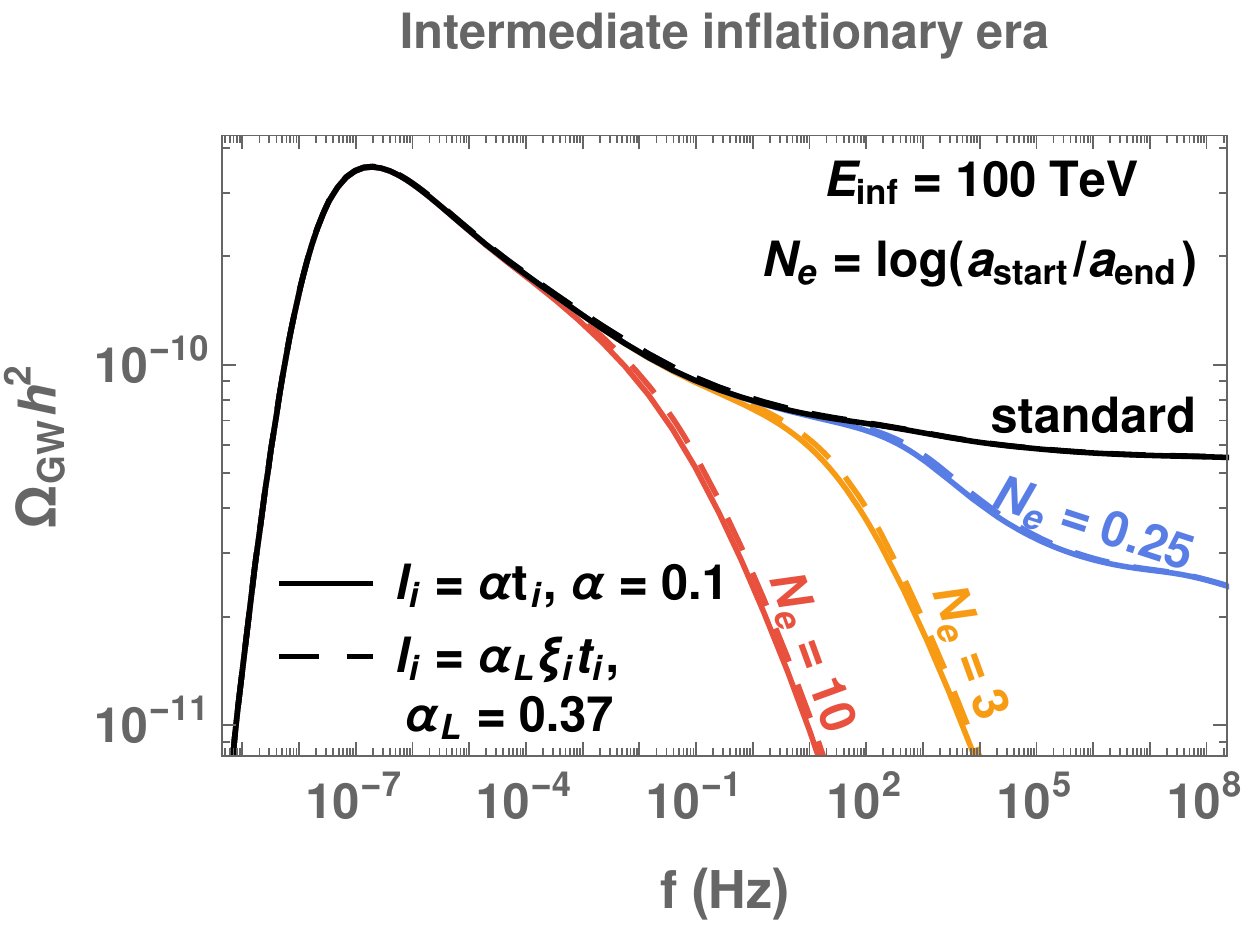}}}
			\raisebox{0cm}{\makebox{\includegraphics[height=0.365\textwidth, scale=1]{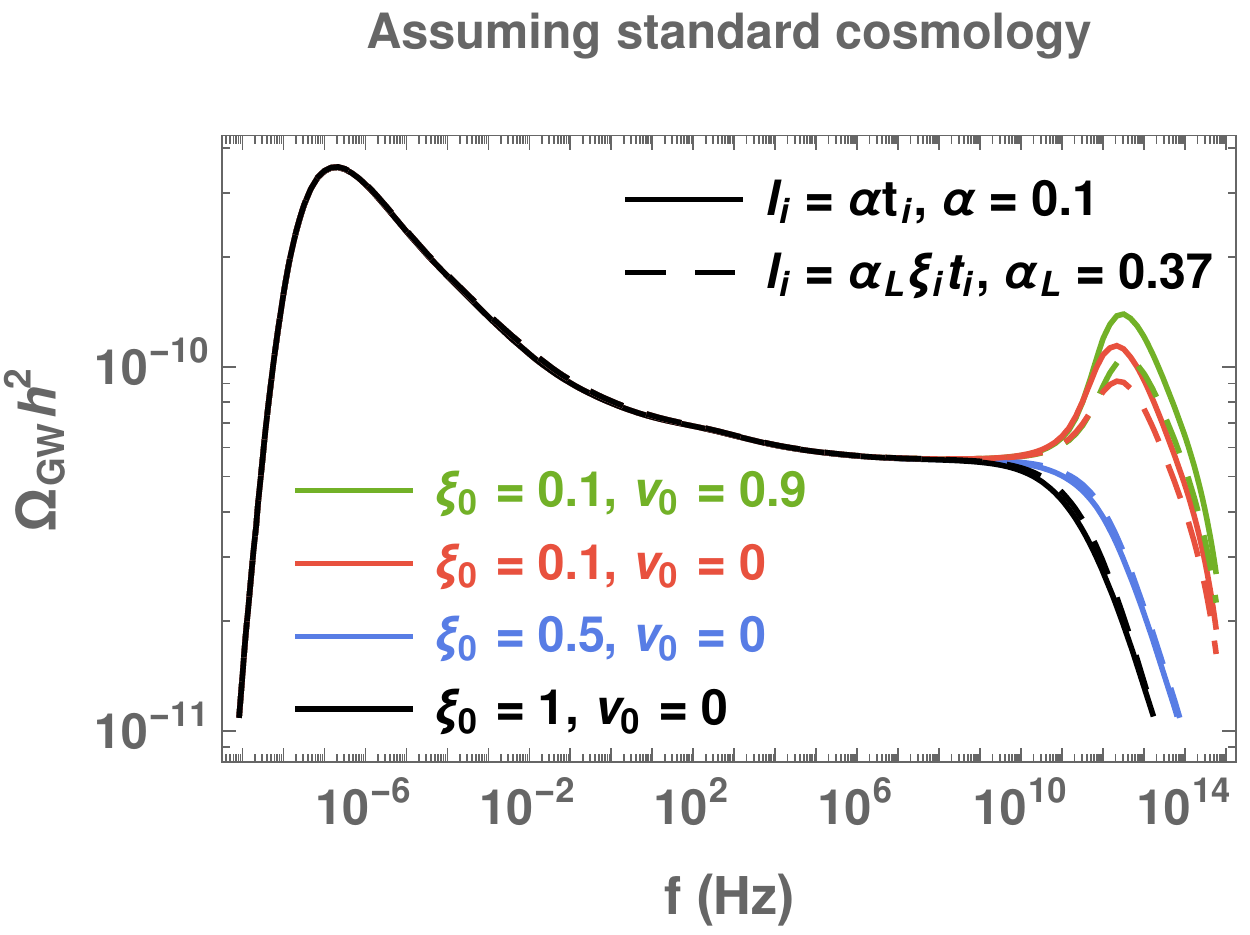}}}
			\hfill
			\caption{\it \small  Comparisons of GW spectra obtained from terminology I and II assuming (top-left) long-lasting non-standard eras, (top-right) intermediate matter era, and (bottom) standard cosmology with various initial conditions of the network. As opposed to the prescription I (plain lines and Eq.~\eqref{eq:loop_size_horizon}), the prescription II (dashed lines and Eq.~\eqref{eq:loop_size_xi}) allows to analytically track the effect of a change of cosmology on the loop-size at formation $l_i$. However, the impact on the spectrum is rather mild.}
			\label{fig:alpha_change}
		\end{figure}
		
\begin{table}[h!]
\centering
\begin{tabular}{c|ccc}
& \begin{tabular}[c]{@{}c@{}}terminology I\\ $\alpha_\textrm{I}$\end{tabular} & \begin{tabular}[c]{@{}c@{}}terminology II\\ $\alpha_\textrm{II}=\alpha_L\xi$\end{tabular} & $\frac{\Omega_{\textrm{GW,II}}}{\Omega_{\textrm{GW,I}}}=\sqrt{\frac{\alpha_\textrm{II}}{\alpha_\textrm{I}}}$ \\ \\[-1em]\hline\\[-0.75em]
radiation & 0.1                                           & (0.37)(0.27)                                           & 1                                                    \\
matter  & 0.1                                           & (0.37)(0.63)                                       & 1.53                                                    \\
kination  & 0.1                                            & (0.37)(0.15)                                            & 0.75                                                     \\[0.25em] \hline 
\end{tabular}
\caption{\it \small Values of loop-size parameter $\alpha_\textrm{I},~\alpha_\textrm{II} \equiv \xi t$ assuming radiation-, matter-, and kination-scaling and ratios between their corresponding GW spectra. Note that the intermediate inflationary scenario provides similar results to that of the matter case. The characteristic length scale $\xi$ is constant during the scaling regime, which makes the comparison between the two terminologies possible.}
\label{table:loopsize_parameters}
\end{table}

\subsection{Impact on the GW spectrum}

In Fig.~\ref{fig:alpha_change}, we compare the impact of the two prescriptions for the loop-size at formation, defined in Eq.~\eqref{eq:loop_size_horizon} and Eq.~\eqref{eq:loop_size_xi}, on the GW spectrum. We show GW spectra assuming intermediate or long-lasting non-standard era, or standard cosmology with various initial conditions of the cosmic-string network. 
The impact of the loop-size at formation on the spectrum mainly comes from the behavior $\Omega_\textrm{GW} \propto \sqrt{\alpha}$, c.f. Eq.~\eqref{eq:lewicki_formula_GW_spectrum_radiation}. Since the actual long-string correlation scale $\xi$ is longer/shorter in matter-/kination-dominated universe respectively, terminology II leads to an enhancement/suppression of the spectrum.
Table \ref{table:loopsize_parameters} displays values of $\alpha$ and expected ratios of amplitudes from two terminologies, for different equations of state.
Because of the technical difficulties in applying terminology II and of the rather small impact of the spectrum, we restrict to terminology I throughout this work, as used in Sec.~\ref{sec:mainAssumptions}.

\section{Sensitivity curves of GW detectors}
\label{app:sensitivity_curves}

\subsection{The signal-to-noise ratio}
The total output of a detector is given by the GW signal plus the noise, $ h(t)+n(t)$ where the level of noise $n(t)$ is measured by its noise spectral density $S_n(f)$ \cite{Caprini:2018mtu}.
\begin{equation}
\left< \tilde{n}^*(f) \tilde{n}(f')  \right> \equiv \delta(f-f')\, S_n(f).
\end{equation}
We define the detector sensitivity $\Omega_{\rm sens}(f)$ as the magnitude of the SGWB energy density which would mimick the noise spectral density $S_n(f)$
\begin{equation}
\Omega_{\rm sens}(f) = \frac{2\pi^2}{3H_0^2}\,f^3\,S_n(f).
\end{equation}
The capability of an interferometer to detect a SGWB of energy density $\Omega_{\rm GW}(f)$ after an observation time $T$ is measured by the signal-to-noise ratio (SNR) \cite{Maggiore:1999vm}
\begin{equation}
\label{eq:SNR_def_Sam}
{\rm SNR} = \sqrt{T \int_{f_{\rm min}}^{f_{\rm max}} df \, \left[ \frac{\Omega_{\rm GW}(f)}{\Omega_{\rm sens}(f)}\right]^2}.
\end{equation}
\subsection{The power-law integrated sensitivity curve}
Assuming a power law spectrum
\begin{equation}
\label{eq:SGWB_power_law_Sam}
\Omega_{\rm GW}(f) = \Omega_\beta\left( \frac{f}{f_{\rm ref}}\right)^\beta,
\end{equation}
with spectral index $\beta$, amplitude $\Omega_\beta$ and reference frequency $f_{\rm ref}$, we deduce from Eq.~\eqref{eq:SNR_def_Sam} the amplitude $\Omega_\beta$ needed to reach a given SNR after a given observation time $T$
\begin{equation}
\Omega_\beta = \frac{{\rm SNR}}{\sqrt{T}}\left( \int_{f_{\rm min}}^{f_{\rm max}} df \left[ \frac{h^2}{h^2\Omega_{\rm sens}(f)} \left( \frac{f}{f_{\rm ref}} \right)^\beta \right]^2  \right)^{-1/2},
\end{equation}
which upon re-injecting into Eq.~\eqref{eq:SGWB_power_law_Sam} gives
\begin{equation}
h^2 \Omega_{\rm GW}(f) = f^\beta \frac{{\rm SNR}}{\sqrt{T}}\left(\int_{f_{\rm min}}^{f_{\rm max}} df \left[ \frac{f^\beta}{h^2\Omega_{\rm sens}(f)} \right]^2 \right)^{-1/2}.
\end{equation}
For a given pair $({\rm SNR}, \, T)$, one obtains a series in $\beta$ of power-law integrated curves. One defines the power-law integrated sensitivity curve $\Omega_{PI}(f)$ as the envelope of those functions \cite{Thrane:2013oya}
\begin{equation}
\Omega_{PI}(f) \equiv \underset{\beta}{\rm max}  \left[ f^\beta \frac{{\rm SNR}}{\sqrt{T}} \left( \int_{f_{\rm min}}^{f_{\rm max}} df \left[ \frac{f^\beta}{h^2\Omega_{\rm sens}(f)}  \right]^2 \right)^{-1/2}   \right].
\end{equation}
Any SGWB signal $\Omega_{\rm GW}(f)$ which lies above $\Omega_{PI}(f)$ would gives a signal to noise ratio $>{\rm SNR}$ after an observation time $T$.
\subsection{Results}
For the purpose of our study, we computed the power-law integrated sensitivity curve $\Omega_{PI}(f)$, starting from the noise spectral density in \cite{Hild:2010id} for ET, \cite{Evans:2016mbw} for CE and \cite{Yagi:2011wg} for BBO/DECIGO. 
For pulsar timing arrays EPTA, NANOGrav and SKA, we directly took the sensitivity curves from \cite{Breitbach:2018ddu}. The signal-to-noise ratio can be improved by using cross-correlation between multiple detectors, e.g. LIGO-Hanford, LIGO-Livingston, VIRGO but also KAGRA which may join the network at the end of run O3, which began on the 1$^{\rm st}$ of April 2019, or LIGO-India which may be operational for run O5 \cite{Aasi:2013wya}. We computed the SNR for LIGO from the expression \cite{Thrane:2013oya}
\begin{equation}
\textrm{SNR} = \left[ 2T \int_{f_{\rm min}}^{f_{\rm max}} df \frac{\Gamma^2(f) S_{\rm h}^2(f)}{S_{n}^1(f)S_n^2(f)}  \right]^{1/2},
\end{equation}
where $S_n^1$ and $S_n^2$ are the noise spectral densities of the detectors in Hanford and in Livingston for the runs \href{https://dcc.ligo.org/LIGO-T1500293/public}{O2}, \href{https://dcc.ligo.org/LIGO-T1800044/public}{O4} or \href{https://dcc.ligo.org/LIGO-T1800042-v4/public}{O5} and $\Gamma(f)$ is the \href{https://dcc.ligo.org/public/0022/P1000128/026/figure1.dat}{overlap function} between the two LIGO detectors which we took from \cite{Abadie:2011fx}. The GW power spectral density $S_h(f)$ is related to the GW energy density through
\begin{equation}
S_h(f) = \frac{3H_0^2}{2\pi^2}\frac{\Omega_{\rm GW}(f)}{f^3}.
\end{equation}

We fixed the signal-to-noise ratio $\rm SNR = 10$ and the observational time $T =$ 268 days for LIGO O2, 1 year for LIGO O4 and O5, and 10 years for other sensitivity curves.

As this paper was completed, Ref.~\cite{Mingarelli:2019mvk} appeared, where the sensitivity curves may differ from us by a factor of order 1.
		
		\medskip
\small
\bibliographystyle{JHEP}
\bibliography{VOS}

\end{document}